\documentclass[a4paper,11pt]{article}
\pdfoutput=1
\usepackage{jheppub}

\usepackage{ifpdf}
\usepackage[usenames,dvipsnames]{xcolor}
\usepackage{slashed}
\usepackage[T1]{fontenc}
\usepackage[ansinew]{inputenc}
\usepackage{amsmath}
\usepackage{amssymb}
\usepackage{booktabs}
\usepackage{siunitx}
\usepackage{graphicx}
\usepackage{color}
\definecolor{darkblue}{cmyk}{0.9,0.9,0,0}
\definecolor{darkgreen}{rgb}{0,0.55,0}
\usepackage{simpler-wick}
\usepackage[export]{adjustbox}
\usepackage{mathtools}
\usepackage{minitoc}
\usepackage{mathtools}
\usepackage{simpler-wick}
\usepackage[normalem]{ulem}
\usepackage{enumitem}

\makeatletter
\DeclareRobustCommand\widecheck[1]{{\mathpalette\@widecheck{#1}}}
\def\@widecheck#1#2{
    \setbox\z@\hbox{\m@th$#1#2$}
    \setbox\tw@\hbox{\m@th$#1
       \widehat{
          \vrule\@width\z@\@height\ht\z@
          \vrule\@height\z@\@width\wd\z@}$}
    \dp\tw@-\ht\z@
    \@tempdima\ht\z@ \advance\@tempdima2\ht\tw@ \divide\@tempdima\thr@@
    \setbox\tw@\hbox{
       \raise\@tempdima\hbox{\scalebox{1.2}[-1]{\lower\@tempdima\box
\tw@}}}
    {\ooalign{\box\tw@ \cr \box\z@}}}
\makeatother

\newcommand{\comment}[1]{}

\newcommand{\tf}{\mathbf{T}}
\newcommand{\nf}{\mathbf{N}}
\newcommand{\bbf}{\mathbf{B}}

\newcommand{\beq}{\begin{equation}}
\newcommand{\eeq}{\end{equation}}
\newcommand{\beqq}{\begin{equation*}}
\newcommand{\eeqq}{\end{equation*}}
\newcommand\beqa{\begin{eqnarray}}
\newcommand\eeqa{\end{eqnarray}}
\newcommand\beqaa{\begin{eqnarray*}}
	\newcommand\eeqaa{\end{eqnarray*}}
\newcommand\bea{\begin{array}}
	\newcommand\eea{\end{array}}

\def\XXint#1#2#3{{\setbox0=\hbox{$#1{#2#3}{\int}$ }
		\vcenter{\hbox{$#2#3$ }}\kern-.5\wd0}}

\def\XXint#1#2#3{{\setbox0=\hbox{$#1{#2#3}{\int}$}
		\vcenter{\hbox{$#2#3$}}\kern-.5\wd0}}

\newcommand{\nn}{\nonumber}

\newcommand{\neqa}{\nonumber\end{eqnarray}}
\newcommand{\la}[1]{\label{#1}}

\def\tr{{\rm tr~}}

\newcommand{\hs}{\frac{\sqrt{3}}{2}}
\renewcommand{\d}{\partial}

\newcommand{\<}{{\langle}}
\renewcommand{\>}{{\rangle}}

\newcommand{\cC}{{\cal C}}

\newcommand{\cL}{{\cal L}}

\newcommand{\re}{\relax{\rm I\kern-.18em R}}

\renewcommand{\sp}{p\hspace{-.40em}/}

\def\su2{{SU(2)}}

\def\[{\left[}
\def\]{\right]}

\def\s{\sigma}

\def\({\left(}
\def\){\right)}
\def\[{\left[}
\def\]{\right]}

\def\<{\langle}
\def\>{\rangle}

\def\cO{{\cal O}}

\def\cC{{\cal C}}
\def\cW{{\cal W}}
\def\cP{{\cal P}}
\def\cM{{\cal M}}

\def\s*{\ *_{\!\!\!\!\!\!\!\!\!\,_{\,_\text{\scriptsize{sym}}}}}
\def\hs*{\ \hat{*}_{\!\!\!\!\!\!\!\!\!\,_{\,_\text{\scriptsize{sym}}}}}
\def\d{\partial}

\def\i2{\frac{i}{2}}

\def\spi{\relax{\rm \pi\kern-0.5em /}}
\def\sA{\relax{\rm A\kern-0.5em /}}
\def\sp{\relax{\rm p\kern-0.5em /}}
\def\sd{\relax{\rm \d\kern-0.5em /}}
\def\sk{\relax{\rm k\kern-0.5em /}}
\def\sn{\relax{\rm n\kern-0.5em /}}
\def\sl{\relax{\rm l\kern-0.5em /}}
\def\sP{\relax{\rm P\kern-0.7em /}}
\def\sBethe{\relax{\rm \Bethe\kern-0.5em /}}

\def\One{\mathbb{I}}

\def\be#1\ee{\begin{equation}\begin{aligned}
#1
\end{aligned}
\end{equation}}

\newcommand{\ii}{\mathrm{i}}
\newcommand{\dd}{\mathrm{d}}
\newcommand{\vv}{\mathtt{V}_\alpha}
\newcommand{\vvv}{\widetilde{\mathtt{V}}}

\newcommand{\te}{\tilde{\epsilon}}

\numberwithin{equation}{section}

\usepackage{varioref}
\usepackage{makeidx}
\usepackage{cleveref}
\makeindex

\title{\boldmath Line operators in Chern-Simons-Matter theories and Bosonization in Three Dimensions II -\\ Perturbative Analysis and All-loop Resummation}

\author[a]{Barak Gabai}
\author[b]{Amit Sever}
\author[b]{and De-liang Zhong}

\affiliation[a]{Jefferson Physical Laboratory, Harvard University, Cambridge, MA 02138 USA}
\affiliation[b]{School of Physics and Astronomy, Tel Aviv University, Ramat Aviv 69978, Israel}

\abstract{We study mesonic line operators in Chern-Simons theories with bosonic or fermionic matter in the fundamental representation. In this paper, we elaborate on the classification and properties of these operators using all loop resummation of large $N$ perturbation theory.
We show that these theories possess two conformal line operators in the fundamental representation. One is a stable renormalization group fixed point, while the other is unstable. They satisfy first-order chiral evolution equations, in which a smooth variation of the path is given by a factorized product of two mesonic line operators. The boundary operators on which the lines can end are classified by their conformal dimension and transverse spin, which we compute explicitly at finite 't Hooft coupling. We match the operators in the bosonic and fermionic theories. Finally, we extend our findings to the mass deformed theories and discover that the duality still holds true.}

\begin{document} 
\maketitle
\flushbottom

\vspace{1.0cm}

\dosecttoc[e]

\newpage

\section{Introduction and Discussion}

Three-dimensional conformal field theories obtained by coupling Chern-Simons (CS) gauge theory to scalars or fermions in the fundamental representation have rich and fascinating dynamics. First, there is overwhelming evidence that the bosonic and fermionic theories are equivalent 
\cite{Sezgin:2002rt, Klebanov:2002ja, Giombi:2009wh,		Benini:2011mf, Giombi:2011kc, Aharony:2011jz, Maldacena:2011jn,		Maldacena:2012sf, Chang:2012kt, Jain:2012qi, Aharony:2012nh,		Yokoyama:2012fa, Gur-Ari:2012lgt, Aharony:2012ns, Jain:2013py,		Takimi:2013zca, Jain:2013gza, Yokoyama:2013pxa, Bardeen:2014paa,		Jain:2014nza, Bardeen:2014qua, Gurucharan:2014cva, Dandekar:2014era,		Frishman:2014cma, Moshe:2014bja, Aharony:2015pla, Inbasekar:2015tsa,		Bedhotiya:2015uga, Gur-Ari:2015pca, Minwalla:2015sca,		Radicevic:2015yla, Geracie:2015drf, Aharony:2015mjs,		Yokoyama:2016sbx, Gur-Ari:2016xff, Karch:2016sxi, Murugan:2016zal,		Seiberg:2016gmd, Giombi:2016ejx, Hsin:2016blu, Radicevic:2016wqn,		Karch:2016aux, Giombi:2016zwa, Wadia:2016zpd, Aharony:2016jvv,		Giombi:2017rhm, Benini:2017dus, Sezgin:2017jgm, Nosaka:2017ohr,		Komargodski:2017keh, Giombi:2017txg, Gaiotto:2017tne,		Jensen:2017dso, Jensen:2017xbs, Gomis:2017ixy, Inbasekar:2017ieo,		Inbasekar:2017sqp, Cordova:2017vab, GuruCharan:2017ftx, Benini:2017aed,		Aitken:2017nfd, Argurio:2018uup, Jensen:2017bjo, Chattopadhyay:2018wkp,		Turiaci:2018nua, Choudhury:2018iwf, Karch:2018mer, Aharony:2018npf,		Yacoby:2018yvy, Aitken:2018cvh, Aharony:2018pjn, Dey:2018ykx, Skvortsov:2018uru, Argurio:2019tvw, Armoni:2019lgb,		Chattopadhyay:2019lpr, Dey:2019ihe, Halder:2019foo, Aharony:2019mbc,		Li:2019twz, Jain:2019fja, Inbasekar:2019wdw, Inbasekar:2019azv,		Jensen:2019mga, Kalloor:2019xjb, Ghosh:2019sqf, Argurio:2020her, Inbasekar:2020hla,		Jain:2020rmw, Minwalla:2020ysu, Jain:2020puw, Mishra:2020wos,		Jain:2021wyn, Jain:2021vrv, Gandhi:2021gwn, Gabai:2022snc,Mehta:2022lgq,Jain:2022ajd}.
This so-called 3d bosonization duality maps the weak coupling limit of one theory to the strong coupling limit of the other. As a result, it is both intriguing and challenging to prove. 
Second, a lot of evidence suggests that these CFTs are holographic duals to parity-breaking versions of Vasiliev's higher-spin theory \cite{Vasiliev:1992av,Giombi:2011kc,Aharony:2011jz,Aharony:2012nh,Klebanov:2002ja,Sezgin:2003pt,Chang:2012kt,Leigh:2003gk}. They can therefore be used to understand, and maybe even derive, some aspects of holography. 
 
One class of fundamental operators in any conformal theory is line operators. They can either be closed, like Wilson loops, or open, like Wilson lines ending on fundamental and anti-fundamental fields. We denote the latter as {\it mesonic line operators}. A generic line operator experiences renormalization group (RG) flows on the line. It can end at either a trivial or a non-trivial conformal line operator. 
In large $N$ gauge theories, we expect the conformal lines in the fundamental representation to be the most elementary ones, having the lowest defect entropy. 
Other conformal line operators can be obtained from them by taking direct sums and perturbing by relevant line operators.\footnote{One such example is discussed in section \ref{RGsection}.}

Here, we study mesonic line operators in the fundamental representation in CS-matter conformal gauge theories. A brief summary of our findings has been published in \cite{short}. In this long companion paper, we analyze the theory using all loop perturbation theory. We focus on the planar limit, in which the gauge group rank $N$ and the CS level $k$ are sent to infinity while keeping $\lambda=N/k$ fixed. This limit does not distinguish between versions of the bosonization duality that differ by half-integer shifts of the Chern-Simons level $k$ and by the gauge group being $SU(N)$ or $U(N)$.\footnote{For a complete list of the dualities,
see e.g., \cite{Benini:2011mf,Aharony:2015mjs,Aharony:2016jvv,Hsin:2016blu,Komargodski:2017keh,Seiberg:2016gmd,Karch:2016sxi,Murugan:2016zal}. In this paper, we shall restrict to the $SU(N)/U(N)$ version.}

In each theory, we find two conformal line operators in the fundamental representation. One is stable, and the other has a single relevant deformation. The latter triggers an RG flow on the line that ends at either the stable line operator or a trivial topological anyonic line, depending on the sign of its coefficient.\footnote{The change in the defect entropy along this flow is computed in \cite{Ivri} and is found to be in agreement with the general theorem of \cite{Cuomo:2021rkm}.} This RG flow picture is summarized in section \ref{RGsection}.

Next, we classify the fundamental (``right'') and anti-fundamental (``left'') boundary operators on which the conformal lines can end. A straight conformal line operator realizes a one-dimensional $SL(2,{\mathbb R})$ subgroup of the conformal symmetry, as well as $U(1)$ transverse rotations symmetry around the line. The boundary operators are uniquely characterized by their $SL(2,{\mathbb R})$ conformal dimension and transverse spin. We find four towers of boundary operators, two fundamental and two anti-fundamental. All the operators at the right (left) end of the line
have the same anomalous spin, equal to $\lambda/2$, ($-\lambda/2$).  
Operators in the two fundamental (anti-fundamental) towers have opposite anomalous dimensions, equal to $\pm\lambda/2$. For each tower, 
there is unique $SL(2,{\mathbb R})$ primary operator with the lowest conformal dimension. The remaining operators in the tower can be obtained from it using path derivatives. 
Some combinations of path derivatives lead to the same boundary operator. This relation between boundary operators is the quantum equivalent of the classical equation of motion.

To perform the all-loop analysis, we define an object called the {\it line integrand}. It is related to the expectation value of the mesonic line operator through line integrations. This integrand is shown to satisfy a simple recursion relation on a straight line. Then, this relation is solved using the appropriate boundary condition. The bosonic theory is considered in section \ref{bossec}, whereas the fermionic theory is considered in sections \ref{fersec} and \ref{sec:condFer}. We find a perfect match between the mesonic line operators in the two theories. 

\begin{figure}[t]
\centering
\includegraphics[scale=0.6]{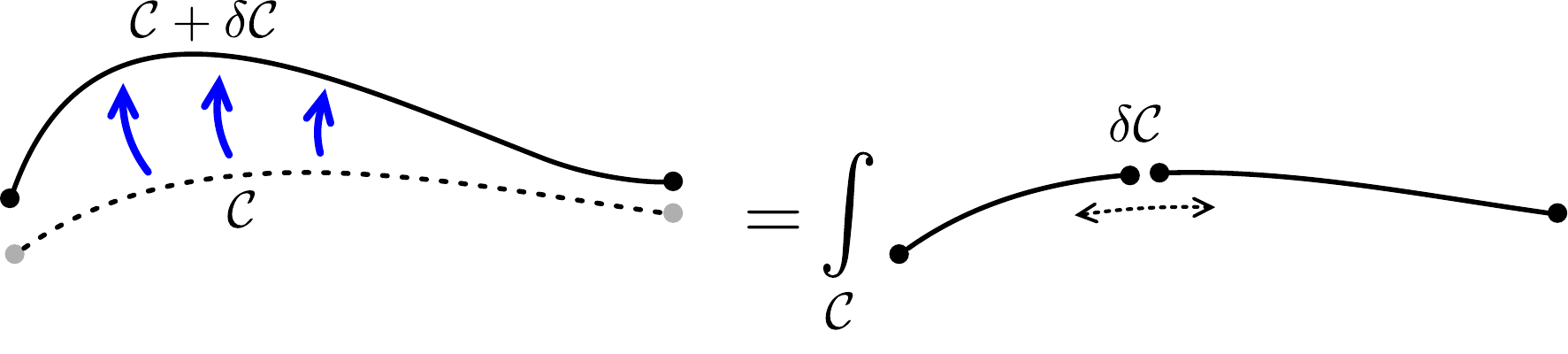}
\caption{The {\it evolution equation} (\ref{ee}) relates a small smooth deformation of a conformal mesonic line operator to an integrated product of two mesonic line operators.}
\label{fig:evolution}
\end{figure}

The classification of the operators that live on the line follows from that of the boundary operators. This is because in the planar limit, the former operators factorize into a product of right and left boundary operators. 
One operator on the line, the displacement operator, always exists. It dictates smooth deformations of the line. 
The fact that this operator factorizes into a product of right and left boundary operators leads to the so-called {\it evolution equation}. 
This operator equation relates any smooth variation of the conformal mesonic line operator to a factorized product of two mesonic line operators, see figure \ref{fig:evolution}. We use this factorization property together with our all-loop result for the expectation value of the mesonic line operators to extract the two-point function of the displacement operator on a straight line or a closed circular loop.

To prove the bosonization duality, one must also match the 
connected correlation functions between the mesonic line operators and with the local 
operators. We expect that the properties of the mesonic lines listed above, together with the trivial spectrum of single trace operators, would be sufficient to complete the proof. 
In \cite{bootstrap}, we take the first step in that direction by showing that these properties are sufficient to determine the expectation value of the mesonic line operators along an arbitrary smooth path. 
Here, in section \ref{sec:bootstrap}, we present a different {\it formal} bootstrap along the line of \cite{Migdal:1983qrz}. We show that for any line operator (conformal or not), the corresponding boundary equation and evolution equation are sufficient to reconstruct the perturbative expansion of the line.

In section \ref{sec:susy}, we show how the mesonic line operators studied in this paper can be embedded into $1/2$-BPS superconformal mesonic line operators in the ${\cal N}=2$ CS-matter theory. The lift to the ${\cal N}=2$ theory is such that the planar expectation value of the operators is the same as in the non-supersymmetric theory. This lift enables us to express the relationship discovered between the anomalous dimension and the anomalous spin of the boundary operators as the boundary BPS condition for the superconformal symmetry that is preserved by the line. The drawback of this section is that the analysis in it is formal. Our explicit computation in sections \ref{bossec}-\ref{sec:condFer} reveals which parts of the SUSY analysis hold in the quantum theory.

In summary, the fundamental conformal line operators realize a very simple yet non-trivial one-dimensional CFT. Based on our experience with other versions of the AdS/CFT correspondence, we expect them to be essential for deriving the duality with Vasiliev's higher-spin theory. 

We end this paper in section \ref{sec:MassDeform}, where we extend our results to the case of CS theory coupled to massive matter fields. We find a similar structure and show that the duality continues to hold for these theories.

\section{The Conformal Line Operators in the Bosonic Theory}\la{bossec}

The ``free scalar'' CFT is defined by coupling $SU(N)$ CS theory at level $k$ to a complex boson in the fundamental representation\footnote{Here the central dot represents the contraction of color indices. For instance, $\phi^\dagger \cdot \phi = \phi^\dagger_j \phi^j$ and $A_\mu \cdot \phi = A_\mu^I (T^I)^i_{\ j}\phi^j$, where $i,j$ are color indices.} 
\beq\label{Sbos}
S_E^\text{bos}=S_{CS}+\int \dd^3 x\,\Big[D_{\mu}\phi^\dagger\cdot D^{\mu}\phi+{\lambda_6\over N^2}(\phi^\dagger\cdot\phi)^3\Big]\ ,
\eeq
where $D_\mu\phi=\d_\mu\phi-\ii A_\mu\cdot \phi$, $D_\mu\phi^\dagger=\d_\mu\phi^\dagger+\ii \phi^\dagger \cdot A_\mu $, and
\beq\la{CSact}
S_{CS}=\frac{\ii k}{4\pi}\int \dd^3x\,\epsilon^{\mu\nu\rho}\, \tr\! (A_{\mu} \partial_\nu A_\rho - \tfrac{2\ii}{3} A_\mu A_\nu A_\rho)\,.
\eeq
In the planar limit $N$ is sent to infinity, holding $\lambda={N\over k}$ and $\lambda_6$ fixed.\footnote{Here, we have assumed the convention where $k$ is the renormalized level that arises, for instance, when the theory is regularized by dimensional reduction. In this convention, $\lambda \in [-1,1]$.} At large $N$ the coupling $\lambda_6$ is a free parameter.\footnote{At finite $N$ it flows to the ``free scalar'' CFT fixed point \cite{Aharony:2011jz}.} This CFT has two relevant operators, the mass $\phi^\dagger\cdot\phi$ and the double trace ${\lambda_4\over N}\(\phi^\dagger\cdot\phi\)^2$. Turning on the double trace deformation generates a flow that can be fine-tuned to end at a non-trivial fixed point, called ``critical scalar'' 
CFT, see \cite{Aharony:2012nh} for details.

The primary focus of this article is on conformal line operators in the fundamental representation that end on fundamental and anti-fundamental boundary operators. 
A fundamental line operator that exists in any gauge theory is a Wilson line. This operator however may not be conformal. When constructing a conformal line operator, we must consider all relevant and marginal operators on the line. 
If they exist, they can generate an RG flow on the line. A fixed point of such flow and a zero of the beta-function on the line can be reached by fine-tuning the coefficients of these operators. 

In the planar limit, color-singlet operators decouple from lines in the fundamental representation, so we only have to consider operators in the adjoint representation. 
For the critical boson CFT (\ref{Sbos}) the adjoint operator with the lowest classical dimension, equal to one, is the bi-scalar $\phi\phi^\dagger$. Hence, we must generalize the Wilson line operator and consider instead the combination
\beq\label{Wbos}
{\cal W}^\alpha[\cC,n]\equiv\Big[{\cal P}e^{\ii\int\limits_\cC\(A\cdot \dd x+\ii \alpha{2\pi\over k}\phi \phi^\dagger|\dd x|\)}\Big]_n\,,
\eeq
where $\cC$ is some smooth path, $n$ is the framing vector \cite{Witten:1988hf}, and $\alpha$ is a free parameter that we have to tune so that the operator is conformal.

To determine the critical values of $\alpha$, we consider the mesonic line operator
\beq \label{M10}
M_{10}\equiv\phi^\dagger(x_1){\cal W}^\alpha[\cC,n]\phi(x_0)\,,
\eeq
with the path $\cC$ taken to be a straight line along the $x^3$ direction between $x_0$ and $x_1$. Here, $x_s=x(s)$ with $s\in[0,1]$ is some parametrization of the straight line. A simple choice that we use is
\beq\la{straight}
x_s=x_0+s\,x_{10}\,,
\eeq
with $x_{st}=x_s-x_t$. For convenience, we will denote $|x_{10}|$ by $x$. 
As long as $\alpha$ is not of order $N$, the planar expectation value of $M_{10}$ is insensitive to the triple and double trace couplings, $\lambda_6$ and $\lambda_4$.\footnote{Which are of order one at the fixed points.} Hence, we can safely ignore these multi-trace interactions, and our computation applies equally to the ``critical'' and ``free'' scalar CFTs. 

To simplify the computation it is convenient to choose a gauge and regularization scheme that are correlated with the direction of the line. We use light-cone gauge in the plane perpendicular to the line, $A_-=0$. Here, $x^\pm=(x^1\pm \ii x^2)/\sqrt2$ are light-cone coordinates with the flat Euclidean metric
\beq
\label{eq:metric}
\dd s^2 = (\dd x^1)^2 +(\dd x^2)^2+(\dd x^3)^2 = 2\, \dd x^+ \dd x^-+(\dd x^3)^2\,.
\eeq
With this choice of gauge, the $A\wedge A\wedge A$ self-interaction of the gauge field vanishes and the CS action takes the quadratic form 
\beq
S_{CS}=\frac{k}{4\pi}\int \dd^3x\, \tr\!(A_+ \partial_- A_3 - A_3 \partial_- A_+) \, .
\eeq
At leading order in the 't Hooft large $N$ limit, where scalar and the ghost loops can be ignored, the gluon propagator in this gauge does not receive corrections. It takes the form
\beq \label{Aprop}
\<A_\mu^I(p)A_\nu^J(-q)\>=(2\pi)^3 \delta^3(p+q)G_{\mu\nu}(p)\delta^{IJ}\,,
\eeq
where $I$ and $J$ denotes $SU(N)$ color indices in the adjoint representation, $A = A^I T^I$. The $SU(N)$ generators, $T^I$, are normalized in the standard way, $\tr T^I T^J = \delta^{IJ}/2$. The non-zero components of $G_{\mu\nu}(p)$ are
\beq\label{eqn-LC-gluonProp}
G_{3+}(p)=-G_{+3}(p)=-\frac{4\pi \ii}{k} \frac{1}{p^+}\,.
\eeq

With this gauge choice, there are only two types of divergent diagrams. One is loop corrections to the scalar self-energy and the other is divergences that result from line integration and are localized on the line. To regularize these UV divergences we found it convenient to work in a new scheme in which the scalar propagator is deformed as
\beq\la{deformedprop}
\<\phi^\dagger_i(p)\phi^j(q)\>^\text{tree}_\epsilon\equiv(2\pi)^3\delta^3(p+q)\,{\delta_i^j\over p^2}\times e^{\ii \epsilon\, p_3}\,.
\eeq
The additional factor of $e^{\ii\epsilon p_3}$ is a point splitting-like regularization and is correlated with the direction of the line that we will take to point along the third direction. As we will show, it is sufficient to regulate the UV divergences.

\subsection{Self-energy}
\label{sec:SEscalar}

The self-energy, denoted by $\Sigma(p;\lambda)$, is the sum over all two-point one-particle irreducible (1PI) diagrams. In our regularization scheme (\ref{deformedprop}), the corresponding full scalar propagator is\footnote{In favor of cleaner expressions, we have omitted contributions to the self-energy with $\lambda_4\left(\phi^\dagger \cdot \phi\right)^2$ or $\lambda_6 \left(\phi^\dagger \cdot \phi\right)^3$ interactions. These always correspond to attaching bubbles in different positions onto scalar propagators. Because such contributions cannot depend on the momenta of the scalar propagator they are attached to, they can be trivially absorbed into a shift of the mass counter-term.}
\beq
\<\phi^\dagger_i(p)\phi^j(q)\>_\epsilon=(2\pi)^3\delta^3(p+q)\,{\delta_i^j\over p^2+e^{\ii\epsilon\, p_3}(m_{c.t.}^2-\Sigma(p))}\, e^{\ii \epsilon\, p_3}\,,
\eeq
where $m_{c.t.}^2$ is a mass counterterm. In \cite{Aharony:2012nh} the self-energy was computed using dimensional regularization in the direction $x^3$, and a cutoff on the momentum in the transverse plane. Their result for the self-energy turns out to be zero. We will show now that the same conclusion holds also in our deformed scalar propagator scheme (\ref{deformedprop}).

\begin{figure}[t]
    \centering
    \includegraphics[width= 1 \textwidth ]{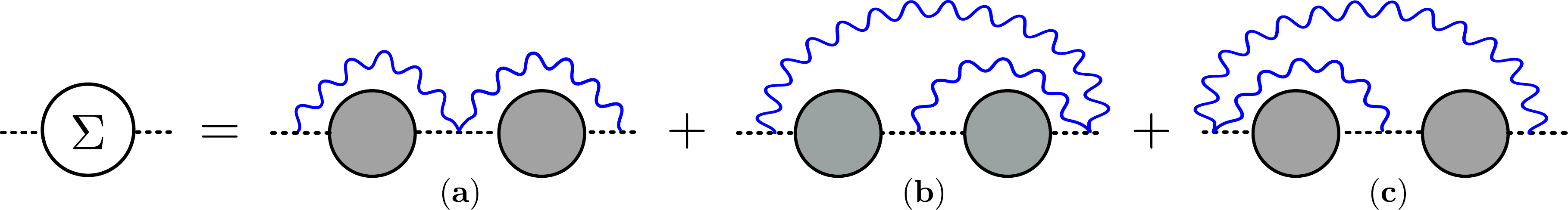}
    \caption{The Schwinger-Dyson equation for scalar self-energy $\Sigma(p)$. Here the filled blob represents the exact propagator $\mathbf{\Delta}(p)$, such that $\mathbf{\Delta}^{-1}(p) = p^2 -\Sigma(p)$.}
    \label{fig:self-energy}
\end{figure}

The self-energy is subject to the Schwinger-Dyson equation plotted in figure \ref{fig:self-energy}. 
Note that the diagram with a single gluon exchange vanishes because the gluon propagator is anti-symmetric, (for more details see the discussion at the end of section \ref{scalarseappendix}). Our regularization prescription preserves the rotational symmetry in the transverse plane and is therefore consistent with the ansatz $e^{\ii \epsilon\,p_3}(\Sigma(p)-m_{c.t.}^2) = f(\lambda) p_\perp^2$, where $p_\perp$ is the magnitude of the momentum in the transverse plane.\footnote{It can be shown that the self-energy vanish in any regularization scheme that is consistent with the ansatz $\<\phi^\dagger_i(p)\phi^i(-p)\>\propto1/
\(k^2-p_\perp^2f(\lambda)\)$.} With this ansatz, diagram ($a$) in figure \ref{fig:self-energy} becomes
\begin{align}
-(a)=&\,\delta_i^j\Big[2\pi\lambda\int\!\!{\dd^3k\over(2\pi)^3}{(k+p)^+\over(k-p)^+}{e^{\ii\epsilon k_3}\over k^2-f(\lambda)k_\perp^2}\Big]^2=\Big[\pi\lambda\int{\dd^2k_\perp\over(2\pi)^2}{(k+p)^+\over(k-p)^+}{e^{-\sqrt{1-f(\lambda)}k_\perp\epsilon}\over k_\perp\sqrt{1-f(\lambda)}}\Big]^2\nn\\
=&\Big[{\lambda\over2}\int \dd k_\perp{2\Theta(k_\perp-p_\perp)-1\over\sqrt{1-f(\lambda)}}e^{-\sqrt{1-f(\lambda)}k_\perp\epsilon}\Big]^2 =\[{\lambda(1-2e^{-\sqrt{1-f(\lambda)}p_\perp\epsilon})\over2(1-f(\lambda))\epsilon}\]^2\,,
\end{align}
where in the first step we have performed the $k_3$ integration by closing the contour on the upper half plane. In the second step we have evaluated the angular integral in the transverse space and in the last step we have performed the remaining radial integral. Similarly, the diagrams ($b$) and ($c$) evaluate to
\beq
(b)+(c)={\lambda^2\over(1-f(\lambda))^2\epsilon^2}e^{-\sqrt{1-f(\lambda)}p_\perp\epsilon}(e^{-\sqrt{1-f(\lambda)}p_\perp\epsilon}-1)\,.
\eeq
The sum of them provides us with the Schwinger-Dyson equation for $\Sigma$, which reads
\beq
\Sigma(p)=e^{-\ii\epsilon p_3}f(\lambda)p_\perp^2+m_{c.t.}^2=(a)+(b)+(c)=-{\lambda^2\over4(1-f(\lambda))^2\epsilon^2}\ ,
\eeq
and therefore
\beq \label{massb}
m^2_{c.t.}=-{\lambda^2\over4\epsilon^2}\ ,\qquad f(\lambda)=0\,.
\eeq
Here $m^2_{c.t.}$ stands for the mass counter-term. By tuning it appropriately, we can reach the conformal fixed points, and the scalar self-energy is zero.

\subsection{The Anomalous Spin}
\label{sec:anomspin}

The transverse spin at the endpoints of the mesonic line operators (\ref{M10}) can receive quantum corrections. These corrections result in an overall factorized \textit{framing factor},
\beq \label{eqn-ff-int}
\mathtt{framing\ factor}= \exp\Big[ \ii \, \frac{\lambda}{2} \int_{\mathcal{C}_{10}} \dd s\, (n_s \times \dot{n}_s) \cdot e_s \Big] \, ,
\eeq
where $e_s \equiv \dot{x}_s/|\dot{x}_s|$ is the unit tangent vector at point $x_s$, and the path $\mathcal{C}_{10}$ is oriented from $x_0$ to $x_1$. Here, the \textit{framing vector} $n_s$ is a unit normal vector, ($n_s \cdot e_s = 0$). The framing phase factor (\ref{eqn-ff-int}) measures the total angle by which the framing vector rotates within the normal plane along $\mathcal{C}_{10}$.
Up to a framing independent phase, this factor can be written as\footnote{Such framing-independent phases factor can be absorbed by wave function renormalization of the boundary operators.} 
\beq \la{ff}
\mathtt{framing\ factor}= \exp \Big[\frac{\lambda}{2} \int_{\mathcal{C}_{10}} \dd \log n_s^+ \Big] =(n_1^+)^{\gamma_s^{(L)}}(n_0^+)^{\gamma_s^{(R)}}\, ,
\eeq
where $n_0$ and $n_1$ are the boundary values of the framing vector and the $n^+$ component is with respect to a local transverse plane. For simplicity of the presentation, we will use the form (\ref{ff}). 

As we rotate the boundary operator at the end of the line, we also drag the endpoint of the framing vector with it. Correspondingly, the framing factor (\ref{ff}) leads to an anomalous spin, equal to $\gamma_s^{(L/R)}$, of the boundary operators. This contribution is independent of $\alpha$. Its derivation applies to any line operator, particularly the straight line in the third direction we focus on. In section \ref{sec:susy}, we will derive $\gamma_s^{(L)}=-\gamma_s^{(R)}=\lambda/2$ using supersymmetry. In appendix \ref{apd:FramingFactor}, we perform further perturbative checks of this result in Lorentz and lightcone gauge. In the case where the two endpoints of the line are pointing in the third direction, we reproduce \eqref{ff}. When this is not the case, we find that the perturbative result in lightcone gauge is not Lorentz covariant. We leave this issue for future study.

Note that the dependence on the integer tree-level boundary spin $s_{R/L}$ can also be accounted for using the boundary framing vector as
\beq\la{sf}
\mathtt{spin\ factor}=(n_1^+)^{s_L}(n_0^+)^{s_R}\,.
\eeq
For simplicity, in the following sections, we will assume that the framing and spin factors, \eqref{ff} and \eqref{sf}, are trivial.

\begin{figure}[t]
    \centering
    \includegraphics[width=0.5 \textwidth]{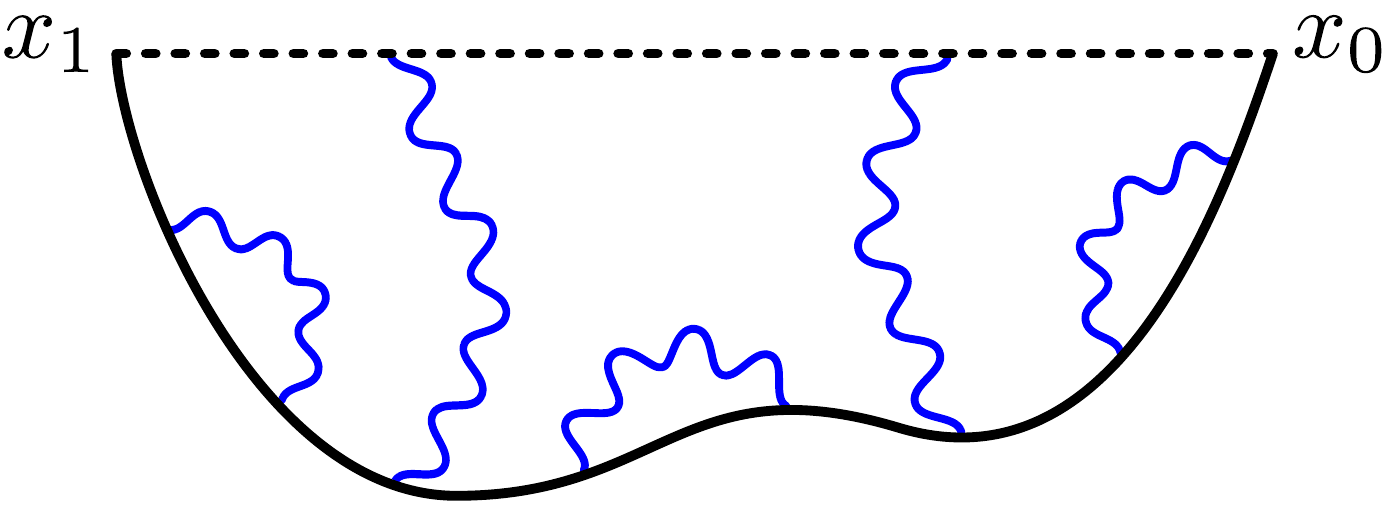}
    \caption{An example of a planar diagram with three gluon exchanges (wiggled blue lines) on the Wilson line (black solid curved line) and two gluon exchanges between the Wilson line and the scalar propagator (dashed straight line). Only the gluon exchanges on the Wilson line contribute to the anomalous spin.}\label{fig:Aspin}
\end{figure}

\subsection{One-loop}
\label{sec:1loop}
\begin{figure}[t]
    \centering
    \includegraphics[width= 0.9 \textwidth ]{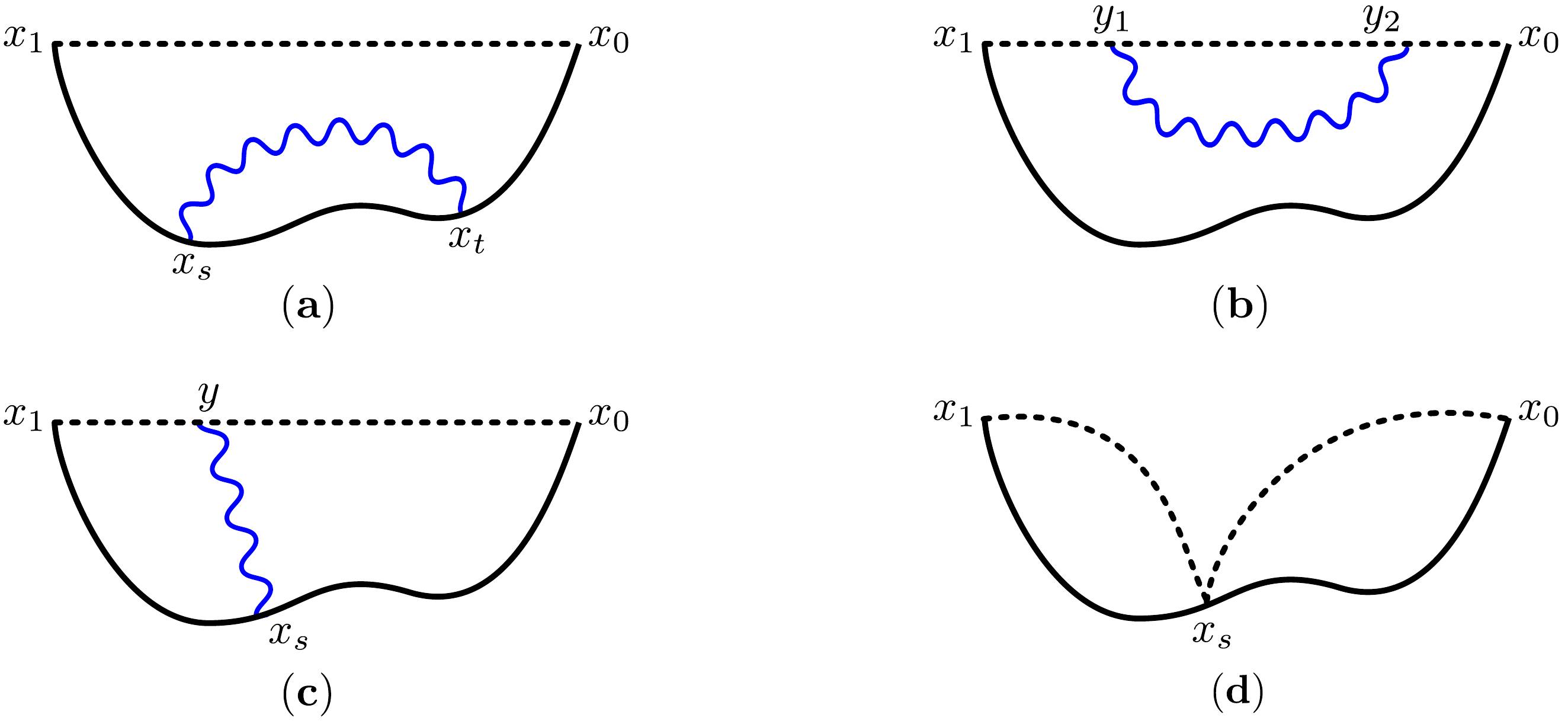}
    \caption{All one-loop diagrams that contribute to the expectation value of the mesonic operator (\ref{Wbos}). The dashed line represents the scalar propagator and the solid black line represents the Wilson line. Diagram ({\bf d}) has the scalar bi-linear inserted on the line.}
    \label{fig:1loop}
\end{figure}

The vacuum expectation value of the mesonic line operator (\ref{M10}) takes the form
\beq\la{Mform}
\<M_{10}\>=\sum_L\cM_L={1\over4\pi|x_{10}|}\(1+\cO(\lambda)\)\,.
\eeq
There are four diagrams that contribute at order $\lambda$, see figure \ref{fig:1loop}. Diagrams (a) and (b) are the one-loop gluon exchange and scalar self-energy that we have considered to all loops above. 
The remaining diagram, (c) and (d), can be computed at once. They are equal to
\beq\la{candd}
\mathtt{(c)+(d)} =-2\pi \lambda|x| 
\int\limits_0^1\! \dd s \! \int\! \frac{\dd^3 k_1}{(2\pi)^3}\frac{e^{\ii k_1^3 x\, (t-s) }}{k_1^2} \! \int \frac{\dd^3 k_0}{(2\pi)^3}\frac{e^{\ii k_0^3 x\, s }}{k_0^2}  \Big[\frac{(k_0+k_{1})^+}{(k_0-k_{1})^+} + \alpha \Big]e^{\ii(k_0^3+k_1^3)\epsilon}\, ,
\eeq
where $t = 1$. The advantage of introducing the $t$ variable is that by using it, the regulator can absorb into a shift, $s \mapsto s- \epsilon/x$, $t \mapsto t - 2\epsilon/x$. 
The resulting integral to consider looks then like the unregularized one, but the exponents are now \textit{strictly positive}, $t-s\ge\tilde\epsilon>0$ and $s\ge\tilde\epsilon>0$, where $\tilde\epsilon\equiv\epsilon/x$.

Let us explain how such momentum integrals are evaluated. Consider for example the integral over $k_0$ in (\ref{candd}). It can be written as $I(k_1;s)+\alpha\, I(0;s)$, with
\beq
I(p;s)\equiv\int \frac{\dd^3 k }{(2\pi)^3} e^{\ii s k^3 x} \frac{1}{k^2} \frac{k^+ + p^+}{k^+ - p^+}\ , \qquad (s>0)\, .
\eeq
We can compute the $k^3$ integral by closing the couture in the upper half-plane, and picking the residue at $k^3=\ii |k_\perp|$,
\beq\la{k3int}
I(p;s) = \int \frac{\dd^2 k_\perp}{(2\pi)^2} e^{- s |x| |k_\perp|} \frac{1}{2|k_\perp|}  \frac{k^+ + p^+}{k^+ - p^+}\, .
\eeq
The remaining two-dimensional transverse integral can be done in polar coordinates
\beq
I(p;s) = \int\limits_0^\infty \frac{|k_\perp|d|k_\perp|}{2\pi}\frac{e^{-|x| |k_\perp|}}{2|k_\perp|} \int\limits_0^{2\pi} \frac{\dd \theta}{2\pi} \frac{e^{\ii \theta} + \sqrt{2} p^+/|k_\perp|}{e^{\ii \theta} - \sqrt{2} p^+/|k_\perp|}\,,
\eeq
where $k^\pm = |k_\perp|e^{\ii \theta}/\sqrt{2}$. Taking $z\equiv e^{\ii \theta}$, the angular integral becomes a contour integral around the unit circle. It gives
\beq\la{angularint}
\int \frac{\dd \theta}{2\pi} \frac{e^{\ii \theta} + \sqrt{2} p^+/|k_\perp|}{e^{\ii \theta} - \sqrt{2} p^+/|k_\perp|} = \oint\limits_{|z|=1} \frac{\dd z}{2\pi \ii z} \frac{z + \sqrt{2} p^+/|k_\perp|}{z - \sqrt{2} p^+/|k_\perp|} = 2 \Theta(|k_\perp|-|p_\perp|) - 1\, ,
\eeq
where $\Theta(x)$ is the step function.
The remaining radial integral is then straightforward to evaluate, and we find
\beq
I(p;s) = \frac{1}{4\pi x\,s} \Big(2 e^{-|p_\perp|x\,s} - 1 \Big)\, .
\eeq

All integrals encountered in this paper will be computed by this method. In particular, we find that the integrals in \eqref{candd} evaluate to
\beq\label{bpc}
\begin{aligned}
\mathtt{(c)+(d)} & 
=\frac{\lambda}{8\pi |x| t}  \int\limits_{\tilde{\epsilon}}^{1+\tilde{\epsilon}}\dd s\, \(\frac{1-\alpha }{t-s}-\frac{\alpha +1}{s}\) \Bigg|_{t= 1+2\tilde{\epsilon}}
= \frac{\lambda\alpha}{4\pi |x|} \log \tilde\epsilon + \mathcal{O}(\tilde\epsilon)\, .   
\end{aligned}
\eeq

This logarithmic divergence comes from the region of integration near the endpoints. It corresponds to an anomalous dimension of the two boundary operators. We see that at one loop order there are no divergences at the bulk of the line and therefore no beta function for the coupling $\alpha$. The two scalars at the endpoints have the smallest tree-level dimension. Hence, they are primaries of the straight line $SL(2,{\mathbb R})$ conformal symmetry and their one-loop anomalous dimension has to be the same. From \eqref{bpc} we find that it is equal to $\Delta_\phi=\Delta_{\bar\phi}=1/2+\lambda\alpha/2+\cO(\lambda^2)$.\footnote{At the technical level, we can shift the divergent region of integration between the two sides by adding a total derivative.}

For completeness, in appendix \ref{apd:oneLoopGen} we present a closed-form expression for the one-loop expectation value of the mesonic line operator (\ref{M10}) along an arbitrary smooth path.

\subsection{Higher Loops}
\label{sec:higher}

As for one-loop, at any loop order, we can divide the diagrams into two classes. The first class consists of the scalar self-energy corrections and the gluon exchange on the line that we have considered above. The second class consists of diagrams of type (c) and (d) in figure \ref{fig:1loop}. These are the gluon exchange between the line and the scalar propagator (c), as well as the contraction of the scalar propagator with the adjoint bi-scalar insertion on the line (d). 
As in the one-loop case \eqref{candd}, these two types of sub-diagrams can be grouped together into a generalized vertex $\vv$ on the line as
\beq
\alpha\,\phi(k_{j+1})\phi^\dagger(k_j)e^{\ii (k_{j+1}-k_j)x\, t_j} \quad\rightarrow\quad \vv (k_j,k_{j+1})\phi(k_{j+1})\phi^\dagger(k_j)e^{\ii(k_{j+1}-k_j) x\,t_j}\,,
\eeq
where $x\,t_j$ is the insertion point on the line and 
\beq 
\vv(k_j,k_{j+1})=\alpha+{(k_j+k_{j+1})^+\over(k_j-k_{j+1})^+}\,.
\eeq
Hence, at $L$-loop order we have to compute the ladder-type diagram
\begin{align}\label{higherloop}
\cM_L=&\,  (-2\pi\lambda)^L \int\limits_0^1 \dd t_L \int\limits_0^{t_L} \dd t_{L-1} \cdots \int\limits_0^{t_2} \dd t_1 \int \prod_{j=0}^L \frac{\dd^3 k_j}{(2\pi)^3}\times\\
&\frac{e^{\ii k_0^3x(t_1+\tilde\epsilon)}}{k_0^2} \vv(k_0,k_1)\frac{e^{\ii k_1^3x(t_{21}+\tilde\epsilon)}}{k_1^2}\ \cdots\ \frac{e^{\ii k_{L-1}^3x(t_{L L-1}+\tilde\epsilon)}}{k_{L-1}^2}\vv(k_{L-1},k_L)\frac{e^{\ii k_L^3x(t_{L+1 L}+\tilde\epsilon)}}{k_L^2}\,.\nn
\end{align}
where we introduce a shorthand notation $t_{jk} \equiv t_j -t_k$. As for the one-loop case, the last variable $t_{L+1}$ is \textit{not} being integrated over and will be set to one at the very end. For simplicity, the ordered measure will be denoted by $\int_0^1 \mathcal{P} \prod_{j=1}^L \dd t_j$ for the rest of the paper.

\subsection{Two-loop Analysis}

The first hint for the RG fixed point value of the constant $\alpha$ comes from the two-loop analysis. At two-loop order the integral \eqref{higherloop} becomes
\begin{align}\la{eq:2loopNaiveBosonic}
\mathtt{2\! -\! loop} = (-2\pi \lambda)^{2}&\int\limits_{\tilde{\epsilon}}^{1+2\tilde{\epsilon}} \dd s \int\limits_{s+\tilde{\epsilon}}^{1+\tilde\epsilon} \dd t \int \frac{\dd^3 k_{2}}{(2\pi)^3} \frac{e^{\ii k_2^3 x \, (u-t)}}{k_{2}^2}
\int \frac{\dd^3 k_1}{(2\pi)^3}\frac{e^{\ii k_1^3 x\, (t-s) }}{k_1^2} \\
&\times \Big[\frac{(k_1+k_{2})^+}{(k_1-k_{2})^+} + \alpha \Big] \int \frac{\dd^3 k_0}{(2\pi)^3}\frac{e^{\ii k_0^3 x\, s }}{k_0^2}  \Big[\frac{(k_0+k_{1})^+}{(k_0-k_{1})^+} + \alpha \Big] \,. \nn 
\end{align}
For notational simplicity, we have relabelled variables as $(t_1,t_2,t_3)=(s,t,u)$ 
after absorbing the regulator in a shift of these positions as $t_j \rightarrow t_j - j \tilde\epsilon$.  

It is rather straightforward to compute this two-loop integral using the integration technique in (\ref{k3int})-(\ref{angularint}). The result is
\beq \la{2loopal}
\mathtt{2\! -\! loop}= \frac{ \lambda^{2}}{8\pi x}\int\limits_{\tilde\epsilon}^{1+\tilde\epsilon} \dd s \int\limits_{s+\tilde\epsilon}^{1+2\tilde\epsilon} \dd t \[ \frac{1-\alpha }{u (s-u) (t-u)}+\frac{(\alpha +1) (2 t-\alpha  u-u)}{2 s t u (t-u)}+\frac{\alpha ^2-1}{2 t (s-t) (t-u)}\]\,.
\eeq
We see that unless $\alpha^2 = 1$, the third term leads to a logarithmic divergence on the line, at $s=t$. Hence, at the conformal fixed points of the line $\alpha^2=1+\cO(\lambda)$. In the next section, we will show that $\alpha^2=1$ is the exact fixed point equation to all loop order.

By plugging $\alpha=\pm1$ into \eqref{2loopal} we arrive at
\beq\la{2loop}
\begin{aligned}
    \mathtt{2\! -\! loop}=& \frac{ \lambda^{2}}{8\pi x}\int\limits_{\tilde\epsilon}^{1+\tilde\epsilon} \!\! \dd s \int\limits_{s+\tilde\epsilon}^{1+2\tilde\epsilon} \dd t \[ \frac{1-\alpha }{u (s-u) (t-u)}+\frac{\alpha +1}{s t u}\] \Big|_{u=1+3\tilde{\epsilon}}\\
=&{1\over4\pi x}\times \frac{\lambda^2}{2}\,\(\log^2\tilde\epsilon-{\pi^2\over6}\)+\cO(\tilde\epsilon)\,.
\end{aligned}
\eeq

We conclude that the anomalous dimension of the boundary scalars is not corrected at two loops and is given by $\Delta_\phi=\Delta_{\bar\phi}=1/2\pm\lambda/2+\cO(\lambda^3)$.

\subsection{All Loop Resummation}\la{allloopsec}

At $L$-loop order we have to evaluate the integral \eqref{higherloop}. We define the corresponding $L$-loop \textit{line integrand}, $\mathtt{B}^{(L)}_\alpha$, to be the result after doing all bulk integration, but not the line integrations
\beq\label{eqn-higherloop-rel}
{\cal M}_L=\int\limits_0^1 \mathcal{P}\prod_{j=1}^L\dd t_j\,\mathtt{B}^{(L)}_\alpha(\{t_j+j\tilde\epsilon\}_{j=1}^L;t_{L+1}+(L+1)\tilde{\epsilon})
\, ,
\eeq
Here, the line integrand is defined with a shift of $t_j$ by $j\tilde\epsilon$. After absorbing all $\tilde{\epsilon}$ dependencies of the line integrand 
into an opposite shift $t_j\to t_j-j\tilde\epsilon$, the $\mathtt{B}^{(L)}_\alpha$ becomes independent of $\tilde\epsilon$. 
The last variable $t_{L+1}$ is a \textit{free parameter} and will be set to its face value $1+(L+1)\tilde{\epsilon}$ (or $1$ if we shift back) in the very end. For example, at tree level we have
\beq \label{eqn-Rec-SeedBos000}
\mathtt{B}^{(0)}_{\alpha}(t_1)={1\over4\pi x\, t_1}\,,
\eeq
where at tree level $t_1=1$, but we keep it for later use. 
In \eqref{bpc} we have found that
\beq 
\mathtt{B}^{(1)}_{\alpha=1}(t_1;t_2)= - \frac{\lambda}{4\pi x\, t_1\, t_{2}}\,,\qquad\text{and}\qquad
\mathtt{B}_{\alpha=-1}^{(1)}(t_1;t_2) = \frac{\lambda}{4\pi x\, (t_{2}-t_1) t_{2}}\,,
\eeq
where $t_2=1+2\tilde\epsilon$, and in (\ref{2loop}) we have found that
\beq
\mathtt{B}^{(2)}_{\alpha=1}(t_1,t_2;t_3)= \frac{\lambda^2}{4\pi x\, t_1\, t_{2}\,t_3}\,,\qquad\text{and}\qquad
\mathtt{B}_{\alpha=-1}^{(2)}(t_1,t_2;t_3) = \frac{\lambda^2}{4\pi x\, (t_{3}-t_2)(t_{3}-t_1) t_{3}}\,,
\eeq
where $t_3=1+3\tilde\epsilon$. Explicitly, for $L>0$ loops the integrand is given by
\beq \label{preintegrand}
\mathtt{B}^{(L)}_\alpha(\{t_j\}_{j=1}^L;t_{L+1})\equiv \int{\dd^3 k_L\over(2\pi)^3}{\dd^3 k_0\over(2\pi)^3}\ {1\over k_0^2}\,\Phi^{(L)}_\alpha(\{t_j\}_{j=1}^L; t_{L+1}| k_0,k_L)
\,{1\over k_L^2}\,,
\eeq
where 
\begin{figure}[t]
\centering
\includegraphics[width=0.8\textwidth]{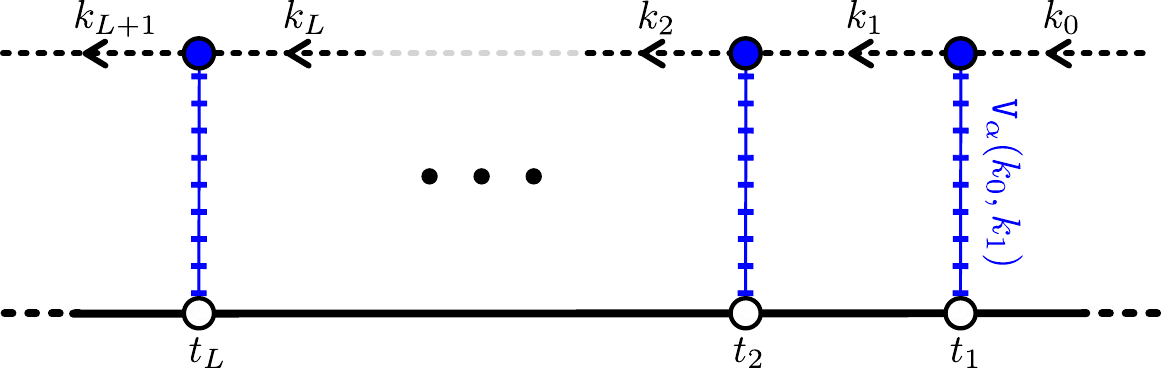}
\caption{In the lightcone gauge, the only diagram that contributes to the expectation value of the mesonic operator is given by the ladder-type diagrams shown in this figure. Here, the blue line and the filled blue blob represent the effective vertex function \eqref{vbosonic}. This vertex is the sum of the gluon-scalar-scalar derivative vertex connected by a gluon propagator to the line, and the bi-scalar insertion at the same point on the line.}\label{vprop}
\end{figure}
\beq\label{eqn-IntegrandNUniVersal}
\Phi^{(L)}_\alpha(\{t_j\}_{j=1}^L;t_{L+1}|k_0,k_L)\equiv(-2\pi \lambda x)^L e^{\ii k_L^3  x \, t_{L+1\, L}}\!\! \int\!\prod_{i=1}^{L-1} \frac{\dd^3 k_i}{(2\pi)^3}\frac{1}{k_i^2}\prod_{j=0}^{L-1} e^{\ii k_j^3 x\,t_{j+1\, j}} \vv(k_j,k_{j+1})\,.
\eeq
Here, $t_{L+1} =1+(L+1)\tilde\epsilon$, $t_0=0$, and 
\beq\label{vbosonic}
 \vv(k_j,k_{j+1})={2\over(k_j-k_{j+1})^+}\times\left\{\begin{array}{ll}k_j^+&\quad\alpha=+1\\ k_{j+1}^+&\quad\alpha=-1
 \end{array}\right. \,.
\eeq
In (\ref{preintegrand}) we have introduced the function $\Phi^{(L)}_\alpha$ that, as oppose to $\mathtt{B}^{(L)}_\alpha$, also depends on $k_0$ and $k_L$. We denote it by {\it pre-integrand}. As we will see below, both functions satisfy the same recursion relation. The reason for introducing the pre-integrand is that it will stay the same also for boundary operators that include derivatives.

\subsection*{The Universal Recursion Relation}
To derive a recursion relation for $\mathtt{B}^{(L)}_\alpha$ (and $\Phi^{(L)}_\alpha$) we 
perform the integration over $k_1$. For both $\alpha=1$ and $\alpha=-1$ this integral takes the form
\begin{align}\label{eqn-Rec-k1int}
\int \frac{\dd^3 k_1}{(2\pi)^3}  \frac{k_1^+\,e^{\ii k_1^3 x\, t_{21}}}{k_1^2(k_0-k_1)^+(k_1-k_2)^+} =&\frac{1}{(k_0-k_2)^+} \int \frac{\dd^3 k_1}{(2\pi)^3} \frac{e^{\ii k_1^3 x\, t_{21}}}{k_1^2} \left(\frac{k_2^+}{k_1^+-k_2^+} -\frac{k^+_0}{k_1^+-k_0^+} \right)\nn \\
=& \frac{1}{4\pi x t_{21}} \frac{1}{(k_0-k_2)^+} \left(e^{-|k_{2\perp}|x\,t_{21}} -e^{-|k_{0\perp}|x\,t_{21}} \right) \,,
\end{align}
where in the first step we have performed a partial fraction for $k_1$ and in the second we have evaluated the integral using polar coordinates.

As for the one-loop case (\ref{candd}), all integrals over $k_j^3$ in (\ref{eqn-higherloop-rel}) and (\ref{eqn-IntegrandNUniVersal}) can be closed on the upper half plane, where they only have a simple pole due to the $1/k^2_j$ factor. Therefore, upon integration, we can identify $\ii |k_{j \perp}|$ with $k^3_j$. 
Under this identification, the exponential term in \eqref{eqn-Rec-k1int} combines with $e^{\ii k_2^3 x\, t_{32}}\simeq e^{-|k_{2\perp}|x\, t_{32}}$ or $e^{\ii k_0^3 x\, t_{10}}\simeq e^{-|k_{0\perp}|x\, t_{10}}$ in (\ref{eqn-IntegrandNUniVersal}), 
to induce the recursion relation
\beq\label{eqn-Rec-Univ}
\boxed{
\mathtt{B}^{(L)}_{\alpha}(\{t_j\}_{j=1}^{L};t_{L+1}) = \frac{\lambda}{t_{21}} \Big[\mathtt{B}^{(L-1)}_{\alpha}( t_2,t_3,\ldots,t_{L};t_{L+1}) -\mathtt{B}^{(L-1)}_{\alpha}( t_1,t_3,\ldots,t_{L};t_{L+1}) \Big]}\, ,
\eeq
for $L > 1$, and
\beq\label{eqn-Rec-SeedBos00}
\mathtt{B}^{(1)}_1(t_1;t_2)=
- \frac{\lambda}{t_1}\, \mathtt{B}^{(0)}(t_{2})\,,\qquad
\mathtt{B}^{(1)}_{-1}(t_1;t_2)=
+ \frac{\lambda}{t_{2}-t_1}\, \mathtt{B}^{(0)}(t_{2})\,.
\eeq

With the initial conditions \eqref{eqn-Rec-SeedBos00}, we can now solve the recursion relations \eqref{eqn-Rec-Univ}. All the higher loop integrands dress the tree level one, as
\beq\label{eqn-Rec-solSca00}
\begin{aligned}
\mathtt{B}^{(L)}_1(\{t_j\}_{j=1}^{L};t_{L+1}) & = \mathtt{B}^{(0)}(t_{L+1})\times \prod_{j=1}^L \frac{-\lambda}{t_j}  = \frac{(-\lambda)^{L}}{4\pi x\, t_{L+1}} \prod_{j=1}^{L} \frac{1}{t_j}\,, \\ 
\mathtt{B}^{(L)}_{-1}(\{t_j\}_{j=1}^{L};t_{L+1}) &=  \mathtt{B}^{(0)}(t_{L+1})\times \prod_{j=1}^L \frac{\lambda}{t_{L+1}- t_j} = \frac{\lambda^{L}}{4\pi x\, t_{L+1}} \prod_{j=1}^{L} \frac{1}{t_{L+1}-t_j}\,. 
\end{aligned}
\eeq
where $t_{L+1} = 1+ (L+1)\tilde{\epsilon}$.

One can now shift all variables back, $t_j \rightarrow t_j + j \tilde{\epsilon}$ to restore the $[0,1]$ range of integration. 
After that, one can set $t_{L+1} = 1$. The final results are
\beq \label{eqn-Rec-solSca01}
\begin{aligned}
\mathtt{B}^{(L)}_1(\{t_j+j\tilde\epsilon\}_{j=1}^L;1+(L+1)\tilde{\epsilon}) & = \frac{(-\lambda)^{L}}{4\pi x} \frac{1}{1+(L+1)\tilde{\epsilon}} \prod_{j=1}^{L} \frac{1}{t_j+j\tilde{\epsilon}}\,, \\
\mathtt{B}^{(L)}_{-1}(\{t_j+j\tilde\epsilon\}_{j=1}^L;1+(L+1)\tilde{\epsilon}) & = \frac{\lambda^{L}_b}{4\pi x\, } \frac{1}{1+(L+1)\tilde{\epsilon}} \prod_{j=1}^{L} \frac{1}{1-t_j+(L+1-j) \tilde{\epsilon}}\,.
\end{aligned}
\eeq

Note that for $t_j \neq 0,1$ ($\forall j$), $\mathtt{B}^{(L)}_{\alpha=\pm1}$ are finite, even when some $t_j$s coincide. 
Hence, at all loop orders, the line operators with $\alpha=\pm1$ are indeed conformal. 

Moreover, these are the only perturbative fixed points. To see this, we note that the bi-scalar beta function takes the form
\beq\la{betafunc}
\beta_\alpha={\lambda\over2}(1-\alpha^2)\times f(\lambda,\alpha)\,,
\eeq
where $f(\lambda,\alpha)=1+\cO(\lambda)$. The prefactor in (\ref{betafunc}) is the two-loop result. It is of order $\lambda^1$ because we have stripped out one power of $1/k$ in front of the bi-scalar in (\ref{Wbos}). If there are other zeros of the beta function, they must come from infinity as we increase $\lambda$ 
and are, therefore, non-perturbative. 

The exact beta function is scheme-dependent. It can be obtained using the Callan-Symanzik equation for the defect entropy or the mesonic line operator expectation value, see \cite{Ivri}.\footnote{For the mesonic line operator the Callan-Symanzik equation takes the form
\beq
\(\beta_\alpha\d_\alpha-\epsilon\d_\epsilon+2\gamma_\phi\)\log\<M_{10}\>=0\,,\nn
\eeq
where $\gamma_\phi$ is the anomalous dimension of the scalar boundary operator.}

Note also that no ``cosmological constant'' term of the form $|dx|/\epsilon$ is generated. Hence, there is no need to tune the coefficient of a corresponding counter-term. This is in contrast to the case of the closed loop operator at order $1/N$ or the case of the condensed fermion line operator that we study in section \eqref{sec:condFer}.

\paragraph{Anomalous Dimensions.}

In appendix \eqref{apd:diffEqnPert}
we have developed an differential equation method to resum the $L$-loop integrals, and the final results for the expectation of our mesonic operators are
\beq \label{m10vev}
\langle M_{10} \rangle ={1\over4\pi x}\times
\begin{dcases}
{1\over\Gamma (1-\lambda)}\Big({\epsilon\,e^{\gamma_E}\over x}\Big)^\lambda\,, \quad\ \  \alpha = +1, \\[5pt]
{1\over\Gamma (1+\lambda)}\Big({\epsilon\,e^{\gamma_E}\over x}\Big)^{-\lambda}\,, \quad \alpha = -1\,. \\
\end{dcases}
\eeq
where $\gamma_E$ is the Euler's constant and we assumed that the framing vector is trivial.

\subsection{Operators with Derivatives}
\label{sec:bosOpWder}

All boundary operators other than the scalar, $\phi$, and $\phi^\dagger$, include derivatives. The recursion relations described above can be generalized to them in a straightforward way. Note first that the boundary operators are uniquely fixed by their tree-level dimension and transverse spin, with no mixing. At tree level, a complete basis of them is given by\footnote{Here $n$ is an integer, not to be confused with the framing vector in (\ref{ff}).}
\beq\label{eq:opsWeConsider}
\cO_{R,\,\text{tree}}^{(n,s)}= \left\{\begin{array}{lc}\d^n_3\d^s_+\phi &\quad s\ge1\\[5pt]
\d^n_3\d^{-s}_-\phi &\quad s\le0\end{array}\right.\ ,\qquad \cO_{L,\,\text{tree}}^{(n,s)}= \left\{\begin{array}{lc}\d^n_3\d^s_+\phi^\dagger &\quad s\ge0\\[5pt]
\d^n_3\d^{-s}_-\phi^\dagger &\quad s\le-1\end{array}\right.\,.
\eeq
Operators with mixed $\partial_+$ and $\partial_-$ derivatives can be converted into $\partial_3^2$ by using the equations of motion, so there is no need to consider them separately. An infinite straight line preserves an $SL(2,{\mathbb R})\times U(1)$ subgroup of the three-dimensional conformal symmetry. 
We see from (\ref{eq:opsWeConsider}) that the boundary operators are uniquely characterized by two numbers, their $SL(2,{\mathbb R})$ conformal dimension and their $U(1)$ spin in the transverse plane to the line. Moreover, the operators of minimal twist, $\cO^{(0,s)}_{R/L}$, are $SL(2,{\mathbb R})$ primaries. The descendants are obtained from the primaries by acting with the $SL(2,{\mathbb R})$ raising generator. Their form depends on the conformal frame (the points $x_0$ and $x_1$). Since we will let the endpoints vary and do not keep the line straight, in (\ref{eq:opsWeConsider}) we have used a simple, frame dependent, classification of the descendants -- by the number of longitudinal derivatives. 

When the interaction is turned on, the partial differentials above are replaced by the corresponding path derivatives.\footnote{The path derivatives pick the operator that multiplies the boundary value of a smooth deformation parameter, with the framing vector kept constant and perpendicular to the direction of the deformation.} Different ordering are related by the equations of motion. We choose to order the longitudinal path derivatives ($\delta_{x_3}$) last. For $\alpha=1$ we have
\beq\label{eq:opsWeConsider2}
\cO_{R}^{(n,s)}= \left\{\begin{array}{lc}\delta^n_{ x^3_R}\delta^{s-1}_{x^+_R}\cO_{R}^{(0,1)} &\quad s\ge1\\[5pt]
\delta^n_{x^3_R}\delta^{-s}_{x^-_R}\cO_{R}^{(0,0)} &\quad s\le0\end{array}\right.\ ,\qquad \cO_{L}^{(n,s)}= \left\{\begin{array}{lc}\delta^n_{ x^3_R}\delta^s_{x^+_R}\cO_{L}^{(0,0)} &\quad s\ge0\\[5pt]
\delta^n_{x^3_R}\delta^{-s-1}_{x^-_R}\cO_{L}^{(0,-1)}&\quad s\le-1\end{array}\right.\,.
\eeq
Here, the use of equal signs instead of a proportionality relation is a relative choice of normalization. At the bottom of these four towers we have the boundary operators
\beq\la{bottom}
\{\cO_L^{(0,0)},\cO_L^{(0,-1)}\}\quad\text{and}\quad\{\cO_R^{(0,0)},\cO_R^{(0,1)}\}\quad\text{for}\quad\alpha=1\,.
\eeq 
We have chosen to group them in this way because, as we will find next, operators in the two different towers on the right or left have opposite anomalous dimensions. Similarly, for $\alpha=-1$ the four bottom boundary operators on top of which all other operators are obtained by taking path derivatives are
\beq\la{bottomm1}
\{\cO_L^{(0,0)},\cO_L^{(0,1)}\}\quad\text{and}\quad\{\cO_R^{(0,0)},\cO_R^{(0,-1)}\}\quad\text{for}\quad\alpha=-1\,.
\eeq

Because the line is conformal at finite $\lambda$, the $SL(2,{\mathbb R})\times U(1)$ symmetry is not broken and the minimal twist operators remain $SL(2,{\mathbb R})$ primaries. The descendants are trivially handled by taking derivatives of the result with respect to $x$
\beq\label{longiDer}
\<\mathcal{O}_L^{(n_L,s)}(x_1)\,\cW_{10}\,\mathcal{O}_R^{(n_R,s)}(x_0)\> = \(\frac{\dd}{\dd x_1^3}\)^{n_L} \(\frac{\dd}{\dd x_0^3}\)^{n_R} \< \mathcal{O}_L^{(0,s)}(x_1)\,\cW_{10}\,\mathcal{O}_R^{(0,s)}(x_0)\>\,,
\eeq
where $\cW_{st}$ is the line operator (\ref{Wbos}) along a straight line between $x_s=x(s)$ and $x_t=x(t)$. 
Hence, in the rest of this paper, we will focus on the operators with $n_L=n_R=0$.

Our derivation of the anomalous spin applies to any boundary operator. The finite coupling spins of the left and right operators are therefore given by
\beq\label{spins}
\mathfrak{s}_L=s_L+\lambda/2\ ,\qquad \mathfrak{s}_R=s_R-\lambda/2\,.
\eeq

The $SL(2,{\mathbb R})\times U(1)$ symmetry of the straight line fixes the expectation value of the straight mesonic line operators with primaries boundary operators to take the form
\beq\label{conformalM3}
\< \mathcal{O}_L^{(0,s_L)}(x_1)\,\cW_{10}\,\mathcal{O}_R^{(0,s_R)}(x_0)\>\ \propto\ 
{\(n_L^+n_R^-\)^{\mathfrak{s}_L}\over|x_L-x_R|^{2\Delta^{(0,s_L)}_L}}\,\delta_{\mathfrak{s}_L+\mathfrak{s}_R,0}\,.
\eeq

For $s_{L/R}\ne0$ the bi-scalar condensate in (\ref{Wbos}) does not contribute to the expectation value. That is because the bi-scalar is a factorized product of right and left scalars. Each of them cannot absorb a non-zero integer transverse spin. As a result, for $s_{L/R}\ne0$ the expectation value is independent of $\alpha$.

We will now generalize the computation of the boundary dimension above to the case with non-zero tree level spin, $\Delta_L^{(0,s)}=\Delta_R^{(0,-s)}$. We will also compute the (scheme dependent) normalization prefactor.

\subsubsection*{The Case of $s_R>0$}

For $s_R=s>0$ and in light-cone gauge ($A_-=0$), the two boundary operators in (\ref{conformalM3}) are  
\beq\la{sRg0}
\mathcal{O}_L^{(0,-s)} = 
\partial_{-}^s \phi^\dagger(x_1)\ , \qquad \mathcal{O}_R^{(0,s)} = 
\(\partial_{+}-\ii A_+(x_0)\)^s  \phi(x_0)\, .
\eeq

The $x_1$ derivatives are ordinary derivatives; therefore, they can be replaced by powers of the corresponding conjugate momenta. On the other hand, at $x_1$ there is also the option of emitting a gluon. 
In the planar limit, that gluon has to be absorbed before the first ladder. 

The sum of all the diagrams in which the gluon is absorbed or absent can be recast into a modification of the first scalar propagator in (\ref{higherloop}), $1/k_0^2\rightarrow\mathbf{D}_+^{(s)}(k_0)/k_0^2$. It is equal to the sum of all planar diagrams connecting $D_+^s \phi$ to the first scalar in the ladder, without interacting with the Wilson line, (see figure \ref{fig:dseed} for example). A diagram in which the $A_+$ at the endpoint connects to the Wilson line vanishes identically due to the rotational symmetry of the straight line. The effective vertex, $\mathbf{D}_+^{(s)}(k_0)$, is a polynomial of degree $s$ in $k_+$ and $1/k_-$. 
It is a meromorphic function of $k_0^3$ and contains contributions up to $(s+1)$-loops. 
For the sake of notational simplicity, we have suppressed it's depends on $\epsilon$ and $\lambda$. 
The precise form of $\mathbf{D}_+^{(s)}(k_0)$ can be computed case by case and is independent of $\alpha$. In the next subsection, we will compute it explicitly for $s=1$. It is, however, irrelevant for computing the anomalous dimension. To see this, for now, we 
factor it out and treat it as if it was independent of $\lambda$. Instead of (\ref{preintegrand}) for the case with no derivatives, the 
$L>0$ ladder integrand now takes the form
\beq\la{Bwithd}
\mathtt{B}^{(L,s)}(\{t_j\}_{j=1}^{L};t_{L+1})= \int \frac{\dd^3 k_0}{(2\pi)^3}  \frac{\dd^3 k_L}{(2\pi)^3} \frac{(\ii k_L^+)^s}{k_L^2}
\Phi^{(L)}_{\alpha}(\{t_j\}_{j=1}^L;t_{L+1}|k_0,k_L)\times{\mathbf{D}_+^{(s)}(k_0)\over k_0^2}\, .
\eeq

Similarly, the tree level result \eqref{Mform} is generalized to
\beq
\mathtt{B}^{(0,s)}(t_{L+1})= \int \frac{\dd^3 k_0}{(2\pi)^3} e^{\ii k_0^3 x\, t_{L+1}}(\ii k_0^-)^s\times \frac{\mathbf{D}_+^{(s)}(k_0)}{k_0^2}\,.
\eeq

The factor of $\mathbf{D}_+^{(s)}(k_0)/k_0^2$ does not affect the $k_1$ integration in \eqref{Bwithd} and therefore this integrand satisfies the same recursion relation \eqref{eqn-Rec-Univ}. It only differs in the recursion seed, given by the $L=1$ integrand that we now focus on. 
The $k_{L=1}$ integral can be done explicitly, (without the knowledge of $\mathbf{D}_+^{(s)}(k_0)$). Using
\beq\label{eq:genIntOneSide}
\int \frac{\dd^3 k_1}{(2\pi)^3} \frac{(k_1^+)^a}{k_1^2(k_0-k_1)^+} e^{\ii k_1^3 x t} = - \frac{(k_0^+)^{a-1}}{4\pi x t} e^{- |k_{0\perp}|x\,t},
\eeq
and the expression \eqref{eqn-IntegrandNUniVersal} for $\Phi^{(1)}_{\alpha}(t_1;t_{L+1}|k_0,k_2)$, we find that 
\beq\label{B1ps}
\mathtt{B}^{(1,s)}(t_1;t_{2})=
\frac{\lambda}{t_{2}-t_1}\int \frac{\dd^3 k_0}{(2\pi)^3}\,e^{\ii k_0^3 x t_2}\,(\ii\, k^+_0)^s\, \frac{\mathbf{D}_+^{(s)}(k_0)}{k_0^2}=\frac{\lambda }{t_{2}-t_1} \mathtt{B}^{(0,s)}(t_{2})\,,
\eeq
where we have identified $k_0^3$ with $\ii |(k_0)_\perp|$.

We see that the $L$-ladder integral is still dressing the zero-ladder one, as for the operator with no derivatives (\ref{eqn-Rec-SeedBos00}). 
As predicted above, it is also independent of $\alpha$. Therefore, the recursion seeds for all $\mathtt{B}^{(L,s)}$, are of the type of the no-derivative operators with $\alpha = -1$, (\ref{eqn-Rec-SeedBos00}). The corresponding anomalous dimension of the boundary operators are also the same and are given by
\beq\la{Dbossg0}
\Delta^{(0,s)}_R=\Delta^{(0,-s)}_L=s+(1-\lambda)/2\,,\qquad s>0\,.
\eeq

\subsubsection*{The Case of $s_R<0$}
For $s_R=-s<0$, we are computing the two-point function between,
\beq
\mathcal{O}_L^{(0,s)} = D_+^s \phi^\dagger(x_1)\,, \quad \text{and} \quad \mathcal{O}_R^{(0,-s)} = 
\partial_-^s\phi(x_0)\, .
\eeq
We see that the rules of the left and right operators are interchanged in compared to (\ref{sRg0}). The recursion relation remains the same, with the seed given by
\beq\la{B1ms}
\mathtt{B}^{(1,-s)}(t_1;t_2)=  \frac{\lambda}{t_1}\int \frac{\dd^3 k_1}{(2\pi)^3}e^{\ii k_1^3  x \, t_2} \frac{\overline{\mathbf{D}}_+^{(-s)}(k_1)}{k_1^2}\,   (-\ii k_1^+)^s = \frac{\lambda}{t_1} \mathtt{B}^{(0,-s)}(t_2) \,,
\eeq
where the vertex $\overline{\mathbf{D}}_+^{(-s)}(k_1)$ modifies the leftmost propagator.

As before, the $k_1$ integral is independent of $t_1$, which leads to the same type of recursion seed as for the $\alpha=1$ case. Correspondingly, we have
\beq\la{Dbossl0}
\Delta^{(0,-s)}_R=\Delta^{(0,s)}_L=s+(1+\lambda)/2\,,\qquad s>0\,.
\eeq

\subsubsection*{Wave Function Normalization Factor}

The overall wave function normalization prefactor in (\ref{conformalM3}) is scheme dependent. Yet, it can combine with the other factors into a scheme-independent physical quantity, such as the two-point function of the cement operator that we consider in section \ref{sec:2ptDisBoson}. Due to the recursion relation, the proportionality prefactor in (\ref{conformalM3}) is inherited from the proportionality prefactor of the seed in (\ref{B1ps}) and (\ref{B1ms}), which by itself can be identified as the zero-ladder result in (\ref{eqn-higherloop-rel}) with boundary derivatives
\begin{align}
\mathtt{B}^{(0,s)}(t_1)=&\int \frac{\dd^3 q}{(2\pi)^3}e^{\ii q^3 x\, t_1}\,(\ii\, q^+)^s\, \frac{\mathbf{D}_+^{(s)}(q)}{q^2}\ ,\label{B01}\\
\mathtt{B}^{(0,-s)}(t_1)=&\int \frac{\dd^3 q}{(2\pi)^3}e^{\ii q^3  x \, t_1} \frac{\overline{\mathbf{D}}_+^{(-s,\alpha)}(q)}{q^2}\,   (-\ii q^+)^s\,.\label{B02}
\end{align}
We now compute these factors for the case of $s=1$ because it is the one that will be relevant for the displacement operator. 

\begin{figure}[t]
\centering
\includegraphics[width=0.85\textwidth]{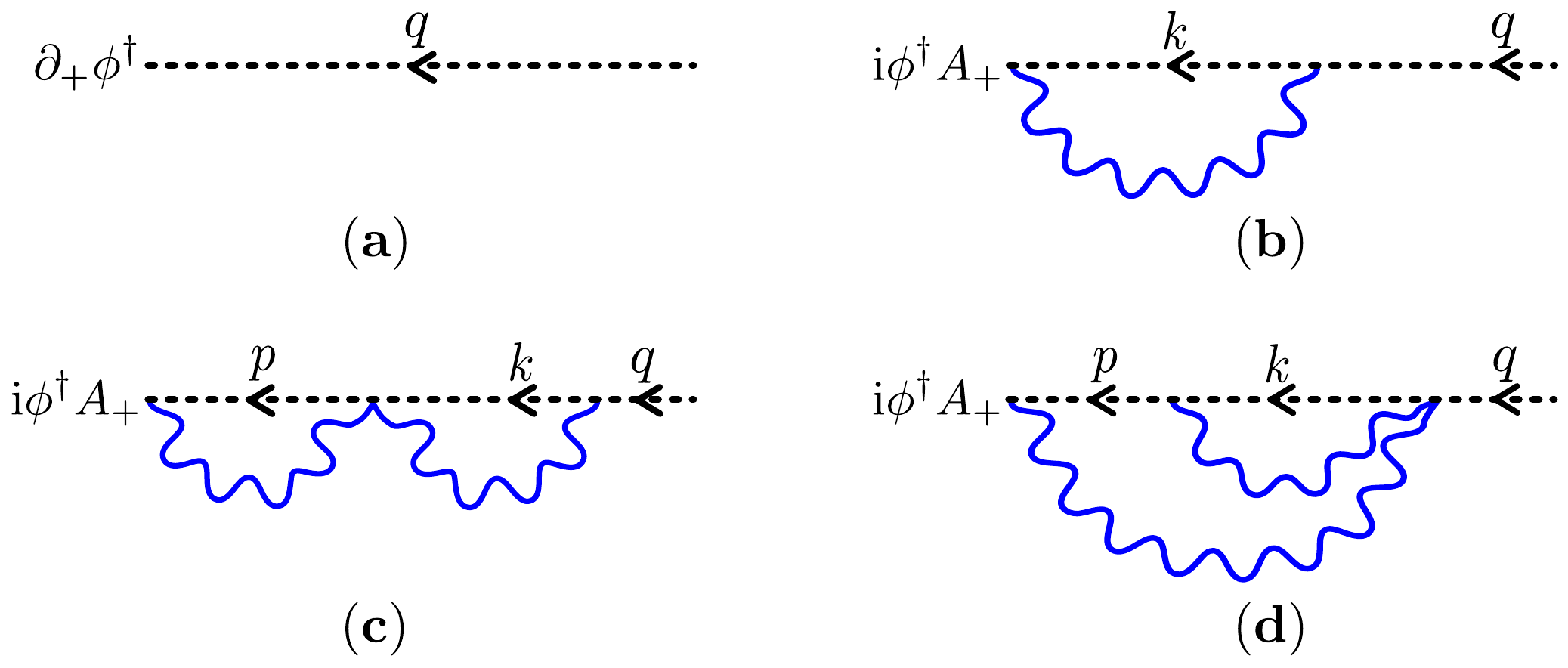}
\caption{The four diagrams that contribute to the modified propagator $\overline{\mathbf{D}}_+^{(-1,\alpha)}(q)$. Note that in lightcone gauge, the diagram in which the gluon is absorbed by the straight Wilson line vanishes identically. 
}
\label{fig:dseed}
\end{figure}

For $s_R=-1$ we have to compute \eqref{B02}. The vertex $\overline{\mathbf{D}}_+^{(-s,\alpha)}(q)$ is given by the sum of the four diagrams in figure \ref{fig:dseed}.      
The first diagram, \ref{fig:dseed}.a, is the tree-level propagator. It leads to the following contribution to $\<M_{10}^{(1,-1)}\>$
\begin{equation} \label{eqn-Dseed-a}
\mathtt{Figure}\, \ref{fig:dseed}.a = \int \frac{\dd^3 q}{(2\pi)^3} \frac{e^{\ii q^3 x\, t_1}}{q^2} q^+ q^- = \frac{1}{4 \pi (x\,t_1)^3}\,.
\end{equation}
The second diagram, \ref{fig:dseed}.b, is a contraction between the gauge field in the covariant derivative $D_+\phi^\dagger$ and the $A\!-\!\phi\!-\!\phi^\dagger$ interaction vertex,
\begin{equation} \label{eqn-Dseed-b}
\begin{aligned}
    \mathtt{Figure}\, \ref{fig:dseed}.b & = -2\pi \ii \lambda\int\frac{\dd^3 k}{(2\pi)^3} \frac{e^{\ii k^3 \epsilon}}{k^2} \int \frac{\dd^3 q}{(2\pi)^3}\frac{e^{\ii q^3 x\, t_1}}{q^2} q^+ \frac{k^3+q^3}{(q-k)^+} \\
 & = \frac{\lambda}{2} \frac{3t_1 + \te}{4\pi x^3 t_1^2 (t_1 + \te)^2} = \frac{3\lambda}{2} \frac{1}{4\pi x^3 t_1^3} + \cO(\te)\, .
\end{aligned}
\end{equation}
The third and the fourth diagrams, \ref{fig:dseed}.c and \ref{fig:dseed}.d, come from adding a quartic interaction to the diagram. They are given by
\begin{align}
	\mathtt{Figure}\, \ref{fig:dseed}.c &= (2\pi \lambda)^2\int\!\!\frac{\dd^3 p}{(2\pi)^3} \frac{e^{\ii p^3 \epsilon}}{p^2} \int\!\! \frac{\dd^3 k}{(2\pi)^3} \frac{e^{\ii k^3 \epsilon}}{k^2}  \int\!\! \frac{\dd^3 q}{(2\pi)^3}\frac{e^{\ii q^3 x\, t_1}}{q^2} \frac{(q+k)^+}{(q-p)^+} \frac{q^+}{(q-k)^+}\,, \\
	\mathtt{Figure}\, \ref{fig:dseed}.d&= (2\pi \lambda)^2 \int\!\! \frac{\dd^3 p}{(2\pi)^3} \frac{e^{\ii p^3 \epsilon}}{p^2} \int\!\! \frac{\dd^3 k}{(2\pi)^3} \frac{e^{\ii k^3 \epsilon}}{k^2}  \int\!\! \frac{\dd^3 q}{(2\pi)^3}\frac{e^{\ii q^3 x\, t_1}}{q^2} \frac{(k+p)^+}{(p-k)^+} \frac{q^+}{(p-q)^+}\,.\label{eqn-Dseed-d}
\end{align}
Summing the two we arrive at
\beqa \label{eqn-Dseed-cd}
\mathtt{Figures}\, \ref{fig:dseed}.c+ \ref{fig:dseed}.d
&=&2(2\pi \lambda)^2 \int\!\frac{\dd^3 p}{(2\pi)^3} \frac{e^{\ii p^3 \epsilon }}{p^2} \int\! \frac{\dd^3 k}{(2\pi)^3} \frac{e^{\ii k^3 \epsilon}}{k^2}  \int\! \frac{\dd^3 q}{(2\pi)^3}\frac{e^{\ii q^3 x \, t_1}}{q^2} \frac{k^+}{(p-k)^+} \frac{q^+}{(k-q)^+}\nn\\
&=&\frac{\lambda^2}{2}  \frac{1}{4 \pi x^3\, t_1 (t_1+\te) (t_1 + 2\te)} = \frac{\lambda^2}{2}  \frac{1}{4 \pi x^3\, t_1^3} + \cO(\te) \, .
\eeqa
The expression for the vertex $\overline{\mathbf{D}}_+^{(-1,\alpha)}(q)$ can be straightforwardly deduced from the expressions above by performing all integrations except for the $q$ integration. We decide not to present them here because the result is not very illuminating.

The final expression for $\mathtt{B}^{(0,-1)}(t)$ is then obtained by taking the sum of all the diagrams. Noticing that it is dressing the $L$-ladder line integral, which is only logarithmic divergent. Hence, it suffices to expand $\mathtt{B}^{(0,-1)}$ to $\mathcal{O}(\tilde{\epsilon}^0)$, 
\beq
\mathtt{B}^{(0,-1)}(t)= \frac{1}{4\pi (x\, t)^3}\times \frac{(\lambda+1)(\lambda+2)}{2} + \mathcal{O}(\tilde{\epsilon}) \,.
\end{equation}
Similarly,
\begin{equation}
\mathtt{B}^{(0,1)}(t)= \frac{1}{4\pi (x\, t)^3}\times \frac{(\lambda-1)(\lambda-2)}{2} + \mathcal{O}(\tilde{\epsilon}) \,. 
\end{equation}
As for the no-derivative case, the final $L$-ladder solution is dressing the expression $\mathtt{B}^{(0,1)}(1+(L+1)\tilde{\epsilon})$. 
Expanding again to $\mathcal{O}(\tilde{\epsilon})$, we conclude that
\begin{align} \label{M1m1prefactor}
\<M_{10}^{(1,-1)}\>&=\frac{\epsilon^{-\lambda}}{x^{-\lambda}} \frac{e^{-\gamma_E  \lambda }}{\Gamma (1+\lambda)}\times \mathtt{B}^{(0,-1)}(1)\(1+\cO(\tilde\epsilon)\)\,,\\
\<M_{10}^{(-1,1)}\>&=\frac{\epsilon^{\lambda}}{ x^{\lambda}} \frac{e^{\gamma_E  \lambda}}{\Gamma (1-\lambda)}\times \mathtt{B}^{(0,1)}(1)\(1+\cO(\tilde\epsilon)\)\,.\nn
\end{align}

\subsection{The Boundary Equation}
\label{sec:EOMBOS}

The scalar equation of motion $D_\mu D^\mu\phi(x)=0$ holds as an operator equation, up to contact terms. The scalar $\phi$ is however not a gauge invariant operator. Instead, it appears at the end of the line operator in (\ref{Wbos}), which couples to the bi-scalar in the exponent. 
Hence, when the bi-scalar approaches the endpoint, it may lead to contact terms and one may wonder what is the fate of the scalar equation of motion 
at the quantum level. At tree level, the equation of motion relates two different looking boundary operators, $\d_3^2\phi$ and $\d_+\d_-\phi=\d_-\d_+\phi$. This relation reduces the number of independent boundary operators. As we turn on the coupling $\lambda$, the number of boundary operators cannot change. Therefore, some form of the boundary equation of motion must hold in the quantum theory. In this section, we fix this operator equation. We first fix its form and then compute the relative coefficient. Finally, we give an alternative derivation of it that is based on formal manipulations of the path integral.

For special paths where the endpoint of the line coincides with a midpoint of the line or with the other endpoint, there are other sources to the boundary equation. Here, we only consider smooth non-self intersecting paths where these terms are not sourced. In section \ref{sec:bootstrap} we consider them formally.  

\subsubsection{The Equation Form}
\label{sec:eqFromBos}

We may first ask what operators can be related to each other by an operator equation in such a way that is consistent with our results from the previous sections. 

Let $\alpha=1$ and consider the operator $\mathcal{O}_R^{(n,-s)}$ with $s\ge1$. Taking its minus or longitudinal path derivatives results in the operators $\mathcal{O}_R^{(n,-s-1)}$ or $\mathcal{O}_R^{(n+1,-s)}$ correspondingly, (\ref{eq:opsWeConsider2}). On the other hand, taking its path derivative in the plus direction result in an operator of tree level spin $1-s$ and tree level dimension $\Delta_\text{tree}=n+s+1$. The only candidate operator is $\mathcal{O}_R^{(n+2,1-s)}$. Because this operator has the same anomalous dimension as $\mathcal{O}_R^{(n,-s)}$, a relation of the form $\delta_+\mathcal{O}_R^{(n,-s)}\propto \mathcal{O}_R^{(n+2,1-s)}$ is consistent at the quantum level. Moreover, since this is the only consistent quantum generalization of the boundary equation of motion, it must be satisfied. 

On the contrary, for $s=0$ the operators $\mathcal{O}_R^{(n,0)}$ and $\mathcal{O}_R^{(n,1)}$ have opposite anomalous dimensions and it does not make sense to identify $\delta_+\mathcal{O}_R^{(n,0)}$ with $\mathcal{O}_R^{(n+1,1)}$. Instead, the only valid relation involving spin one is $\mathcal{O}_R^{(n+2,1)}\propto \delta_-\mathcal{O}_R^{(n,2)}$. 

The same considerations extend to higher spins as well as the left boundary operators. In total, we have\footnote{Correspondingly, we expect the null cusp anomalous dimension to be trivial for $+\to-$ cusp, but nontrivial for $-\to+$ cusp.}
\beq\la{boundaryeq}
\mathcal{O}_R^{(n+2,s)}\propto\left\{\begin{array}{ll}
\delta_-\mathcal{O}_R^{(n,s+1)}&\quad s\ge1\\
\delta_+\mathcal{O}_R^{(n,s-1)}&\quad s\le0
\end{array}\right.\,,\quad \mathcal{O}_L^{(n+2,s)}\propto\left\{\begin{array}{ll}
\delta_-\mathcal{O}_L^{(n,s+1)}&\quad s\ge0\\
\delta_+\mathcal{O}_L^{(n,s-1)}&\quad s\le-1
\end{array}\right.\,,\quad\alpha=1\,.
\eeq
These boundary operator equations are complete because, in the free theory, they exactly cover all the operator relations that follow from the scalar equation of motion. 

Below, we will show that the proportionality factor does not receive quantum corrections and therefore takes its tree-level value, $-2$. Few comments are in order
\begin{itemize}
\item Note that operators of the form $\delta_+ \cO_L^{(n,-1)}$, $\delta_- \cO_L^{(n,0)}$, $\delta_- \cO_R^{(n,1)}$ or $\delta_+\cO_R^{(n,0)}$ do \textit{not} exist in the quantum theory. In other words, if we start with the operator $\cO_L^{(n,-1)}$ and perform a smooth deformation of the line, $x(\cdot)\to x(\cdot)+v(\cdot)$, we will not have a term linear in $v_L^+$. This fact is confirmed by a perturbative computation below and is directly related to the chiral form of the evolution equation, which is the subject of the next section.

\item If we start with the boundary operator $\cO_L^{(0,0)}$, then at second order in the smooth deformation, the unique boundary operator that multiplies $v_L^-v_L^+$ is $\cO_L^{(2,0)}$. It can be reached by either, first perturbing in the plus direction and then in the minus or the other way around. Correspondingly, when we try to evaluate $\delta_- \delta_+ \cO_L^{(n,-1)}$ in perturbation theory, we find that it does exist and equals to $-{1\over2}\cO_L^{(n+2,-1)}$, even though $\delta_+ \cO_L^{(n,-1)}$ does not exist. For more details, we refer the reader to \cite{bootstrap}.

\item For $\alpha=-1$, the anomalous dimensions of the $s=0$ operators are flipped. 
In that case, the relations that are consistent with the spectrum of boundary operators are
\beq\la{boundaryeqm1}
\mathcal{O}_R^{(n+2,s)}\propto\left\{\begin{array}{ll}
\delta_-\mathcal{O}_R^{(n,s+1)}&\quad s\ge0\\
\delta_+\mathcal{O}_R^{(n,s-1)}&\quad s\le-1
\end{array}\right.\,,\quad \mathcal{O}_L^{(n+2,s)}\propto\left\{\begin{array}{ll}
\delta_-\mathcal{O}_L^{(n,s+1)}&\quad s\ge1\\
\delta_+\mathcal{O}_L^{(n,s-1)}&\quad s\le0
\end{array}\right.\,,\quad\alpha=-1\,.
\eeq
\end{itemize}

\subsubsection{Fixing the Relative Coefficient}

To fix the relative coefficient in (\ref{boundaryeq}), (\ref{boundaryeqm1}) in our choice of normalization (\ref{eq:opsWeConsider}), (\ref{eq:opsWeConsider2}) we compute the expectation value of a straight line operator with these operators at its end. Consider first the equation
\beq\label{derinrel}
\(\delta_+\cO_L^{(0,-s-1)}\equiv\delta_+\delta_-\cO_L^{(0,-s)}\)\ \propto\ \(\cO_L^{(2,-s)}\equiv\delta_3^2\cO_L^{(0,-s)}\)\,,\qquad s\ge1\,,
\eeq
that applies for both $\alpha=1$ and $\alpha=-1$. We attach either of these left boundary operators to the straight line operator, with the boundary operator $\cO_R^{(0,s)}$ at the right end. 
We then express it in terms of the corresponding line integrand as
\beq\label{eqn-higherloop-relbeq}
\<\widehat\cO_L\cW_\alpha\cO_R^{(0,s)}\>_{L-\text{ladder}}=\int\limits_0^1 \mathcal{P}\prod_{j=1}^L\dd t_j\,\mathtt{B}^{(L,s,\widehat\cO_L)}_\alpha(\{t_j+j\tilde\epsilon\}_{j=1}^L;t_{L+1}+(L+1)\tilde{\epsilon})
\, ,
\eeq
where $\widehat\cO_L$ is either $\delta_+\delta_-\cO_L^{(0,-s)}$ or $\delta_3^2\cO_L^{(0,-s)}$. The explicit form of the line integrand is
\beq\la{intgrandO}
\mathtt{B}^{(L,s,\widehat\cO_L)}(\{t_j\}_{j=1}^{L};t_{L+1})= \int\!\! \frac{\dd^3 k_0}{(2\pi)^3}  \frac{\dd^3 k_L}{(2\pi)^3} \frac{\mathbf{V}_{\widehat\cO_L}(k_L)}{k_L^2}
\Phi^{(L)}_{\alpha}(\{t_j\}_{j=1}^L;t_{L+1}|k_0,k_L) \frac{\mathbf{D}_+^{(s)}(k_0)}{ k_0^2}\,,
\eeq
where the vertex $\mathbf{V}_{\widehat\cO_L}(k_L)$ is meromorphic in $k_L^3$. Here, the pre-integrand, $\Phi^{(L)}_{\alpha}$, is given in (\ref{eqn-IntegrandNUniVersal}) and $\mathbf{D}_+^{(s)}$ in (\ref{eqn-Dseed-a})-(\ref{eqn-Dseed-d}). The computations of the line integrand 
is almost identical to the one with derivatives in \eqref{Bwithd}. The only difference is that the leftmost (generalized) propagator in \eqref{Bwithd}, $(\ii k_L^+)^s/k_L^2$, is modified to a new function that we label as $\mathbf{V}_{\widehat\cO_L}(k_L)/k_L^2$. The pre-integrand $\Phi$ remains the same as in (\ref{eqn-IntegrandNUniVersal}), hence, \eqref{eqn-higherloop-relbeq} satisfies the same recursion relation. To check \eqref{derinrel} and confirm that the relative coefficient does not receive loop correction it is therefore sufficient to 
show that 
\beq\la{zerovertex}
2\mathbf{V}_{\delta_+ \delta_- \cO_L^{(0,-s)}}(k_L)+\mathbf{V}_{\delta_3^2 \cO_L^{(0,-s)}}(k_L)=0\,,
\eeq
where the equality is understood to hold under the integration in (\ref{intgrandO}).

In terms of the fields, the two operators take the form
\begin{align}
\delta_+\delta_-\cO_L^{(0,-s)}&= \partial_+\big( \partial_-^{s+1} \phi^\dagger\big) + \ii \big( \partial_-^{s+1} \phi^\dagger\big) A_+ \, ,\label{dpdmO}\\
\delta_3^2\cO_L^{(0,-s)}&= \partial_3^2 \big( \partial_-^s \phi^\dagger\big) + \ii \partial_3 \big( \partial_-^s \phi^\dagger \mathbb{A}_3\big) + \ii \partial_3 \big( \partial_-^s \phi^\dagger \big) \mathbb{A}_3 + \ii^2 \big( \partial_-^s \phi^\dagger \big) \mathbb{A}_3^2 \, ,
\end{align}
where $\mathbb{A}_3 = A_3 + {2\pi \ii \alpha\over k} \phi \phi^\dagger$ is the generalized connection. 
The diagrams that contribute to the vertex $\mathbf{V}_{\delta_+ \delta_- \cO_L^{(0,-s)}}$ are the same as the ones in figure \eqref{fig:dseed}. The only difference is that the additional minus derivatives in (\ref{dpdmO}) bring down a factor of $(\ii p^+)^s$, where $p$ is the momentum of the scalar. 
The final result that is inferred from \eqref{eqn-Dseed-a}, \eqref{eqn-Dseed-b}, and \eqref{eqn-Dseed-cd}, is
\beq\label{eqn-BdEOM-VPM}
\begin{aligned}
	2\mathbf{V}_{\delta_+ \delta_- \cO_L^{(0,-s)}}(q) & = 2\ii^2 q^+ q^- (\ii q^+)^s + 4\pi \lambda\int\!\! \frac{\dd^3 k}{(2\pi)^3} e^{\ii k^3 \epsilon} \frac{(\ii k^+)^{s+1}}{k^2} \frac{(q+k)^3}{(q-k)^+} \\
	& \quad + (4\pi \lambda)^2 \int\!\frac{\dd^3 p}{(2\pi)^3} e^{\ii p^3 \epsilon} \frac{(\ii p^+)^{s+1}}{p^2} \int\! \frac{\dd^3 k}{(2\pi)^3} \frac{e^{\ii k^3 \epsilon}}{k^2} \frac{k^+}{(p-k)^+} \frac{1}{(k-q)^+}\, .
\end{aligned}
\eeq

\begin{figure}[t]
\centering
\includegraphics[width=0.9\textwidth]{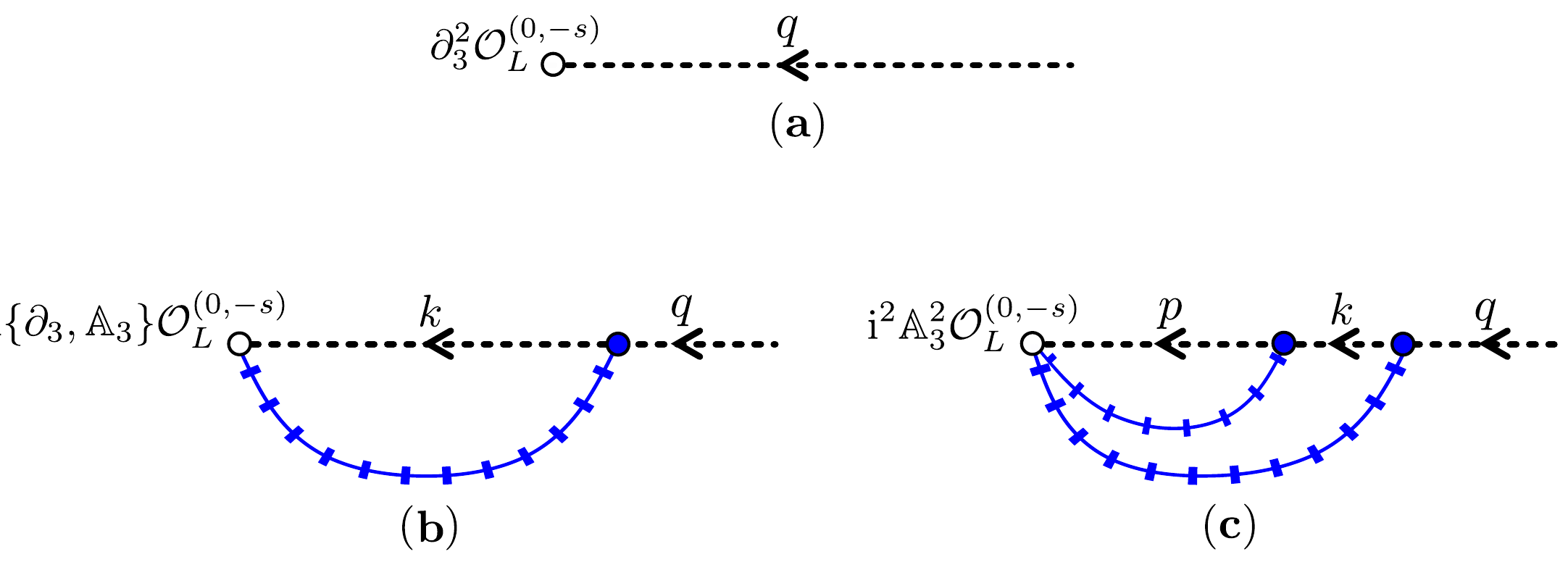}
\caption{The three diagrams that contribute to the vertex $\mathbf{V}_{\delta_3^2 \cO_L^{(0,-s)}}$. Here, the gauge field is understood to act on $\mathcal{O}_L^{(0,-s)}$ from the right. For instance, $\{\partial_3, \mathbb{A}_3\} \mathcal{O}_L = \partial_3 \big( \mathcal{O}_L \mathbb{A}_3\big) + \partial_3 \big( \mathcal{O}_L \big)\mathbb{A}_3$. The blue line and the filled blue blob are the same as in figure \ref{vprop}, and represent the effective vertex function \eqref{vbosonic}.
}
\label{fig:dseedeom}
\end{figure}
The diagrams that contributes to the vertex $\mathbf{V}_{\delta_3^2 \cO_L^{(0,-s)}}$ are shown in figure \ref{fig:dseedeom}. They give
\beq\label{eqn-BdEOM-V33}
\begin{aligned}
\mathbf{V}_{\delta_3^2 \cO_L^{(0,-s)}}(q) & = \ii^2 (q^3)^2 (\ii q^+)^s - 2\pi \lambda\int\!\! \frac{\dd^3 k}{(2\pi)^3} e^{\ii k^3 \epsilon} \frac{(\ii k^+)^{s}}{k^2} \ii (q+k)^3\,\vv(q,k) \\
& \quad - (2\pi \lambda)^2 \int\!\frac{\dd^3 p}{(2\pi)^3} e^{\ii p^3 \epsilon} \frac{(\ii p^+)^s}{p^2} \int\! \frac{\dd^3 k}{(2\pi)^3} \frac{e^{\ii k^3 \epsilon}}{k^2} \vv(q,k)\vv(k,p)\, ,
\end{aligned}
\eeq
where $\vv$ is defined in \eqref{vbosonic}. 

We can still identify $\ii q_3$ with $|q_\perp|$ upon integration, because (\ref{eqn-BdEOM-VPM}) and (\ref{eqn-BdEOM-V33}) are meromorphic in $q^3$.  
Upon this identification, the terms of order 
$\mathcal{O}(\lambda^0)$ in \eqref{eqn-BdEOM-VPM} and \eqref{eqn-BdEOM-V33} manifestly cancel each other. From (\ref{eq:genIntOneSide}) it follows that we can identify $(k^+)^{s+1}/(k-q)^+$ with $(q^+)^{s+1}/(k-q)^+$. Upon this identification, the terms of order 
$\mathcal{O}(\lambda^1)$ and $\mathcal{O}(\lambda^2)$ in \eqref{eqn-BdEOM-VPM} and \eqref{eqn-BdEOM-V33} also cancel each other.
Therefore, we conclude that 
\beq\la{beqb}
\cO_L^{(2,-s)}+2\delta_+\cO_L^{(0,-s-1)}=0\,.
\eeq
Note that the individual integrals in (\ref{eqn-BdEOM-VPM}) and (\ref{eqn-BdEOM-V33}) develops a power divergence at $\epsilon\to0$, which mean that these composite operators has counter terms. These however cancel out in the combination (\ref{beqb}).

Similarly, it follows that all the proportionality factors in (\ref{boundaryeq}) and (\ref{boundaryeqm1}) do not receive quantum corrections and are equal to $-2$.

 \subsection{The Line Evolution Equation}
\label{sec:bosevo}
As for the boundary operators, also the bulk of the line is subject to an operator equation. One way of presenting it is as follows.  
Under a small smooth deformation of the path $x(\cdot)\mapsto x(\cdot)+v(\cdot)$, the change in the line operator (\ref{Wbos}) can be expressed in terms of the displacement operator ${\mathbb D}_\mu(s)$ as
\beq\la{displacementop}
\delta\cW=\int\! \dd s\,|\dot x(s)|\,v^\mu(s)\,\cP\[{\mathbb D}_\mu(s)\cW\]\,,
\eeq
where the deformation is parameterized such that $v(s)$ is a normal vector. This equation can be taken as a definition of the displacement operator. In order to define the deformation of the line operator properly, we also need to specify how the framing vector transforms. The dependence on the framing vector is however topological and does not contribute to the displacement operator. We find that
\beq\la{Dpm1}
{\mathbb D}_+=-4\pi\lambda\, \cO_R^{(0,1)}\,\cO_L^{(0,0)}\,,\qquad
{\mathbb D}_-=-4\pi\lambda\, \cO_R^{(0,0)}\,\cO_L^{(0,-1)}\,,\qquad\text{for}\quad\alpha=+1\,,
\eeq
and
\beq\la{Dpmm1}
{\mathbb D}_+=+4\pi\lambda\, \cO_R^{(0,0)}\,\cO_L^{(0,1)}\,,\qquad
{\mathbb D}_-=+4\pi\lambda\, \cO_R^{(0,-1)}\,\cO_L^{(0,0)}\,,\qquad\text{for}\quad\alpha=-1\,,
\eeq
where it is understood that the framing vector, being continuous, is the same on the right and left. This factorized form of the 
operator leads to a closed equation for the mesonic line operators. We label them using the shorthand notation
\beq \label{Mev}
M_{st}^{(s_L,s_R)}[x(\cdot)]\equiv \cO_L^{(0,s_L)}\cW_{st}[x(\cdot)]\cO_R^{(0,s_R)}\,,
\eeq
where $x(\cdot)$ is a smooth path between $x_L=x(s)$ and $x_R=x(t)$. In this notations, equation (\ref{displacementop}) takes the form
\begin{align} \label{ee}
\delta M_{10}^{(s_L,s_R)}[x(\cdot)]=
&-4\pi\lambda\!\int\limits_0^1\! \dd s|\dot x_s|\!\[v^+_sM_{1s}^{(s_L,1)}M_{t0}^{(0,s_R)}\!+v^-_sM_{1s}^{(s_L,0)}M_{t0}^{(-1,s_R)}\]\\
&+\[\text{boundary terms}\]\,,\nn
\end{align}
where $u_s\equiv u(s)$. The boundary terms can be determined by the consistency of the equation, \cite{bootstrap}. We call this equation {\it the evolution equation} because it allows us to evolve the mesonic line operators (\ref{Mev}) as the path is smoothly deformed.

Note that the variation of the line also contains contributions from self-crossing points. Here, we restrict our consideration to paths that do not self-cross. Because the mesonic line operator is open, any shape can be reached from any other without going through such points.

In this section, we present two alternative derivations of this equation. In section \ref{sec:bootstrap} we show that the boundary and bulk equations are sufficient to systematically determine the perturbative expansion of the line expectation value.

\subsubsection{Derivation I}\la{derivation1}

It follows from (\ref{displacementop}) that ${\mathbb D}$ has a transverse spin equal to one. For a conformal line operator that is well defined along an arbitrary smooth path, it also follows from (\ref{displacementop}) that ${\mathbb D}$ is a dimension two $SL(2,\mathbb{R})$ primary. Reversing the logic, the existence of dimension two spin one primary operator on a fixed line can be used to define the deformed operator as (\ref{displacementop}).

In our case of the conformal line operators (\ref{Wbos}) with $\alpha=\pm1$ we find that there is a unique dimension two spin one primary operator, given by (\ref{Dpm1}), (\ref{Dpmm1}). The proportionality factor in (\ref{Dpm1}), (\ref{Dpmm1}) depends on our convention and does not follow from this dimensional argument alone. We will compute it explicitly in the next section in our normalization convention (\ref{m10vev}), (\ref{eq:opsWeConsider2}). In \cite{bootstrap} we bootstrap it from the consistency of the evolution equation and the spectrum of boundary operators.

\subsubsection{Derivation II}

The second derivation is a Schwinger-Dyson type equation that uses the definition of the operator in terms of fields, (\ref{Wbos}), inside the path integral. This is a rather formal derivation because it does not take into account renormalization of the operator. For $\alpha=\pm1$ however, we have seen that the operator is conformal, with no need for other counter terms to be adjusted. Hence, for this case, the corresponding Schwinger-Dyson type equation is on more firm ground. Indeed, we will find that it matches with (\ref{Dpm1}), (\ref{Dpmm1}) that we have argued for above and fix the proportionality coefficient in our normalization convention, (\ref{m10vev}), (\ref{eq:opsWeConsider2}).

The conformal line operator (\ref{Wbos}) can be written as
\beq\la{gconnection}
\cW=\cP\,e^{\ii\int_\cC{\mathbb A}\cdot\dd x}\,,\qquad\text{with}\quad{\mathbb A}_\mu=A_\mu+\ii\alpha{2\pi\over k}\phi\phi^\dagger{\dd x_\mu\over|\dd x|}\,.
\eeq
Under a small smooth deformation of the path $x(\cdot)\mapsto x(\cdot)+v(\cdot)$, the change in the line operator (\ref{gconnection}) can be written as
\beq \label{displacementF}
\delta\cW=\ii\int\! \dd s\,\dot x(s)^\mu\,v^\nu(s)\,\cP\[{\mathbb F}_{\mu\nu}(s)\cW\]\,.
\eeq
If ${\mathbb A}_\mu$ would have been a connection that is a local function of spacetime, then ${\mathbb F}_{\mu\nu}$ would have been given by the standard expression, $\d_\mu{\mathbb A}_\nu-\d_\nu{\mathbb A}_\mu-\ii\big[{\mathbb A}_\mu,{\mathbb A}_\nu\big]$. However, the factor $\dd x_\mu/|\dd x|$ in (\ref{gconnection}) is only defined on the line and therefore $\d_\mu{\mathbb A}_\nu$ is not well defined. Instead, the computation of (\ref{displacementF}) requires the expansion of this factor in $v$ and performing integration by parts. We find that    
\beq \label{mathbbF}
\dot x^\mu\,v^\nu\,{\mathbb F}_{\mu\nu}= \dot x^\mu\,v^\nu\[F_{\mu\nu}+\ii\alpha\frac{2\pi}{k} \dot{x}_\mu\Big(D_\nu(\phi\,\phi^\dagger)-\phi\,\phi^\dagger{\ddot x_\nu\over|\dot x|^2}\Big)\]\,,
\eeq
where we have used the fact that $v$ is normal to $\dot x$.

The gauge field equation of motion reads
\beq\la{Aeom}
(F_{\mu \nu})^i_{\ j} = \frac{2\pi}{k} \epsilon_{\mu \nu \rho}\[\phi^i(D^\rho \phi)^\dagger_j  -  (D^\rho \phi)^i\phi^\dagger_j\]\,,
\eeq
where $i$ and $j$ are color indices. This equation holds inside (\ref{displacementF}) up to contact terms that are present at self-crossing points. We will not write these contact terms explicitly because they will not be relevant for our consideration, (they can be found for example in \cite{Migdal:1983qrz}). By plugging (\ref{Aeom}) into (\ref{mathbbF}) we arrive at 
\begin{align} \label{displacementeom}
\delta\cW=\frac{2\pi\ii}{k}\int\! \dd s\,\dot x(s)^\mu\,v^\nu(s)\,\cP\Big[\Big(&\epsilon_{\nu\mu\rho}[\phi(D^\rho \phi)^\dagger-(D^\rho \phi)\phi^\dagger] + \ii \alpha [ \phi(D_\nu \phi)^\dagger + (D_\nu\phi)\phi^\dagger]\nn\\
-&\ii\alpha\phi\,\phi^\dagger\ddot x_\nu/|\dot x|^2\Big)\cW\Big]+\text{[self crossing contact term]}\,.
\end{align}
These terms nicely combine into 
\beq \label{eqn-displacementeOp}
v^\mu{\mathbb D}_\mu=\frac{4\pi\ii}{k}\left\{\begin{array}{ll}-v^+\phi \Big[\big(D_-+\ii\,\ddot x_-/|\dot x|^2\big)\phi\Big]^\dagger-v^-\Big[\big(D_-+\ii\,\ddot x_-/|\dot x|^2\big)\phi\Big]\phi^\dagger&\quad\alpha=+1\\
+v^-\phi \Big[\big(D_++\ii\,\ddot x_+/|\dot x|^2\big)\phi\Big]^\dagger+v^+\Big[\big(D_++\ii\,\ddot x_+/|\dot x|^2\big)\phi\Big]\phi^\dagger&\quad\alpha=-1
\end{array}\right.
\eeq

For a straight line, $\ddot x=0$, and this equation is manifestly the same as (\ref{Dpm1}), (\ref{Dpmm1}), with the spinning operators related to each other by path derivatives (\ref{eq:opsWeConsider2}). 
The case where the line is not straight can be obtained from the straight one by applying a conformal transformation. 
In that way, the term with $\ddot x$ in (\ref{eqn-displacementeOp}) is generated 
from the derivative of scalar field conformal factor. 

\subsection{The Two-point Function of the Displacement Operator} \label{sec:2ptDisBoson}

We would like to understand the dependence of the mesonic line operators expectation value of the path. We start with a straight mesonic line operator \eqref{conformalM3} with $s_R=-s_L$, so that it has a non-zero expectation value. We then deform the path as $x(s)\to x(s)+v(s)$, where $v(s)$ is a smooth normal vector. Due to the transverse rotation symmetry on the straight line, the first-order deformation has zero expectation value. At second order, the first non-trivial contribution appears as the expectation value of the line operator with two displacement operators insertions 
\beq\la{predistpf}
v(s)^\mu v(t)^\nu\,\< \mathcal{O}_L\,\cW_{1s}\,{\mathbb D}_\mu(s)\,\cW_{st}\,{\mathbb D}_\nu(t)\,\cW_{t0}\,\mathcal{O}_R\>\,.
\eeq
By plugging the displacement operator in (\ref{Dpm1}) we find that for $\alpha=1$, a non-zero expectation value is obtained for $s_R=-s_L=0$ or $s_R=-s_L=1$ only, being the bottom operators in (\ref{bottom}). 
In the former case only the $(\mu,\nu)=(-,+)$ term in (\ref{predistpf}) contributes and in the latter case only the $(\mu,\nu)=(+,-)$ term contributes. In either of these instances, we define the two point-function of the displacement operator to be given by the normalization independent factor, $\Lambda$, in the ratio
\beq \la{distpf}
{\< \mathcal{O}_L\,\cW_{1s}\,{\mathbb D}_\mu(s)\,\cW_{st}\,{\mathbb D}^\mu(t)\,\cW_{t0}\,\mathcal{O}_R\>\over\< \mathcal{O}_L\,\cW_{10}\,\mathcal{O}_R\>}={\Lambda\over x_{st}^4}\({x_{10}x_{st}\over x_{1s}x_{t0}}\)^{2\Delta}\,,
\eeq
where $\Delta$ is the conformal dimension of $\cO_{R/L}$. We find that it is given by the combination
\beq\la{MtoLambda}
\Lambda=(4\pi\lambda)^2x_{10}^4\<M_{10}^{(0,0)}\>\<M_{10}^{(-1,1)}\>\qquad\text{for}\qquad\alpha=1\,.
\eeq
Similarly, for $\alpha=-1$, using the form of the displacement operator in (\ref{Dpmm1}), we find
\beq\la{MtoLambdam1}
\Lambda=(4\pi\lambda)^2x_{10}^4\<M_{10}^{(0,0)}\>\<M_{10}^{(1,-1)}\>\qquad\text{for}\qquad\alpha=-1\,.
\eeq

Instead of deforming the straight mesonic line operator, we can also deform the closed circular loop, which is conformal to the infinite straight line. In that case (\ref{distpf}) is replaced by
\beq\la{distpfcirc}
\<\tr\(\cW_{\infty s}\,{\mathbb D}_\mu(s)\,\cW_{st}\,{\mathbb D}^\mu(t)\,\cW_{t-\infty}\)\>={\Lambda\over  x_{st}^4}\,.
\eeq
The coefficient $\Lambda$ is the same, given by the combinations (\ref{MtoLambda}) and (\ref{MtoLambdam1}) for $\alpha=1$ and $\alpha=-1$ correspondingly.

In \cite{bootstrap} 
we show that the two-point of the displacement operator is a function of $\Delta$, the dimension of either the left or right operators in the factorized displacement operator (\ref{Dpm1}), (\ref{Dpmm1}). Hence, $\Lambda(\Delta)=\Lambda(2-\Delta)$, in agreement with the fact that the same factor is obtained using either $s_R=0$ or $s_R=1$ for $\alpha=1$, ($s_R=0$ or $s_R=-1$ for $\alpha=-1$) in (\ref{distpf}). By plugging the scheme dependent coefficients in (\ref{m10vev}), (\ref{M1m1prefactor}) into the combination (\ref{MtoLambda}) we find that
\beq\la{fDD}
\Lambda(\Delta)=-\frac{(2\Delta-1)(2\Delta-2)(2\Delta-3)\sin(2\pi\Delta)}{2\pi}\,,
\eeq
where for $\alpha=1$ we have $\Delta=(1+\lambda)/2$ or $\Delta=(3-\lambda)/2$, and $\Delta=(1-\lambda)/2$ or $\Delta=(3+\lambda)/2$ for $\alpha=-1$.
Note that here we do not deal with the integrations over $s$ and $t$ in (\ref{predistpf}). They are treated in \cite{bootstrap}, where we bootstrap the relation (\ref{fDD}) from the evolution equation, the boundary equation, and the spectrum of boundary operators.

\subsection{Implication of High Spin Conserved Charges}\la{iohscc}

Note that the left and right boundary operators in the displacement operator have non-zero anomalous dimensions and anomalous spins. These, however, cancel out in the combinations in (\ref{Dpm1}) and (\ref{Dpmm1}). Likewise, we have protected operators at any non-zero integer spin. The existence of such tilt operators\footnote{The arguments about tilt operators having protected dimension date back to \cite{Bray:1977tk}; a modern treatment is given in \cite{Cuomo:2021cnb,Padayasi:2021sik,Herzog:2017xha, Cuomo:2022xgw}.} 
follows from the higher spin symmetry of CS-matter theories, that is only broken at order $1/N$.\footnote{Higher spin symmetry also leads to integrated constraints, 
see \cite{bootstrap}.} Namely, the action of a spin $S\ge2$ conserved charge on the line generates a protected $SL(2,{\mathbb R})$ primary operator of dimension $\Delta=S$ and transverse spin $\mathfrak{s}=\pm(S-1)$. To represent the charge, we wrap the line by a two-dimensional surface and integrate the spin $S$ conserved charge $J^{(S)}$ over it as
\beq\la{charge}
Q_v^{(S)}=\int \dd^2\Sigma^\mu J_{\mu\,\nu_1\nu_2\ldots\nu_{S-1}}^{(S)}v^{\nu_1}v^{\nu_2}\dots v^{\nu_{S-1}}\,,
\eeq
where $v$ is a constant vector normal to the line. We can then shrink the surface on the line and express the result as an operator ${\mathbb O}$ integrated on the line with transverse spin $S-1$ as
\beq
Q_v^{(S)}\cW=\ii\int\! \dd s\,|\dot x_s|\(v^{\nu_1}\dots v^{\nu_{S-1}}\)\,\cP\[{\mathbb O}_{\nu_1\ldots\nu_{S-1}}(s)\cW\]\,.
\eeq
For example, the action of the conserved energy-momentum tensor induce the insertion of the displacement operator as in (\ref{displacementop}). 

In addition, to the local conserved charges we can also use in (\ref{charge}) their (three dimensional) conformal descendant to generates $SL(2,{\mathbb R})$ protected primary line operators. 
Explicitly, using the operators $D_+^n J_{\mu\,++\ldots+}$ and $D_-^n J_{\mu\,--\ldots-}$, for any transverse spin $s\ne 0 $ we generate $|s|$ primary operators of dimension $\Delta=|s|+1$. For $\alpha=1$ they take the form
\beq\la{Dpm1S}
{\mathbb D}^{(S)}= \cO_R^{(0,S-n)}\,\cO_L^{(0,n)}\,,\qquad
{\mathbb D}^{(-S)}=\cO_R^{(0,-n)}\,\cO_L^{(0,n-S)}\quad\text{with}\quad n=0,1,\ldots,S-1\,.
\eeq
Similarly, for $\alpha=-1$ they take the form
\beq\la{Dpmm1S}
{\mathbb D}^{(S)}= \cO_R^{(0,n)}\,\cO_L^{(0,S-n)}\,,\qquad
{\mathbb D}^{(-S)}=\cO_R^{(0,n-S)}\,\cO_L^{(0,-n)}\quad\text{with}\quad n=0,1,\ldots,S-1\,.
\eeq

As we go to $1/N$ order, we expect all these protected operators except the displacement operator to develop non-zero anomalous dimensions.

\section{The Conformal Line Operators in the Fermionic Theory}\la{fersec}

The ``free fermion'' theory is defined by coupling a fundamental fermion to Chern-Simons theory as
\beq\la{Sfer}
S_E^\text{fer}=S_{CS}+\int \dd^3 x\,\bar\psi\cdot \gamma^\mu D_{\mu}\psi\,,
\eeq
where $D_\mu\psi=\d_\mu\psi-\ii A_\mu \cdot \psi$ and $S_{CS}$ is given in \eqref{CSact} with gauge group $SU(N)$ or $U(N)$. Here, $\psi$ is a two-component spinor and the Euclidean $\gamma$ matrices are taken to be the Pauli matrices, $\gamma^\mu = \sigma^\mu$.\footnote{
The explicit form of the matrix representation of fermion and $\gamma$ matrices are given by \begin{equation*}
\bar{\psi}\gamma\psi = (\bar{\psi}_-,\bar{\psi}_+)
\begin{pmatrix}
\gamma_{+-} & \gamma_{++} & \\
\gamma_{--} & \gamma_{-+} & 
\end{pmatrix} 
\begin{pmatrix}
\psi_+ \\
\psi_- 
\end{pmatrix},
\end{equation*}
where $\psi_\pm$ and $\bar\psi_\pm$ carry tree level spin $\pm{1\over2}$ around the third direction correspondingly.
}
After tuning the mass to zero, the theory (\ref{Sfer}) is conformal. This CFT can also be obtained from the so-called critical (Gross-Neveu) CS-fermion theory by a double trace deformation. The flow between the two theories is expected to be the fermionic dual description of the flow between the boson and the critical boson theories mentioned at the beginning of section \ref{bossec}. We consider the CFT (\ref{Sfer}) in the planar limit, $N\to\infty$ with $\lambda=N/k$ fixed. As for the scalar case, the contribution of the double trace deformation to the expectation value of the mesonic line operators is $1/N$ suppressed. Therefore, the difference between the fermion and the critical fermion theories will not be relevant for our consideration.  

The simplest line operator in the fundamental representation is the standard Wilson line,
\beq \label{WLop}
W[x(\cdot)]=\cP e^{\ii \int A_\mu\dot x^\mu \dd s}\,.
\eeq
In the free theory (at $\lambda=0$) there is no adjoint operator of dimension one or less. Hence, at least for some range of $\lambda$, the Wilson line operator (\ref{WLop}) is conformal.\footnote{This is in contrast with the free bosonic theory, where the adjoint bi-scalar has free dimension equal to one.} In this section we analyze the mesonic line operators that are obtained by stretching the Wilson line along an arbitrary smooth path between right (fundamental) and left (anti-fundamental) boundary operators
\beq
M=\cO_LW[x(\cdot)]\cO_R\,.
\eeq
At tree level and for a line in the third direction, a basis of these boundary operators is
\beq
\label{eq:opsWeConsider_fer}
\cO_{R,\,\text{tree}}^{(n,s)}= \left\{\begin{array}{lc}\d^n_{ x^3_R}\d^s_{x^+_R}\psi_+ &\quad s > 0\\[5pt]
	\d^n_{x^3_R}\d^{-s}_{x^-_R}\psi_- &\quad s < 0\end{array}\right.\ ,\qquad \cO_{L,\,\text{tree}}^{(n,s)}= \left\{\begin{array}{lc}\d^n_{ x^3_L}\d^{s-1/2}_{x^+_L}\bar{\psi}_+ &\quad s>0\\[5pt]
	\d^n_{x^3_L}\d^{-s+1/2}_{x^-_L}\bar \psi_- &\quad s<0\end{array}\right.\,,
\eeq
where $\psi_\pm$ and $\bar\psi_\pm$ carry tree level spin $\pm{1\over2}$ around the third direction correspondingly.
At loop level, these operators are renormalized. They receive anomalous dimension, anomalous spin, and wave function renormalization that we compute in this section to all loop orders. In what follows, we will use the shorthand notation
\beq\la{Mfer}
M^{(s_L,s_R)}_{st}\equiv\cO_L^{(0,s_L)}W_{st}[x(\cdot)]\cO_R^{(0,s_R)}\,,
\eeq
where now $s_L$ and $s_R$ take half-integer values, and $x(\cdot)$ is a smooth path between $x_L=x(s)$ and $x_R=x(t)$.

As in the bosonic case, we work in light cone gauge $A_-=0$. In this gauge, the gluon propagator takes the form (\ref{eqn-LC-gluonProp}), and the ghosts decouple in the planar limit. To regularize these UV divergences we introduce a new regularization scheme, similar to (\ref{deformedprop}), in which the exact fermion propagator 
\beq\la{ferprop}
\langle \bar{\psi}_i(p) \psi^j(q) \rangle
=(2\pi)^3 \delta^3(p+q)\delta^j_i\,\Delta(p)\,,
\eeq
is deformed to take the following tree-level form
\beq\label{eq:ferReg}
\Delta_0(p) = 
\frac{1}{\ii \slashed{p} + \mathcal{O}(\epsilon)} \times e^{\ii \epsilon p_3} \,.
\eeq
As in the scalar case, the factor of $e^{\ii\epsilon p_3}$ is a point splitting-like regularization and is correlated with the direction of the line that we will take to point along the third direction. The $\mathcal{O}(\epsilon)$ term in the denominator of the deformed tree-level propagator is properly chosen, such that the exact propagator takes the simple form\footnote{Here, we follow the convention of \cite{Giombi:2011kc}, the actual self-energy is given by  $-\Sigma(p)$. For later convenience, we use non-standard fermionic propagator \eqref{ferprop} instead of the standard one $\langle \psi(p) \bar \psi(q) \rangle=(2\pi)^3\delta^3(p+q)\,\tilde{\Delta}(p)$. They are related by $\Delta(p) = - \tilde{\Delta}(-p)$.} 
\beq\label{eq:exPropFer}
\Delta(p) \equiv \frac{\Delta_0(p)}{1+\Sigma(p)\Delta_0(p)} = \frac{1}{\ii \slashed{p} + \Sigma(p) + M_{c.t.}} e^{\ii \epsilon p_3}\, .
\eeq
Here $\Sigma = \ii \gamma^\mu \Sigma_\mu + \mathbb{I}\, \Sigma_0$ denotes (minus) the self-energy, which equals the sum of all 1PI diagrams, and $M_{c.t.}$ is a mass counterterm. 
By plugging (\ref{eq:ferReg}) into (\ref{eq:exPropFer}), the $\mathcal{O}(\epsilon)$ corrections of the deformed tree level fermion propagator can be expressed in terms of the self-energy as 
\beq \label{deformedpropf2}
\Delta_0^{-1}(p) = (\ii \slashed{p} + M_{c.t.})e^{-\ii \epsilon p_3} + (e^{-\ii \epsilon p_3} -1) \Sigma(p) \, .
\eeq
The explicit form of the self-energy $\Sigma$, and thus $\Delta_0$, will be solved in section \ref{sec:fer:selfenergy} using the Schwinger-Dyson equation.

\subsection{The Anomalous Spin}
\label{sec:anomspinf}

The anomalous spin is derived in section \ref{sec:susy} using supersymmetry arguments, (see appendix \ref{apd:FramingFactor} for perturbative checks). It turns out to be identical to the one for mesonic line operators in the bosonic theory and results in the overall framing factor in (\ref{ff}). In addition, the tree-level half-integer spin $s$ of the boundary fermions can be accounted for through the dependence on the boundary framing vector as $(n_{R/L}^\pm)^{|s|}$. In total, the dependence on the boundary framing vector reads
\beq\label{fff}
(\mathtt{spin\ factor})\times(\mathtt{framing\ factor})=(n_L^+)^{s_L}(n_R^+)^{s_R}\times\(2n_L^+ n_R^-\)^{\lambda\over2}=(n_L^+)^{\mathfrak{s}_L}(n_R^+)^{\mathfrak{s}_R}\,,
\eeq
where the total boundary spins, $\mathfrak{s}_{R/L}$, are given in (\ref{spins}). For simplicity, in what follows we will choose the framing and normalization such that this factor is trivial.

\subsection{Self-energy} \label{sec:fer:selfenergy}
\begin{figure}[t]
    \centering
    \includegraphics[width= 0.6 \textwidth ]{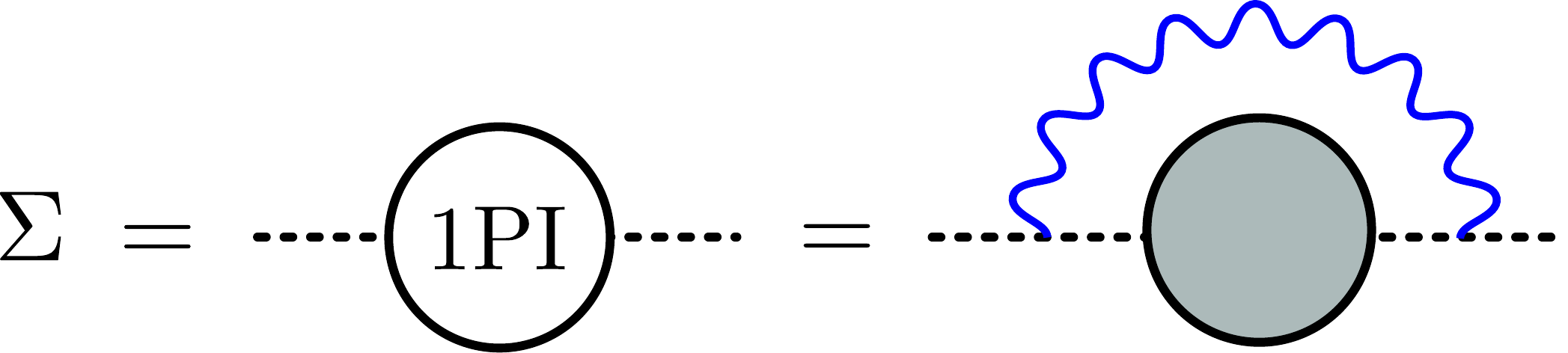}
    \caption{The planar fermion self energy $\Sigma(p)$ satisfies a simple Schwinger-Dyson equation.}
    \label{fig:self-energy-fer}
\end{figure}

In the regularization scheme (\ref{eq:ferReg}) the exact fermion propagator takes the form \eqref{eq:exPropFer}.
The self-energy, $\Sigma(p)$, is subject to the Schwinger-Dyson equation in figure \ref{fig:self-energy-fer}. Explicitly, this equation reads\footnote{Here, we have performed a shift $\Sigma_0 \rightarrow \Sigma_0 - M_{c.t}$.}
\beq \label{SDeq}
\Sigma(p)\equiv \ii \Sigma_\mu(p)\gamma^\mu  + \(\Sigma_0(p)-M_{c.t.}\) \mathbb{I}=2\pi\ii\lambda\int{\dd^3 q\over(2\pi)^3}\gamma^\mu\Delta(q)\,\gamma^\nu G_{\mu\nu}(p-q)\,,
\eeq
where the gluon propagator $G_{\mu\nu}$ is given in (\ref{eqn-LC-gluonProp}) and $M_{c.t.}$ is the bare mass (or mass counterterm). 
To solve this equation we first use that $\Delta(p)$ can be written as
\beq
\Delta(p)= \frac{-\ii \gamma^\mu p_\mu - \Sigma_\mu(p) \gamma^\mu  + \Sigma_{0}(p) \mathbb{I}}{\(p_\mu +  \Sigma_\mu(p)\)^2 + \Sigma_0(p)^2} e^{\ii \epsilon p_3}\,.
\eeq
Next, we 
plug the following ansatz 
for the denominator of the exact propagator\footnote{In appendix \ref{sec:ferselfEnergyDiff}, we present a different way of solving the Schwinger-Dyson equations using differential equation techniques.}
\beq \label{Sigmaguess}
\(p_\mu + \Sigma_\mu(p)\)^2 +\Sigma_0(p)^2 = p^2\,.
\eeq
That ansatz, which will be satisfied by the exact solution below, is directly correlated with the choice of regularization scheme \eqref{eq:exPropFer} and \eqref{deformedpropf2}. For such a function the Schwinger-Dyson equation (\ref{SDeq}) simplifies to
\beq
\Sigma(p) = 4\pi \ii \lambda \int \frac{\dd^3 q}{(2\pi)^3} \frac{\gamma^+ \Sigma_0(q) + \ii (q^+ + \Sigma_-(q))}{q^2} \frac{e^{\ii \epsilon q_3}}{(p-q)^+} \, . 
\eeq
Comparing the matrix structure on both sides, we see that $\Sigma_- = \Sigma_3 = 0$. The other two non-vanishing components satisfy the following equations,
\begin{align} 
\Sigma_0(p) - M_{c.t.}=&-4\pi \lambda \int \frac{\dd^3 q}{(2\pi)^3} \frac{1}{q^2} \frac{ q^+}{(p-q)^+} e^{\ii \epsilon q_3}=\lambda\frac{e^{-\epsilon |p_\perp|}}{\epsilon}\,, \label{eqn-Fer-Selfenergy01} \\ 
\Sigma_+(p)=& + 4\pi \lambda \int \frac{\dd^3 q}{(2\pi)^3}\frac{\Sigma_0(q)}{q^2} \frac{1}{(p-q)^+}e^{\ii \epsilon q_3}\, . \label{eqn-Fer-Selfenergy02}
\end{align}
The equations \eqref{eqn-Fer-Selfenergy01} and \eqref{eqn-Fer-Selfenergy02}, together with \eqref{Sigmaguess} completely fixes the exact propagator. In particular, $M_{c.t}= -\lambda/\epsilon$.

To summarize, the exact propagator \eqref{eq:exPropFer} takes the form
\beq\la{sigmatodelta}
\Delta(p) =\frac{-\ii \gamma^\mu p_\mu - \ii \Sigma_+(p) \gamma^+  + \Sigma_{0}(p) \mathbb{I}}{p^2}e^{\ii \epsilon p_3} \,.
\eeq
In the matrix basis,
\beq \label{propdecomp}
\Delta(q)=\Delta_\mu(q)\gamma^\mu+\Delta_0(p){\mathbb I}\,,
\eeq
we have
\beq \label{eq:exactprop}
\begin{aligned}
\Delta_3(q) & = \frac{e^{\ii q_3 \epsilon}}{q^2} (-\ii q_3)\,,\\
\Delta_-(p) & = \frac{e^{\ii q_3 \epsilon}}{q^2} (-\ii q_-)\,, \\
\Delta_+(p) & 
= \frac{e^{\ii q_3 \epsilon}}{q^2} (-\ii q_+) \Big[1  -\lambda^2 \frac{(1-e^{-\epsilon |q_\perp|})^2}{\epsilon^2 |q_\perp|^2}\Big] = \frac{e^{\ii q_3 \epsilon}}{q^2} (-\ii q_+) (1-\lambda^2) + \cO(\epsilon)\,,\\
\Delta_0(q) & = \frac{e^{\ii q_3 \epsilon}}{q^2} \times \lambda \frac{e^{-\epsilon |q_\perp|}-1}{\epsilon}\qquad\qquad\qquad = \frac{e^{\ii q_3 \epsilon}}{q^2} \times (-\lambda |q_\perp|)  + \mathcal{O}(\epsilon)\,.
\end{aligned}
\eeq
For the computations in this section, the $\cO(\epsilon)$ expansion above 
would be sufficient.

Note that the exact propagator satisfies the following equation
\begin{equation}\label{eq:propExactProp}
\Delta_0^2-\Delta_\mu\Delta^\mu=(\Delta_0+\Delta_3) (\Delta_0-\Delta_3)- 2\Delta_+ \Delta_-=\frac{e^{2\ii q_3 \epsilon}}{q^2}\,,
\end{equation}
which is \textit{exact} in $\epsilon$. The right-hand side is a contact term that we will keep track of, but will not contribute to the line integration. This equation can be viewed as the compatibility of our regularization scheme with the equation of motion. We will use it in section \ref{sec:condFer} for deriving the quantum version of the boundary equation. By plugging it into the square of (\ref{sigmatodelta}), we see that (\ref{Sigmaguess}) is indeed satisfied.

\subsection{All Loop Resummation} \label{sec:ferAllLoop}

Consider first the operator
\beq \label{Mf}
\big(M^f\big)_{ab}=\bar\psi_b\, W[x(\cdot)]\psi_a\,.
\eeq
where $a,b$ are spinor indices. We denote the spin $\pm{1\over2}$ component by $a,b=\pm$ correspondingly. In the basis where $\gamma^\mu=\sigma^\mu$ and the line is along the third direction, we have $\psi=(\psi_+,\psi_-)^\intercal$ and $\bar\psi=(\bar\psi_-,\bar\psi_+)$. In this basis, any matrix $M$ which acts on the spinor polarizations takes the form $M = \begin{pmatrix} M_{+-} & M_{++} \\ M_{--} & M_{-+} \\ \end{pmatrix}.$ We take the path $x(\cdot)$ to be a straight line between $x^3=0$ and $x^3=x>0$. Due to the rotation symmetry of the straight line, the only non-zero components of (\ref{Mf}) are $(M^f)_{+-}$ and $(M^f)_{-+}$. As in the bosonic case, we introduce the $L$-point line integrand as
\beq\label{eqn-higherloop-rel-fer}
\<M^f\>=\sum_L\cM_L^f\,,\quad\text{with}\quad {\cal M}^{f}_{L}=\int\limits_0^1 \mathcal{P}\prod_{j=1}^L\dd t_j\,\mathtt{F}^{(L)}(\{t_j+j\tilde\epsilon\}_{j=1}^L;t_{L+1}+(L+1)\tilde{\epsilon})
\, .
\eeq
We call it $L$-point integrand because the exact fermionic propagator (\ref{eq:exPropFer}) receives loop corrections, and thus the number $L$ of gluon insertions on the line is no longer equal to the loop order. Note that ${\cal M}^{f}_{L}$ and $\mathtt{F}^{(L)}$ are $2$ by $2$ matrices. The integrand is again defined with a shift of $t_j$ by $j\te$ as explained below (\ref{eqn-higherloop-rel}), and the last variable $t_{L+1}$ is a free parameter that will be set to its face value at the end. However, unlike the bosonic integrand $\mathtt{B}^{(L)}$, not all dependence on of $\mathtt{F}^{(L)}$ on $\tilde{\epsilon}$ is explicit in the these shifts. Namely, the shift accounts for the factor of $e^{\ii p_3\tilde\epsilon}$ in the numerator of the fermion propagator (\ref{eq:exactprop}), 
but not the more detailed dependence on $\te$. 
Yet, we will find that this dependence is of order $O(\tilde{\epsilon})$. It will factor out of the $t$-integrals, and will not affect the final result. 
 
\begin{figure}[t]
\centering
\includegraphics[width=0.7\textwidth]{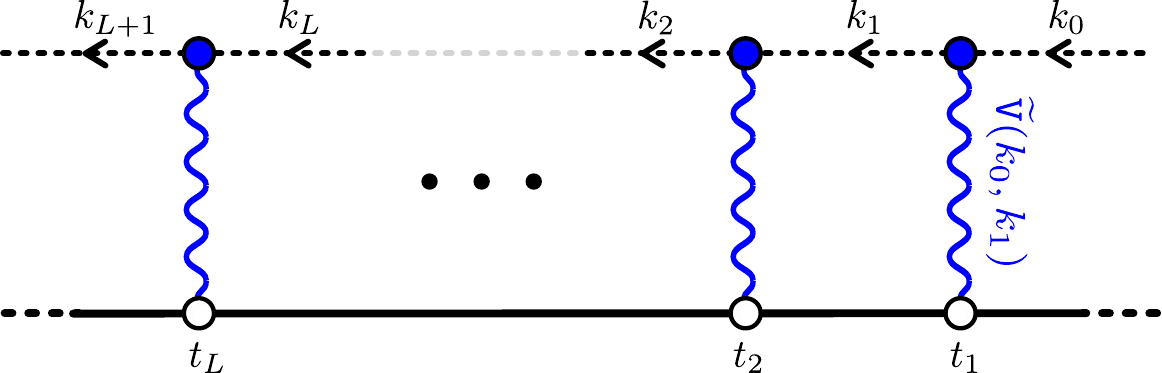}
\caption{\small Ladders of gluon exchange diagrams between the Wilson line and the fermion line. Here, the dashed lines are exact fermion propagators and the solid black line represents the Wilson line. These are the only diagrams that remains after resuming the loop corrections to the fermion propagator and the gluon exchange on the Wilson line.}
\label{vprop_fer}
\end{figure}

The only diagrams that contribute to $\<M^f\>$, other than the gluon exchange on the line that we have considered in section \ref{sec:anomspinf}, are ladders of gluon exchange diagrams between the Wilson line and the fermion line, see figure \ref{vprop_fer}.
As in the bosonic case, the corresponding line integrand takes the form 
\begin{align}\label{higherloopFer0}
\mathtt{F}^{(L)}(\{t_j\}_{j=1}^L;t_{L+1})=&\,  (-2\pi\ii \lambda x)^L\int \prod_{j=0}^{L}\frac{\dd^3 k_j}{(2\pi)^3} e^{\ii k_j^3\, x\, t_{j+1\, j}} \\
& \times \widetilde\Delta(k_0) \vvv(k_0,k_1)\widetilde\Delta(k_1)\vvv(k_1,k_2) \cdots\widetilde\Delta(k_{L-1}) \vvv(k_{L-1},k_L)\widetilde\Delta(k_L)\,,\nn
\end{align}
where the effective interaction vertex is the matrix
\beq\la{Vfer}
\vvv(k_j,k_{j+1})=\frac{\gamma^+}{(k_j-k_{j+1})^+}\,,
\eeq
and 
\beq\la{tildeDelta}
\widetilde\Delta(p)=\Delta(p)\,e^{-\ii \epsilon p_3} =\frac{-\ii \gamma^\mu p_\mu - \ii \Sigma_+(p) \gamma^+  + \Sigma_{0}(p) \mathbb{I}}{p^2}\,.
\eeq
We now show that the fermion integrand (\ref{higherloopFer0}) satisfies the same recursion relation as the bosonic integrand, (\ref{eqn-Rec-Univ}). 

First, we note that $\vvv\propto\gamma^+$ and therefore the matrix product structure in (\ref{higherloopFer0}) simplifies drastically. When inserting the decomposition of the exact fermion propagator (\ref{propdecomp}), (\ref{eq:exactprop}) into (\ref{higherloopFer0}), we have 
\beq
\gamma^+\widetilde\Delta(k) \gamma^+ = \widetilde\Delta_-(k) \gamma^+ \gamma^- \gamma^+ = \frac{-\ii k^+}{k^2} \gamma^+ \gamma^- \gamma^+ \, . 
\eeq
Consequently, 
\beq
\gamma^+\widetilde\Delta(k_1) \gamma^+\widetilde\Delta(k_2) \gamma^+ \cdots \gamma^+\widetilde\Delta(k_{L-1}) \gamma^+ = (\gamma^+ \gamma^-)^{L-1} \gamma^+ \prod_{j=1}^{L-1}\widetilde\Delta_-(k_j) = 2^{L-1} \gamma^+ \prod_{j=1}^{L-1} \frac{-\ii k^+_j}{k^2_j} \, .\nn
\eeq
When combining this expression with the denominator of the effective vertex (\ref{Vfer}), we see that the line integrand becomes almost identical to the bosonic one. Explicitly, we have
\begin{align} \la{preintegrand-fer}
\mathtt{F}^{(L)}(\{t_j\}_{j=1}^L;t_{L+1})=& \int{\dd^3 k_L\over(2\pi)^3}{\dd^3 k_0\over(2\pi)^3}\,\Phi^{(L)}_{\alpha=-1}(\{t_j\}_{j=1}^L; t_{L+1}| k_0,k_L)
\times\widetilde\Delta(k_0) \,{\gamma^+\over k_L^+}\widetilde\Delta(k_L)\\
=&\int{\dd^3 k_L\over(2\pi)^3}{\dd^3 k_0\over(2\pi)^3}\,\Phi^{(L)}_{\alpha=+1}(\{t_j\}_{j=1}^L; t_{L+1}| k_0,k_L)
\times\widetilde\Delta(k_0) \,{\gamma^+\over k_0^+}\widetilde\Delta(k_L)\,,\nn
\end{align}
where $\Phi^{(L)}_{\alpha}$ is the bosonic pre-integrand (\ref{eqn-IntegrandNUniVersal}). Because the bosonic integrand satisfied the recursion relation (\ref{eqn-Rec-Univ}), so does the fermionic integrand $\mathtt{F}^{(L)}$.

\subsection*{Recursion Seed and All Orders Solution}
To solve the recursion relation we need to evaluate the recursion seed, $\mathtt{F}^{(0)}(t_1)$ (which is the exact propagator in position space), and $\mathtt{F}^{(1)}(t_1,t_2)$. Using the techniques described in the section \ref{bossec} (see subsection \ref{sec:1loop} for example) and the form of the fermion propagator (\ref{eq:exactprop}), we find that
\beq
\label{eq:F0}
\big(\mathtt{F}^{(0)}(t)\big)_{-+} = -\frac{1}{4 \pi x^2\,t^2}\(1+{\lambda\,t\over t+\tilde{\epsilon}}\)\,,\qquad \big(\mathtt{F}^{(0)}(t)\big)_{+-}= \frac{1}{4 \pi x^2\,t^2}\(1-{\lambda\,t\over t+\tilde{\epsilon}}\)\,.
\eeq
Similarly, we have
\begin{align}
\big(\mathtt{F}^{(1)}(t_1;t_2)\big)_{-+} =-{\lambda\over t_1}\big(\mathtt{F}^{(0)}(t_2)\big)_{-+}\,,\qquad \big(\mathtt{F}^{(1)}(t_1;t_2)\big)_{+-} ={\lambda\over t_{21}}\big(\mathtt{F}^{(0)}(t_2)\big)_{+-}\,.
\end{align}
These are precisely the relations satisfied by the scalar integrand $\mathtt{B}^{(1)}_\alpha$ in (\ref{eqn-Rec-SeedBos00}) for $\alpha=1$ and $\alpha=-1$ correspondingly. The solution of the recursion is therefor also the same and is given by (\ref{eqn-Rec-solSca00}) with $\mathtt{B}^{(L)}\to\mathtt{F}^{(L)}$. In these expressions, the function $\mathtt{F}^{(0)}(t_{L+1})$ only appears as an overall factor, 
with $t_{L+1} = 1+ (L+1)\tilde{\epsilon}$. Because the $t_j$ integrals only produce logarithmic divergences, we can simply set $t_{L+1} = 1$, (see appendix \ref{apd:diffEqnPert} for more details). Therefore, up to an overall factor, the result is the same as in the bosonic case and is given by
\beq \label{fwfrf}
\begin{split}
    \<M^{({1\over2},-{1\over2})}\>
&=-{1\over4\pi x^2}\Big({\epsilon\, e^{\gamma_E}\over x}\Big)^\lambda\frac{1+\lambda}{\Gamma (1-\lambda)}+\cO(\te)\,,\\
\<M^{(-{1\over2},{1\over2})}\>
&= + {1\over4\pi x^2}\Big({\epsilon\, e^{\gamma_E}\over x}\Big)^{-\lambda}\frac{1-\lambda}{\Gamma (1+\lambda)} + \cO(\te)\,.
\end{split}
\eeq

\subsection{Operators with Derivatives}

At tree level, a complete basis of boundary operators is given in (\ref{eq:opsWeConsider_fer}). These operators are uniquely classified by their tree-level dimension and spin. When the interaction is turned on, they receive anomalous dimension, anomalous spin, and wave function renormalization. We now generalize our computation of these corrections for $s=\pm{1\over2}$ to the rest of the boundary operators, with $|s|>{1\over2}$. The anomalous spin, considered in section \ref{sec:anomspinf}, is independent of the boundary operator and therefore reminds the same, (\ref{spins}).

As in the bosonic case (\ref{eq:opsWeConsider}), we keep a simple labeling of the $SL(2,{\mathbb R})$ descendant operators, with $n>0$, by the number of longitudinal path derivatives ($\delta_{x_3}$) acting on the $n=0$ primaries, (\ref{longiDer}). Hence, in what follows it is sufficient to consider the operators with $n=0$. 

Recall that the $SL(2,{\mathbb R})\times U(1)$ symmetry of the straight line fixes the expectation value of the straight mesonic line operators to take the form (\ref{conformalM3}). The only difference is that in the fermionic case $s_R=-s_L\equiv s$ takes half-integer values. We denote the corresponding line integrand (\ref{eqn-higherloop-rel-fer}) by $\mathtt{F}^{(L,s)}$. In this convention 
the operators without boundary derivatives are $\mathtt{F}^{(L)}_{-+}=\mathtt{F}^{(L,{1\over2})}$ and $\mathtt{F}^{(L)}_{+-}=\mathtt{F}^{(L,-{1\over2})}$.

\subsubsection*{The Case $s_R>0$}
For $s=s_R>0$, the left and right boundary operators are,
\beq
\mathcal{O}_L^{(0,-s)} = \d_-^{s-1/2} \bar{\psi}_-(x_1)\,, \qquad \mathcal{O}_R^{(0,s)} = 
D_+^{s-1/2}\psi_+(x_0)\, .
\eeq
As in the bosonic case, the gluon in the covariant derivative has to be absorbed into the matter propagator before the first ladder. Hence, the form of the $L$-point integrand remands the same with only a modification of the first and the last fermion propagators
\beq\la{Fwithd}
\mathtt{F}^{(L,s)}
=\int{\dd^3 k_L\over(2\pi)^3}{\dd^3 k_0\over(2\pi)^3}\ \,\Phi^{(L)}_{-1}(\{t_j\}_{j=1}^L; t_{L+1}| k_0,k_L)\[{\mathbf{D}_+^{(s)}(k_0)\over k_0^2}{\gamma^+\over k_L^+}\widetilde\Delta(k_L)(\ii k_L^+)^{s-{1\over2}}\]_{+-}\,.
\eeq
Here, the components of the modified propagator $\mathbf{D}_+^{(s)}$ are polynomials of maximal degree $s+1/2$ in $k^-$ and $1/k^+$, and are analytic in $k_3$. Their exact form will not be relevant for the computation of anomalous dimensions. 

It follows that the integrand $\mathtt{F}^{(L,s)}$ satisfies the same recursion relation (\ref{eqn-Rec-Univ}). To construct the relevant solution we then have to compute the seed. We have 
\begin{align}\la{F1ls>0}
\mathtt{F}^{(0,s)}_{+-}(t_{1}) =&\int{\dd^3 k\over(2\pi)^3}\,e^{\ii k^3 x\, t_1 } \[{\mathbf{D}_+^{(s)}(k)\over k^2}{(\ii k^+)^{s-1/2}}\]_{+-}\,,\\
\mathtt{F}^{(1,s)}_{+-}(t_1;t_{2}) = &\int{\dd^3 k_1\over(2\pi)^3}{\dd^3 k_0\over(2\pi)^3}\ \,\Phi^{(1)}_{-1}(t_1; t_{2}| k_0,k_1)\[{\mathbf{D}_+^{(s)}(k_0)\over k_0^2}{\gamma^+\over k_1^+}\widetilde\Delta(k_1)(\ii k^+)^{s-{1\over2}}\]_{+-}\,.\nn
\end{align}
Using the matrix identity
\beq
\Big[\mathbf{D}_+^{(s)}(k) \gamma^+ \widetilde\Delta(k_1)\Big]_{+-} = \sqrt{2} \Big[\mathbf{D}_+^{(s)}(k) \Big]_{+-} \Big[\widetilde\Delta(k_1)\Big]_{--}
\eeq
and \eqref{eq:genIntOneSide} to integrate $k_1$ we find that the resulting $k_0$ integral is identical to $\mathtt{F}^{(0,s)}(t_2)$ up to an overall factor,
\beq\la{F1ms}
\mathtt{F}_{+-}^{(1,s)}(t_1;t_{2})= \frac{\lambda}{t_{2\, 1}} \mathtt{F}_{+-}^{(0,s)}(t_{2})\,.
\eeq
This is the same relation as for $\mathtt{B}_{-1}$ in (\ref{eqn-Rec-SeedBos00}). The anomalous dimension is therefore the same and is given by
\beq\la{Deltasg0}
\Delta^{(0,s)}_R=\Delta^{(0,-s)}_L=\Big({1\over2}+s\Big)-{\lambda\over2}\,,\qquad s>0\,.
\eeq

\subsubsection*{The Case $s_R<0$}
For $s_R=-s<0$, the two boundary operators are
\beq
\mathcal{O}_L^{(0,s)} = D_+^{s-1/2} \bar{\psi}_+(x_1)\,, \quad \text{and} \quad \mathcal{O}_R^{(0,-s)} =
\partial_-^{s-1/2}\psi_-(x_0)\, .
\eeq
In this case, the $L$-point line integrand takes the form
\beq\la{Fwithdb}
\mathtt{F}^{(L,-s)}_{-+}
=\int{\dd^3 k_L\over(2\pi)^3}{\dd^3 k_0\over(2\pi)^3}\ \,\Phi^{(L)}_{-1}(\{t_j\}_{j=1}^L; t_{L+1}| k_0,k_L)\[(\ii k_0^+)^{s-{1\over2}}\widetilde\Delta(k_0){\gamma^+\over k_L^+}{\overline{\mathbf{D}}_+^{(s)}(k_0)\over k_0^2}\]_{-+}\,.
\eeq
It, therefore, satisfies the same recursion relation as before. By repeating similar steps to the positive right spin case, we get
\beq\la{F1mms}
\mathtt{F}^{(1,-s)}_{-+}(t_1;t_2)= \frac{\lambda}{t_1} \mathtt{F}_{-+}^{(0,-s)}(t_2)\,.
\eeq
This is the same relation as for $\mathtt{B}_{-1}$ in (\ref{eqn-Rec-SeedBos00}). Hence, 
\beq\la{Deltasl0}
\Delta^{(0,-s)}_R=\Delta^{(0,s)}_L=\Big(s+{1\over2}\Big)+{\lambda\over2}\,,\qquad s>0\,.
\eeq

\subsubsection*{Wave Function Normalization Factor}

\begin{figure}[t]
\centering
\includegraphics[width=0.85\textwidth]{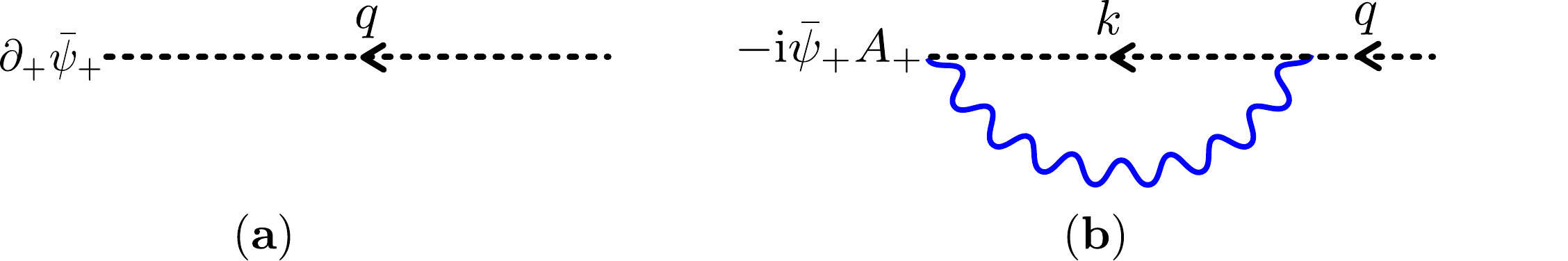}
\caption{The two diagrams that contribute to the modified propagator $\overline{\mathbf{D}}_+^{(1)}(q)$.}
\label{fig:ferderseed}
\end{figure}

As for the low spin boundary operators (\ref{fwfrf}), we can systematically compute the wave function renormalization factor for operators with $|s|>1$. This is not needed for the study of the Wilson line operator in this section. For the next section, however, we will need the wave function renormalization factor of the operator $M^{({3\over2},-{3\over2})}$, (\ref{Mfer}). 
To compute its expectation value, we need the form of the modified propagator $\overline{\mathbf{D}}_+^{({3\over2})}$, or equivalently, the seed $\mathtt{F}^{(0,-{3\over2})}_{-+}(t_1)$ in (\ref{Fwithdb}). 

There are two diagrams contributing to the vertex $\[\overline{\mathbf{D}}_+^{(s)}(q)\]_{-+}$, see figure \ref{fig:ferderseed}. 
The diagram in figure \ref{fig:ferderseed}.a has the derivative of the fermion $\d_+\bar{\psi}_+$ at the left end while the diagram in \ref{fig:ferderseed}.b has a gluon contraction between the fermion propagator and the covariant derivative, $\bar{\psi}_+A_+$. Their contribution to $\mathtt{F}^{(0,-{3\over2})}_{-+}(t_1)$ read
\begin{align}
\mathtt{Figure}\, \ref{fig:ferderseed}.a =& \qquad\int \frac{\dd^3 q}{(2\pi)^3} {e^{\ii q^3 x\, t_1}}\widetilde\Delta_{-+}(q) q^+ q^- =-\frac{3(1+\lambda)}{4 \pi (x\,t_1)^4} + \cO(\te)\,,\la{eqn-ferDseed-a}\\
\mathtt{Figure}\, \ref{fig:ferderseed}.b=& 4\pi \lambda\int \frac{\dd^3 q}{(2\pi)^3} \frac{\dd^3 k}{(2\pi)^3} {e^{\ii q^3 x\, t_1}}  \frac{[\Delta(q)\gamma^3\widetilde\Delta(k)]_{-+} }{(q-k)^+}  q^+=-{\lambda(\lambda+1)(5+\lambda)\over8\pi (x t_1)^4}+\cO(\te)\,,\nn
\end{align}
where we used (\ref{eq:genIntOneSide}) to perform the integrations. By summing these two contributions we get
\beq
\mathtt{F}^{(0,-{3\over2})}_{-+}(t_1) = -\frac{1}{2}\frac{(1+\lambda)(2+\lambda)(3+\lambda)}{4 \pi(t_1x)^4} +\cO(\te)\,.
\eeq
Through the recursion relations (\ref{F1ms}) and (\ref{eqn-Rec-Univ}), the factor of $1/t_1^4$ in $\mathtt{F}^{(0,{3\over2})}_{-+}$ result in a factor of $1/t_{L+1}^4$ for $\mathtt{F}^{(L,{3\over2})}_{-+}$. Because $t_{L+1}=1+\cO(\te)$ is not being integrated, that factor does not lead to power divergences. We can therefore safely set $t_{L+1}=1$ at the end to obtain
\beq\la{M1m1prefactor_fer}
\<M^{({3\over2},-{3\over2})}\> =-{1\over4\pi x^4}\(\frac{\epsilon\,e^{\gamma_E}}{x}\)^{\lambda}{(1+\lambda)(2+\lambda)(3+\lambda)\over2\Gamma (1-\lambda)}\,.
\eeq

\subsection{Boundary and Evolution Equations}
\label{sec:ferevoandbdry}

The relation between the 't Hooft coupling of the fermionic and bosonic theories reads \cite{Aharony:2012nh}
\beq\label{bosfermap}
\lambda_\text{bos}=1+\lambda_\text{fer}\,,\qquad\text{for}\qquad k_\text{bos}>0\,.
\eeq
Under this map, both the transverse spins (\ref{spins}) and the conformal dimensions (\ref{Deltasg0}), (\ref{Deltasl0}) of the boundary operators match those of the $\alpha=1$ line operator of the bosonic theory, see (\ref{Dbossg0}) and (\ref{Dbossl0}). In particular, the four bottom operators map as
\beq\label{tab_comp}
\begin{tabular}{llllcc} \toprule
{Fermionic} & {Tree}& {Bosonic}  & {Tree}&$\mathfrak{s}$  & $\Delta$\\ 
\midrule\midrule
$\ \cO_R^{(0,-{1\over2})} \quad$ & $\ \psi_-\qquad$ & $\ \cO_R^{(0,0)}\quad$  & $\ \phi$&$-{\lambda_\text{bos}\over2}$ & $\frac{1+\lambda_\text{bos}}{2}$\\
\midrule
$\ \cO_R^{(0,{1\over2})} \quad$ & $\ \psi_+\qquad$ & $\ \cO_R^{(0,1)}\quad$  & $\ \d_+ \phi\quad$&${2-\lambda_\text{bos}\over2}$ & $\frac{3-\lambda_\text{bos}}{2}$\\
\toprule
$\ \cO_L^{(0,{1\over2})} \quad$ & $\ \bar\psi_+\qquad$ & $\ \cO_L^{(0,0)}\quad$  & $\ \phi^\dagger\quad$&${\lambda_\text{bos}\over2}$& $\frac{1+\lambda_\text{bos}}{2}$ \\
\midrule
$\ \cO_L^{(0,-{1\over2})} \quad$ & $\ \bar\psi_-\qquad$ & $\ \cO_L^{(0,-1)}\quad$  & $\ \d_-\phi^\dagger\quad$&${\lambda_\text{bos}-2\over2}$ & $\frac{3-\lambda_\text{bos}}{2}$\\
\bottomrule\\
\end{tabular}
\eeq

Since the form of the boundary and evolution equation follow from this spectrum, they also agree. In terms of the fermionic labeling of the operators, the boundary equation reads
\beq\la{boundaryeqs}
\mathcal{O}_{L/R}^{(n+2,s)}\propto\left\{\begin{array}{ll}
\delta_-\mathcal{O}_{L/R}^{(n,s+1)}&\quad s\ge+{1\over2}\\
\delta_+\mathcal{O}_{L/R}^{(n,s-1)}&\quad s\le-{1\over2}
\end{array}\right.\,,
\eeq
and the displacement operator is 
\beq\la{Dis_fer}
{\mathbb D}_+\propto\cO_R^{(0,{1\over2})}\,\cO_L^{(0,{1\over2})}\,,\qquad
{\mathbb D}_-\propto\cO_R^{(0,-{1\over2})}\,\cO_L^{(0,-{1\over2})}\,.
\eeq
It remains to fix the relative coefficients. 

\subsubsection*{Evolution Equation Coefficient}
Consider the evolution equation first. Under a smooth deformation of the path $x(\cdot)\mapsto x(\cdot)+v(\cdot)$, the Wilson line operator (\ref{WLop}) transform as
\beq\la{displacementF_fer}
\delta\cW=\ii\int\! \dd s\,\dot x^\mu(s)\,v^\nu(s)\,\cP\[F_{\mu\nu}(s)\cW\]\,.
\eeq
The gauge field equation of motion reads
\beq\la{Aeom_fer}
(F_{\mu \nu})^i_{\ j} = -\frac{2\pi}{k} \epsilon_{\mu \nu \rho} \psi^i \gamma^\rho \bar \psi_j\,,
\eeq
where $i$ and $j$ are color indices. We can replace $F_{\mu\nu}$ in (\ref{displacementF_fer}) away from self-crossing points, which we generally avoid in our analysis. In this way, we arrive precisely at the displacement operator (\ref{Dis_fer}). The relative coefficient on the right-hand side is $-2\sqrt{2}\pi\lambda$. In terms of the notation (\ref{Mfer}) the corresponding evolution equation takes the form
\begin{align} \label{ee_fer}
\delta M_{10}^{(s_L,s_R)}[x(\cdot)]=&-2\sqrt{2}\pi\lambda\!\int\limits_0^1\! \dd s|\dot x_s|\!\[v^+_s M_{1s}^{(s_L,\frac{1}{2})}M_{t0}^{(\frac{1}{2},s_R)}\!+v^-_s M_{1s}^{(s_L,-\frac{1}{2})}M_{t0}^{(-\frac{1}{2},s_R)}\]\\
&+\[\text{boundary terms}\]\,.\nn
\end{align}

\subsubsection*{Boundary Equation Coefficient}

It remains to fix the relative coefficient in the boundary equation (\ref{boundaryeqs}). Working in the normalization implied by (\ref{eq:opsWeConsider_fer}), we now show that it is equal to $-2$. To evaluate it, we consider a straight line and place the operator $\cO_R^{(0,s)}$ with $s\ge1/2$ at its right end. At the left end, we place either of the operators that appear in the boundary equation (\ref{boundaryeqs}) for $\cO_L$. 
The corresponding line integrand takes the form (\ref{Fwithd}) with the only difference being that the leftmost fermion propagator, $\widetilde\Delta(k_L)(\ii k_L^+)^{s-{1\over2}}$ in (\ref{Fwithd}), is modified to a new effective propagator that we denote by $\widetilde\Delta(k_L)\mathbf{V}_{\widehat\cO_L}(k_L)$.
Only the $(--)$ component of this modified propagator appears in the corresponding line integrand. This is because it is sandwiched between $\gamma^+$ and the minus polarization of $\bar\psi$, see (\ref{Fwithd}). 
Hence, to confirm that the relative coefficient in (\ref{boundaryeqs}) is indeed equal to $-2$, it is sufficient to check that
\beq\la{boudaryeqf}
	\[\Delta(q)\(\mathbf{V}_{\delta_+\cO_L^{(0,-s-1)}}(q)+2\mathbf{V}_{\cO_L^{(2,-s)}}(q)\)\]_{--}=0\,,
\eeq
where the equality is understood to hold under the integration, (\ref{Fwithd}). 

In terms of fields, we have
\begin{align}
	\delta_+\cO_L^{(0,-s-1)}&= \partial_+\big( \partial_-^{s+1/2} \bar{\psi}_-\big) + \ii \big( \partial_-^{s+1/2} \bar{\psi}_-\big) A_+ \, ,\label{dpdmOFer}\\
	\cO_L^{(2,-s)}&= \partial_3^2 \big( \partial_-^{s-1/2} \bar{\psi}_-\big) + \ii \partial_3 \big( \partial_-^{s-1/2} \bar{\psi}_- {A}_3\big) + \ii \partial_3 \big( \partial_-^{s-1/2} \bar{\psi}_- \big) {A}_3 + \ii^2 \big( \partial_-^{s-1/2} \bar{\psi}_- \big) {A}_3^2 \, .\nn
\end{align}
The two diagrams that contribute to $\delta_+\cO_L^{(0,-s-1)}$ are the same as in (the reflection of) figure \ref{fig:ferderseed}. The only differences are the additional $\d_-^s$ derivatives. They sum to 
\begin{align}\label{eqn-FerEOM-PM} 
\mathbf{V}_{\delta_+\cO_L^{(0,-s-1)}}(q) =&\, \ii q^-(\ii q^+)^{s+1/2}-2\pi \ii \lambda\int\!\! \frac{\dd^3 k}{(2\pi)^3}\frac{(\ii k^+)^{s+1/2}}{(q-k)^+}\gamma^3\Delta(k)\, . 
\end{align}

\begin{figure}[t]
	\centering
	\includegraphics[width=0.9\textwidth]{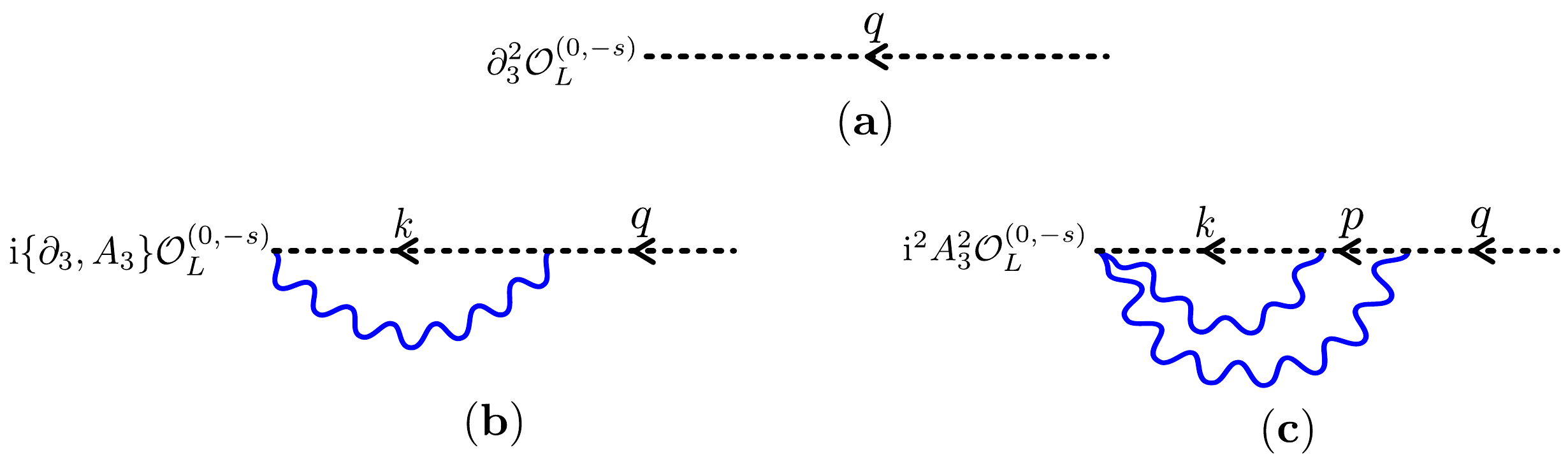}
	\caption{The three diagrams that contributes to the vertex $\mathbf{V}_{\delta_3^2 \cO_L^{(0,-s)}}(q)$. This vertex enters the computation of $\<\delta_3^2 \cO_L^{(0,-s)}\cW\cO_R^{(0,s)}\>$. 
 }\label{fig:dseedeom_fer}
\end{figure}

Similarly, the diagrams that contributes to the vertex $\mathbf{V}_{\delta_3^2 \cO_L^{(0,-s)}}$ are shown in figure \ref{fig:dseedeom_fer}. 
They give
\begin{align}\label{eqn-BdEOM-V33Fer}
&\mathbf{V}_{\delta_3^2 \cO_L^{(0,-s)}}(q)=-(q^3)^2 (\ii q^+)^{s-1/2} \\
&+2\pi\ii \lambda\int\!\! \frac{\dd^3 k}{(2\pi)^3}  {(\ii k^+)^{s-1/2}}  \(\frac{\ii(k+q)^3}{(q-k)^+}+2\pi\ii \lambda\int\! \frac{\dd^3 p}{(2\pi)^3} \frac{\gamma^+}{(q-p)^+}\frac{\Delta(p)}{(p-k)^+}\)\gamma^+\Delta(k)\,.\nn
\end{align}

As before, apart from the on-shell pole and the $e^{\ii\epsilon q^3}$ factor in $\Delta(q)$, the dependence on $q^3$ is polynomial. Hence, we can safely identify $\ii q^3$ with $-|q_\perp|$ under the integral. Upon this identification, the first term in \eqref{eqn-FerEOM-PM} explicitly cancels against the first term in \eqref{eqn-BdEOM-V33Fer}. To compare the rest we first perform the $p$-integration for the second line of (\ref{eqn-BdEOM-V33Fer}) to obtain
\beq\label{V33Fer}
-2\pi\ii \lambda\int\!\! \frac{\dd^3 k}{(2\pi)^3} \frac{{(\ii k^+)^{s-1/2}}}{(q-k)^+} \Big( |q_\perp| + |k_\perp| + \lambda \frac{e^{-\epsilon |q_\perp|}-e^{-\epsilon |k_\perp|}}{\epsilon} \Big)
\gamma^+\Delta(k)\, .
\eeq
We proceed by looking at the $(--)$ component in (\ref{boudaryeqf}). For \eqref{V33Fer}, the relevant matrix element is 
\beq
\Big[\Delta(q)\gamma^+ \Delta(k)\Big]_{--} = \sqrt{2} [\Delta(q)]_{--}[\Delta(k)]_{--}\,, \quad \text{with} \quad [\Delta(k)]_{--} = - \frac{\ii \sqrt{2}}{k^2} k^+ e^{\ii k^3 \epsilon}\, .
\eeq
Similarly, the relevant matrix element of \eqref{eqn-FerEOM-PM} is
\beq \label{eqn-FerEOM-PM-IntMedStep}
\Big[\Delta(q)\gamma^3 \Delta(k)\Big]_{--} = [\Delta(q)]_{--}[\Delta(k)]_{+-} - [\Delta(q)]_{-+}[\Delta(k)]_{--}\, .
\eeq
Akin to the above, we replace $\ii k^3\to-|k_\perp|$ under integration. After doing so, the $(+-)$ and $(-+)$ elements of the exact fermionic propagator \eqref{eq:exactprop} take the simple form
\beq
[\Delta(k)]_{\pm \mp} = \frac{e^{-|k_\perp|\epsilon}}{k^2} \Big(\lambda \frac{1-e^{-\epsilon|k_\perp|}}{\epsilon} \pm |k_\perp| \Big)\, .
\eeq

Lastly, the integrals over $k$ 
are of the form (\ref{eq:genIntOneSide}) or its $|k_{\perp}|$ derivatives. That means that we can identify 
$k^+$ with $q^+$ in the numerator of (\ref{eqn-FerEOM-PM}) and (\ref{V33Fer}). 
After this identification, the $|k_\perp|$ dependent
term in \eqref{V33Fer} exactly cancels against the $[\Delta(k)]_{+-}$ term in \eqref{eqn-FerEOM-PM-IntMedStep}. Similarly, the 
$|q_\perp|$ dependent term in \eqref{V33Fer} cancels against the $[\Delta(q)]_{-+}$ term in \eqref{eqn-FerEOM-PM-IntMedStep}. 
As a result, we conclude that
\beq \label{eqn-FerBEOM}
\cO_L^{(2,-s)}+2\delta_+\cO_L^{(0,-s-1)}=0\,,
\eeq
as an operator equation. Note that the individual integrals in (\ref{eqn-FerEOM-PM}) and (\ref{V33Fer}) develop a power divergence at $\epsilon\to0$, which is to be subtracted when constructing each of the two boundary operators individually. These however cancel out in the combination (\ref{beqb}). An analogous derivation holds for the rest of the boundary equations in (\ref{boundaryeqs}). 

\subsection{Summary}
\label{sec:sumfer}

To summarize, we see that under the map between the 't Hooft coupling of the bosonic and fermionic theories (\ref{bosfermap}) the spectrum of conformal dimension and the boundary spins match with those of the $\alpha=1$ line operator in the bosonic theory, see table (\ref{tab_comp}). It follows that the form of the evolution and boundary equations also match. In \cite{bootstrap} we show that these are sufficient to fix the expectation values of these mesonic line operators along an arbitrary smooth path. As we deform the path away from a straight line one, the first non-trivial and physical quantity we encounter is the two-point function of the displacement operator, $\Lambda(\Delta)$ in (\ref{distpf}). Using the results from the previous sections we can explicitly check that it indeed matches. From (\ref{ee_fer}) we have 
\beq\la{MtoLambda_fer}
\Lambda=(2 \sqrt{2} \pi\lambda)^2x^4\<M^{(\frac{1}{2},-\frac{1}{2})}\>\<M^{(-\frac{1}{2},\frac{1}{2})}\>\,.
\eeq
By plugging in values from (\ref{fwfrf}), we reproduce (\ref{fDD}) with $\Delta=1\pm\lambda_\text{fer}/2$, in agreement with the duality.

\section{The Condensed Fermion Operator} \label{sec:condFer}

For the duality between the bosonic and fermionic theories to hold, we should be able to identify a conformal line operator in the fermionic theory that is the dual of the $\alpha=-1$ operator of the bosonic theory. 

To understand what type of operator we are looking for, consider first the scalar $\phi$ at the right end of the $\alpha=1$ line operator. At tree level, it has dimension one-half and zero spin. At the maximal coupling, $\lambda_\text{bos}=1$, its dimension and transverse spin exactly match those of a free fermion, $\lambda_\text{fer}=0$, equal to one and minus one-half correspondingly, see table (\ref{tab_comp}). On the other hand, if we place it at the end of the $\alpha=-1$ line operator, then the dimension and transverse spin reach zero and minus one-half at $\lambda_\text{fer}=0$ correspondingly. This looks puzzling at first because, in the free fermionic theory, there is no field of zero dimension. The free fermion, however, has dimension one; therefore it makes sense to integrate it along the line. The corresponding endpoint of integration indeed has dimension zero. In order for the endpoint of the line to also have transverse spin equal to minus one-half, we should translate the spin of the integrated fermion to the end. This can be done by attaching it to a spin parallel transport along the line as
\beq\la{topspintrans}
\int\limits_0\dd s|\dot x_s|\psi_-(x_s)\quad\rightarrow\quad \int\limits_0\dd s|\dot x_s|\Big[\cP e^{\ii \int\limits_0^s\dd t\,\Gamma_\mu\dot x^\mu} \cdot
\psi(x_s) \Big]_-\,.
\eeq
The spin connection $\Gamma_\mu\dd x^\mu$ has to be both, reparametrization invariant and topological (shape-independent). It turns out that in order to have such a topological connection, one must introduce a framing vector $n(s)$ along the line. The construction of this unique connection is detailed in appendix \ref{sec:topoPT}. The result is
\beq\label{toptransport}
\Gamma_\mu\dot x^\mu= {1 \over2}\epsilon_{\mu\nu\rho}\(\dot{n}^\mu  n^\nu e^\rho\, e\cdot \gamma-e^\mu\dot e^\nu\gamma^\rho\) \, ,
\eeq
where we recall that $e^\mu = \dot{x}^\mu/|\dot{x}^\mu|$. Similarly, on the left end, we can have the conjugate fermion component $\bar\psi_+$ integrated at tree level with the same fermion parallel transport. 

At the loop level, we must attach a Wilson line to the fundamental fermion. As seen in the previous section, such a fermion attached to a Wilson line has a non-zero anomalous dimension. Therefore, it does not make sense to integrate it on a line. Instead, one can think of the fermion attached to the spin transport at one end and a Wilson line on the other as an analog of the scalar bi-linear in the bosonic theory. As in the bosonic case, the conformal line operator is obtained by adding it to the line action and tuning its coefficient to the conformal fixed point. In the case at hand, we have two such operators, $\psi_-$ attached to a topological spin transport from the left and a $\bar\psi_+$ attached to a topological spin transport from the right. To construct the line operator, we start from a direct product of the Wilson line and an empty line parameterized by the topological spin transport 
\beq \label{Mfc}
\widetilde M_0\equiv\(\begin{array}{cc}\bar\psi_+&0\\0&1\end{array}\)\cdot\mathcal{P}e^{\ii\int \dd x\cdot \widetilde A_0}\cdot \(\begin{array}{cc}\psi_-&0\\0&1\end{array}\)
\,,\qquad\text{with}\qquad(\widetilde A_0)_\mu=\Big(\begin{array}{cc}A_\mu & \\& \Gamma_\mu + \frac{\lambda}{2} \partial_\mu \log n^+\end{array}\Big)\,.
\eeq
Here, we recall that the lower diagonal component $\dd \log n^+$ is the local framing factor \eqref{ff}.
The upper left component of this $2\times2$ matrix is the mesonic line operator that we have considered in the previous section (\ref{Mf}), $M_{-+}=\(\widetilde M_0\)_{\uparrow \uparrow}$. We label its downright component as
\beq
\widetilde M^{({1\over2},-{1\over2})}_0\equiv\(\widetilde M_0\)_{\downarrow \downarrow}\,.
\eeq
The dependence of $\widetilde M_0$ on the framing vector is topological and is given by (\ref{fff}) with $\mathfrak{s}_R=-\mathfrak{s}_L=(1+\lambda)/2$. Namely, under a change in the framing vector $n(s)\to\tilde n(s)$, the line operator (\ref{Mfc}) transforms as
\beq\la{transn}
\widetilde M_0[\tilde n]=\({\tilde n_L^+\over n_L^+}\)^{-{1+\lambda\over2}}\({\tilde n_R^+\over n_R^+}\)^{1+\lambda\over2}\,\times\,\widetilde M_0[n]\,.
\eeq
This transformation property of the $\uparrow\uparrow$ component is the same as in (\ref{fff}). The fact that the topological spin transport, given by the $\downarrow\downarrow$ component, transforms in this way is proven in appendix \ref{sec:topoPT}.  

We then deform $\widetilde A_0$ in (\ref{Mfc}) as 
\beq\la{condencedfcon}
\widetilde A_0\ \rightarrow\ \widetilde A=\widetilde A_0+\ii\sqrt{4\pi b\over k}\(\begin{array}{cc}0&P_- \psi\\
\bar\psi P_-&0\end{array}\)
+\frac{f}{\epsilon}\(\begin{array}{cc}0&0\\0&P_-\end{array}\)\,,
\eeq
\begin{figure}[t]
    \centering
    \includegraphics[width=0.8\textwidth]{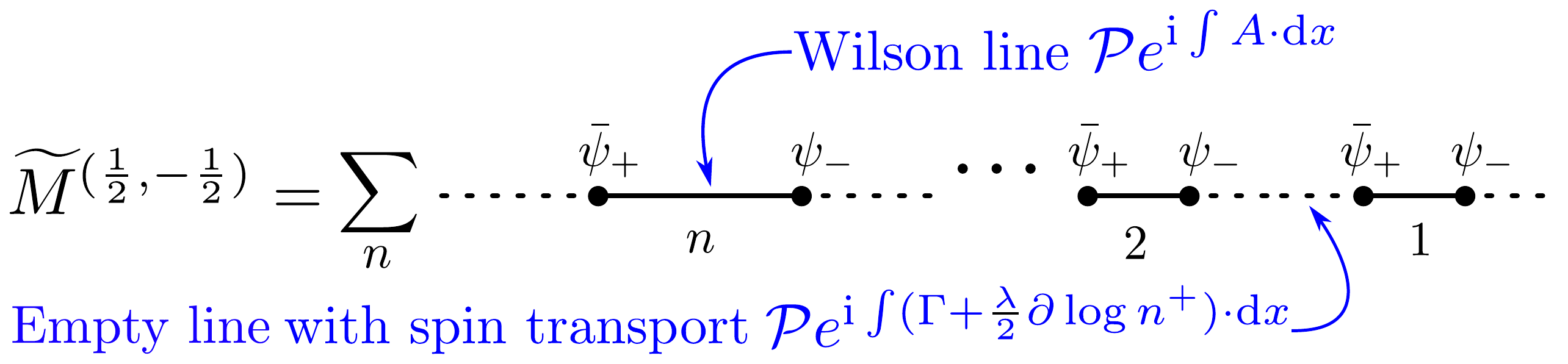}
    \caption{The condensed fermionic line operator is defined perturbatively by integrating fundamental and anti-fundamental fermions along the path, (\ref{condencedfcon}). Neighboring ordered pairs of fundamental and anti-fundamental are connected by Wilson lines. In between these pairs, we have a topological transport of the transverse spin with the connection (\ref{toptransport}). We call these segments ``\textit{empty}'' because they do not have fields inserted on them.}
    \label{fig:condensedFer}
    \end{figure}
where $P_{\pm}^\mu \equiv(e^\mu \pm \gamma^\mu)/2$ 
is a projector that projects the fermion to its local $\pm$ components with respect to the tangential direction of the line. Here, $b$ is the coefficient that we have to fix so that the corresponding line operator
\beq\la{condensedfer}
\widetilde M^{({1\over2},-{1\over2})}=\(\widetilde M\)_{\downarrow \downarrow}=\(\mathcal{P}e^{\ii\int \dd x\cdot \widetilde A}\)_{\downarrow \downarrow}\,,
\eeq
is conformal. 
The last term in (\ref{condencedfcon}), proportional to $f/\epsilon$, is a scheme dependent counterterm that is needed to cancel the power divergences that come about when two conjugate fermions come close together. Finally, this deformation does not affect the dependence on the framing vector and therefore the boundary spins remind the same and are equal to $\mathfrak{s}_R=-\mathfrak{s}_L=(1+\lambda)/2$. In figure \ref{fig:condensedFer} the form of the expansion of this operator is illustrated. The $n$'th term in this figure is given by $n$ pairs of conjugate fermions integrated along the line. We have Wilson lines along the filled parts between the fundamental and anti-fundamental fermions. Along the empty parts we have the topological spin $(1+\lambda)/2$ transport. 

In the following, we fix the coefficients $b$ and $f$ and repeat the all-loop resummation for the corresponding conformal operator. We find that it is indeed dual to the $\alpha=-1$ line operator in the bosonic theory. 

We have also considered the condensation of the positive spin component of fermions $\psi_+$. As for the $\psi_-$ case, we also find a non-trivial fixed point. However, it corresponds to a non-unitary conformal line operator. Hence, we will not present it here. 

\subsection{All-loop Resummation}
\begin{figure}[t]
    \centering
    \includegraphics[width=12cm]{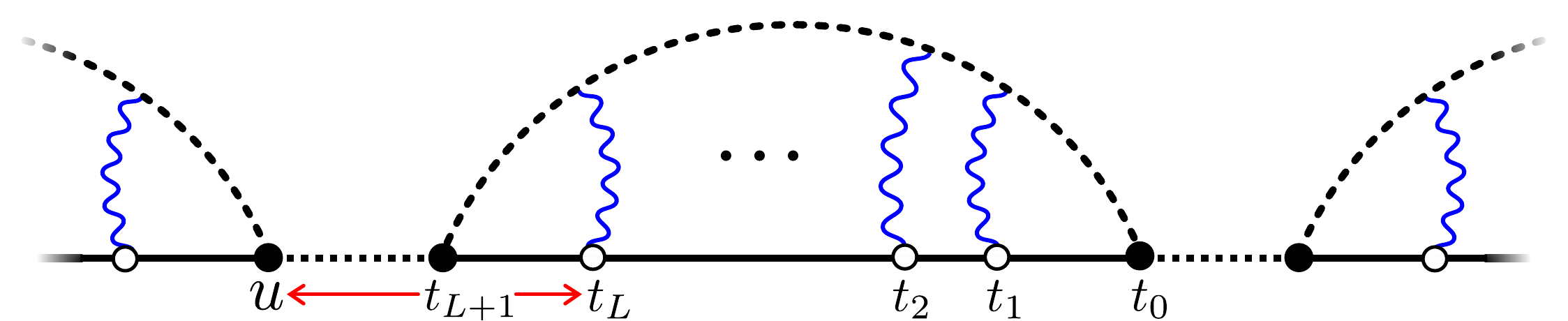}
    \caption{One segment of the $L$-ladder fermion-antifermion pair insertion. Once integrating the last endpoint $t_{L+1}$, we arrive at the same integrand as for the line operators in the bosonic theory with $\alpha=1-2b$.}
    \label{fig:condensed-Identity}
\end{figure}

As before, we take the path to be a straight line between $x^3=0$ and $x^3=x>0$. Any diagram that contributes to the expectation value has some ordered set of fermion anti-fermion pairs integrated along the line, together with their corresponding counter-terms, see figure \ref{fig:condensed-Identity}. 
By performing all bulk integration for each such pair, we associate to it the line integrand $\mathtt{F}^{(L)}_{-+}$ in (\ref{eqn-higherloop-rel-fer}). Recall that this line integrand satisfies the same recursion relation as the bosonic line integrand $\mathtt{B}^{(L)}_{\alpha=1}$ in (\ref{eqn-Rec-Univ}) and (\ref{eqn-Rec-SeedBos00}). Hence, as in (\ref{eqn-Rec-solSca00}), it takes the form
\begin{equation}\label{eq:FLexplicit}
\mathtt{F}^{(L)}(\{t_j\}_{j=1}^L;t_{L+1})= \mathtt{F}^{(0)}(t_1)\times \prod_{j=1}^{L}{-\lambda\over t_j}= -\frac{1}{4 \pi x^2}\(\frac{1}{t_{L+1}^2} + \frac{\lambda}{t_{L+1}(t_{L+1}+\te)}\) \prod_{j=1}^{L}{-\lambda\over t_j}\,,
\end{equation}
where we used the form of $\mathtt{F}^{(0)}(t)$ in (\ref{eq:F0}) and for compactness of the expression we have suppressed the shift of $t_j$ by $j\te-t_0$, see (\ref{eqn-higherloop-rel-fer}). To proceed, we open the parenthesis in (\ref{eq:FLexplicit}) and integrate only the first term (of order $\lambda^L$) in $t_{L+1}$ between $t_L$ and the insertion point of the $\psi_-$ in the next fermion pair that we denote by $u$, see figure \ref{fig:condensed-Identity},
\begin{equation}
    \int\limits_{t_L}^{u}\frac{\dd t_{L+1}}{(t_{L+1}-t_0+(L+1)\te)^2} = -\frac{1}{u-t_0+(L+1)\te}+\frac{1}{t_{L}-t_0+(L+1)\te}\,.
\end{equation}
For $L>0$ the second contribution precisely cancels the term of order $\lambda^L$ in $\mathtt{F}^{(L-1)}$ (the second term in the parenthesis of (\ref{eq:FLexplicit})).
For $L=0$, it gives a power divergence that we can tune the counter term to cancel, by setting $f=\lambda\, b$ in (\ref{condencedfcon}).

We remain with the contribution of the upper limit of integration. At this upper limit, there are no longer empty regions on the line. Instead, we have two types of points. One that corresponds to a gluon insertion
\beq \label{condensed_gluon_exchange}
\centering
\includegraphics[valign=c,width=60mm]{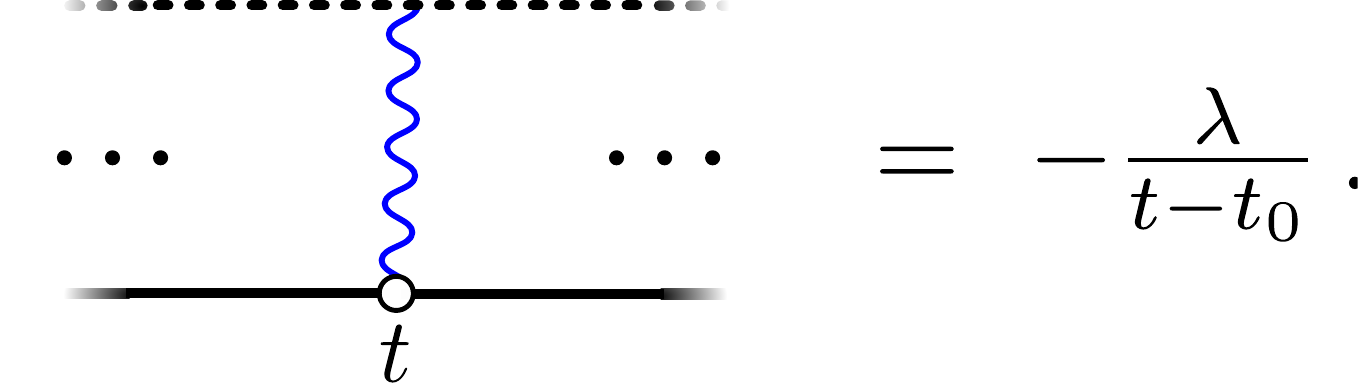}
\eeq
The other corresponds to the upper limit of a $\bar\psi$ integration
\beq \label{condensed_psib_endpoint}
\includegraphics[valign=c,width=60mm]{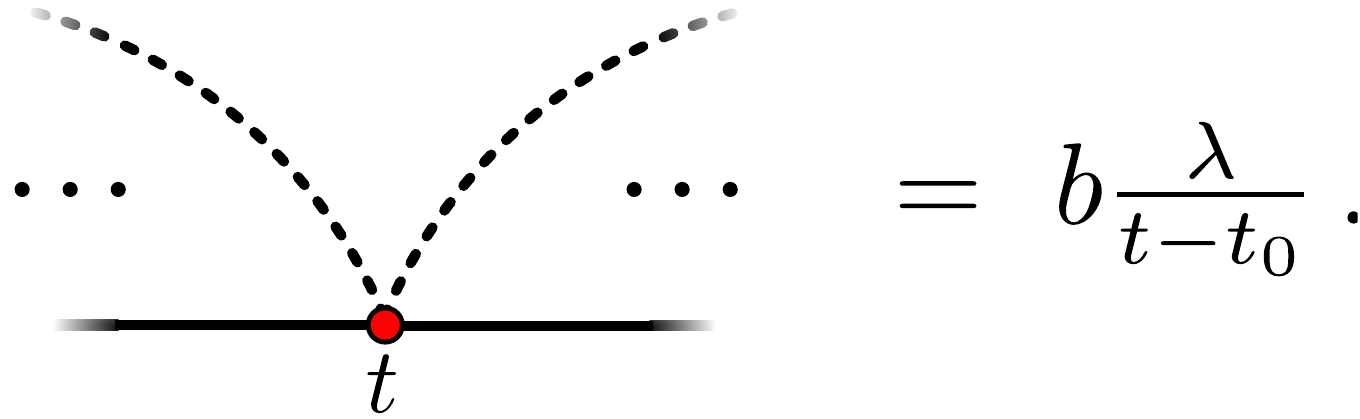}
\eeq
Here, $t_0$ is a point inside the line, were a $\psi_-$ is inserted. Upon integration over $t$, each of these two contributions leads to a bulk logarithmic divergence and corresponding bulk RG flow. To cancel these two divergences and have an RG fixed point, we must set $b=1$ (of course $b=0$ is also a trivial solution). The same also applies to all $t_i>t$, with either $t_0\to t$ for (\ref{condensed_psib_endpoint}) or that $t_0$ remands the same for (\ref{condensed_gluon_exchange}).

The perturbative expansion now takes the form
\beq\label{eqn-higherloop-rel-fer-cond}
    \<\widetilde{M}^{({1\over2},-{1\over2})}\>= \sum_{L=0}^\infty\widetilde\cM_L^f\,,\quad\text{with}\quad {\widetilde{\cal M}}^{f}_{L}=\int\limits_0^1 \mathcal{P}\prod_{j=1}^L\dd t_j\,\widetilde{\mathtt{F}}^{(L)}(\{t_j+j\tilde\epsilon\}_{j=1}^L;t_{L+1}+(L+1)\tilde{\epsilon})
\, .
\eeq
where the $L$ loop term is an $L$ points line integrand $\widetilde{\mathtt{F}}_b^{(L)}$, similar to (\ref{eqn-higherloop-rel}) and (\ref{eqn-higherloop-rel-fer}). Each integration point can be of the type (\ref{condensed_gluon_exchange}) or (\ref{condensed_psib_endpoint}), and we have to sum over all the corresponding possibilities. As before, the argument of the lines integrand appears with a shift of $t_j$ by $j\te$, so that $t_j$ is integrated exactly up to $t_{j+1}$, and $t_{L+1}=1+(L+1)\tilde{\epsilon}$, (see explanation below (\ref{eqn-higherloop-rel})). 

\begin{figure}[t]
    \centering
    \includegraphics[width=0.95\textwidth]{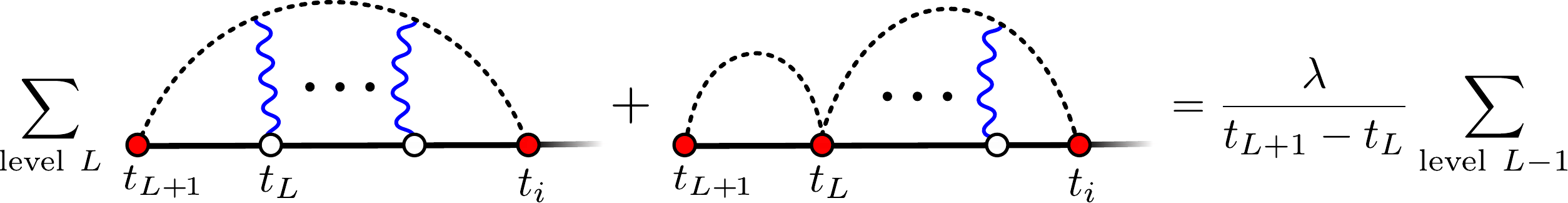}
    \caption{The line integrand $\widetilde F^{(L)}$ satisfies a simple recursion relation (\ref{fig:condesedPluckerId}). It is derived by summing up the two vertex possibilities at the last integration point $t_L$. The vertex can either be a gluon insertion (\ref{condensed_gluon_exchange}) or the $\bar{\psi}$ vertex insertion (\ref{condensed_psib_endpoint}). The sum of the two yields the one lower loop integrand $\widetilde F^{(L-1)}$, times $\lambda/(t_{L+1}-t_L)$.}
    
\label{fig:condesedPluckerId}
\end{figure}

By setting $b=1$ and iterating the 
relation 
\beq
{1\over(s-t)(t-u)}+{1\over(t-u)(u-s)}+{1\over(u-s)(s-t)}=0\,,
\eeq
from the left as, (see figure \ref{fig:condesedPluckerId}), 
\begin{align}
    \widetilde{\mathtt{F}}^{(L)}(\{t_{j}\}_{j=1}^L; t_{L+1})=&\sum_{{\rm level}\  L} \(\frac{\lambda}{(t_{L+1}-t_L)(t_L-t_i)} - \frac{\lambda}{(t_{L+1}-t_i)(t_L-t_i)}\) \times \dots\nn \\
    =&\sum_{{\rm level}\ L} \frac{\lambda}{(t_{L+1}-t_L)( t_{L+1}-t_i)}  \times \dots\\
    =&{\lambda\over t_{L+1}-t_L}\times\widetilde{\mathtt{F}}^{(L-1)}(\{t_{j}\}_{j=1}^{L-1}; t_{L+1})\,,\nn
\end{align}
we arrive at
\begin{equation}
    \widetilde{\mathtt{F}}^{(L)}(\{t_{j}\}_{j=1}^L; t_{L+1}) = \lambda^L \prod_{j=1}^{L} \frac{1}{t_{L+1}-t_j}\,. 
\end{equation}
This is precisely of the form (\ref{eqn-Rec-solSca00}) that we have already encountered in the bosonic case. We, therefore, conclude that\footnote{It is interesting to note that in our regularization scheme, for any value of $b$, the following relation holds
\beq
\<\delta_{x_0^3} \widetilde{M}_b^{({1\over2},-{1\over2})}\> = -4 \pi \lambda \<M\>|_{\alpha=1-2b}\,.
\eeq
This match along the RG flow is however only a technical coincidence.}
\begin{equation}
\label{eq:expValCondEmpty}
    \<\widetilde{M}^{({1\over2},-{1\over2})}\> = 4 \pi x \<M\>\big|_{\alpha=-1} + \cO(\te) = {1 \over \Gamma (1+\lambda)}\Big({\epsilon\,e^{\gamma_E}\over x}\Big)^{-\lambda} + \cO(\te)\,,
\end{equation}
where here $\lambda$ is the same parameter on both sides of the first equality, not to be confused with the relation between the couplings of the bosonic and fermionic theories (\ref{bosfermap}).

\subsection{The Four Towers of Operators}

We denote the empty boundary operators considered above in accordance with their tree-level spin and the number of longitudinal derivatives as
\beq\la{t11}
\widetilde\cO_L^{(0,{1\over2})}=\text{empty left}\,,\qquad\widetilde\cO_R^{(0,-{1\over2})}=\text{empty right}\,,\qquad\text{and}\qquad\widetilde\cO^{(n,s)}=\delta_3^n\widetilde\cO^{(0,s)}\,.
\eeq
In particular, by taking a single longitudinal path derivative we arrive at the operators
\beq\la{t12}
\widetilde\cO_L^{(1,{1\over2})}=
-\sqrt{{4\pi\over k}}\bar\psi_+ - \frac{\lambda}{\epsilon}\,\qquad\text{and}\qquad\widetilde\cO_R^{(1,-{1\over2})}=
\sqrt{{4\pi\over k}}\psi_- + \frac{\lambda}{\epsilon}\,.
\eeq
The anomalous dimension of these descendant operators is the same as that of (\ref{t11}), equal to $-\lambda/2$. It is opposite from the anomalous dimension of the same fermion components at the boundary of a standard Wilson line, $\cO^{(0,{1\over2})}_L=\bar\psi_+$ and $\cO^{(0,-{1\over2})}_R=\psi_-$ in (\ref{tab_comp}). Namely, the deformation of the Wilson line action by the fermion condensation in (\ref{condencedfcon}) has the effect of flipping their anomalous dimension. 
On the other hand, this deformation does not affect the anomalous dimension of all boundary operators other than $\widetilde\cO_L^{(n,{1\over2})}$ and $\widetilde\cO_R^{(n,-{1\over2})}$. That is because all these operators are orthogonal to the fermions in (\ref{condencedfcon}) on a straight line, and therefore are insensitive to them. 
Their anomalous dimension and anomalous spin remain the same as if they were placed at the boundary of the standard Wilson line and are given in (\ref{fff}), (\ref{Deltasg0}), and (\ref{Deltasl0}). Altogether, we have four towers of boundary operators that are built on top of the operators
\beq\la{fourbottomcf}
\{\widetilde\cO_R^{(0,-{1\over2})},\widetilde\cO_R^{(0,-{3\over2})}\}\,\qquad\text{and}\qquad\{\widetilde\cO_L^{(0,{3\over2})},\widetilde\cO_L^{(0,{1\over2})}\}\,.
\eeq
Operators in each tower are related to each other by path derivatives. They have the same anomalous dimension and anomalous spin. This spectrum of boundary operator match exactly the one of the $\alpha=-1$ line operator in the bosonic theory, provided that the 't Hooft couplings of the two descriptions are related as in (\ref{bosfermap}). In particular, the four bottom operators (\ref{fourbottomcf}) and (\ref{bottomm1}) map to each other as
\beq\label{tab_comp_cond}
\begin{tabular}{llllcc} \toprule
	{Fermionic} & {Tree}& {Bosonic}  & {Tree}&$\mathfrak{s}$  & $\Delta$\\ 
	\midrule\midrule
	$\ \widetilde\cO_R^{(0,-{1\over2})} \quad$ & $\ \rm{Empty}\qquad$ & $\ \cO_R^{(0,0)}\quad$  & $\ \phi$&$-{\lambda_\text{bos}\over2}$ & $\frac{1-\lambda_\text{bos}}{2}$\\
	\midrule
	$\ \widetilde\cO_R^{(0,-{3\over2})} \quad$ & $\ \d_-\psi_-\qquad$ & $\ \cO_R^{(0,-1)}\quad$  & $\ \d_- \phi\quad$&$-{2+\lambda_\text{bos}\over2}$ & $\frac{3+\lambda_\text{bos}}{2}$\\
	\toprule
	$\ \widetilde\cO_L^{(0,{1\over2})} \quad$ & \ \rm{Empty} & $\ \cO_L^{(0,0)}\quad$  & $\ \phi^\dagger\quad$&${\lambda_\text{bos}\over2}$& $\frac{1-\lambda_\text{bos}}{2}$ \\
	\midrule
	$\ \widetilde\cO_L^{(0,{3\over2})} \quad$ & $\ \d_+\bar\psi_+\qquad$ & $\ \cO_L^{(0,1)}\quad$  & $\ \d_+\phi^\dagger\quad$&${\lambda_\text{bos}+2\over2}$ & $\frac{3+\lambda_\text{bos}}{2}$\\
	\bottomrule
\end{tabular}\,.
\eeq

\subsection{The Evolution Equation}

As before, there is a unique spin one, dimension two primary displacement operator. It is given by
\beq\la{Dis_fer_cond}
{\mathbb D}_+\propto\widetilde\cO_R^{(0,-{1\over2})}\,\widetilde\cO_L^{(0,{3\over2})}\,,\qquad
{\mathbb D}_-\propto\widetilde\cO_R^{(0,-{3\over2})}\,\widetilde\cO_L^{(0,{1\over2})}\,.
\eeq
Using the identification in (\ref{tab_comp_cond}), this is precisely the form of the displacement operator of the line operator in the bosonic theory with $\alpha=-1$, (\ref{Dpmm1}). As in the computation of the evolution equations appearing in previous sections \ref{sec:bosevo} and \ref{sec:ferevoandbdry}, the dependence on the framing vector is topological and does not contribute.

To fix the relative coefficient in (\ref{Dis_fer_cond}) we now vary the condensed fermion operator explicitly. As before, this is a formal functional variation of the operator that ignores renormalization. We have four types of contributions, coming from the functional variation 
of $\Gamma$, $P_-$, $A$, $\psi$, and $\bar\psi$ in \eqref{condensedfer}. We consider each one of these in turn. Without loss of generality, we specify to a straight line. 

\paragraph{Variation of the Topological Phase} 
The topological phase $\int \dd \log n^+$ in the lower component of $\widetilde A$ (\ref{Mfc}) combines with the same factor for the Wilson line in the upper component (\ref{eqn-Bos-ASpinFactor}) to an overall topological phase factor (\ref{fff}). Hence, it does not contribute to a smooth internal variation of the path.

\paragraph{Variation of $\Gamma$} 
Along the empty parts of the line, we have the topological spin transport 
\beq
\label{eq:PT}
U(s,t)\equiv\cP \exp{\Big[\ii \int\limits^{x_s}_{x_t}\dd x^\mu\,\Gamma_\mu}\Big]\, ,
\eeq
It is topological by construction. Hence, its variation is a boundary term
\beq
\delta U(s,t)= V(s)U(s,t)-U(s,t)V(t)\,.
\eeq
As we show in  equation \eqref{eqn-apd-PT-var02}, for a straight line 
\beq
V(s)\propto(\dot{v}_s \times e_s)\cdot \gamma\,.
\eeq
This term is sandwich between $\bar\psi(s) P_-$ and the parallel transport of $P_-\psi(t)$. For a straight line however, ${\[(\dot{v}_s \times e_s)\cdot\gamma\]_{-+}}=0$, so this term does not contribute. In other words, the two fermion components in (\ref{condencedfcon}) have an opposite transverse spin and therefore cannot contract with a spin one variation.

\paragraph{The Variation of the Wilson Line} Second, we have the variation of the Wilson line. As we explained before (\ref{Aeom_fer}), it amounts to inserting the field strength,
\beq\la{WLvar}
    \ii\, \dot x^\nu v^\mu F_{\nu\mu}\quad\rightarrow\quad\frac{2\pi \ii}{k} \dot{x}^\nu v^\mu \epsilon_{\nu\mu \rho} \bar \psi \gamma^\rho \psi\,.
\eeq

\paragraph{The Variation of $P_-$} 

The third contribution is the variation of the projector 
\begin{equation}
    \delta_v\, \dot x\cdot P_- = \dot v \cdot \(e \One - \gamma\)/2\,.
\end{equation}
For a straight line, $\dot v\cdot e=0$, and therefore the identity component vanishes. 
We remain with
\beq
    \dot x\cdot\(\begin{array}{cc}0&\delta_{v}P_-\psi(s)\\\bar \psi(s)\delta_vP_-&0\end{array}\)=-\frac{1}{2}\sqrt{\frac{4\pi}{k}}\dot v_s^\mu\,\(\begin{array}{cc}0&-\gamma_\mu\,\psi(s)\\-\bar \psi(s)\, \gamma_\mu&0\end{array}\)\,.
\eeq
This operator is inserted on the line operator $\mathcal{P}\exp(\ii\int \dd x\cdot \widetilde A)$ and integrated over $s$. Next, we perform integration by parts to remove the derivative from $\dot{v}_s$. Using the relation (\ref{t12}) (without regularization), we arrive at
\beq\la{Pmvar}
\arraycolsep=6.4pt\def\arraystretch{1.5}
-v_s^\mu\,\(\!\!\begin{array}{cc}\frac{2\pi}{k}\bar{\psi}^a\[\gamma_\mu,\dot x_s\cdot P_-\]^b_a \psi_b&\sqrt{\pi\over k}\gamma_\mu \dot x_s \cdot D \psi\,\\
\sqrt{\pi\over k}\,\dot x_s \cdot D \bar\psi \gamma_\mu &0\end{array}\!\!\)\, .
\eeq
The commutator term evaluates to
\beq
\[v^\mu\gamma_\mu,\dot x_\nu P_-^\nu\] = \ii \dot{x}^\nu v^\mu \epsilon_{\nu \mu \rho} \gamma^{\rho}\,.
\eeq
It exactly cancels with the Wilson line variation (\ref{WLvar}). Note that for this cancellation to happen, it is crucial that $b=1$.

\paragraph{The Variation of $\psi$ and $\bar \psi$}  The fourth and final contribution is the variation of the fermions in (\ref{condencedfcon}),
\beq\la{eq:4thcont}
\(\begin{array}{cc}0&P_{-} \delta_v\psi\\ \delta_v\bar\psi P_-&0\end{array}\)=
\(\begin{array}{cc}0&P_{-} v \cdot D\psi\\
v \cdot D \bar\psi P_{-}&0\end{array}\)\,.
\eeq

Because of the spin of the condensated fermion, the following fermion equations of motion are not affected by the condensation
\begin{align} \label{eqn-EOMferCondensedUnaffected}
& D_3\psi_++\sqrt{2}D_+\psi_-=0\,, \quad D_3\bar\psi_-+\sqrt{2}D_-\bar\psi_+=0\,.
\end{align}
Using these and the relations
\beq
\gamma^-P_-=P_-\gamma^+=0\,,\qquad\gamma^+P_-=\gamma^+\,,\qquad P_-\gamma^-=\gamma^-\,,
\eeq
we see that one of the off-diagonal terms in (\ref{Pmvar}) is zero while the other cancels out with the corresponding off-diagonal term in (\ref{eq:4thcont}). We end up with 
\begin{align} \la{docf}
{\mathbb D}_+&=\sqrt{4\pi\over k}\(\begin{array}{cc}0&0\\ D_+\bar\psi_+&0 \end{array}\)= \sqrt{4\pi \lambda}\,\widetilde\cO_R^{(0,-{1\over2})}\,\widetilde\cO_L^{(0,{3\over2})}\,,\\
{\mathbb D}_-&=\sqrt{4\pi\over k}\(\begin{array}{cc}0&D_-\psi_-\\0&0 \end{array}\)= \sqrt{4\pi\lambda}\,\widetilde\cO_R^{(0,-{3\over2})}\,\widetilde\cO_L^{(0,{1\over2})}\,,
\end{align}
in agreement with (\ref{Dis_fer_cond}).

\subsection{The Boundary Equation}
The same logic as for other line operators (see \ref{sec:EOMBOS} and \ref{sec:ferevoandbdry}
) leads to the following form of the boundary equation for right operators,
\beq \label{boundaryeqs_cond}
\widetilde{\mathcal{O}}_{R}^{(n+2,s)}
\propto\left\{\begin{array}{ll}
	\delta_-\widetilde{\mathcal{O}}_{R}^{(n,s+1)}&\quad s\ge-{1\over2}\\
	\delta_+\widetilde{\mathcal{O}}_{R}^{(n,s-1)}&\quad s\le-{3\over2}
\end{array}\right.\,,
\eeq
and for left operators,\footnote{Here, the framing is always taken perpendicular to the line and to the direction of deformation, as explained in the footnote above (\ref{eq:opsWeConsider2}).}
\beq
\widetilde{\mathcal{O}}_{L}^{(n+2,s)}
\propto\left\{\begin{array}{ll}
	\delta_-\widetilde{\mathcal{O}}_{L}^{(n,s+1)}&\quad s\ge {3 \over 2}\\
	\delta_+\widetilde{\mathcal{O}}_{L}^{(n,s-1)}&\quad s\le {1 \over 2}
\end{array}\right.\,.
\eeq

Next, we derive the proportionality constant $-2$ for right operators (\ref{boundaryeqs_cond}). The derivation for left operators follows the same steps. Recall that 
the expectation value of straight mesonic line operators with $s_R \neq -1/2$ are unaffected by the condensation. Hence, for these boundary equations the factor of $-2$ carries over from the result of the previous section \ref{sec:ferevoandbdry}. 
It remains to show that
\beq\la{becf}
\widetilde\cO_R^{(2,-{1\over2})}=-2\delta_-\cO_R^{(0,{1\over2})}\,.
\eeq
In appendix \ref{cfbeapp} we verify (\ref{becf}) by an explicit computation.

\subsection{The Two-point Function of the Displacement Operator}

By plugging the displacement operator in (\ref{docf}) into (\ref{distpf}), we arrive at
\beq\la{MtoLambda_fer_cond}
\Lambda= 4 \pi \lambda\, x^4\<\widetilde{M}^{(\frac{1}{2},-\frac{1}{2})}\>\<\widetilde{M}^{(\frac{3}{2},-\frac{3}{2})}\> = 4 \pi \lambda\, x^4\<\widetilde{M}^{(\frac{1}{2},-\frac{1}{2})}\>\<ֿ\widetilde{M}^{(\frac{3}{2},-\frac{3}{2})}\> \,.
\eeq
Using the explicate expressions for these expectation values in (\ref{eq:expValCondEmpty}) and (\ref{M1m1prefactor_fer}), we reproduce (\ref{fDD}) with $\Delta=-\lambda_\text{fer}/2$ or $\Delta=2+\lambda_\text{fer}/2$, in agreement with the duality.

\section{Comments About Line RG Flow at Large $N$}\la{RGsection}

We have found that there are two conformal line operators in the fundamental representation. They correspond to two fixed points in a one-parameter family. In the bosonic description, this parameter is the coefficient $\alpha$ in front of the scalar bi-linear in the line action (\ref{Wbos}). 
In this section, we study the RG flow which takes place in this one-parameter family of lines. 

An RG flow on the line is triggered by 
local operators in the line action. In the planar limit, these operators factorize into a product of right and left boundary operators as
\beq
\mathcal{O}_{\text{inner}} = \mathcal{O}_R \times \mathcal{O}_L\, \qquad\text{with}\qquad 
\Delta(\mathcal{O}_{\text{inner}}) = \Delta(\mathcal{O}_R) + \Delta(\mathcal{O}_L) + \cO(1/N)\, .  
\eeq

Consider first the $\alpha=1$ fixed point. 
The operator on the line with the minimal conformal dimension is the bi-scalar. 
Its conformal dimension is
\beq
\Delta(\mathcal{O}_R^{(0,0)}\times\mathcal{O}_L^{(0,0)}) = 1 +  \lambda\ge1 \qquad\text{for}\qquad\alpha=1\, .
\eeq
Hence, the $\alpha=1$ fixed point is stable. 

In contrast, when $\alpha=-1$, the scalar bi-linear with dimension $\Delta=1-\lambda\le1$ is relevant. It is the only relevant deformation of the $\alpha=-1$ fixed point, whose coefficient grows along the RG flow. The sign of the relevant coupling determines the direction of that flow. Given that the $\alpha =1$ is the only stable fixed point in this one-parameter family, there is a definite flow toward this point. If the sign of the relevant coupling is reversed,  
the magnitude of the bi-scalar coefficient will continue to grow. 
To understand where this flow ends, we turn to the fermionic description. 

\begin{figure}[t]
\centering
\includegraphics[width=0.7\textwidth]{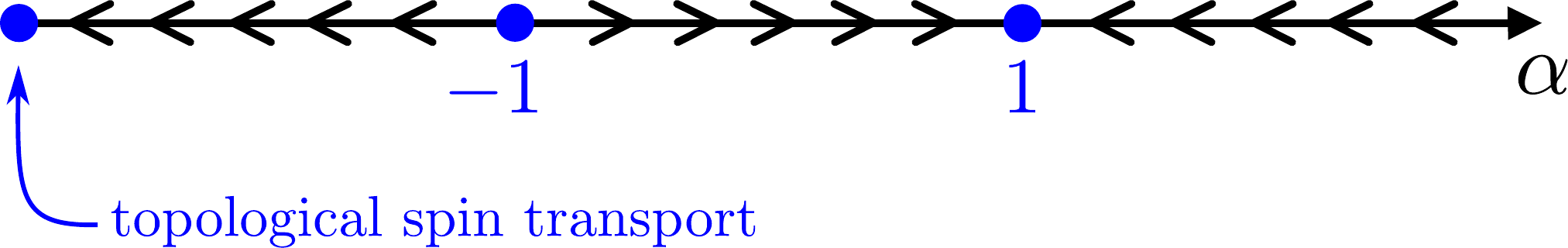}
\caption{The coefficient of the scalar bi-linear $\alpha$ in (\ref{Wbos}) undergoes an RG flow on the line. The fixed point at $\alpha=1$ is stable while the fixed point at $\alpha=-1$ is unstable. Starting from the unstable fixed point, two flows exist. 
One moves in the direction of the stable fixed point, while the other moves in the other direction toward a topological spin transport that is essentially trivial 
(\ref{toptransport}).}\label{flowfig}
\end{figure}

The fermionic dual of the $\alpha=-1$ operator is the condensed fermion operator (\ref{condencedfcon}) with $b=1$. When expended in perturbation theory, the corresponding line is decomposed from ``empty'' segments with the topological spin connection (\ref{toptransport}) and ``filled'' segments with a Wilson line between a pair of fundamental and anti-fundamental fermions, see figure \ref{fig:condensedFer}. The duals of the fundamental scalar boundary operators are the empty endpoints. Deforming the bosonic line action by the scalar bi-linear is consequently equivalent, to leading order in the deformation parameter, to modifying the coefficient in front of $P_-$ in the bottom right component of the connection $\widetilde A$ in (\ref{condencedfcon}). This modification changes the weight of the empty segments. It thus generates an RG flow on the line in which the empty segments grow or shrink accordingly. This flow ends when the line is either completely filled or completely empty. The wholly filled line is the Wilson line operator, dual to the $\alpha=1$ operator in the bosonic description. The completely empty line is an almost trivial line, with only the topological spin connection (\ref{toptransport}). Figure \ref{flowfig} shows the RG flow picture that emerges. Another support for this RG picture comes from the beta function for the scalar bi-linear. In section \ref{allloopsec} we have found that it is proportional to $\beta_\alpha\propto(1-\alpha^2)$.

Curiously, the $\alpha=-1$ unstable line operator is the endpoint of a different RG flow. To describe it, we work in the fermionic description and start with the direct sum of the Wilson line and an anyonic parallel transport, defined by the connection $\tilde{A}_0$ in (\ref{Mfc}). As a direct sum of conformal lines, it is conformal as well. The line operator 
\begin{equation}
\cO_\text{inner}= -{\ii\over\sqrt k} \(\begin{array}{cc}0&P_- \psi\\
    \bar\psi P_-&0\end{array}\)\,,
\end{equation}
is relevant and therefore can be used to trigger an RG flow on the line. It ends when the coefficient of this operator reaches the value $\sqrt{4\pi}$, which is the condensed fermion conformal line operator (\ref{condencedfcon}).\footnote{Note that the boundary entropy $g$ of a direct sum of two lines is $g_{1\oplus2}=\log\(e^{g_1}+e^{g_2}\)$. Hence, the direct sum of the Wilson line operator with the trivial parallel transport gives an operator with an higher $g$-function, in agreement with the $g$-theorem \cite{Cuomo:2021rkm}.}

Finally, the same RG flow picture \ref{flowfig} has appeared in \cite{Aharony:2022ntz} when studying an RG flow on a line of the four-dimensional QED in the large charge limit. It also appeared in the study of spin impurities in \cite{Cuomo:2022xgw}. It would be interesting to understand how these cases are related. 

\section{Perturbative Bootstrap of the Line Integrand} \label{sec:bootstrap}

In this section, we demonstrate how the 
evolution and boundary equations of a mesonic line operator can be used to recursively reconstruct its perturbative expansion. The drawback of the approach presented here is that the manipulations are formal. This is because we do not perform the line integrations, which are the only place where divergences, regularization, and renormalization of the line come about. In particular, the framing regularization of the line is not taken into account. Hence, these approach can be applied to any line operator, conformal or not. For the sake of simplicity of the presentation, we simply set $\alpha=0$ in \eqref{Wbos} and adjust the form of the displacement operator accordingly. 
We will show explicitly how the resulting formal solution reproduces the result of perturbation theory at one loop order.

To reconstruct the perturbative expansion of the line, we introduce 
a set of functions
$\mathtt{B}^{(L)}_{\mu_1 \ldots \mu_m}$. 
They are generalizations of the line integrand in (\ref{preintegrand}) that also includes the gluon-gluon exchange on the line. We bootstrap them recursively from the corresponding evolution and boundary equations. 
In spirit, the manipulations in this section are in line with the original Makeenko-Migdal solution to the four-dimensional loop equation \cite{Makeenko:1979pb}. 

While working formally, the main point of this section is that the evolution and boundary equations are very restrictive constraints. Motivated by this, in \cite{bootstrap} we present a physical non-perturbative bootstrap of the renormalized mesonic line operators. Indeed, we find there that the evolution and boundary equations combined with the spectrum of the boundary operators determine the expectation values uniquely.  

\subsection{Setup}

We consider the mesonic line operator in the bosonic theory that is formally defined as
\beq \label{Mev2}
M[\cC_{10}]\equiv
\< \phi^\dagger(x_1)\, W[\mathcal{C}_{10}]\phi(x_0)\>=\sum_{L=0}^\infty\lambda^LM^{(L)}[\cC_{10}]\,.
\eeq
Here, $\cC_{10}$ is some arbitrary smooth path between $x_0$ and $x_1$ and $W$ is the standard Wilson line operator in (\ref{WLop}). We start by writing an ansatz for the perturbative solution of the form
\beq \label{anzats}
M^{(L)}[\mathcal{C}_{10}] \equiv{1\over4\pi}
\sum_{m=0}^{2L} \int_{s_1>s_2>\dots>s_m}\hspace{-40pt}\dd s_1\dots \dd s_m\ \dot x_{s_1}^{\mu_1} \cdots\dot x_{s_m}^{\mu_m}\, \mathtt{B}^{(L)}_{\mu_1 \ldots \mu_m}(x_{s_1},\dots,x_{s_m})\ ,
\eeq
where the $\mathtt{B}^{(n)}_{\mu_1 \ldots \mu_m}(s_1,\dots,s_m)$ is a dimension $m+1$ function of $(x_{s_1},\dots,x_{s_m})$ as well as $x_0$ and $x_1$, but is \textit{independent} of the path $\mathcal{C}$. We think of these points as the $L$-loop contribution from $m$ gluons inserted on the line. However, we will not use Feynman diagrams directly. The relation between $\mathtt{B}^{(L)}_{\mu_1 \ldots \mu_m}$ and the integrand $\mathtt{B}_{\alpha=0}^{(L)}$ in \eqref{eqn-higherloop-rel} will not be relevant for us.\footnote{It involves fixing a light cone gauge and performing some of the line integrations, which correspond to the gluon exchanges on the line.}

The gauge symmetry of the CS theory manifests itself in an ambiguity in the form of the ansatz \eqref{anzats}. Namely, \eqref{anzats} is invariant under simultaneous transformations 
\be \label{gaugered}
&\mathtt{B}^{(L)}_{\mu_1 \ldots \mu_m}\rightarrow \mathtt{B}^{(L)}_{\mu_1 \ldots \mu_m}+{\d\over\d x^{\mu_j}_{s_j}}\Omega_{\mu_1,\dots\mu_{j-1}\mu_{j+1}\dots\mu_m}(x_{s_j}|x_{s_1},\dots,x_{s_{j-1}},x_{s_{j+1}},\dots,x_{s_m})\ ,\\
&\mathtt{B}^{(L)}_{\mu_1 \ldots \mu_{m-1}}\rightarrow \mathtt{B}^{(L)}_{\mu_1 \ldots \mu_{m-1}}+\Omega_{\mu_1,\dots\mu_{j-1}\mu_{j+1}\dots\mu_m}(x_{s_{j-1}}|x_{s_1},\dots,x_{s_{j-1}},x_{s_{j+1}},\dots,x_{s_m})\\
&\hspace{108pt}-\Omega_{\mu_1,\dots\mu_{j-1}\mu_{j+1}\dots\mu_m}(x_{s_{j+1}}|x_{s_1},\dots,x_{s_{j-1}},x_{s_{j+1}},\dots,x_{s_m})\ ,
\ee
with $s_0=0$ and $s_{m+1}=1$. Note that this ambiguity is more than the original CS gauge symmetry because we can use different functions, $\Omega$, for different segments of the line.

Our aim is to demonstrate that up to this gauge ambiguity, $\mathtt{B}^{(L)}$ is uniquely fixed by the boundary and evolution equations. In other words, $\mathtt{B}^{(L)}$ in (\ref{anzats}) is fixed by the scalar and gauge field equations of motion restricted to the line. For this, it is crucial to include all contact terms, which are sources when the line has a self-crossing point or when the two endpoints coincide. 

Consider the boundary equation first. This equation takes the form\footnote{It originated from the following gauge non-invariant Schwinger-Dyson equation: 
\begin{equation*}
 0 = \int[\mathcal{D} A][\mathcal{D} \phi]
 \,  \frac{\delta}{\delta \phi^i(x_1)}\Big[e^{-S} \,  (\mathcal{P}e^{\ii\int_{\cC_{10}} A\cdot\dd x})^{i}_{\ j} \phi(x_0)^j \Big] \, ,  
\end{equation*}
where repeated color indices $i,j$ are being summed over.}
\beq\label{boundaryvar}
\Box_{x_0}M[\mathcal{C}_{10}]=\Box_{x_1}M[\mathcal{C}_{10}]=
-\delta^3(x_0-x_1)\times{1\over N}\tr W[\mathcal{C}_{10}]\ ,
\eeq
where $\Box=\{\delta_+,\delta_-\} + \delta_3^2$ is the boundary Laplacian. The right-hand side in (\ref{boundaryvar}) is a contact term that is sourced when the two endpoints coincide. It is proportional to the closed Wilson loop that is obtained by joining the endpoints together. This term was not relevant before when we classified the boundary operators at finite coupling. 

Second, we consider the \textit{evolution equation}. It is obtained by setting $\alpha =0$ in \eqref{displacementF}-\eqref{displacementeom} and reads
\begin{figure}[t]
\centering
\includegraphics[width=0.9\textwidth]{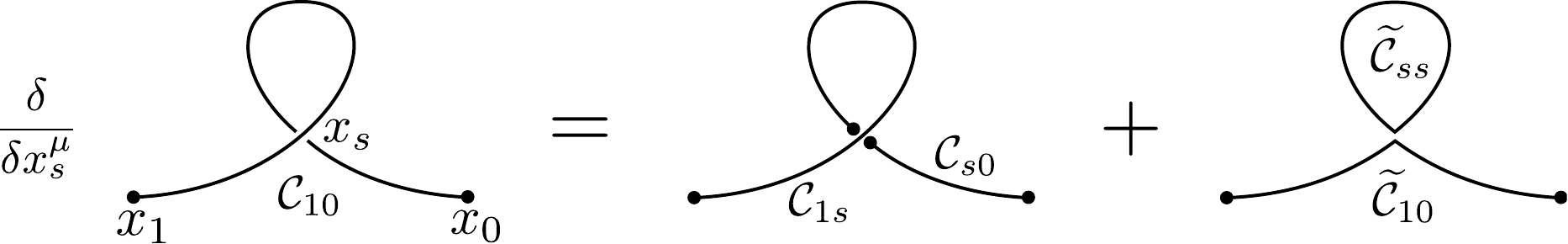}
    \caption{The evolution equation for a mesonic line operator with a Wilson line along the path ${\cal C}_{10}$, is given in (\ref{middelvar}). Local functional variation 
    of the path result in the insertion of the field strength operator.  
    When substituting the equation of motion $F= J$, an additional contact term contributes at self-crossing points. In the planar limit, both terms factorize into a product of two lines. In particular, the self crossing term factorize into a product of a closed Wilson loop $W[\widetilde{\mathcal{C}}_{ss}]$ times a mesonic line along the path $\widetilde{{\cal C}}_{10}$ in the figure.}
    \label{fig:loopeqnBulk}
\end{figure}
\begin{align}\label{middelvar}
{\delta M[\mathcal{C}_{10}]\over\delta x^\mu_s}
=2 \pi \ii \lambda\,\epsilon_{\mu\nu\rho} \dot x^\nu_s\Big[&\Big({\delta\over \delta x^\rho_s}M[\mathcal{C}_{1s}]\Big) M[\mathcal{C}_{s0}]-M[\mathcal{C}_{1s}] \Big({\delta\over\delta x^\rho_s}M[\mathcal{C}_{s0}]\Big)\\
&-
\int\limits_{\mathcal{C}_{10}} \dd x_t^\rho\, \delta^3(x_s-x_t) M[\widetilde{\mathcal{C}}_{10}]\times{1\over N}\tr W[\widetilde{\mathcal{C}}_{ss}]\Big] +{\cal O}(1/N)\, .\nn
\end{align}
Here the term in the second line is the contact term that is sourced when the contour has a self-crossing point, similar to (\ref{eqn-Bootstrap-loopEqn}). At such point, the operator factorizes into the product of a closed Wilson loop times a mesonic line, see figure \ref{fig:loopeqnBulk}. 

The closed Wilson loop on the right-hand side of (\ref{boundaryvar}) and (\ref{middelvar}) satisfies a loop equation that is the same as the one satisfied by the mesonic line operator in (\ref{middelvar}). The only difference is that instead of an open line between fundamental and anti-fundamental scalars, we have a closed loop with a color trace. As a result, the first term in (\ref{middelvar}) becomes $1/N$ suppressed and we remained with
\beq\la{eqn-Bootstrap-loopEqn}
\frac{\delta}{\delta x_s^\mu} \tr W[\mathcal{C}_{10}] = -{2 \pi \ii\over k} \int \limits_{\mathcal{C}_{10}} \dd x_t^\rho\, \delta^3(x_s-x_t)\,\tr W[\widetilde{\mathcal{C}}_{st}]\times\tr W[\widetilde{\mathcal{C}}_{ts}] + \cO(1/N)\,.
\eeq
Here, the loop at any of the self-crossing point $x_s$ splits into two \textit{disjoint closed loops}, denoted by $\widetilde{\mathcal{C}}_{ts}$ and $\widetilde{\mathcal{C}}_{st}$, respectively. This equation can also be derived by plugging (\ref{boundaryvar}) into (\ref{middelvar}). 

\subsection{Tree Level}

At tree level, \eqref{anzats} is independent of the shape of the path, so the boundary Laplacian $\Box_x$ reduces to the normal Laplacian $\partial^2_x$. The boundary equation \eqref{boundaryvar} then implies that
\beq\la{treeM}
\mathtt{B}^{(0)}={1\over|x_0-x_1|}\,,
\eeq
where we used the three dimensional Green's function $\partial^2_x \frac{1}{|x|}= -4\pi \delta^3(x)$. One may worry about additional constant and linear solutions to (\ref{treeM}). These however are ruled out by translation symmetry and the fact that $\mathtt{B}^{(0)}$ has dimension one. 

\subsection{One Loop}

At loop level, the boundary Laplacian, which is decomposed from path derivatives $\Box=\{\delta_+,\delta_-\} + \delta_3^2$, also act in the line integrations. 
To map the boundary equation (\ref{boundaryvar}) into a constraint on $\mathtt{B}^{(L)}$, 
it is convenient to use the following identities 
\begin{align}\label{eq:varLemma0}
&\Box_x  \int\limits_x^{x_1} \dd x^\mu_s \, \Gamma_\mu(x, x_s) = \int\limits_x^{x_1} \dd x^\mu_s \, \partial^2_x \Gamma_\mu(x, x^\mu_s)  - 2 \d_u^\mu \Gamma_\mu(u,x)  \Big|_{u = x} \,,\nn\\
&\Box_x \int\limits_{x}^{x_1} \dd x_s^\mu \int\limits_x^{x_s} \dd x^\nu_t \, \Gamma_{\mu \nu}(x, x_s, x_t)=  \int\limits_{x}^{x_1} \dd x_s^\mu\Big(\int\limits_x^{x_s} \dd x^\nu_t \, \partial^2_x \Gamma_{\mu \nu}(x, x_s, x_t)-2\partial^\sigma_{y} \Gamma_{\mu \sigma}(y, x_s, x) \Big|_{y=x}\Big)\nn\\
&\qquad\qquad\qquad\qquad\qquad\qquad\qquad+ \Gamma_{\mu}^{\ \mu}(x, x, x) \, , \\
&\Box_x \int\limits_{x}^{x_1} \dd x_s^\mu \int\limits_x^{x_s} \dd x^\nu_t \int\limits_x^{x_t} \dd x^\rho_u \, \Gamma_{\mu \nu \rho}(x, x_s, x_t, x_u)= \int\limits_{x}^{x_1} \dd x_s^\mu \int\limits_x^{x_s} \dd x^\nu_t \int\limits_x^{x_t} \dd x^\rho_u \,\partial^2_x  \Gamma_{\mu \nu \rho}(x, x_s, x_t, x_u)\nn\\
&\qquad\qquad\qquad\qquad\qquad\qquad+\int\limits_{x}^{x_1}\dd x_s^\mu\Big(\Gamma_{\mu \nu}^{\ \ \nu}(x, x_s, x, x)-2 \int\limits_x^{x_s} \dd x^\nu_t \,\partial^\sigma_y\Gamma_{\mu \nu \sigma}(y, x_s, x_t, x) \Big|_{y=x}\Big) \, ,\nn
\end{align}
where we recall that $\partial^2_x$ denotes the three dimensional Laplacian. The last equation also applies to multiple line integration which depends on four or more integration variables. 

At one loop order we have to solve for the three functions $\mathtt{B}^{(1)}$, $\mathtt{B}^{(1)}(x)$, and $\mathtt{B}^{(1)}(x_s,x_t)$, that also depend on $x_0$ and $x_1$ but not on the path $\cC$. We plug the ansatz \eqref{anzats} into the box equation \eqref{boundaryvar} and evaluate both sides at order $\lambda$. 
Using (\ref{eq:varLemma0}) with $x=x_0$ we get
\begin{align}\la{box1l}
\Box_{x_0} M^{(1)}[\mathcal{C}_{10}] = & \frac{\lambda}{4\pi}\Big[\partial^2_{x_0} \mathtt{B}^{(1)}-2 \partial_{x_0}^\mu \mathtt{B}_\mu^{(1)}(x_0) + \int\limits_0^1 \dd s \, \dot{x}_s^\mu \, \partial^2_{x_0} \mathtt{B}_\mu^{(1)}(x_s) \\
& \qquad + \eta^{\mu \nu}\mathtt{B}^{(1)}_{\mu \nu}(x_0, x_0) -2 \int\limits_0^1 \dd s \, \dot{x}_s^\nu \partial_{x_0}^\mu \mathtt{B}^{(1)}_{\nu \mu}(s,0) +\int\limits_{s>t}\! \dd s\,\dd t\,\dot{x}_s^\mu\dot{x}_t^\nu\, \partial^2_{x_0} \mathtt{B}^{(1)}_{\mu\nu}(x_s,x_t)\Big]\nn\\
=& -\delta^3(x_0-x_1)\times{1\over N}\tr W[\mathcal{C}_{10}] \Big|_{\mathcal{O}(\lambda)}\,.\nn
\end{align}

This equation should hold for an arbitrary shape of the path $\mathcal{C}_{10}$. At the level we work at, we must treat the expectation value of the closed Wilson loop in the last line on the same footing as we treat the mesonic line operator. Namely, 
we make the following ansatz for it
\beq
{1\over N}\tr W[\mathcal{C}_{10}] \Big|_{\mathcal{O}(\lambda)}
=\frac{\lambda}{2} \int\limits_{s>t}\! \dd s\,\dd t\,\dot{x}_s^\mu \dot{x}_t^\nu\, \mathtt{G}_{\mu\nu}(x_s, x_t) + \mathcal{O}(\lambda^2) \ ,\
\eeq
where terms with three or more integration points are suppressed by powers of the coupling. Plugging the ansatz into \eqref{eqn-Bootstrap-loopEqn}, to leading order in $\lambda$, the equation that determines $\mathtt{G}_{\mu \nu}$ is simply
\beq
{\d\over\d x_s^\mu}\mathtt{G}_{\nu \rho}(x_s,x_t)-{\d\over\d x_s^\nu}\mathtt{G}_{\mu \rho}(x_s,x_t) = -2\pi \ii \lambda\,\epsilon_{\mu \nu \rho}\, \delta^3(x_s-x_t)\,.
\eeq
where the equation is understood to hold up to total derivatives.\footnote{This equation originates from the integral equation $\oint_{\mathcal{C}_{10}} \dd x^\rho_t (\mathtt{LHS}-\mathtt{RHS})=0$, so the solution is up to a total derivative term that vanishes for a closed loop.}
We can identify $\mathtt{G}_{\mu_1 \mu_2}$ with the gauge \textit{dependent} gluon propagator in position space
\beq\la{Apropx}
\<A_\mu^I(x)A_\nu^J(y)\>={1\over k}\mathtt{G}_{\mu\nu}(x-y)\delta^{IJ}\,.
\eeq

By comparing the double line integration in (\ref{box1l}), we conclude that up to total derivatives, $\mathtt{B}^{(1)}_{\mu\nu}(x_s,x_t)$ is the gluon propagator multiplies a free scalar propagator $1/|x_{01}|$,
\beq\la{GammatoG}
\mathtt{B}^{(1)}_{\mu\nu}(x_s,x_t)= \frac{1}{2|x_{01}|} \mathtt{G}_{\mu \nu}(x_s, x_t)+{\d\over\d x_s^\mu} F_\nu(x_s,x_t)+{\d\over\d x_t^\nu}\tilde F_{\mu}(x_s,x_t)\,.
\eeq
We can always use the redundancy (\ref{gaugered}) to set $F=\tilde F=0$. This is equivalent to redefining $\mathtt{B}^{(1)}_\mu$.

At this point, it is convenient to resolve the ambiguity \eqref{gaugered} by fixing a gauge. A choice that is equivalent to Lorentz gauge in the CS theory is to pick
\beq\label{AAprop}
\d_{x_{s_j}}^{\mu_j}
\mathtt{B}^{(n)}_{\mu_1 \ldots \mu_m}(x_{s_1},\dots,x_{s_m})=0\,.
\eeq
From (\ref{GammatoG}) it follows that $\mathtt{G}_{\mu\nu}$ has to satisfy the same gauge constraint. This gauge condition still allows a linear gauge transformation such as $\Omega(x)=v_\mu x^\mu$. Demanding that the $\mathtt{B}^{(L)}$ decays at infinity require $v^\mu=0$, and no gauge redundancy is left, up to a constant $\Omega$ that decouples from $M$. 
In this gauge, we have that 
\beq\la{GtoB2}
\mathtt{B}^{(1)}_{\mu\nu}(x_s,x_t)= \frac{1}{2|x_{01}|} \mathtt{G}_{\mu \nu}(x_s, x_t)=  \frac{\ii}{2|x_{01}|}\epsilon_{\mu \nu \rho} \partial^\rho_{x_s} \frac{1}{|x_s- x_t|}\,.
\eeq

After plugging (\ref{GtoB2}) into the boundary equation (\ref{box1l}), it reduces to 
\beq
\partial^2_{x_0} \mathtt{B}^{(1)}_\mu(x_s) = 2 \d_{x_0}^\nu \mathtt{B}^{(1)}_{\mu \nu}(x_s,x_t)\Big|_{t=0}\ ,\qquad
\partial^2_{x_0} \mathtt{B}^{(1)} = 2 \d_{x_0}^\mu \mathtt{B}^{(1)}_\mu(x_t) \Big|_{t=0} \, .
\eeq

Next, we move to the evolution equation \eqref{middelvar}. By plugging \eqref{anzats} into it and imposing the Lorentz-type gauge (\ref{AAprop}) we get for the left-hand side
\begin{align}\la{evolutionGamma}
{\delta M[\mathcal{C}_{10}]\over\delta x^\mu_s}=&{\lambda\over4\pi}\dot{x}_s^\nu\Big[{\d\over\d x_s^\mu} \mathtt{B}^{(1)}_\nu(x_s)-{\d\over\d x_s^\nu} \mathtt{B}^{(1)}_\mu(x_s)\\
+&\int\limits_0^1 \dd t\,\dot x^{\rho}_{t}\Big({\d\over\d x_s^\mu}\mathtt{B}^{(1)}_{\nu\rho}(x_s,x_t)-{\d\over\d x_s^\nu}\mathtt{B}^{(1)}_{\mu\rho}(x_s,x_t)\Big)\Big]+\cO(\lambda^2)\,,\nn
\end{align}
where we used that $\mathtt{B}^{(1)}_{\mu \nu}(x,y) = \mathtt{B}^{(1)}_{\nu \mu}(y,x)$ and our Lorentz gauge choice (\ref{AAprop}). The term $\mathtt{B}^{(1)}_{\mu \nu}(x_s, x_t)$ in (\ref{GtoB2}) satisfies the Bianchi identity up to a contact term when $x_s, x_t$ coincides,
\beq \label{Bianchi}
{\d\over\d x_s^\mu} \mathtt{B}^{(1)}_{\nu \rho}(x_s, x_t)+{\d\over\d x_s^\nu} \mathtt{B}^{(1)}_{\rho\mu}(x_s, x_t) + {\d\over\d x_s^\rho}\mathtt{B}^{(1)}_{\mu \nu}(x_s, x_t) = -\frac{2\pi \ii}{|x_{01}|} \epsilon_{\mu \nu \rho} \,  \delta^{(3)}(x_s -x_t) \, .
\eeq
By plugging (\ref{Bianchi}) into (\ref{evolutionGamma}) and using that $\d_{x_s}^\rho\mathtt{B}^{(1)}_{\mu \nu}(x_s, x_t)=-\d_{x_t}^\rho\mathtt{B}^{(1)}_{\mu \nu}(x_s, x_t)$ and $\mathtt{B}^{(1)}_{\mu \nu}(x_s, x_t)=-\mathtt{B}^{(1)}_{\nu\mu}(x_s, x_t)$, we express the second line of (\ref{evolutionGamma}) as a sum of contact and a boundary terms
\begin{align}\la{secondline}
&\int\limits_0^1 \dd t\,\dot x^{\rho}_{t}\Big({\d\over\d x_s^\mu}\mathtt{B}^{(1)}_{\nu\rho}(x_s,x_t)-{\d\over\d x_s^\nu}\mathtt{B}^{(1)}_{\mu\rho}(x_s,x_t)\Big)\\
= & \,\mathtt{B}_{\mu \nu}^{(1)}(x_s,x_1) - \mathtt{B}_{\mu \nu}^{(1)}(x_s,x_0)-\frac{2\pi \ii}{|x_{01}|} \epsilon_{\mu \nu \rho}   \int\limits_0^1 \dd t\,\dot x^{\rho}_{t}\delta^{(3)}(x_s -x_t)\,.\nn
\end{align}

The right-hand side of the evolution equation (\ref{middelvar}) is already of order $\lambda$. Hence, we only need to plug in the tree-level expressions (\ref{treeM}) there. We get
\beq\la{RHS1loop}
\mathtt{RHS}=
\lambda\(\frac{\ii}{2}\epsilon_{\mu \nu \rho} \dot{x}_s^\nu V^\rho(x_s)-2\pi \ii\, \epsilon_{\mu \nu \rho} \dot{x}_s^\nu \int \dd x_t^\rho \delta^3(x_s-x_t) \frac{1}{4\pi |x_{10}|}\)+\cO(\lambda^2)\,,
\eeq
where
\beq
V^\rho(x_s)\equiv\frac{1}{|x_{0s}|}\d_{x_s}^\rho \frac{1}{|x_{1s}|}-\frac{1}{|x_{1s}|} \d_{x_s}^\rho\frac{1}{|x_{0s}|}\,.
\eeq

The contact terms term in (\ref{secondline}) and (\ref{RHS1loop}) cancel against each other. The evolution equation then reduces to 
the following relation between functions of points, with no dependence on the path $\mathcal{C}_{10}$
\beq
{\d\over\d x_s^\mu} \mathtt{B}^{(1)}_\nu(x_s)-{\d\over\d x_s^\nu} \mathtt{B}^{(1)}_\mu(x_s) +\mathtt{B}_{\mu \nu}^{(1)}(x_s,x_1) - \mathtt{B}_{\mu \nu}^{(1)}(x_s,x_0) = \frac{\ii}{2}\epsilon_{\mu \nu \rho} V^\rho(x_s) \, .
\eeq
Acting on both sides with $\partial^\mu_{x_s}$ and using the gauge condition (\ref{AAprop}), this equation reduces to
\beq
\partial^2_{x} \mathtt{B}^{(1)}_\nu(x) = \frac{\ii}{2} \epsilon_{\mu \nu \rho} \partial^\mu_{x} V^\rho(x) \,.
\eeq
The rotational covariant solution to this Laplace's equation that decays at infinity is
\beq\la{B1mu}
\mathtt{B}^{(1)}_\mu(x) = \frac{\ii}{2} \int \frac{\dd^3 y}{4\pi |x-y|}  \epsilon_{\mu \nu \rho} \partial^\nu_{y} V^\rho(y) \, .
\eeq

It reminds to solve for $\mathtt{B}^{(1)}$, which is a function of $x_0$ and $x_1$ only. The remaining non-trivial constraint from the boundary equation (\ref{box1l}) is
\beq
\partial^2_{x_0} \mathtt{B}^{(1)} = 2 \d_{x_0}^\mu \mathtt{B}^{(1)}_\mu(x_t) \Big|_{t=0}\,.
\eeq
From the explicit form of $\mathtt{B}^{(1)}_\mu(x)$ in (\ref{B1mu}) one can check that $\d_{x_0}^\mu \mathtt{B}^{(1)}_\mu(x_t) \Big|_{t=0} =0$. Therefore, we have to solve $\partial^2_{x_0} \mathtt{B}^{(1)} = 0$. 
The only dimension one translation and rotation symmetric solution is 
$\mathtt{B}^{(1)} = 0$.

We conclude that at one-loop order, the combination of the evolution equation and the box equation determine the solution up to the gauge freedom (\ref{gaugered}). The result agrees with the explicit 1-loop calculation in the Lorentz gauge, presented in the appendix \ref{apd:oneLoopGen}. We can proceed systematically like this and reconstruct the full formal perturbative expansion of the mesonic line operator (\ref{Mev2}). Turning on a non-zero value for $\alpha$ adds terms to the equations, but does not lead to any ambiguity in their solution. 

\section{Lift to Superconformal Line Operators in the ${\cal N}=2$ CS-matter Theory}
\label{sec:susy}

For all of the conformal line operators we have considered, the anomalous dimension and anomalous spin of the boundary operators were the same up to a sign. This section shows that this relation follows from supersymmetry, even though the theories we have considered are not supersymmetric. In short, the reason for this is the following. Every line operator we have considered has a lift into a different supersymmetric line operator in the ${\cal N}=2$ CS-matter theory. The lift has the property that the expectation value of the corresponding mesonic line operator in the supersymmetric and non-supersymmetric CS-matter theories is identical to the leading order in $1/N$.  
At the same time, the primary boundary operators are BPS operators of the supersymmetry that is preserved by the line. The three-dimensional transverse spin becomes an R-symmetry of the one-dimensional super-conformal algebra. Hence, the BPS condition of the boundary operators relates their conformal dimension to their transverse spin. 

One implication of our study for the ${\cal N}=2$ CS-matter theory is that the lift of the $\alpha=-1$ operator is very different from the lift of the $\alpha=1$ one. To our surprise, only the latter was explicitly considered before. 

It is important to note that the analysis in this section is formal because it is done at the level of the classical action, and we do not renormalize the operators. Classically, the line operators we consider preserve a one-dimensional $\mathfrak{osp}(2|2)$ superconformal symmetry. However, the line supersymmetry is allowed to be anomalous, and we can have a non-trivial RG flow on the line.\footnote{See \cite{Bianchi:2016vvm} for an example of a line operator with anomalous supersymmetry.} The number of broken supersymmetries is related to the number of dimensions $3/2$ fermionic operator on the line \cite{Agmon:2020pde}.\footnote{We thank Yifan Wang and Nathan Agmon for illuminating discussions on this point.} Our analysis in sections \ref{bossec}-\ref{sec:condFer} shows that the lines studied in section \ref{apm1susysec} are not anomalous. At the same time, the family of lines described in section \ref{sec:SUSY:condensed} has only two isolated superconformal fixed points.

For each of these superconformal line operators, we consider the BPS boundary operators on which the line can end. At the formal level, these boundary operators preserve one-half of the line supersymmetry.
Provided that the line is superconformal at the quantum level, the BPS boundary operators are guaranteed to remain BPS also at the quantum level. This is because they do not have other boundary operators to combine into long multiples. In summary, most of the formal analysis in this section holds true also at the quantum level.

\subsection{The $\alpha=\pm1$ Superconformal Line Operators}\la{apm1susysec}

We consider the ${\cal N}=2$ CS-matter theory of a single chiral superfield in the fundamental representation, \cite{Kapustin:2009kz,Hama:2010av,Jain:2012qi}. After integrating out all auxiliary fields, the action of the theory takes the form $S_{{\cal N}=2}=S_{CS}+S_{m}$ where the CS action is given in (\ref{CSact}) and
\beq \la{N2act}
S_{m}\!=\!\int\!\dd^3 x\left[D_{\mu}\phi^\dagger D^{\mu}\phi-\ii  \bar\psi \gamma^{\mu}D_{\mu}\psi
-\frac{4\pi\ii}{k}(\phi \phi)(\bar\psi \psi)-\frac{2\pi\ii}{k}(\bar\psi \phi)(\phi^\dagger \psi)+
\frac{4\pi^2}{k^2} (\phi^\dagger\phi)^3\right]\,.
\eeq
Here, in this section, we use the standard SUSY convention for the kinetic term of the fermions. It differs from the normalization we have used so far (\ref{Sfer}) by a factor of $-\ii$. The dictionary between the two conventions is
\beq\label{eqn-SUSY-fermiCompDict}
\psi^1_\text{here} = \sqrt{\ii}\, \psi_+\,, \quad \psi^2_\text{here} =  \sqrt{\ii}\, \psi_-\,, \quad \bar{\psi}^1_\text{here} = -\sqrt{\ii}\, \bar{\psi}_+\,, \quad  \bar{\psi}^2_\text{here} = \sqrt{\ii}\,\bar{\psi}_-\, .
\eeq

The theory (\ref{N2act}) is expected to be self-dual \cite{Giveon:2008zn,Benini:2011mf,Kapustin:2011gh,Willett:2011gp}. In the planar limit, the CS level and the 't Hooft coupling are related to the dual ones as $k\to-k$ and $\lambda-{\rm sign}(k)\to\lambda$. The theory (\ref{N2act}) posses a three dimensional ${\cal N}=2$ superconformal symmetry. In appendix \ref{susyappendix} we uniformize our notations for the algebra and its realization with the existing literature cited there. Readers who are familiar with it can skip this appendix, which otherwise is needed for reading this section.

Consider the modified Wilson line operator \eqref{Wbos} or equivalently \eqref{Wbos2}, along a straight line in the third direction. A general supersymmetry transformation can be parameterized using two constant Grassmannian spinors, $\epsilon$ and $\bar\epsilon$, as
\beq
\delta = \epsilon\, Q + \bar{\epsilon}\, \bar{Q}\,.
\eeq
The condition that the line stays invariant under this transformation reads
\footnote{The same equation holds for superconformal transformations. Taking $\epsilon = x_3 \gamma^3 \epsilon_c$ and using \eqref{eqn-SUSY-spinorId2}, one finds that the constraint equation for $\epsilon_c$ reads $(\alpha \mathbb{I} + \gamma_3) \epsilon_c = 0$. Since $\alpha^2 = 1$, this is identical to \eqref{eqn-SUSY-SUSYLineCondition}. One can also verify that the solution is compatible with the reality condition  
\eqref{eqn-SUSY-RC-QS}.}
\beq \label{eqn-SUSY-SUSYLineCondition}
\bar{\epsilon}^a \varepsilon_{ab} (\gamma_3 + \alpha \mathbb{I})^b_{\ c} = 0\qquad\text{and}\qquad(\gamma_3 + \alpha \mathbb{I})^b_{\ c} \epsilon^c = 0\,.
\eeq
Since $\det(\gamma_3+\alpha\One)=1-\alpha^2$, for the line to be invariant we must have $\alpha=\pm1$ and half of the SUSY variations must vanish. We have
\beq
\delta= \epsilon^2 Q_2 + \bar{\epsilon}^1 \bar{Q}_1\qquad\text{for}\quad\alpha=+1\,,
\eeq
and
\beq
\delta=\epsilon^1 Q_1 + \bar{\epsilon}^2 \bar{Q}_2\qquad\text{for}\quad\alpha=-1\,.
\eeq
In either of these cases, the line brakes half of the supersymmetry, leaving two supercharges unbroken. 
The line also preserves their conjugate 
superconformal charges. 

The bosonic symmetries of the straight line are generated by the conformal  $\mathfrak{sl}(2,\mathbb{R})$ algebra and the $\mathfrak{so}(2) \simeq \mathfrak{u}(1)$ transverse rotations. Together with the two supercharges and the two superconformal charges that preserve the line, 
they generate the line superconformal group $\mathfrak{osp}(2|2)$. The non-trivial commutation relations that we will need below are the ones between the $Q$'s and the $S$'s. Using \eqref{eqn-SUSY-QScomm}, we find 
\beq \label{eqn-SUSY-BPSalphappP1}
\begin{aligned}
\{Q_2,\bar{S}^2 \} & = -\ii D +(\mathcal{R}+M_{12})\\
\{S^1,\bar{Q}_1 \} & = - \ii D-(\mathcal{R}+M_{12})
\end{aligned}\qquad\text{for }\alpha=+1\,,
\eeq
and
\beq \label{eqn-SUSY-BPSalphaM1}
\begin{aligned}
\{Q_1,\bar{S}^1 \} & = - \ii D +(\mathcal{R}- M_{12})\\
\{S^2,\bar{Q}_2 \} & = - \ii D-(\mathcal{R}-M_{12})
\end{aligned}\qquad\text{for }\alpha=-1\,.
\eeq
Here, $D$ is the dilatation generator, ${\cal R}$ is the three-dimensional R-charge generator, and $M_{12}$ generates rotations in the transverse plane to the line, see appendix \ref{susyappendix} for details.

Next, we study the actions of this superconformal symmetry of the line on the boundary operators.

\subsection{Boundary Operators of the $\alpha=\pm1$ Superconformal Lines}

The two superconformal line operators, (\ref{Wbos}) with $\alpha=\pm1$, can end on operators in the fundamental/anti-fundamental representation. Among these, we would like to understand what are the boundary operators that are super-conformal primaries, preserving one-half of the straight line super and superconformal charges. Depending on the sign of $\alpha$, these boundary operators are annihilated by one supercharge and one superconformal charge in (\ref{eqn-SUSY-BPSalphappP1}) or (\ref{eqn-SUSY-BPSalphaM1}) correspondingly. The right hand side of these equations then relates their conformal dimension $\Delta$ (eigenvalue of $\ii D$) to their transverse spin $\mathfrak{s}$ (eigenvalue of $-M_{12}$). 

It is sufficient to consider the four bottom operators of the lowest dimension and transverse spin. All the other operators in the towers are related to these by path derivatives. Correspondingly, their dimension and transverse spin are related to the bottom operator in the tower by an integer shift.

At weak coupling, the boundary operators of the lowest dimension are the scalar and fermion fields. We now study their superconformal transformation. For convenience, we place them at the origin. A general superconformal transformation can be parameterized by $\epsilon = \epsilon_s + x^\mu \gamma_\mu \epsilon_c$ (see \eqref{eqn-SUSY-KillingVecSol}), with $\epsilon_s, \epsilon_c$ being constant spinors. The corresponding variation of the fundamental fields takes the form
\beq \label{eqn-SUSY-commEpsBar}
\delta_{\epsilon} (\cdot) = [\epsilon_s Q, \cdot] + \ii [\epsilon_c S, \cdot]\,,
\eeq
with a similar equation for the conjugate transformation.

\subsubsection{Boundary Operators of the $\alpha =1$ Line}

The $\alpha=1$ line is preserved by the supercharges $Q_2, \bar{Q}_1$ and their conjugate superconformal charges $\bar{S}^2, S^1$. These generate the following transformation of the fields at the origin, (see \eqref{eqn-SUSY-matter})
\beq \label{eqn-SUSY-a1}
\begin{aligned}
\delta \phi & = -\bar{\epsilon}^1_s \psi^2\,, \quad & & &
\delta \phi^\dagger & = \epsilon^2_s \bar{\psi}^1\,, \\
\delta \psi^1 & = \ii \sqrt{2} \epsilon^2_s D_+ \phi\,, \quad & & &
\delta \bar{\psi}^1 & = \ii \bar{\epsilon}^1_s \mathbb{D}_3 \phi^\dagger + \ii \bar{\epsilon}_c^1 \phi^\dagger\,, \\
\delta \psi^2 & = -\ii \epsilon^2_s \mathbb{D}_3 \phi + \ii \epsilon_c^2 \phi\,, \quad & & &
\delta \bar{\psi}^2 & = \ii \sqrt{2} \bar{\epsilon}^1_s D_- \phi^\dagger\,,
\end{aligned}
\eeq
where ${\mathbb D}_3 \phi=(\partial_3-\ii A-\sigma) \phi$ is the \textit{full} line covariant derivative in the third direction. Here, the auxiliary field $\sigma \simeq \phi \phi^\dagger$ and the auxiliary field $F$ has been integrated out and set to zero.

From this transformation rule, we learn that $\{\phi,\phi^\dagger,\psi^1,\bar\psi^2\}$ are superconformal primaries, annihilated by $\bar{S}^2$ and $S^1$, because their transformations are \textit{independent} of $\epsilon_c$ and $\bar{\epsilon}_c$. In addition, each of those primary fields is annihilated by one supercharge, so from the line perspective they are all $1/2$ BPS operators. Concretely, $\phi$ and $\bar{\psi}^2$ are annihilated by $Q_2$, because their transformations are \textit{independent} of $\epsilon_s^2$. Similarly, $\phi^\dagger$ and $\psi^1$ are annihilated by $\bar{Q}_1$. We can therefore use \eqref{eqn-SUSY-BPSalphappP1} to relate their dimension to the transverse spin.

Explicitly, the scalar $\phi$ is invariant under $Q_2$ and $\bar{S}^2 = (Q_2)^\dagger$. The vanishing of the 
$\{Q_2,\bar{S}^2\}$ anticommutator \eqref{eqn-SUSY-BPSalphappP1} gives,\footnote{
Note that when acting on fields, $M_{12}$ gives \textit{minus} the spin. 
Explicitly, we have $[\ii D,\mathcal{O}] =\Delta_O \mathcal{O}$, $[\mathcal{R},\mathcal{O}] = r_\mathcal{O} \mathcal{O}$, and $[M_{12},\mathcal{O}] = - M_{12}^R \mathcal{O}$, 
where $M_{12}^R$ is the representation matrix of the Lorentz representation $R$. For spinors, $
M_{12}^R = -\frac{\ii}{4}[\gamma_1,\gamma_2] = \frac{1}{2} \gamma^3$, see \cite{Minwalla:1997ka} for more details. 
Note also that the fermionic spin is obtained by acting with the spin generators \textit{from the left}, for both $\psi$ and $\bar{\psi}$.}
\beq \label{eqn-SUSY-a1-BPSCond1}
\Delta_\phi = \frac{1}{2} - \mathfrak{s}(\phi)\,.
\eeq
Similar relations are found for the rest of the $1/2$ BPS primaries, they are given by
\beq \label{eqn-SUSY-a1-BPSCond2}
\Delta_{\phi^\dagger} = \frac{1}{2} + \mathfrak{s}(\phi^\dagger)\,, \qquad \Delta_{\psi^1} = \frac{1}{2} + \mathfrak{s}(\psi^1)\,, \qquad  \Delta_{\bar{\psi}^2} = \frac{1}{2} - \mathfrak{s}(\bar{\psi}^2)\, .
\eeq

The fields $\psi^2$ and $\bar{\psi}^1$ on the contrary are descendent of $\phi$ and $\phi^\dagger$ respectively, (\ref{eqn-SUSY-a1}). The scaling dimension of the SUSY parameters $\bar\epsilon_s^1$, $\epsilon_s^2$ is $-1/2$ and their transverse spin is $+1/2$ and $-1/2$ correspondingly. Hence, it follows from (\ref{eqn-SUSY-a1}) that 
\beq
\begin{aligned}
\Delta_{\psi^2} &=\Delta_\phi+{1\over2}\,, & \qquad \mathfrak{s}(\psi^2) &=\mathfrak{s}(\phi)-{1\over2}\,,\\
\Delta_{\bar\psi^1}&=\Delta_{\phi^\dagger}+{1\over2}\,,& \qquad \mathfrak{s}(\bar\psi^1)&=\mathfrak{s}(\phi^\dagger)+{1\over2}\,.
\end{aligned}
\eeq
Once combined with (\ref{eqn-SUSY-a1-BPSCond1}), (\ref{eqn-SUSY-a1-BPSCond2}), we find that
\beq\la{psi2barpsi1}
\Delta_{\psi^2}={1\over2}-\mathfrak{s}(\psi^2)\,,\qquad\Delta_{\bar\psi^1}={1\over2}+\mathfrak{s}(\bar\psi^1)\,.
\eeq

In addition to those relations, the two bottom bosonic operators with spin, $\mathcal{O}_L^{(0,-1)} = D_- \phi^\dagger$ and $\mathcal{O}_R^{(0,1)} = D_+ \phi$ are descendants of the primary fermion $\bar{\psi}^2$ and $\psi^1$ respectively, (\ref{eqn-SUSY-a1}). Therefore, their dimension and spin are also related as
\beq
\Delta_{D_+ \phi} = \frac{1}{2} + \mathfrak{s}(D_+ \phi), \qquad \Delta_{D_- \phi^\dagger} = \frac{1}{2} - \mathfrak{s}(D_- \phi^\dagger) \, .
\eeq

\subsubsection{Boundary Operators of the $\alpha=-1$ Line}

The $\alpha=-1$ line is preserved by the supercharges $Q_1$, $\bar{Q}_2$ and their conjugate superconformal charges $\bar{S}^1$, $S^2$. The corresponding field transformations at the origin read
\beq \label{eqn-SUSY-a2}
\begin{aligned}
\delta \phi & = \bar{\epsilon}^2_s \psi^1\,, \quad & & &
\delta \phi^\dagger & = -\epsilon^1_s \bar{\psi}^2\,, \\
\delta \psi^1 & = \ii \epsilon^1_s \mathbb{D}_3 \phi + \ii \epsilon_c^1 \phi\,, \quad & & &
\delta \bar{\psi}^1 & = \ii \sqrt{2} \bar\epsilon^2_s D_+ \phi^\dagger\,, \\
\delta \psi^2 & = \ii \sqrt{2} \epsilon^1_s D_- \phi\,, \quad & & &
\delta \bar{\psi}^2 & = -\ii \bar{\epsilon}^2_s \mathbb{D}_3 \phi^\dagger + \ii \bar{\epsilon}^2_c \phi^\dagger \,,
\end{aligned}
\eeq
where the covariant derivative now takes the form $\mathbb{D}_3 = \partial_3 - \ii (A_3 + \ii \sigma)$.

One can verify that $\phi, \phi^\dagger$ and $\psi^2,\bar{\psi}^1$ are superconformal primaries and are $1/2$-BPS operators. The relevant $1/2$-BPS conditions \eqref{eqn-SUSY-BPSalphaM1} now read 
\beq
\Delta_\phi = \frac{1}{2} + \mathfrak{s}(\phi)\,, \quad \Delta_{\phi^\dagger} = \frac{1}{2} - \mathfrak{s}(\phi^\dagger)\,, \quad \Delta_{\psi^2} =\frac{1}{2}- \mathfrak{s}(\psi^2)\,, \quad \Delta_{\bar{\psi}^1} = \frac{1}{2}+\mathfrak{s}(\bar{\psi}^1)\,.
\eeq
Comparing with the $\alpha =1$ case, see \eqref{eqn-SUSY-a1-BPSCond1} and \eqref{eqn-SUSY-a1-BPSCond2}, the sign of the spin term for bosons flips.

The fermion components $\psi^1$ and $\bar{\psi}^2$ on the other hand, are SUSY \textit{descendants} of $\phi$ and $\phi^\dagger$ respectively. Their conformal dimensions and transverse spins are related to those of the primaries as
\beq
\begin{aligned}
\Delta_{\psi^1} &=\Delta_\phi+{1\over2}\,, & \qquad \mathfrak{s}(\psi^1) &=\mathfrak{s}(\phi)+{1\over2}\,,\\
\Delta_{\bar\psi^2}&=\Delta_{\phi^\dagger}+{1\over2}\,,& \qquad \mathfrak{s}(\bar\psi^2)&=\mathfrak{s}(\phi^\dagger)-{1\over2}\,.
\end{aligned}
\eeq
Once combined with (\ref{eqn-SUSY-BPSalphaM1}), we find that
\beq\la{psi1barpsi2}
\Delta_{\psi^1}={1\over2}+\mathfrak{s}(\psi^1)\,,\qquad\Delta_{\bar\psi^2}={1\over2}-\mathfrak{s}(\bar\psi^2)\,.
\eeq

In addition to those relations, the two bottom bosonic operators with spin, $\mathcal{O}_L^{(0,1)} = D_+ \phi^\dagger$ and $\mathcal{O}_R^{(0,-1)} = D_- \phi$ are descendants of the primary fermion $\bar{\psi}^1$ and $\psi^2$ respectively, (\ref{eqn-SUSY-a2}). Therefore, their dimension and spin are also related as
\beq
\Delta_{D_- \phi} = \frac{1}{2} - \mathfrak{s}(D_- \phi)\,, \qquad \Delta_{D_+ \phi^\dagger} = \frac{1}{2} + \mathfrak{s}(D_+ \phi^\dagger) \, .
\eeq

\subsubsection{Summary}
In total, we find that the sign of $\alpha$ only enters the scalar relations
\beq
\Delta_\phi = \frac{1}{2}-\alpha\, \mathfrak{s}(\phi)\,, \qquad \Delta_{\phi^\dagger} = \frac{1}{2}+\alpha\, \mathfrak{s}(\phi^\dagger)\,,
\eeq
while the fermionic relations always take that form (\ref{psi2barpsi1}), (\ref{psi1barpsi2}). It follows that the anomalous dimension and the anomalous spin of all these boundary operators are the same (up to a sign), and that only the bosons are sensitive to the sign of $\alpha$. 

These results are in agreement with our explicit computations of the previous sections. However, as discussed above, due to potential anomalies it is not \textit{a priori} guaranteed that the classical analysis in this section also holds in the quantum theory. Using our results from sections \ref{bossec} and \ref{fersec} we can show that such anomalies do not occur. To show this we note that for any spacetime supersymmetry that is broken by the line, there is a protected fermionic operator on the line of dimension $3/2$ and transverse spin $1/2$, \cite{Agmon:2020pde}. Similarly to the protected operators discussed in section \ref{iohscc}, this operator is responsible for infinitesimal deformations of the line in the direction of the broken symmetry. In the case of the operators (\ref{Wbos}), we have two such operators 
\beq\la{Sigmaalpha1}
\Sigma_-=\phi\times\bar\psi_-\,,\qquad\Sigma_+=\psi_+\times\phi^\dagger\qquad\text{for}\qquad\alpha=1\,,
\eeq
and
\beq\la{Sigmaalpham1}
\widetilde\Sigma_+=\phi\times\bar\psi_+\,,\qquad\widetilde\Sigma_-=\psi_-\times\phi^\dagger\qquad\text{for}\qquad\alpha=-1\,.
\eeq
These two superconformal primary operators on the line (which are also BPS) correspond to the two broken supersymmetries. As the displacement operator, they are constructed such that the right and left anomalous dimensions cancel each other. The fact that there are only two such operators at finite coupling (and not four or more) implies that the conformal line operator only breaks two out of the four spacetime supercharges (at least at large $N$). 

From the spectrum of the boundary operators, we expect that the $\alpha=1$ be mapped back to itself under the Fermion-Boson duality. Flipping the sign of $\alpha$ has the effect of flipping the sign of the anomalous dimension for the boundary bosons. It however does not affect the anomalous dimension of the boundary fermions. Hence, under Fermion-Boson duality, the $\alpha=-1$ line operator cannot be mapped to itself.

Like the line operators, (\ref{Sigmaalpha1}), (\ref{Sigmaalpham1}), and the displacement operator, we also have protected bosonic operators at any nonzero integer spin and protected fermionic operators at any half integer spin. Their existence follows from the existence of high spin conserved charges and super charges that are only broken at order $1/N$, see section \ref{iohscc}.

\subsection{The Supersymmetric Condensed Fermion Line}
\label{sec:SUSY:condensed}

The condensed fermionic line operator considered in section \ref{sec:condFer} also has a lift into a $1/2$-BPS line operator in the ${\cal N}=2$ CS-matter theory. This supersymmetric line operator was considered before in the context of quiver gauge theories, with the fermion in the bi-fundamental representation \cite{Drukker:2009hy,Ouyang:2015iza,Lee:2010hk,Mauri:2018fsf, Drukker:2019bev}.\footnote{The construction of such fermionic type BPS line operators was first considered in the $\mathcal{N}=6$ ABJM theory \cite{Drukker:2009hy}.} 
The lift of the condensed fermionic line operator is obtained by taking the rank of one of the gauge groups to one. This limit was discussed previously in \cite{Drukker:2020opf}.

The starting point of the construction, following \cite{Drukker:2020opf}, is to embed the $\alpha = 1$ BPS Wilson line discussed in the previous sections into the upper diagonal of a $(2|2)$ super-connection,
\beq
\mathcal{L}_0 = 
\begin{pmatrix}
A_\mu \dot{x}^\mu - \ii \sigma |\dot{x}| &\quad 0 \\
0 &\quad 0 \\
\end{pmatrix}\, .
\eeq
This connection is preserved by $Q_2$ and $\bar{Q}_1$. Next, one constructs a family of (formally) SUSY line operators that are parameterized by a complex number $u$ as\footnote{
Note that the only physical parameter is the product of the coefficients in front of the fermion and the conjugate fermion in $\mathcal{L}_u$. Here, we have chosen them to be equal.}
\beq \label{eqn-SUSY-Ldeformed}
\cL_0\quad\rightarrow\quad\cL_u \equiv 
\mathcal{L}_0- \begin{pmatrix}
u^2 \, \phi \phi^\dagger & \ii u\, \psi^2  \\
\ii u\, \bar{\psi}^1 & u^2 \, \phi^\dagger \phi
\end{pmatrix} |\dot{x}|\,.
\eeq
To see that $\cL_u$ indeed leads to an $1/2$-BPS line operator, one note that
\beq
\cL_u=\cL_0-\(\ii \mathcal{Q}_- \mathcal{G}_{u}+\mathcal{G}_{u}^2\)|\dot x|\qquad\text{with}\qquad\mathcal{G}_{u}= 
\begin{pmatrix}
0 & u\, \phi \\
u\, \phi^\dagger & 0
\end{pmatrix}\,,
\eeq
where $\mathcal{Q}_\pm = Q_2 \pm \bar{Q}_1$. 
Under a SUSY variation, $\cL_u$ transforms into a total derivative
\beq \label{eqn-SUSY-condensedF-BdrTerm}
\mathcal{Q}_- \mathcal{L}_{u}= -\mathcal{D}_{\mathcal{L}_{u}} \mathcal{G}_{u}\,, \qquad
\mathcal{Q}_+ \mathcal{L}_{u} = -\mathcal{D}_{\mathcal{L}_{u}}
\begin{pmatrix}
0 & u \phi  \\
-u \phi^\dagger & 0
\end{pmatrix}\,,
\eeq
where $\mathcal{D}_{\mathcal{L}_{u}} $ is the covariant derivative with respect to the deformed connection $\cL_u$.

To see that $\cL_u$ is also superconformal we consider the conjugate charges $\mathcal{S}^\pm = (\mathcal{Q}_\pm)^\dagger = \bar{S}^2 \pm S^1$. Similarly to (\ref{eqn-SUSY-condensedF-BdrTerm}), their action on the deformed connection reads\footnote{More explicitly, notice the action of $\mathcal{S}$ on the deformed connection reads,
\beq
\begin{aligned}
\mathcal{S}^\pm \mathcal{L}_u & = \mathcal{S}^\pm \mathcal{L}_0 - \ii \mathcal{S}^\pm \mathcal{Q}_- \mathcal{G}_u - \{ \mathcal{S}^\pm \mathcal{G}_u, \mathcal{G}_u \}\, . \\
\end{aligned}
\eeq
The first term vanishes because the original $\alpha =1$ line is superconformal. For the second term, using the field transformation rules \eqref{eqn-SUSY-matter}, one finds 
\beq
- \ii\, \mathcal{S}^\pm \mathcal{Q}_- \mathcal{G}_u = -\ii
\begin{pmatrix}
0 & \pm u (x_s^3 \mathbb{D}_3 \phi + \phi) \\
-u (x_s^3 \mathbb{D}_3 \phi^\dagger + \phi^\dagger ) & 0 \\
\end{pmatrix}\, .
\eeq
The terms in the bracket nicely combine into a total derivative. For the last term, one notices that
\beq
\mathcal{S}^\pm \mathcal{G}_u = \pm \ii x_s^3 \mathcal{Q}_\pm \mathcal{G}_u \, .
\eeq
Summing over all the contributions leads to \eqref{eqn-SUSY-condensedF-actionS}.}
\beq \label{eqn-SUSY-condensedF-actionS}
\mathcal{S}^- \mathcal{L}_u = \ii\, \mathcal{D}_{\mathcal{L}_u} \Big[ x_s^3
\mathcal{G}_u
\Big], \qquad \mathcal{S}^+ \mathcal{L}_u = -\ii\, \mathcal{D}_{\mathcal{L}_u} \Bigg[ x_s^3
\begin{pmatrix}
0 & u \phi \\
-u \phi^\dagger & 0 \\
\end{pmatrix}
\Bigg] \, .
\eeq

The family of line operators $\{W_u=\cP\,\exp\int\cL_u\dd s\,,\ u\in{\mathbb C}\}$ are all formally superconformal. Changing $u$, however, corresponds to a deformation of $\cL_u$ by a classically marginal operator that preserves supersymmetry. We should therefore ask which, if any, of these operators is also superconformal in the quantum theory. To answer this question, we note that at leading order in $1/N$, the fermions do not contribute to the expectation value of the mesonic line operator 
\beq
M=(\phi^\dagger\ 0)\cdot\cP e^{\ii \int \cL_u \dd s}\cdot\begin{pmatrix}
\phi\\0\end{pmatrix}\,.
\eeq
It, therefore, takes the same value as the expectation value of the operator (\ref{M10}) in the bosonic CS-matter theory. This picks up the points
\beq
u^2 ={4\pi \ii \over k}\,,\qquad\text{and}\qquad u^2 = 
0\,,
\eeq
as the two quantum conformal operators, in correspondence with the $\alpha=-1$ and $\alpha=1$ operators respectively. 

As discussed above, to check whether the $u=\sqrt{4\pi\ii/k}$ line is superconformal at the quantum level, we have to count the protected fermionic operators on the line of dimension $3/2$ and transverse spin $\pm1/2$. At the tree level, there are six such operators. Two of them consist of the empty endpoints
\beq\la{twoop}
\begin{pmatrix}
D_- \phi \\
0 \\
\end{pmatrix} \times (0 \ 1)\,, \qquad \text{and} \qquad 
\begin{pmatrix}
0 \\
1 \\
\end{pmatrix} \times 
(D_+ \phi^\dagger\ 0) \, .
\eeq
As discussed in the next section, the anomalous dimensions and spins of these two operators cancel between their left and right factors. Hence, they remain of dimension $3/2$ and transverse spin $\pm 1/2$ at the quantum level. The other four take the form

\beq\la{fourop}
\begin{pmatrix}
 \widetilde{\cO}_R\\
 0 \\
\end{pmatrix} \times 
(\widetilde{\cO}_L\ 0)\,,\qquad\text{with}\qquad(\widetilde{\cO}_R,\widetilde{\cO}_L)\in\{(\phi,\bar{\psi}_\pm),\,(\psi_\pm,\phi^\dagger)\}\,.
\eeq
The anomalous dimensions of the right and left factors of these operators sum up, so they will obtain non-zero anomalous dimensions. We therefore conclude that the $u=\sqrt{4\pi\ii/k}$ line is a superconformal $1/2$-BPS line at the quantum level.

The reason for the anomalous dimensions of the operators in (\ref{fourop}) can be understood as follows. The condensation of the fermions has the effect of flipping the anomalous dimension of the two fermion components that are being condensed while leaving the other two components the same as for the normal Wilson line operator of section \ref{fersec}. Namely, the anomalous dimensions of operators $\psi_- (\psi^2)$ and $\bar{\psi}_+ (\bar{\psi}^1)$ flips the sign, while the anomalous dimensions of $\psi_+ (\psi^1)$ and $\bar{\psi}_- (\bar{\psi}^2)$ remains the same. After this flip, all fermion components on the condensed line have the same anomalous dimension, equal to $-\lambda/2$. Similarly, the condensation flips the sign of the scalar's anomalous dimension. This is because the upper-diagonal component of $\cL_u-\cL_0$ in \eqref{eqn-SUSY-Ldeformed} flips the sign of the coefficient in front of the bi-scalar, from $\alpha=1$ in $\cL_0$ to $\alpha=-1$ in $\cL_u$ with $u={\sqrt{4\pi\ii/k}}$. After this flip, the scalar's anomalous dimension is equal to $-\lambda/2$.

\subsubsection{Boundary Operators of the Condensed Fermion Line}

The lift of the tree-level boundary operators in table (\ref{tab_comp_cond}) to superconformal line construct from $\cL_u$ is
\beq\la{bocondenced}
\widetilde\cO_R^{(0,-{1\over2})}\to\begin{pmatrix}0\\1\end{pmatrix}\,,\quad
\widetilde\cO_R^{(0,-{3\over2})}\to
\begin{pmatrix} \d_-\psi_- \\ 0
\end{pmatrix}\,,\quad
\widetilde\cO_L^{(0,{1\over2})}\to\(0\ 1\)\,,\quad
\widetilde\cO_L^{(0,{3\over2})}\to\( \d_+\bar\psi_+\ 0\)\,.
\eeq

In appendix \ref{cfboapp}, we analyze the superconformal properties of these operators. We find that the operators $\widetilde\cO_R^{(0,-{1\over2})}$ and $\widetilde\cO_L^{(0,{1\over2})}$ are $1/2$-BPS superconformal primaries while the operators $\widetilde\cO_R^{(0,-{3\over2})}$ and $\widetilde\cO_L^{(0,{3\over2})}$ are superconformal descendants. The corresponding relations between their conformal dimensions and transverse spins are derived in the appendix. They read
\beq \label{eqn-SUSY-CondFer-Half}
\Delta(\widetilde{\cO}_R^{(0,-\frac{1}{2})}) = \frac{1}{2} +\mathfrak{s}(\widetilde{\cO}_R^{(0,-\frac{1}{2})})\,, \qquad \Delta(\widetilde{\cO}_L^{(0,\frac{1}{2})}) = \frac{1}{2}- \mathfrak{s}(\widetilde{\cO}_L^{(0,-\frac{1}{2})})\,,
\eeq
and
\beq \label{eqn-SUSY-CondFer-3Half}
\Delta(\widetilde\cO_R^{(0,-{3\over2})})=\frac{1}{2}-\mathfrak{s}(\widetilde\cO_R^{(0,-{3\over2})})\,,\qquad \Delta(\widetilde\cO_L^{(0,{3\over2})})=\frac{1}{2}+\mathfrak{s}(\widetilde\cO_L^{(0,{3\over2})})\,.
\eeq

\section{Mesonic Line Operators in the Mass Deformed CS-matter Theories}
\label{sec:MassDeform}

The recursion techniques used in this paper can also be applied to the mass-deformed version of the CS-matter theories studied in this paper. These theories are defined in the same way as the ones studied in the other sections, but with the physical mass not tuned to zero. This mass, $m$, is defined as the location of the pole in the exact propagator. Because it is the only dimensionful parameter in the theory, all non-zero masses are equivalent. 

In this section, we compute the expectation values of the mesonic line operators along a straight line of length $x$ for the massive case. We find that even though the $3d$ theory is gapped, for $\alpha=\pm1$, there is no RG flow on the line. Moreover, the mesonic line operators in the fermionic and bosonic theories are consistent with the bosonization duality, provided that we identify the dimensionless parameter $m^2x^2$ between the two theories.

\subsection{Bosonic Case}

\paragraph{Massive Boson Propagator}

For massive case, we chose the regularization scheme
\beq \label{deformedprop02}
\<\phi^\dagger_i(p)\phi^j(q)\>^\text{tree}_\epsilon\equiv(2\pi)^3\delta^3(p+q)\,{\delta_i^j\over p^2 +m^2}\times e^{\ii \epsilon\, p_3}\,.
\eeq
Thus, the exact propagator reads
\beq
\<\phi^\dagger_i(p)\phi^j(q)\>_\epsilon=(2\pi)^3\delta^3(p+q)\,{\delta_i^j\over p^2+m^2+e^{\ii\epsilon\, p_3}(m_{c.t.}^2-\Sigma(p))}\, e^{\ii \epsilon\, p_3}\, ,
\eeq
where $m$ is the physical mass. The Schwinger-Dyson equation determining $\Sigma$ is still given in figure \ref{fig:self-energy}. One can also include the $\lambda_4$ and $\lambda_6$ type vertex. Diagrammatically, they are given by the standard $\lambda \phi^4$ and $\lambda \phi^6$ bubble-type vertices. Importantly, in the planar limit those contributions are \textit{independent} of the momenta of the self-energy. Therefore, in any scheme, the effect of $\lambda_4$ and $\lambda_6$ on the self-energy can be absorbed into a shift of the mass counter-term. Hence, we shall not take them into account explicitly in the following.

We can now apply the differential equation techniques described in the previous section \eqref{sec:ferselfEnergyDiff}. Notice that the rotational symmetry in the transverse plane determines $e^{\ii \epsilon p_3}(m_{c.t}^2-\Sigma(p) ) \equiv f(p_\perp;\lambda,m) $ to be a function of $|p_\perp|$ only. Taking differentials with respect to $p^-$, one finds,
\beq
\frac{\partial}{\partial p^-} \Sigma(p) = \frac{\partial}{\partial p^-} \Big( (a)+(b)+(c) \Big) = 0\, .
\eeq
This implies $\Sigma(p)$ is not a function of $p^-$. As the $p^-$ dependence of $\Sigma(p)$ is only trough $p_\perp^2 = 2p^+ p^-$, we conclude that $\Sigma(p)$, and thus $f$, are \textit{independent} of $|p_\perp|$.

The computation then simplifies drastically, as one can take $p^\pm =0$ in the Schwinger-Dyson equation. In this limit, diagrams (b) and (c) vanish because the integrands of these diagrams are antisymmetric with respect to exchanging of the two loop momenta. Diagram (a) then evaluates to 
\beq
(a) = -\frac{\lambda^2}{4\epsilon^2} e^{-\epsilon \sqrt{m^2 - f(\lambda, m)}}\, .
\eeq
The Schwinger-Dyson equation now reads,
\beq
\Sigma(p)=e^{-\ii\epsilon p_3}f(\lambda,m)+m_{c.t.}^2=(a)+(b)+(c)= -\frac{\lambda^2}{4\epsilon^2} e^{-2\epsilon \sqrt{m^2 - f(\lambda,m)}}\ ,
\eeq
As there is no momenta dependence on the RHS, one has to take
\beq
f(\lambda,m)=0, \qquad m^2_{c.t.}=-{\lambda^2\over4\epsilon^2} e^{-2\epsilon |m|}\,.
\eeq
One conclude that the exact massive propagator is given by 
\beq
\<\phi^\dagger_i(p)\phi^j(q)\>_\epsilon=(2\pi)^3\delta^3(p+q)\,{\delta_i^j\over p^2+m^2}\, e^{\ii \epsilon\, p_3}\, ,
\eeq
and the tree-level mass $m$ becomes the physical pole mass.

\paragraph{Massive Line Integrand.}

The generalization of the analysis in section \ref{sec:higher} is straightforward. Equation \eqref{higherloop} becomes,
\begin{align}
\label{mass-eqn-IntegrandNUniVersal}
\cM_L=&\,  (-2\pi\lambda)^L \int\limits_0^1 \dd t_L \int\limits_0^{t_L} \dd t_{L-1} \cdots \int\limits_0^{t_2} \dd t_1 \int \prod_{j=0}^L \frac{\dd^3 k_j}{(2\pi)^3}\times\\
&\frac{e^{\ii k_0^3x(t_1+\tilde\epsilon)}}{k_0^2+m^2} \vv(k_0,k_1)\frac{e^{\ii k_1^3x(t_{21}+\tilde\epsilon)}}{k_1^2+m^2}\ \cdots\ \frac{e^{\ii k_{L-1}^3x(t_{L L-1}+\tilde\epsilon)}}{k_{L-1}^2+m^2}\vv(k_{L-1},k_L)\frac{e^{\ii k_L^3x(t_{L+1 L}+\tilde\epsilon)}}{k_L^2+m^2}\,.\nn
\end{align}
This leads to a simple modification of the integral necessary for the recursion \eqref{eqn-Rec-k1int},
\begin{multline}\label{mass-eqn-Rec-k1int}
\int \frac{\dd^3 k_1}{(2\pi)^3}  \frac{k_1^+\,e^{\ii k_1^3 x\, t_{21}}}{(k_1^2+m^2)(k_0-k_1)^+(k_1-k_2)^+} \\
= \frac{1}{4\pi x\ t_{21}} \frac{1}{(k_0-k_2)^+} \left(e^{-\sqrt{k^2_{2\perp}+m^2}x\,t_{21}} -e^{-\sqrt{k^2_{0\perp}+m^2}x\,t_{21}} \right)\,,    
\end{multline}
The identification of $k_j^3$ is with $\ii\sqrt{k_{j\perp}^2+m^2}$. Under this identification, the exponential term in \eqref{mass-eqn-Rec-k1int} combines with $e^{\ii k_2^3 x\, t_{32}}\simeq e^{-\sqrt{k_{2\perp}^2+m^2}x\, t_{32}}$ or $e^{\ii k_0^3 x\, t_{10}}\simeq e^{-\sqrt{k_{2\perp}^2+m^2}x\, t_{10}}$ in (\ref{mass-eqn-IntegrandNUniVersal}), 
to induce exactly the same recursion relation as in the massless case \eqref{eqn-Rec-Univ} and \eqref{eqn-Rec-SeedBos00}. This also leads to the same solution to the recursion, up to a modification of the seed term,
\begin{equation}
    \mathtt{B}_{m^2>0}^{(0)}(t) = \int \frac{\dd^3 k}{(2\pi)^3} \frac{e^{\ii k^3 x t}}{k^2 + m^2} = \frac{e^{-|m| x\, t}}{4 \pi\, x\, t}\ .
\end{equation}
Hence, similarly to \eqref{m10vev}, we find that
\beq \label{mass-m10vev}
\langle M_{10} \rangle_{m^2>0} ={e^{-|m| x}\over4\pi x}\times
\begin{dcases}
{1\over\Gamma (1-\lambda)}\Big({\epsilon\,e^{\gamma_E}\over x}\Big)^\lambda \,, \quad\ \  \alpha = +1, \\[5pt]
{1\over\Gamma (1+\lambda)}\Big({\epsilon\,e^{\gamma_E}\over x}\Big)^{-\lambda}\,, \quad \alpha = -1\,. \\
\end{dcases}
\eeq
We see that there is no RG flow taking place on the line. 

\paragraph{Derivative Operator.}

Following the same logic, the recursion for the derivative operator can also be generalized to the massive case, and the only modification is the recursion seed. 

To verify the duality, we present the result for the recursion seed here. Explicitly, the recursion seed of the operator $\<M_{10}^{(1,-1)}\>$
can be obtained by replacing all propagators in figure \ref{fig:dseed} with the massive propagator. After some algebra, one finds,
\beq
\mathtt{B}^{(0,-1)}(t)= \frac{e^{- |m| t x}}{4\pi (x\, t)^3}\times \frac{(\lambda+1)(\lambda+2+2 |m| t\, x)}{2} + \mathcal{O}(\tilde{\epsilon}) \,.
\end{equation}
The other seed term $\mathtt{B}^{(0,1)}(t)$ is obtained by flipping the sign of $\lambda$. Hence, similar to \eqref{M1m1prefactor}, to leading order in $\te$, we find
\begin{align} 
\<M_{10}^{(1,-1)}\>&=\frac{e^{- |m| x}}{4\pi x^3}\times \(\frac{\epsilon\,e^{\gamma_E  }}{x}\)^{-\lambda} \times  \frac{(1+\lambda)(2+\lambda+2 |m| x)}{2 \Gamma (1+\lambda)} \,,\\
\<M_{10}^{(-1,1)}\>&=\frac{e^{- |m| x}}{4\pi x^3}\times  \(\frac{\epsilon\,e^{\gamma_E  }}{x}\)^{+\lambda} \times \frac{(1-\lambda)(2-\lambda+2 |m| x)}{2 \Gamma (1-\lambda)} \,.\nn
\end{align}

\subsection{Fermionic Case}

The recursion relation can also be applied to the massive fermionic case.

\paragraph{Massive Fermion Propagator}
The symmetry argument that leads to the Schwinger-Dyson equation \eqref{eqn-Fer-Selfenergy01} and \eqref{eqn-Fer-Selfenergy02} still applies to the massive case, so the equations to be solved are still given by these two.
Assuming the exact propagator has a mass pole at $m$, we can generalize the condition \eqref{Sigmaguess} as\footnote{Note that the mass pole condition only determines $|m|$. However, we can reproduce the sign by agreement with the free mass when $\lambda \to 0$. The duality map implies that $m_{\mathrm{bos}}=-m_{\mathrm{fer}}$ \cite{Minwalla:2015sca}.}
\beq 
\(p_\mu + \Sigma_\mu(p)\)^2 +\Sigma_0(p)^2 = p^2 + m^2 \, .
\eeq
The non-vanishing components of the fermionic self-energy then read
\beq
\begin{aligned}
\Sigma_0 & = |m| - \frac{\lambda}{\epsilon}\Big(e^{-\epsilon |m|} -e^{-\epsilon \sqrt{|p_\perp|^2 + m^2} } \Big)\, ,\\
& \simeq |m| - \lambda \Big(\sqrt{|p_\perp|^2 + m^2} -|m| \Big)+ \cO(\epsilon^2) \,, \\
\Sigma_+ 
& = - \frac{\lambda}{2\epsilon^2} \Big(e^{-\sqrt{p_\perp^2 + m^2}}-e^{-|m|\epsilon}\Big) \Big(2|m|\epsilon + \lambda e^{-\sqrt{p_\perp^2 + m^2}}-\lambda e^{-|m|\epsilon}\Big)\\
& \simeq \frac{p^-}{p_\perp^2} \Big[ |m| \lambda \big( \sqrt{p_\perp^2 + m^2} - |m| \big) + \lambda^2 \Big( |m| \big( \sqrt{p_\perp^2 + m^2} - |m| \big) - \frac{p_\perp^2}{2} \Big) \Big]  + \cO(\epsilon^2) \, .
\end{aligned}
\eeq

\paragraph{Massive Line Integrand.}
The fermionic recursion relation stays the same form as for the massless case, for the same reason as for the bosonic case. The first two recursion seed terms now reads
\beq
\begin{aligned}
\big(\mathtt{F}^{(0)}(t)\big)_{-+} & = - \frac{e^{-|m|x t }}{4 \pi x^2\,t^2}\(1+ \frac{\lambda t}{t+ \te} e^{-|m|\epsilon} \)\, , \\ 
\big(\mathtt{F}^{(0)}(t)\big)_{+-} & = + \frac{e^{-|m| x t}}{4 \pi x^2\,t^2}\(1+ 2 |m| t x  - \frac{\lambda t}{t+ \te} e^{-|m|\epsilon}\)\, . \\ 
\end{aligned}
\eeq
When inserting the initial value $t = 1 + L \epsilon$, to leading order, one finds,
\beq
\begin{aligned}
\big(\mathtt{F}^{(0)}(t)\big)_{-+}  & \simeq -\frac{e^{-|m| x}}{4 \pi x^2} (1+\lambda ) + \cO(\epsilon) \, , \\
\big(\mathtt{F}^{(0)}(t)\big)_{+-} & \simeq +\frac{e^{-|m| x}}{4 \pi x^2} (1-\lambda+ 2|m| x)+ \cO(\epsilon)\, . \\ 
\end{aligned}
\eeq
As a consequence, the expectation values of the massive mesonic lines read\footnote{The new $|m|x$ dependence of  $\<M^{(-{1\over2},{1\over2})}\>$ might look confusing, but it is needed for the duality. Recall from the dictionary \eqref{tab_comp} that this operator is dual to the bosonic operator with derivatives. At the tree-level, the massive bosonic operator with derivatives is proportional to 
\begin{equation*}
    \propto \int \frac{\dd^3 p}{(2\pi)^3} e^{\ii p_3 x} \frac{p^+ p^-}{p^2+m^2} \propto \frac{e^{-|m| x}}{x^3} (1+ |m| x)\, ,
\end{equation*}
therefore the dual fermionic operator must involve the $|m|x$ dependence.
}
\beq 
\begin{split}\la{Yukawa2}
    \<M^{({1\over2},-{1\over2})}\>
&=-{e^{-|m| x}\over4\pi x^2}\Big({\epsilon\, e^{\gamma_E}\over x}\Big)^{+\lambda} \ \frac{1+\lambda}{\Gamma (1-\lambda)}+\cO(\te)\,,\\
\<M^{(-{1\over2},{1\over2})}\>
&= + {e^{-|m| x}\over4\pi x^2}\Big({\epsilon\, e^{\gamma_E}\over x}\Big)^{-\lambda}\ \frac{1-\lambda+ 2|m| x}{\Gamma (1+\lambda)} + \cO(\te)\,.
\end{split}
\eeq

\paragraph{Derivative Operators.}

The recursion relation also applies to the derivative cases. The only modification is the recursion seed. 

For instance, one can modify the propagator in $\mathtt{Figure}\, \ref{fig:ferderseed}.a$ and $\mathtt{Figure}\, \ref{fig:ferderseed}.b$ correspondingly. After some algebra, one obtains
\beq
\mathtt{F}^{(0,-{3\over2})}_{-+}(t) = -\frac{e^{-|m| t x}}{4 \pi(t x)^4} \times \frac{(1+\lambda)(2+\lambda)(3+\lambda + 2 |m| x t)}{2} +\cO(\te)\,.
\eeq
Therefore, 
\beq\la{Yukawa3}
\<M^{({3\over2},-{3\over2})}\> =-{e^{-|m| x}\over4\pi x^4}\(\frac{\epsilon\,e^{\gamma_E}}{x}\)^{\lambda}{1\over\Gamma (1-\lambda)} \frac{(1+\lambda)(2+\lambda)(3+\lambda + 2 |m| x)}{2} \, .   
\eeq

\paragraph{Condensed Fermion Operator.}

The derivation of the condensed fermionic operator can also be repeated. First, we observe that in the presence of mass, the definition of the condensed line operator must be modified. Repeating the derivation of the displacement operator \eqref{docf} is one way to see this. All the steps go through as before, with one exception: the fermion equation of motion that are unaffected by the condensation \eqref{eqn-EOMferCondensedUnaffected} must now include the mass term contribution on the right-hand side, leading to
\beq
{\mathbb D}_+ \propto \(\begin{array}{cc}0&m \psi_+\\ D_+\bar\psi_+&0 \end{array}\)\, , \qquad 
{\mathbb D}_- \propto
\(\begin{array}{cc}
0&D_-\psi_-\\
m \bar{\psi}_-&0 
\end{array}\)
\, .
\eeq
The additional off-diagonal term
violates the condition that $\mathbb{D}_\mu$ has a quantum dimension two in the UV, provided that $m$ has dimension one.\footnote{Note that $\psi_+$ has the opposite anomalous dimension than that of $D_+ \bar{\psi}_+$.} Therefore, the naive definition of the line does not result in a good operator, and $m$-dependent terms are expected to be added to the definition of the condensed line to cancel out these unwanted contributions.

The massive modification of operator \eqref{condencedfcon} is simple to implement: merely scale the topological spin connection in the empty part by the total length as
\beq
U_{\mathcal{C}}(s,t) \rightarrow U_{\mathcal{C}}(s,t) \times \exp \Big(-\int\limits_{\mathcal{C}} |m| |\dd x| \Big)\, .
\eeq
Under this modification, each filled or empty component of the condensed fermionic line operator contributes a factor proportional to the length of the corresponding segment.  These factors add up to a factor $\exp(-|m|x)$ proportional to the total length of the condensed line.

The argument in section \ref{sec:condFer} now can be generalized to the massive case. After factorizing out the $\exp(-m t)$ factor, the $L$-ladder integrand now reads,
\footnote{Recall that the $t_j$ terms are shifted by $j+\te$; after shifting back and factoring out the total length dependence, there will be a $L$-dependent exponential remainder factor.}
\begin{multline}
\big(\mathtt{F}^{(L)}(t_{L+1})\big)_{-+} = - \frac{e^{-|m|x (L+1)\epsilon}}{4 \pi x^2}\Big(\frac{1}{(t_{L+1}+(L+1)\te)^2} \\
+ \frac{\lambda e^{-|m|\epsilon} }{(t_{L+1}+(L+1)\te)(t_{L+1}+(L+2)\te)} \Big) \prod_{j=1}^L \frac{-\lambda}{t_j+j \te}\, . 
\end{multline}
Integrating the first term yields two terms, with the term proportional to $t_L$ canceling out the second term in the previous order. The argument in section \ref{sec:condFer} can now be repeated \textit{verbatim}. The value of the contact term coefficient $f$ and the coefficient $b$ now becomes,
\beq
f = \lambda b e^{-|m| \epsilon}, \qquad b = 1\, .
\eeq
The expectation value of the condensed line picks up an additional exponential factor compared with the massless case,
\begin{equation}\la{Yukawa4}
    \<\widetilde{M}^{({1\over2},-{1\over2})}\> = {e^{-|m|x} \over \Gamma (1+\lambda)}\Big({\epsilon\,e^{\gamma_E}\over x}\Big)^{-\lambda} + \cO(\te)\, .
\end{equation}

The form of the displacement operator remains the same as \eqref{docf}. The only change in the derivation is that the unaffected fermion equation of motion \eqref{eqn-EOMferCondensedUnaffected} now depends on the mass term, which was then canceled by the contribution from the deformed parallel transport.

\subsection{Evolution and Boundary Equations}

The mass term does not affect the gluon equations of motion. Therefore, the bosonic and fermionic evolution equations remain the same as in the conformal case and are given by \eqref{displacementop} and \eqref{ee_fer} respectively. The boundary equation, however, is modified in a simple way, as we derive below. 

\paragraph{The Bosonic Boundary Equation.}

Since the tree-level scalar equation of motion is deformed by the mass term, the boundary equations must also be modified. The derivation of the quantum boundary equation generalizes to the massive case in a straightforward way. First, we notice that the identity \eqref{eq:genIntOneSide} generalizes to the massive case as
\beq
\int \frac{\dd^3 k_1}{(2\pi)^3} \frac{(k_1^+)^a}{(k_1^2+m^2)(k_0-k_1)^+} e^{\ii k_1^3 x t} = - \frac{(k_0^+)^{a-1}}{4\pi x t} e^{- \sqrt{k_{0\perp}^2 +m^2}|x\,t}\,.
\eeq
Therefore, all the identifications of the plus component of loop momenta made in sec \eqref{beqb} still applies to the massive case.  

Replacing the scalar propagators in 
\eqref{eqn-BdEOM-VPM} and \eqref{eqn-BdEOM-V33} by the massive propagator and repeating the steps, one finds that the $\cO(\lambda)$ and $\cO(\lambda^2)$ contributions cancels exactly, while the tree-level part is cancelled by adding a vertex proportional to the mass,
\beq
\mathbf{V}_{m^2 \cO_L^{(0,-s)}} = -m^2 (\ii q^+)^s\, .
\eeq
The quantum equations of motion becomes,
\beq \label{zerovertexMass}
2\mathbf{V}_{\delta_+ \delta_- \cO_L^{(0,-s)}}(k_L)+\mathbf{V}_{\delta_3^2 \cO_L^{(0,-s)}}(k_L) + \mathbf{V}_{m^2 \cO_L^{(0,-s)}}=0\,,
\eeq
where the equality is understood to hold under the integration. This implies the quantum equation of motion holds 
\beq \la{beqbMass}
\cO_L^{(2,-s)}+2\delta_+\cO_L^{(0,-s-1)} - m^2 \cO_L^{(0,-s)}=0\,.
\eeq
The form stays the same as for the tree-level equation of motion. The other boundary equation is modified in the same way.

\paragraph{The Fermionic Boundary Equation.}

The fermionic boundary equation is also modified as for the tree-level equation of motion. All the arguments that leads to \eqref{eqn-FerBEOM} can be applied to the massive case, with one minor modification: the identification $q_3 \rightarrow \ii |q_\perp|$ under the integral sign now becomes $q_3 \rightarrow \ii \sqrt{q_\perp^2 +m^2}$. One therefore conclude that
\beq 
\cO_L^{(2,-s)}+2\delta_+\cO_L^{(0,-s-1)}- m^2 \cO_L^{(0,-s)}=0\,.
\eeq

\subsection{Duality Check}

For the massive case, the ratio \eqref{distpf} 
\begin{equation*}
{\< \mathcal{O}_L\,\cW_{1s}\,{\mathbb D}_\mu(s)\,\cW_{st}\,{\mathbb D}^\mu(t)\,\cW_{t0}\,\mathcal{O}_R\>\over\< \mathcal{O}_L\,\cW_{10}\,\mathcal{O}_R\>}\, 
\end{equation*} 
is still a normalization-independent physical quantity that one could match.
However, since we do not have conformal invariance, this ratio does not take a simple form as it does in (\ref{distpf}). 

\paragraph{$\alpha =1$ Bosonic Case.}

For the same reason as for the massless case, for $\alpha=1$, a non-zero expectation value is obtained for $s_R=-s_L=0$ or $s_R=-s_L=1$ only, being the bottom operators in (\ref{bottom}).  In the former case, only the $(\mu,\nu)=(-,+)$ term in (\ref{predistpf}) contributes and the ratio reads ($\Delta =(1+\lambda)/2$)
\begin{multline} \label{MtoLambdaMass1}
{\< \mathcal{O}_L\,\cW_{1s}\,{\mathbb D}_\mu(s)\,\cW_{st}\,{\mathbb D}^\mu(t)\,\cW_{t0}\,\mathcal{O}_R\>\over\< \mathcal{O}_L\,\cW_{10}\,\mathcal{O}_R\>} \\
= {1\over x_{st}^4}\({x_{10}x_{st}\over x_{1s}x_{t0}}\)^{\Delta} {\sin(\pi  \lambda)\over\pi}(1-\lambda ) \lambda (2 +2 |m| x_{st}-\lambda )\, .      
\end{multline}

In the latter case, only the $(\mu,\nu)=(+,-)$ term contributes. The result is ($\Delta =(3-\lambda)/2$)
\begin{multline} \label{MtoLambdaMass2}
{\< \mathcal{O}_L\,\cW_{1s}\,{\mathbb D}_\mu(s)\,\cW_{st}\,{\mathbb D}^\mu(t)\,\cW_{t0}\,\mathcal{O}_R\>\over\< \mathcal{O}_L\,\cW_{10}\,\mathcal{O}_R\>} \\
= {1\over x_{st}^4}\({x_{10}x_{st}\over x_{1s}x_{t0}}\)^{\Delta} {\sin(\pi  \lambda)\over\pi}(1-\lambda ) \lambda \frac{(2 +2 |m| x_{1s}-\lambda )(2 +2 |m| x_{t0}-\lambda )}{(2 +2 |m| x_{10}-\lambda )}\, .     
\end{multline}

\paragraph{Fermionic Case.}

One can also compute the same ratio in fermionic theories. There are two nontrivial contributions. For $\mathcal{O}_L = \bar{\psi}_+$ and $\mathcal{O}_R = \psi_-$, we find 
\begin{multline} \label{MtoLambda_ferMass1}
{\< \mathcal{O}_L\,\cW_{1s}\,{\mathbb D}_\mu(s)\,\cW_{st}\,{\mathbb D}^\mu(t)\,\cW_{t0}\,\mathcal{O}_R\>\over\< \mathcal{O}_L\,\cW_{10}\,\mathcal{O}_R\>} \\
= {1\over x_{st}^4}\({x_{10}x_{st}\over x_{1s}x_{t0}}\)^{\Delta} {\sin (\pi  \lambda) \over\pi}\lambda(\lambda+1)  (1+2|m| |x_{st}|-\lambda)\, , 
\end{multline}
which matches with the \eqref{MtoLambdaMass1}, provided that we identify $m^2 x^2_{st}$ between the two theories.

The other one ratio is obtained by taking $\mathcal{O}_L = \bar{\psi}_-$ and $\mathcal{O}_R = \psi_+$, we find 
\begin{multline} \label{MtoLambda_ferMass2}
{\< \mathcal{O}_L\,\cW_{1s}\,{\mathbb D}_\mu(s)\,\cW_{st}\,{\mathbb D}^\mu(t)\,\cW_{t0}\,\mathcal{O}_R\>\over\< \mathcal{O}_L\,\cW_{10}\,\mathcal{O}_R\>} \\
= {1\over x_{st}^4}\({x_{10}x_{st}\over x_{1s}x_{t0}}\)^{\Delta} {\sin(\pi  \lambda)\over\pi}(1+\lambda ) \lambda \frac{(1 +2 |m| x_{1s}-\lambda )(1 +2 |m| x_{t0}-\lambda )}{(1 +2 |m| x_{10}-\lambda )} \, , 
\end{multline}
which matches with the \eqref{MtoLambdaMass2}, provided that we identify $m^2 x^2$ between the two theories.

\paragraph{Infinite Mass Limit.}

Interestingly, at $x|m| \to \infty$ limit, the ratio \eqref{MtoLambda_ferMass2} becomes,
\beq
\lim_{x|m| \to\infty}{\< \mathcal{O}_L\,\cW_{1s}\,{\mathbb D}_\mu(s)\,\cW_{st}\,{\mathbb D}^\mu(t)\,\cW_{t0}\,\mathcal{O}_R\>\over\< \mathcal{O}_L\,\cW_{10}\,\mathcal{O}_R\>}
= {|m|\over x_{st}^3}\({x_{10}x_{st}\over x_{1s}x_{t0}}\)^{\Delta-1}{\sin(\pi  \lambda)\over\pi}(1+\lambda ) \lambda \, .
\eeq
This result resembles a non-trivial, $\lambda$-dependent conformal result, wherein the dimensions are shifted by half a unit. It suggests that in the IR limit and after dressing out the trivial $\exp(-|m|x)$ factor, we remained with a non-trivial CFT on the line. 
Further investigation of this limit would be of great interest. In addition, it would be intriguing to determine if the unconventional form of the planar S-matrix, as studied in \cite{Jain:2014nza, Mehta:2022lgq, Gabai:2022snc}, is related to the Yukawa-like potential in (\ref{mass-m10vev}), (\ref{Yukawa2}), (\ref{Yukawa3}), and (\ref{Yukawa4}).

\acknowledgments

We thank Yunfeng Jiang for his collaboration during the initial stages of this work. We are grateful to Nathan Agmon, Ofer Aharony, Nissan Itzhaki, Zohar Komargodski, Yifan Wang, and Shimon Yankielowicz for useful discussions. We thank Zohar Komargodski for comments on the manuscript. AS and DlZ are supported by the Israel Science Foundation (grant number 1197/20). BG is supported by the Simons Collaboration Grant on the Non-Perturbative Bootstrap and by the DOE grant DE-SC0007870. BG would like to thank the Galileo Galilei Institute for its hospitality. DlZ would like to thank the support of the Simons Foundation for the Varna Workshop organized by the International Center for Mathematical Sciences in Sofia.

\appendix

\section{The Mesonic Wilson Line in the Bosonic Theory at 1-loop Order} \label{apd:oneLoopGen}

In this section, we present the one-loop 
expectation value of mesonic line operator (\ref{M10}) in the bosonic theory along an arbitrary smooth path $\cC$ between $x_0$ and $x_1$,
\beq
\cM[\cC,n]\equiv\<\phi^\dagger(x_1)\Big[{\cal P}e^{\ii\int\limits_\cC A\cdot \dd x+ \frac{2\pi \ii \alpha}{k} \phi \phi^\dagger |dx|  }\Big]_n\phi(x_0)\>\,.
\eeq
For non-self-intersecting paths and for $\alpha=0$, the one-loop result is finite. For $\alpha\ne0$, it only diverges near the endpoints of the line. Accordingly, the RG flow on the line only starts at two-loop order. The one-loop result, of course, depends on the path. Here, we express it as single and double integrals along the path, with no bulk integration. For $\alpha=0$, we find that the result coincides with the one bootstrapped in section \ref{sec:bootstrap}.

At tree level, the expectation value of the line operator is given by 
the free scalar propagator,
\beq
{\cal M}[\mathcal{C},n]_\text{tree}= \frac{1}{4\pi} \frac{1}{|x_0-x_1|}\,.
\eeq
It is thus independent of the path.

At one-loop, regardless of the choice of the gauge, there are three non-trivial gluon exchange diagrams -- gluon exchange on the Wilson line, gluon exchange on the scalar propagator, gluon exchange between the Wilson line and the scalar propagator, see figure \ref{fig:1loop}. The fourth contribution, \ref{fig:1loop}(d), comes from the bi-scalar deformation and is equal to an integrated product of two free scalar propagators. In the following, we analyze them one by one in Lorentz gauge.

\subsection{Feynman Rules in Lorentz Gauge}
\label{apd:1loopLorentzFeynRules}

We work in Lorentz gauge ($\partial^\mu A_\mu =0$) and evaluate the diagrams in position space directly. The standard gauge-fixing procedure leads to the following gauge-fixing and Faddeev-Popov ghost terms \cite{Guadagnini:1989kr},
\beq
\begin{aligned}
S_{\text{GF}} & = \frac{\ii k}{2\pi} \int \dd^3 x \, \mathrm{tr} A_\mu \partial^\mu B, \qquad
S_{\text{FP}} = \int \dd^3 x \,  \tr{\partial_\mu c^\dagger D^\mu c} \, .
\end{aligned}
\eeq
Here the auxiliary scalar field $B$ is the Lagrange multiplier imposing the gauge condition, and $c$, $c^\dagger$ are complex anti-commuting ghost fields.

The corresponding free propagators are
\beq \label{eqn-Lorentz-FeynRules}
\begin{aligned}
\langle \phi^\dagger_j(x) \phi^i(y) \rangle & = + \frac{1}{4\pi} \delta^{i}_{j} \frac{1}{|x-y|} \, ,\\ 
\langle A_\mu^I(x) A_\nu^J(y) \rangle & = -\frac{\ii}{k} \delta^{IJ} \epsilon_{\mu \nu \rho} \frac{(x-y)^\rho}{|x-y|^3} = \frac{\ii}{k} \delta^{IJ} \epsilon_{\mu \nu \rho} \partial^\rho_x \frac{1}{|x-y|} \, ,\\
\langle A_\mu^I(x) B^J(y) \rangle & = -\frac{\ii}{k} \delta^{IJ} \frac{(x-y)^\mu}{|x-y|^3} \, .\\
\end{aligned}
\eeq

Detailed perturbative analysis shows that there are no loop corrections to the gluon propagator up to two-loops \cite{Chen:1992ee}. 

\subsection{Gluon Exchange on the Wilson line}

The gluon exchange diagram on the Wilson line is plotted in figure \ref{fig:1loop}(a).
We parametrize the smooth non-self-intersecting Wilson line $x_s^\mu=x^\mu(s)$ with $\dot{x}_s^3 > 0$. For definiteness, we set $0<s<1$
unless otherwise stated.
Using the Feynman rules \eqref{eqn-Lorentz-FeynRules}, we obtain the expression\footnote{The factor of one quarter in (\ref{eqn-Lorentz-adiag}) has resulted from the normalization $\tr (T^I T^J)=\frac{1}{2}\delta^{IJ}$ and from converting ordered integrals to full integrals.}
\beq \label{eqn-Lorentz-adiag}
\begin{aligned}
\mathcal{M}_{a} 
=& {\cal M}[\gamma]_\text{tree} \frac{\ii \lambda}{4} \int\limits_0^1 \dd s\int\limits_0^1 \dd t \, \epsilon_{\mu \nu \rho} \frac{(x_s-y_t)^\rho}{|x_s-y_t|^3} \dot{x}_s^\mu \dot{y}_t^\nu.
\end{aligned}
\eeq
As discussed around equation (\ref{eqn-Bos-ASpinFactor}), the Wilson line is defined using framing regularization prescription. At one loop order, it amounts to deforming one of the two integration points in (\ref{eqn-Lorentz-adiag}) as $y_t=x_t+\epsilon\, n_t$, where $n$ is the framing vector and $\epsilon \to 0$, see \cite{Witten:1988hf} for details.

For convenience, we shall now review some basic properties of this integral. We divide the integral above to two distinct finite pieces, the Writhing number and the Twisting number.

\paragraph{Writhing Number.}

The unregularized integral in (\ref{eqn-Lorentz-adiag}) is called the ``Writhing Number''. This expression is finite but not topological.

\paragraph{Twisting Number.}

Although the region of integration in (\ref{eqn-Lorentz-adiag}) where $x_t \rightarrow x_s$ is not divergent, it contains a well-defined finite piece. This so-called ``Twisting Number'' is extracted from the framed integral by integrating only the infinitesimal neighborhood of the collision point.
For computational convenience, let us switch to the arc-length parameterization for the moment by setting $\dd s = \sqrt{\dd x^\mu \dd x_\mu}$. The upper limit of integration is the length of the curve $L_\gamma$. In the vicinity of $t \rightarrow s$, we can approximate $x_t = x_s + (t-s)  \dot{x}_s + \mathcal{O}((t-s)^2)$, 
\beq \label{eqn-apd-relTorsionLorentz}
\begin{aligned}
\mathcal{T}_a=&\lim_{\epsilon,\delta\to0} \int\limits_0^{L_\gamma}\dd s\int\limits_{s-\delta}^{s+\delta} \dd t \, \epsilon_{\mu \nu \rho} \frac{(x_t-x_s + \epsilon n_s)^\rho}{|x_s-x_t- \epsilon n_s|^3} \dot{x}_s^\mu (\dot{x}_t^\nu+ \epsilon \dot{n}_t^\nu)\\
=&\lim_{\epsilon,\delta\to0} \int\limits_0^{L_\gamma} \dd s\int\limits_{s-\delta}^{s+\delta} \dd t \epsilon_{\mu \nu \rho}\, \frac{\epsilon^2 n_s^\rho}{|(t-s)\dot{x}_s+ \epsilon n_s|^3} \dot{x}_s^\mu \dot{n}_s^\nu\\
= &\lim_{\epsilon,\delta\to0} \int\limits_0^{L_\gamma}\dd s\, \epsilon_{\mu \nu \rho} \dot{x}_s^\mu \dot{n}_s^\nu n_s^\rho \int\limits_{s-\delta}^{s+\delta} \dd t  \frac{\epsilon^2 }{[(t-s)^2+ \epsilon^2 ]^{\frac{3}{2}}} = 2 \int\limits_0^{L_\gamma}\dd s\, \epsilon_{\mu \nu \rho} \dot{x}_s^\mu \dot{n}_s^\nu n_s^\rho \ ,
\end{aligned}
\eeq
where the $\epsilon \to 0$ is taken first. Here, we have used that in arc-length parameterization $|\dot{x}_s| = |n_s| \equiv 1$. We see that $\mathcal{T}_a$ depends on the framing regularization. 

The quantity $\mathcal{T}_a$ is also called the \textit{relative torsion} of the curve. Its geometric meaning is discussed in appendix \ref{sec:topoPT}.

\subsection{Scalar Self-energy}\la{scalarseappendix}

The second diagram is a gluon exchange on the scalar propagator (self-energy); see figure \ref{fig:1loop}(b). This 1-loop diagram does not depends on the shape of the Wilson loop. It reads
\be
\mathcal{M}_{b}=&\frac{\ii^3 2\lambda}{(4\pi)^3} \int \dd^3 y_1 \dd^3y_2\,\epsilon_{\mu\nu\rho}\d^\rho_{y_1}{1\over|y_1-y_2|}\times\\
&\[\frac{1}{|x_0-y_1|} (\d_{y_1}^\mu \frac{1}{|y_1 - y_2|}) (\d_{y_2}^\nu \frac{1}{|y_2 - x_1|}) 
- (\d_{y_1}^\mu\frac{1}{|x_0-y_1|}) \frac{1}{|y_1 - y_2|} (\d_{y_2}^\nu \frac{1}{|y_2 - x_1|})\right.\nn \\
&\left.-\frac{1}{|x_0-y_1|}(\d_{y_1}^\mu\d_{y_2}^\nu\frac{1}{|y_1 - y_2|} ) \frac{1}{|y_2 - x_1|} 
+ (\d_{y_1}^\mu \frac{1}{|x_0-y_1|}) (\d_{y_2}^\nu \frac{1}{|y_1 - y_2|} ) \frac{1}{|y_2 - x_1|} \]\nn\\
=&-\frac{\ii 4\lambda}{(4\pi)^3} \int \dd^3 y_1 \dd^3y_2\,\epsilon_{\mu\nu\rho}\d^\rho_{y_1}{1\over(y_1-y_2)^2}(\d_{y_1}^\mu \frac{1}{|x_0-y_1|})  (\d_{y_2}^\nu \frac{1}{|y_2 - x_1|})\nn\\
=&-\frac{\ii 4\lambda}{(4\pi)^3} \int \dd^3 y_1 \dd^3y_2\,\epsilon_{\mu\nu\rho}\d^\rho_{y_1}\[{1\over(y_1-y_2)^2}(\d_{y_1}^\mu \frac{1}{|x_0-y_1|})  (\d_{y_2}^\nu \frac{1}{|y_2 - x_1|})\]=0\,,\nn
\ee
where in the first step, we observe that only one term is non-zero, and in the last step, we have an integral over a total derivative. 

The vanishing of the diagram can be seen more easily from momentum space. If we denote the incoming/outgoing momenta by $p$ and the scalar loop momenta by $q$, the two derivative vertices in \eqref{fig:1loop}(b) contribute a factor of $(p+q)^\mu (p+q)^\nu$, which is contracted with the gluon propagator $\langle A_\mu A_\nu \rangle$. Since the gluon propagator is antisymmetric in $\mu \nu$, the contraction vanishes automatically.

\subsection{Scalar-Wilson Line Gluon Exchange}

The diagram in figure \ref{fig:1loop}(c) is the exchange of a gluon between the scalar propagator and the Wilson line. 
It is given by
\beq
\begin{aligned} \label{eqn-OneLoopGen-GluonMatter}
\mathcal{M}_{c} =
& \frac{\ii \lambda}{32\pi^2} \int\limits_0^1 \dd s\ \dot{x}_s^\nu \int \dd^3y \Big[ \frac{\partial}{\partial y_\mu} (\frac{1}{|x_0-y|}) \frac{1}{|y-x_1|} - \frac{\partial}{\partial y_\mu} (\frac{1}{|x_1-y|}) \frac{1}{|y-x_0|} \Big] \epsilon_{\mu \nu \rho} \frac{y^\rho - x_s^\rho}{|y-x_s|^3},\\
\equiv & -\frac{\ii \lambda}{32\pi^2} \int\limits_0^1 \dd s\,{\cal M}_c(s)\,.
\end{aligned} 
\eeq
The spacetime integral over $y$ in $\mathcal{M}_c(s)$ can be evaluated explicitly,
\beq \label{eqn-Oneloop-McExplicit}
\begin{aligned}
{\cal M}_c(s)=& \dot{x}_s^\nu (\frac{\partial}{\partial x_{0 \mu}}-\frac{\partial}{\partial x_{1 \mu}}) \epsilon_{\mu \nu \rho} \int \dd^3 y \frac{1}{|x_0-y|} \frac{1}{|y-x_1|} \frac{y^\rho - x_s^\rho}{|y - x_s|^3} \\
= & \dot{x}_s^\nu (\frac{\partial}{\partial x_{0 \mu}}-\frac{\partial}{\partial x_{1 \mu}}) \epsilon_{\mu \nu \rho} \int \dd^3 y \frac{1}{|y-x_{0s}|} \frac{1}{|y-x_{1s}|} \frac{y^\rho}{|y|^3} \\
= & 2 \pi \dot{x}_s^\nu \epsilon_{\mu \nu \rho}(\frac{\partial}{\partial x_{0 \mu}}-\frac{\partial}{\partial x_{1 \mu}}) \frac{|x_{0s}|+|x_{1s}|-|x_{01}|}{|x_{0s}||x_{1s}| + x_{0s}\cdot x_{1s}} \Big[\frac{x_{0s}^\rho}{|x_{0s}|} + \frac{x_{1s}^\rho}{|x_{1s}|}\Big] \\
= & \frac{8\pi}{|x_{01}||x_{0s}||x_{1s}|} \frac{1}{|x_{0s}|+|x_{1s}|+|x_{01}|}  \epsilon_{\mu \nu \rho} x_{01}^\mu x_{1s}^\nu \dot{x}_s^\rho\,.
\end{aligned}
\eeq
Here, we have performed the $y$ integration using the following formula \cite{Guadagnini:1989am},
\beq
\int \dd^3 y \frac{y^\mu}{|y|^3|y-a||y-b|} =  2\pi \frac{|a|+|b|-|a-b|}{|a||b|+a\cdot b} \left(\frac{a^\mu}{|a|}+\frac{b^\mu}{|b|} \right) \, .
\eeq

We now show that the integral in \eqref{eqn-Oneloop-McExplicit} is finite for a smooth path. The integrand is finite for $x_s \neq x_0, x_1$; therefore, the only potential divergences are at the boundaries. 
Consider the limit where $x_s \rightarrow x_0$. In arc-length parametrization, we can approximate $x_s$ by
\beq
x_s = x_0 +(s -\frac{s^3 \kappa^2}{6}) \tf + (\frac{s^2 \kappa}{2} + \frac{s^3 \kappa'}{6}) \nf + \frac{s^3 \kappa \tau}{6} \bbf + \mathcal{O}(s^4)\,,
\eeq
where $\tf = \dot{x}_s$, $\nf = \ddot{x}_s$, $\bbf = \tf \times \nf$ are the vierbein (tetrad) of the curve at $x_0$. Here, $\kappa$, $\tau$ are the curvature and torsion at $x_0$, respectively.

Near the boundary, the $\epsilon$-tensor factor in \eqref{eqn-Oneloop-McExplicit} can be approximated as
\beq
\epsilon_{\mu \nu \rho} x_{01}^\mu x_{1s}^\nu \dot{x}_s^\rho \simeq (x_{01}) \cdot \Big((s\tf + s^2 \nf) \times (\tf + s \nf) \Big) = \mathcal{O}(s^2)\,,
\eeq
while the denominator behaves like $|x_{0s}| =\mathcal{O}(s)$. 
Hence, the integrand vanishes near the $x_0$ endpoint. A similar analysis applies at $x_1$.

\subsection{Bi-Scalar Contribution}

The last diagram, figure \ref{fig:1loop}(d), is given by products of two tree-level propagators,
\beq
\mathcal{M}_d = - \frac{\alpha \lambda}{8 \pi} \int\limits_0^1 \dd s \frac{|\dot{x}_s|}{|x_{1s}|\, |x_{s0}|} = - \frac{\alpha \lambda |x_{01}|}{8 \pi} \int\limits_0^1 \dd s \frac{1}{|x_{1s}|\, |x_{s0}|}\, .
\eeq
Near the endpoints, the terms in the denominator behave linearly, $|x_{0s}| \simeq \cO(s), \ |x_{1s}| \simeq \cO(1-s)$, which lead to logarithmic divergences at the boundaries and the corresponding anomalous dimension $\gamma=\alpha\lambda/2$, see section \ref{sec:1loop}.

\section{The Mesonic Wilson Line in the Fermionic Theory at 1-loop Order} \label{apd:oneLoopGenFer}

In this section, we present the one-loop expectation value of mesonic Wilson line operator \eqref{Mf} in the fermionic theory along an arbitrary smooth path $\cC$ between $x_0$ and $x_1$,
\beq
\cM^{f}[\cC,n]\equiv\<\bar\psi(x_1)\Big[{\cal P}e^{\ii\int\limits_\cC A\cdot \dd x}\Big]_n\psi(x_0)\>\,.
\eeq
Now, $\cM^{f}[\cC,n]$ is a $2\times2$ matrix in the fermionic indices. For non-self-intersecting paths, the one-loop result only diverges near the endpoints of the line.

At tree level, the expectation value of the mesonic line operator is given by the free fermion propagator,
\beq \label{apd-1loop-FerMesonTree}
\begin{aligned}
{\cal M}[\mathcal{C}]_\text{tree} =& \frac{1}{4\pi} \gamma_\mu \frac{x^\mu_1 -x_0^\mu}{|x_1-x_0|^3}\, .
\end{aligned}
\eeq
It is thus independent of the path.

At one loop, there are three non-trivial diagrams, regardless of the choice of the gauge.\footnote{It is because diagrams with cubic gluon vertex $A^3$ start at two-loop.} They are the gluon exchange on the Wilson line in figure \ref{fig:1loop}(a), the gluon exchange on the scalar propagator in figure \ref{fig:1loop}(b), and the gluon exchange between the Wilson line and the scalar propagator in figure \ref{fig:1loop}(c). 
We now analyze them in turn. 

\subsection{Gluon Exchange on the Wilson line}

The first diagram \ref{fig:1loop}(a) does not depend on the matter content of the theory. It is therefore the same as in \eqref{eqn-Lorentz-adiag}, with the tree level scalar propagator replaced by the fermionic one, \eqref{apd-1loop-FerMesonTree}.

\subsection{Fermion Self-energy}

The second diagram is a gluon exchange on the fermion propagator (self-energy); see figure \ref{fig:1loop}(b). It has been computed in momentum space in \cite{Giombi:2011kc}. 
Here, we reproduce it directly in position space. The expression is 
\beq
\begin{aligned}
\mathcal{M}_b^{f}=& \frac{\ii^2 N}{(4\pi)^3} \tr(T^I T^J)\, \gamma^{ֿ\alpha}\gamma^\mu \gamma^{\beta} \gamma^\nu \gamma^{\delta}\!\int \dd^3 y_1 \dd^3 y_2\frac{(x_1-y_1)_{\delta}}{|x_1-y_1|^3} \frac{(y_{12})_{\beta}}{|y_{12}|^3} \frac{(y_2-x_0)_{\alpha}}{|x_0-y_2|^3} \langle A_\nu^I(y_1) A_\mu^J(y_2) \rangle \\
\end{aligned}
\eeq

To proceed, we can project $\mathcal{M}_b$ to the complete matrix basis $\{\mathbb{I},\gamma^\mu\}$. For instance,
\beq
\begin{aligned}
\tr \mathcal{M}_b^{f}=& \, - \frac{\lambda}{32\pi^3} \int \dd^3 y_1 \dd^3 y_2 \, \frac{x_1^{\sigma}-y_1^{\sigma}}{|x_1-y_1|^3} \frac{1}{|y_{12}|^4} \frac{(y_2-x_0)_{\sigma}}{|x_0-y_2|^3} \\
=& \, +\frac{\lambda}{32\pi^3} \frac{\partial}{\partial x_1^\sigma} \frac{\partial}{\partial x_{0\sigma}}\int \dd^3 y_1 \dd^3 y_2 \, \frac{1}{|x_1-y_1|} \frac{1}{|y_{12}|^4} \frac{1}{|x_0-y_2|} \\
=& \, -\frac{\lambda}{32\pi^3} \int \dd^3 y_1 \dd^3 y_2 \, \frac{1}{|y_1|} \frac{1}{|y_{12}|^4} \Box_{y_2} \frac{1}{|x_{01}-y_2|} \\
=& \, +\frac{\lambda}{8\pi^2} \int \dd^3 y_1 \, \frac{1}{|y_1|} \frac{1}{|y_{1}-x_{01}|^4} \, .\\
\end{aligned}
\eeq
Using Feynman parametrization to collect the denominators, and dimensional regularization with $d = 3 +\varepsilon$, we get\footnote{In fact, we have to do analytical continuation for $d>4$, which amounts to dropping a linearly divergent term.} 
\beq
\tr \mathcal{M}_b^{f}
=\frac{\lambda}{8\pi^2} \frac{3}{4}\mu^{3-d} \int\limits_0^1 \dd u \frac{u}{\sqrt{1-u}} \frac{(2\pi)^d}{(4\pi)^{\frac{d}{2}}} \frac{\Gamma(\frac{5}{2}-\frac{d}{2})}{\Gamma(\frac{5}{2})} \Big(\frac{1}{u(1-u)x_{10}^2} \Big)^{\frac{5}{2}-\frac{d}{2}}= -\frac{\lambda}{4\pi x_{01}^2} + \cO(\varepsilon) \,.
\eeq

Similarly, we can project $\mathcal{M}_b^{f}$ to the $\gamma$ matrices,
\beq
\begin{aligned}
\tr \mathcal{M}_b^{f} \gamma^\alpha & \propto \int \dd^3 y_1 \dd^3 y_2\, \epsilon^{\alpha \lambda_1 \lambda_3}  \frac{(x_1-y_1)_{\lambda_3}}{|x_1-y_1|^3} \frac{1}{|y_{12}|^4} \frac{(y_2-x_0)_{\lambda_1}}{|x_0-y_2|^3}\\
& \propto \epsilon^{\alpha \lambda_1 \lambda_3} \d^{x_1}_{\lambda_3} \d^{x_0}_{\lambda_1} \int \dd^3 y_1 \dd^3 y_2 \frac{1}{|y_1 -x_1||y_{12}|^4 |y_2 -x_0|} \\
& \propto \epsilon^{\alpha \lambda_1 \lambda_3} \d^{x_1}_{\lambda_3} \d^{x_0}_{\lambda_1} \int \dd^3 y_1 \frac{1}{|y_1 -x_1||y_{1}-x_0|^2} \\
& \propto \epsilon^{\alpha \lambda_1 \lambda_3} \d^{x_1}_{\lambda_3} \d^{x_0}_{\lambda_1} \left( \log x_{10}^2 + \mathrm{regulator} + \mathrm{finite} \right)\\
& = 0 \, .
\end{aligned}
\eeq

To summarize, the 1-loop self-energy is given by
\beq
\mathcal{M}_b^{f} = -\frac{\lambda}{8\pi x_{01}^2} \mathbb{I} \, .
\eeq
This result suggests that the one-loop correction to the fermionic propagator in momentum space is proportional to $1/|p|$, or the 1PI part receives finite correction $\Sigma(p) \propto |p|$, in agreement with \cite{Giombi:2011kc}.

\subsection{Fermion-Wilson Line Gluon Exchange}

The last diagram is the gluon exchange between the fermion propagator and the Wilson line; see figure \ref{fig:1loop}(c). It is given by
\beq
\mathcal{M}_c^{f} = -\frac{\ii^2 N}{2 (4\pi)^2}  \int\limits_0^1 \dd s \, \dot{x}_s^\nu \int \dd^3y\,  \gamma^\sigma \gamma^\mu \gamma^\rho \frac{(x_1-y)_\rho}{|x_1-y|^3} \frac{(y-x_0)_\sigma}{|y-x_0|^3} \langle A_\mu^I(y) A_\nu^I(x_s) \rangle\, .
\eeq

We can compute it by projecting on the complete matrix basis as before. The results are 
\beq
\tr \mathcal{M}_c^{f} = -\frac{\lambda}{8\pi} \int\limits_0^1 \dd s\, \dot{x}_s^\nu  (\delta^\sigma_\lambda \delta^\rho_\nu-\delta^\sigma_\nu \delta^\rho_\lambda ) \partial^\rho_{x_1} \partial^\sigma_{x_0}
\frac{|x_{0s}|+|x_{1s}|-|x_{01}|}{|x_{0s}||x_{1s}| + x_{0s}\cdot x_{1s}} \Big[\frac{x_{0s}^\lambda}{|x_{0s}|} + \frac{x_{1s}^\lambda}{|x_{1s}|}\Big],
\eeq
and
\beq
\tr \gamma^\alpha \mathcal{M}_c^{f} = \frac{\ii \lambda}{8\pi} \int\limits_0^1 \dd s\, \dot{x}_s^\nu   (\delta^{\sigma \mu} \delta^{\rho \alpha} -\delta^{\sigma \rho} \delta^{\mu \alpha}+\delta^{\sigma \alpha} \delta^{\mu \rho})  \epsilon_{\mu \nu \lambda}\partial^\rho_{x_1} \partial^\sigma_{x_0} \frac{|x_{0s}|+|x_{1s}|-|x_{01}|}{|x_{0s}||x_{1s}| + x_{0s}\cdot x_{1s}} \Big[\frac{x_{0s}^\lambda}{|x_{0s}|} + \frac{x_{1s}^\lambda}{|x_{1s}|}\Big]\, .
\eeq

For a straightline, $x_s^\mu = x_0^\mu + s x_{10}^\mu$, we find
\beq
\mathcal{M}_c^{f} \Big|_{x_s = x_0 + s x_{10}} = \frac{\lambda}{8\pi x_{01}^2} \mathbb{I} \int\limits_0^1 \frac{\dd s}{s(1-s)} \, . 
\eeq
Based on the preceding expressions, the one-loop integral can only be logarithmic divergent near the endpoint. Any smooth line is straight near its endpoints, so the divergent behavior is identical to the expression given above.

\section{The Topological Parallel Transport}
\label{sec:topoPT}

In this appendix, we discuss the construction of the topological parallel transport.

\subsection{Construction}
\label{sec:apd:PT:constuction}

\paragraph{Definition.}

The parallel transport $U_\mathcal{C}(s,t)$ preserves the local $\pm$ component of the spinor indices along the segment of a generic curve $\mathcal{C}$. When it does not lead to confusion, we shall omit the subscript $\mathcal{C}$. For definiteness, we choose a parameterization of the curve s.t. the start/end-points are labeled by $t$ and $s$, respectively.

Recall that the local spinor $\pm$ components at any point $x_s$ are defined as the eigenvectors of $e_s \cdot \gamma$, where $e_s=\dot x_s/|\dot x_s|$. As a result, the parallel transport $U_\mathcal{C}(s,t)$ preserves eigenvectors of $\gamma \cdot e$,
\beq \label{eqn-PTconstraint}
(\gamma \cdot e_s) U_\mathcal{C}(s,t) = U_\mathcal{C}(s,t) (\gamma \cdot e_t) \, .
\eeq 

We also demand that if we break the path into two as $\cC=\cC_1\cup\cC_2$ then $U_{\cC_1\cup\cC_2}(s,t)=U_{\cC_1}(s,u)U_{\cC_2}(u,t)$. Hence, we are looking for a solution of the following form,
\beq
U_\mathcal{C}(s,t) = \mathcal{P} e^{\ii \int\limits_t^s \dd \tau\, \Gamma(\tau)} \equiv \mathcal{P} e^{\ii \int\limits_t^s \dd \tau\, \Gamma_\mu \dot{x}^\mu_\tau }\, .
\eeq
Without loss of generality, we take the initial point to be $t=0$. Taking derivatives of \eqref{eqn-PTconstraint} w.r.t $s$, we find the following differential equation for $U_\mathcal{C}(s) \equiv U_\mathcal{C}(s,0)$,
\beq
(\gamma \cdot \dot{e}_s) U_\mathcal{C}(s) + (\gamma \cdot e_s) \frac{\dd U_\mathcal{C}(s)}{\dd s} = \frac{\dd U_\mathcal{C}(s)}{\dd s} (\gamma \cdot e_0) =\frac{\dd U_\mathcal{C}(s)}{\dd s} U_\mathcal{C}(s)^{-1} (\gamma \cdot e_s) U_\mathcal{C}(s),
\eeq
where in the last step we have used the \eqref{eqn-PTconstraint} to translate $\gamma \cdot e_0$ to the point $x_s$. In terms of $\Gamma$, the equation above reads,
\beq \label{eqn-CommExpParallel}
\gamma \cdot \dot{e}_s +\ii (\gamma \cdot e_s) \Gamma_s =\ii \Gamma_s (\gamma \cdot e_s) \, .
\eeq

\paragraph{Solving for the Exponent.}

Let us first discuss the possible structures of $\Gamma_s$. 
It is a scalar matrix function acting on the spinor space. It can therefore be expressed in terms of a complete matrix 
basis $\{\mathbb{I}, \vec{\gamma} \}$. 
To absorb the Lorentz indices of $\vec{\gamma}$, the only way is to contract them with the local orthonormal frame $\{e, \frac{\dot{e}}{|\dot{e}|}, e \times \frac{\dot{e}}{|\dot{e}|}  \}$. Remember that this local orthonormal frame is a complete basis, so there are no longer any independent vector fields at a given point.

As a consequence, $\Gamma$ has to take the following general form,\footnote{Note that the first term is proportional to $\ddot{e} \cdot \gamma$.} 
\beq 
\Gamma_s = \frac{\alpha}{2} \epsilon_{\mu \nu \rho} e_s^\mu \dot{e}_s^\nu \gamma^\rho +  \tilde{\beta} \dot{e} \cdot \gamma + \beta |\dot{e}| e \cdot \gamma + \delta\, \mathbb{I} \, ,
\eeq
where four unknown coefficients $\alpha, \beta, \tilde{\beta},\delta$ were introduced. They can depend on local scalar functions of the curve, such as the curvature and the torsion.\footnote{As well as the relative angle between $\dot{e}$ and the framing vector, if such exists.}

Using the following commutation relations, 
\beq
\begin{aligned}
(\gamma \cdot e_s) (\gamma \cdot e_s) & = (\gamma \cdot \dot{e}_s)(\gamma \cdot \dot{e}_s) = 0, \\
(\gamma \cdot e_s)(\gamma \cdot \dot{e}_s) & = \ii \epsilon_{\mu \nu \rho} e^\mu \dot{e}^\nu \gamma^\rho, \\
(\gamma \cdot e_s) (\ii \epsilon_{\mu \nu \rho} e^\mu \dot{e}^\nu \gamma^\rho) & = \dot{e}_s \cdot \gamma,
\end{aligned}
\eeq
one can verify that
\beq
[\Gamma_s, \gamma \cdot e_s] = \ii \alpha \gamma \cdot \dot {e}_s-2 \tilde{\beta} \epsilon_{\mu \nu \rho} e^\mu \dot{e}^\nu \sigma^\rho.
\eeq
Therefore, we must take $\alpha = -1, \tilde{\beta} = 0$. We conclude that the parallel transport with the desired properties is given by
\beq \label{paralelltransport}
\begin{aligned}
U_\mathcal{C}(s,0) &= \mathcal{P}\exp \int\limits_0^s \dd\tau \left(-\frac{\ii}{2} \epsilon_{\mu \nu \rho} e_\tau^\mu \dot{e}_\tau^\nu \gamma^\rho + \beta |\dot{e}| e \cdot \gamma  + \delta \One\right) \, .
\end{aligned}
\eeq
The coefficients $\beta$ and $\delta$ are not constrained. As we now explain, demanding that the path be topological specifies their form.

\paragraph{General Topological Transport.}

A topological spin transport is, by definition, invariant under smooth deformation of the path with fixed boundaries. The first term in (\ref{paralelltransport}) is, however, not invariant under such a deformation. Instead, it gives a contribution that cannot be canceled by setting $\beta$ to be any local function of the path. That is, a function of the curvature and torsion.\footnote{See also discussions in \cite{Witten:1988hf} about the non-topological nature of Polyakov's spin factor \cite{Polyakov:1988md}.} To overcome this obstruction, one must introduce a framing vector $n$, that we take to be a unit orthogonal vector. In the presence of a framing vector, the form of the topological spin connection is unique, up to the freedom in the choice of $\delta$ in (\ref{paralelltransport}). In the next section we show that this topological connection is given by
\beq
\label{eq:integ-PT}
\Gamma_s = -\frac{1}{2} \Big( (e \times \dot{e}) - \tau_n(s) e \Big) \cdot \gamma + \delta\, \One\,,
\eeq
where $\times$ stands for the cross product for two vectors and
\beq
\tau_n(s) = \frac{\(\dot{n}\times n\)\cdot e}{|e\times n|^2}\,,
\eeq
is the so-called \textit{relative torsion}. In the arc-length parameterization, where $|\dot{x}_s| = |n_s|=1$, it reduces to (\ref{eqn-apd-relTorsionLorentz}). The constant $\delta$ in (\ref{eq:integ-PT}) leads to an overall phase factor. For the rest of this appendix, we set $\delta=0$. In the main text, we include such an overall topological phase factor that we separate from $\Gamma$, see the term $\frac{\lambda}{2}\d_\mu \log n^+$ in $(\widetilde A_0)_\mu$, (\ref{Mfc}).

\subsection{Functional Variation}
\label{sec:funVar-PT}

The topological constraint means that the spin transport is invariant under smooth variations of the path. We now show that $\Gamma_s$ in (\ref{eq:integ-PT}) indeed has this property.

\paragraph{Variation of Vector Fields.}

We shall use the arc-length parameterization throughout the section unless otherwise stated.

We deform the curve $x_s^\mu$ by a vector $v^\mu$, $x_s^\mu \mapsto x_s^\mu + \epsilon\, v^\mu$. To linear order in $\epsilon$ we have
\beq
\begin{aligned}
e^\mu &\quad\rightarrow\quad
e^\mu+\epsilon (\dot v^\mu-(e\cdot\dot v)e^\mu)+\cO(\epsilon^2)\,, \\
\dot{e}^\mu &\quad \rightarrow\quad \dot{e}^\mu + \epsilon \Big(\ddot{v}^\mu - e^\mu [(\dot{e}\cdot \dot{v} + e \cdot \ddot{v})] - (e \cdot \dot{v}) \dot{e}^\mu \Big)+\cO(\epsilon^2)\, .
\end{aligned}
\eeq
Using the linear order result, we find the linear order variation of the first term in \eqref{eq:integ-PT} is
\beq
e \times \dot{e}\quad \rightarrow\quad e \times \dot{e}  + \epsilon \Big( \dot{v}\times \dot{e} - 2 (e \cdot \dot{v}) e\times \dot{e} + e \times \ddot{v} \Big) + \mathcal{O}(\epsilon^2) \, .
\eeq

\paragraph{Variation of the Transport.}

The functional variation of $U(1,0)$ is given by,
\beq
\delta U(1,0) = \frac{\ii}{2}\int\limits_0^1 \dd s\, U(1,s) \Big(\delta \tau_n(s) e+\tau_n(s) \delta e-\delta(e \times \dot{e})\Big) \cdot \gamma\, U(s,0)\,.
\eeq

Under the constraint $e \cdot v =0$, we can trade all derivatives on $v$ appearing in the inner products by derivatives on $e$. For example, using
\beq
0 =\frac{\dd}{\dd s} (e \cdot v)\quad \Rightarrow \quad e \cdot \dot{v} = -\dot{e} \cdot v\, ,
\eeq
we can simplify the second term of the variation of $e \times \dot{e}$,
\beq
\delta(e \times \dot{e} ) = \dot{v}\times \dot{e} + 2 (v \cdot \dot{e}) e\times \dot{e} + e \times \ddot{v} \, .
\eeq
Now, performing IBP for the $e\times \ddot{v}$ term once, we find,
\begin{align}\label{eqn-apd-PT-var02}
\delta U(1,0) 
& =\frac{\ii}{2}\int\limits_0^1 \dd s\, U(1,s) \Big(\delta \tau_n e-\dot{v}\times \dot{e}-2 (v \cdot \dot{e}) e\times \dot{e}+\dot{e} \times \dot{v}-[(e\times \dot{e})\cdot \dot{v}]e\Big) \cdot \gamma\, U(s,0)\nn\\
&- \frac{\ii}{2} \int \dd s\, \frac{\dd}{\dd s} \Big( U(1,s) (e \times \dot{v})\cdot \gamma\, U(s,0)\Big)\,.
\end{align}

Next, we will show that the expression in the first line of \eqref{eqn-apd-PT-var02} vanishes,
\beq \label{eqn-VariationPT-bulk}
\dot{v}\times \dot{e} + 2 (v \cdot \dot{e}) e\times \dot{e} - \dot{e} \times \dot{v} +
[(e\times \dot{e})\cdot \dot{v}]e - \delta \tau_n e 
=0\,.
\eeq
As a result, the variation is the boundary term in the second line. 
The parallel transport described above is, therefore, topological.

\paragraph{Frenet Frame.}

A convenient method to deal with geometric quantities is to use the local orthonormal Frenet frame for the curve, $\{\tf,\nf,\bbf\}$. Here $\tf = e$ is the unit tangent, $\nf = \dot{e}/|\dot{e}|$ is the unit normal, $\bbf= \tf \times \nf$ is the so-called bi-normal. As we move along the path, they rotate as
\beq\label{eqn-apd-PT-Frenet}
\frac{\dd}{\dd s} 
\begin{pmatrix}
\tf \\
\nf \\
\bbf \\
\end{pmatrix}= |\dot{x}_s|
\begin{pmatrix}
0 & \kappa & 0 \\
- \kappa & 0 & \tau \\
0 & -\tau & 0  \\
\end{pmatrix}
\begin{pmatrix}
\tf \\
\nf \\
\bbf \\
\end{pmatrix} \, ,
\eeq
where the two scalar function $\kappa, \tau$ are the \textit{curvature} and the \textit{torsion} of the curve, respectively. Our strategy is to simplify the first four terms in \eqref{eqn-VariationPT-bulk}, then show that the last term cancels the sum of them.

\paragraph{$v$ Dependent Terms.}

The first four terms can be expressed in a compact way. Explicitly, we will show that
\beq\la{nicerelation}
\dot{v}\times \dot{e} + 2 (v \cdot \dot{e}) e\times \dot{e} - \dot{e} \times \dot{v} +
[(e\times \dot{e})\cdot \dot{v}]e = - [(e \times \dot{e}) \cdot \dot{v}] e \, .
\eeq

The normal deformation $v$ can be projected on the Frenet-Serret frame,
\beq
v= a\,\nf+ c\,\bbf\,.
\eeq
The derivative of $v$ can then be computed using (\ref{eqn-apd-PT-Frenet}),
\beq \label{eqn-VariationPT-vDiff}
\dot{v} = - a \kappa \tf +(\dot{a}-c \tau) \nf +(\dot{c}+a\tau) \bbf \equiv - a \kappa \tf + \mathfrak{a} \nf +\mathfrak{c} \bbf.
\eeq
For the $v$-dependence terms in \eqref{eqn-VariationPT-bulk}, we can now compute them term by term, 
\beq
\begin{aligned}
\dot{v} \times \dot{e} & = \kappa \dot{v} \times \nf = - a \kappa^2 \bbf - \mathfrak{c} \tf, \\
 (v \cdot \dot{e}) e\times \dot{e} & = a \kappa^2 \tf \times \nf = a \kappa^2 \bbf, \\
(e\times \dot{e})\cdot \dot{v} & = \kappa \bbf \cdot \dot{v} = \mathfrak{c} \, .
\end{aligned}
\eeq
Hence, we see that
\beq \label{eqn-VariationPT-bulk01}
-\kappa \mathfrak{c} \tf \cdot \sigma = - [(e \times \dot{e}) \cdot \dot{v}] (e \cdot \sigma) \, .
\eeq
These expressions result in the relation (\ref{nicerelation}). 

For plane curves, $v$ only has normal component $c=\dot{c}=\tau =0$ so the sum (\ref{eqn-VariationPT-bulk01}) vanishes (and so does the fifth term of \eqref{eqn-VariationPT-bulk}). Generally, for non-planar curves, they will not vanish. Below, we will show that they cancel.

\paragraph{Geometry of the Relative Torsion.}

Let us first study the geometric meaning of $\tau_n$. The framing vector $n$ is a unit normal vector, so we can expand it onto the Frenet-Serret frame,
\beq\la{rt}
n = \cos \theta_s \nf + \sin \theta_s \bbf.
\eeq

After some simple algebra, one can verify that the relative torsion differs from the geometric torsion by a total derivative of the angle $\theta$,\footnote{Recall that the geometric torsion is defined in $\eqref{eqn-apd-PT-Frenet}$. Explicitly, it is defined by taking the normal $n$ in $\tau_n$ to be the geometric normal $\nf$, $\tau =\tau_{\nf}$.} 
\begin{equation}
    \tau_n = \frac{\(\dot{n}\times n\)\cdot e}{|e\times n|^2} = \tau + \dot{\theta}_s\,.
\end{equation}
From this expression, we learn that the meaning of the ``relative'' is the relative angle between the geometric normal $\nf$ and the framing normal $n$. 

With this geometric meaning in mind, computing the functional variation of $\tau_n$ reduces to computing the variation of the geometric torsion and the relative angle $\dot{\theta}_s$. 

\paragraph{Variation of the Torsion.}

The deformation by a vector $v$ induces the deformation of the unit tangent, which is given by (see equation \eqref{eqn-VariationPT-vDiff})
\beq
\begin{aligned}
\delta \tf & = \mathfrak{a} \nf +\mathfrak{c} \bbf \, .
\end{aligned}
\eeq

Consider the functional variation of the second Frenet-Serret equation. Taking inner product with binormal $\bbf$, we get
\beq
\bbf \cdot \delta\Big( \frac{\dd \nf}{ds} \Big) = \delta(|\dot{x}| \tau) -\kappa \bbf \cdot \delta \tf + \tau \underbrace{\bbf \cdot \delta \bbf}_{=\delta(\bbf \cdot \bbf) =0} \Rightarrow \delta(|\dot{x}| \tau) = \kappa \bbf \cdot \delta \tf + \bbf \cdot \delta \partial_s \nf \, .
\eeq

Notice that the second term is actually a total derivative
\beq
\bbf \cdot \delta (\partial_s \nf) = \frac{\dd}{\dd s} (\bbf \cdot \delta \nf) - \underbrace{\frac{\dd\bbf}{\dd s} \cdot \delta \nf}_{=\tau \nf \cdot \delta \nf = \tau \delta (\nf \cdot \nf)=0} \, .
\eeq
Therefore, the final expression for the functional variation of $\tau |\dot{x}|$ is given by\footnote{The functional variation of $\tau$ should correspond to the variation of the combination $\tau |\dot{x}|$ in a generic parametrization of the line. In arc-length parameterization, $|\dot{x}| =1$, so the variation of the combination is equal to the variation of $\tau$. Refer to the section's concluding paragraph for clarification.}
\beq \label{eqn-PT-varTorsion}
\delta(|\dot{x}| \tau) = \kappa \mathfrak{c} + \frac{\dd}{\dd s} \Big( \bbf \cdot \delta \nf \Big)\,.
\eeq

\paragraph{Variation of the Relative Torsion.}

Since the relative torsion $\tau_n$ differs from the geometric torsion $\tau$ only by $\dot{\theta}$, it suffices to study the variation of $\theta$.
Recall $n$ is kept fixed,
\beq
\cos \theta_s = n \cdot \nf \quad\Rightarrow\quad - \sin \theta_s \delta \theta_s = n \cdot \delta \nf = \sin \theta_s \bbf \cdot \delta \nf\,,
\eeq
or in other words,
\beq \label{eqn-PT-varTorRel}
\delta \theta_s = - \bbf \cdot \delta \nf, \quad \Rightarrow \quad \delta \dot{\theta}_s = - \frac{\dd}{\dd s} \Big(\bbf \cdot \delta \nf \Big)\,.
\eeq

Combining \eqref{eqn-PT-varTorsion} and \eqref{eqn-PT-varTorRel}, we find a simple result,
\beq
\delta \tau_n = \kappa \mathfrak{c}\,.
\eeq
It cancels the remaining bulk term \eqref{eqn-VariationPT-bulk01}, so we conclude that (\ref{eqn-apd-PT-Frenet}) indeed holds true. 

\paragraph{Remark on Functional Variation.} \label{eqn-apd-PT-variationRemark}

We would like to show that functional variation commutes with derivatives. For that aim, consider two smooth curves related by a small variation. The first curve $\cC$ parameterized by $x_s$, and the other $\mathcal{C}^\epsilon$ parameterized by $y_s= x_s + \epsilon v_s$. Here, we have chosen the same parametrization for the two curves. Provided that we use the arc length parametrization of the path $\cC$, the resulting parametrization of $\cC^\epsilon$ is \textit{not} its arc-length one. 
Consequently, one must 
carefully keep track of the functional variations of the geometric quantities. For example, consider the functional variation of the derivative of the tangent vector $\tf$
\beq
\delta \frac{\dd \tf}{\dd s} \equiv \frac{\dd\tf^\epsilon_s}{\dd s} - \frac{\dd\tf_s}{\dd s} = |\dot{x}^\epsilon_s| \kappa^\epsilon_s \nf^\epsilon_s - \kappa |\dot{x}_s| \nf = \delta(\kappa |\dot{x}_s|) \nf + \kappa |\dot{x}_s| \delta \nf \, .
\eeq
For the deformed unit tangent, one has to use the generic Frenet-Serret equation, with $|\dot{x}| \neq 1$. 

To summarize, when taking functional variations of the Frenet-Serret equations, the functional variation of $\kappa,\tau$ is the variation of the combination $|\dot{x}|\kappa, |\dot{x}|\tau$.
With this in mind, we conclude that the functional variation commutes with derivatives.

\subsection{The Dependence on the Framing Vector}

Recall that the dependence of a mesonic line $M^{({1\over2},-{1\over2})}$ on the framing vector $n$, given in (\ref{fff}), is $\(2\, n_L^+n_R^-\)^{\frac{\lambda+1}{2}}$. 
We now show that the same framing factor is also generated by the topological spin connection in (\ref{eq:PT}) and the term $\frac{\lambda}{2}\d_\mu \log n^+$ in $(\widetilde A_0)_\mu$, (\ref{Mfc}). Hence, this dependence continues to hold for the condensed fermion operator (\ref{condensedfer}), which is decomposed from segments with Wilson line and segments with the topological parallel transport, see figure \ref{fig:condensedFer}.

The topological spin connection (\ref{toptransport}) is decomposed from two pieces, the geometric transport $(e \times \dot{e})\cdot\gamma$, and the relative torsion $\tau_n\,e\cdot\gamma$. 
In section \ref{sec:apd:PT:constuction}, we have seen that the geometric transport, as a $\beta = \delta = 0$ case of the generic parallel transport (\ref{paralelltransport}), satisfies (\ref{eqn-PTconstraint}).
Namely, it allows one to transport any terms of the form $e \cdot \gamma$ to the endpoints of the line. Since the relative torsion term is of this form, it can be transported to the end of the line as 
\beq\la{rttransport}
\mathcal{P} e^{\ii \int\limits_0^1 \dd s\, \Gamma_s} =\big[ e^{\frac{\ii}{2}e_s \cdot \gamma\int\limits_0^1 \dd s\, \tau_n}\big]\cdot\big[\mathcal{P} e^{-\frac{\ii}{2}\int\limits_0^1 \dd s\, (e_s \times \dot{e}_s) \cdot \gamma }\big] =\big[\mathcal{P} e^{-\frac{\ii}{2}\int\limits_0^1 \dd s\, (e_s \times \dot{e}_s) \cdot \gamma }\big]\cdot\big[ e^{\frac{\ii}{2}e_t \cdot \gamma\int\limits_0^1 \dd s\, \tau_n}\big]\,.
\eeq
After this, the relative torsion term becomes a pure phase acting on the plus and minus spin components,
\beq
P_\pm(s)\cdot e^{\frac{\ii}{2}e_s \cdot \gamma\int\limits_0^1 \dd s\, \tau_n}
= P_\pm(s) e^{\mp{\ii \over2} \int\limits_t^s \tau_n}, 
\qquad e^{\frac{\ii}{2}e_t \cdot \gamma\int\limits_0^1 \dd s\, \tau_n} \cdot P_\pm(t)= e^{\mp{\ii \over2} \int\limits_t^s \tau_n} P_\pm(t)\, .
\eeq

Consider next a straight line along the third direction. 
We then parameterize the components of the framing vector using polar coordinates as $n^\pm_\tau = e^{\pm\ii \theta_\tau}/\sqrt{2}$. We find
\beq \label{eqn-TorsionRelExpression}
\frac{\ii}{2} \tau_n = \frac{\ii}{2} \epsilon_{\mu\nu\rho} \dot{n}^\mu  n^\nu e^\rho = \frac{\ii}{2} \epsilon_{+-3} \left( \dot{n}^+ n^- -\dot{n}^- n^+  \right) = -\frac{\ii}{2} \dot{\theta}_\tau = -\frac{1}{2}\dd \log n^+_\tau\,,
\eeq
and therefore
\beq \label{eqn-apd-PT-framingFactor1}
e^{\mp{\ii \over2} \int\limits_0^1\dd s\, \tau_n(s)}= \Big( \frac{n_0^+}{n_1^+} \Big)^{\pm\frac{1}{2}} = \Big(2\, n_0^+ n_1^-\Big)^{\pm\frac{1}{2}}\,.
\eeq
Because the spin transport (\ref{rttransport}) is topological, the same result holds for a path that is not straight. 

The spin factor \eqref{eqn-apd-PT-framingFactor1} transports the classical $\pm 1/2$ spin of the fermions to the endpoints. The anomalous part of the spin is taken care of by the term $\frac{\lambda}{2}\d_\mu \log n^+$ in $(\widetilde A_0)_\mu$, (\ref{Mfc}). Together, they result in the framing factor in (\ref{transn}).

\section{Computation of the Perturbative Integral} \label{apd:diffEqnPert}

In this appendix, we show how to re-sum \eqref{eqn-Rec-solSca01} in the small $\te$ limit. 
Namely, we wish to compute the wave function renormalization factor in
\beq
\<M_{01}\>={\tilde c_0(- \lambda)\over4\pi x}\({\epsilon\over x}\)^{\lambda}\,,\qquad\text{for}\quad\alpha= 1 \quad\text{and}\quad \lambda \geq 0\,.
\eeq
The result we derive in this appendix is
\beq\la{c0value}
\tilde c_0(-\lambda)=\frac{e^{\lambda \gamma_E}}{\Gamma(1- \lambda)}\,.
\eeq

Our starting point is the following representation of $\tilde c_0(z)$
\begin{equation}\label{eq:intToBeProved}
    \tilde c_0(z)
     = \lim_{\te \to 0} \sum_{n=0}^{\infty}\Big[\te^{z} \frac{z^n}{1+(n+1) \te} \prod_{j=1}^{n} \ \int\limits_{t_{j-1}}^{1} \frac{\dd t_j}{t_j+j\te} \Big]\,,
\end{equation}
with $t_0=0$. It is obtained by plugging the expression for the line integrand (\ref{eqn-Rec-solSca01}) into (\ref{eqn-higherloop-rel}). We now evaluate $\tilde{c}_0(z)$ for positive $z$. In the end, we will analytically continue it to $z=-\lambda$. 

In the small $\te$ limit, all the integrals are logarithmic divergent, with no power divergences. The proof of this is simple. We have
\footnote{The second integral of (\ref{eq:upperInt}) is symmetric in all arguments, so the path ordering measure $\int \mathcal{P}\prod_j \dd t_j$ reduces to factorized product $1/n!\prod_j \int_0^1 \dd t_j$.}
\beq
\label{eq:upperInt}
0<\tilde c_0(z)\,\tilde\epsilon^{-z}
< \prod_{j=1}^{n} \ \int\limits_{t_{j-1}}^{1} \frac{\dd t_j}{t_j +\te} = \frac{1}{n!}\log^n (1+\te) = \frac{1}{n!} \log^n (\te) + \mathcal{O}(\te)\, .
\eeq
Therefore, we can safely expand the prefactor in (\ref{eq:intToBeProved}) to the leading order, $1/(1+(n+1)\te)\to1$. Moreover, the integrals are well-behaved in the vicinity of the upper limits of integration. Hence, after shifting the integration variables as $t_j\to t_j-j\tilde\epsilon$, we can safely drop the $\tilde\epsilon$ shift in the upper limits of integration. In that way, we arrive at
\begin{equation}
\label{eq:eqwithUpperlim}
	\tilde c_0(z)=\lim_{\epsilon \to 0} \sum_{n=0}^{\infty}\Big[\te^{z} z^n \prod_{j=1}^{n} \ \int\limits_{t_{j-1}+\te}^{1} \frac{\dd t_j}{t_j} \Big]\,.
\end{equation}
Consider the following auxiliary sum, that generalizes the expression inside the parenthesis in (\ref{eq:eqwithUpperlim}) by setting $t_0=s$ instead of $t_0=0$,
\beq\la{Fdef}
    F^{\te}(z,s)\equiv\sum_{n=0}^\infty z^n \[\prod_{j=1}^{n} \ \int\limits_{t_{j-1}+\te}^{1} \frac{\dd t_j}{t_j} \]\Bigg|_{t_0 = s}\,.
\eeq
To evaluate (\ref{eq:intToBeProved}), we need to first take the $s \rightarrow 0$ limit of $F^{\te}(s)$, and only then take the $\te \to 0$
\beq
\tilde c_0(z)=\lim_{\te \to 0}\te^{z}\,F^{\te}(z,0)\,.
\eeq
This function satisfies the differential-difference equation,
\begin{equation}
	({s+\te})\d_s 
	F^{\te}(z,s) = -z {F^{\te}(z,s+\te)}\,. 
\end{equation}
The inverse Fourier transform of $F^{\te}(z,s)$ then satisfies the following differential equation 
\begin{equation}
	(\ii\d_k + \epsilon) (\ii k) \widetilde{F}^{\te}(z,k) = -z\, e^{\ii k\epsilon} \widetilde{F}^{\te}(z,k)\,.
\end{equation}
This equation has an exact solution in terms of the exponential integral
\beq
\widetilde{F}^{\te}(z,k) =d_{\te}(z) \frac{e^{z  \text{Ei}(\ii k \epsilon ) }}{k}e^{\ii k \epsilon}\,,
\eeq
where $d_{\te}(z)$ is a constant of integration, and
\beq \label{eqn-apd-DiffEqn-defEi}
\text{Ei}(x) = -\int\limits_{-x}^\infty \frac{\dd t}{t} e^{-t}\,,
\eeq
is the exponential integral. This integral has a logarithmic branch point at the origin. Hence, to perform the Fourier transform back to $F^{\te}(z,s)$, we need to specify the contour of integration $\cC$ in 
\beq\la{eq:integThatcommutes}
	{F}^{\te}(z,s) = d_{\te}(z) \int\limits_{\mathcal{C}} \frac{\dd k}{2\pi} \frac{\, e^{z\,\text{Ei}(\ii k \te )}}{k} e^{\ii k(s+\te)} \underset{k \rightarrow \frac{k}{\te}}{=} d_{\te}(z) \int\limits_{\mathcal{C}} \frac{\dd k}{2\pi} \frac{\, e^{z\,\text{Ei}(\ii k )}}{k} e^{\ii k\(1+\frac{s}{\te}\)} \,.
\eeq
Different choices of $\cC$ result in different overall prefactors, $d_{\te}(z)$.  Here, we choose the integration contour $\mathcal{C}$ to go below the branch point at $p=0$ and asymptote to the real line.

To fix $d_{\te}(z)$ we note that at $s=1-\te$, the domain of all the integrals in (\ref{Fdef}) shrink to zero and therefore
\beq\la{1tod}
1= F^{\te}(z,1-\te)
=d_{\te}(z) \int\limits_{\mathcal{C}} \frac{\dd k}{2\pi} \frac{\, e^{z\,\text{Ei}(\ii k \te)}}{k} e^{\ii k} \, .
\eeq
In subsection \ref{eqn-apd-DiffEqn-cVal0}, we show that the small $\te$ limit of the integral in (\ref{1tod}) commutes with the integration. Hence, we can expend $\text{Ei}(\ii k\, \te) =\gamma_E + \log(k\,\te) - \ii \pi/2 + \cO(\te)$ inside the integral, 
\begin{equation}\label{eto0}
	\lim_{\te \to 0} \[\te^{z}\, d_{\te}(z)\] = \[
	e^{\gamma_E z} e^{-\ii \pi(z-1)/2}\int\limits_{\mathcal{C}} \frac{\dd k}{2\pi}  k^{z -1} e^{\ii k}\]^{-1} \,.
\end{equation}
By deforming the contour to engulf the branch cut along the positive imaginary axis we arrive at,
\beq\la{eqn-apd-DiffEqn-cVal}
	\lim_{\te \to 0}  \[\te^{z}\, d_{\te}(z)\] = \[2 \ii\,e^{\gamma_E z} \sin(\pi z)\int\limits_0^\infty \frac{\dd \omega}{2\pi}  \omega^{z -1} e^{- \omega}\]^{-1} = -\ii \frac{z  \Gamma (-z )}{e^{\gamma_E  z}} \,,
\eeq
where we used that $z>0$. 
We conclude that
\beq
\tilde c_0(z)=
-\ii \frac{z  \Gamma (-z )}{e^{\gamma_E  z }}\lim_{\te \to 0} \lim_{s\to 0}\int\limits_{\mathcal{C}} \frac{\dd k}{2\pi} \frac{\, e^{z\,\text{Ei}(\ii k )}}{k} e^{\ii k\(1+\frac{s}{\te}\)} \,.
\eeq

We can now set $s=0$. After doing so, the integral does not depend on $\te$ anymore. By encircling the cut as above, we arrive at
\beq\la{backtoc}
\tilde c_0(z)=-\ii \frac{z  \Gamma (-z )}{e^{\gamma_E  z }}\times 2\ii  \sin(\pi z) \int\limits_0^{\infty} \frac{\dd \omega}{2\pi} \frac{\, e^{z\,\text{Ei}(-\omega)}}{\omega} e^{-\omega}\,.
\eeq
Next, we notice that the $\text{Ei}(-\omega)$ independent terms nicely combine to a total differential (see \eqref{eqn-apd-DiffEqn-defEi})
\beq
\frac{\dd \omega}{\omega} e^{-\omega} = \frac{\dd}{\dd \omega} \text{Ei}(-\omega)\,.
\eeq
Hence,
\begin{equation}
\lim_{\omega_0\to0^+}\int\limits_{\omega_0}^\infty \frac{\dd\omega}{2\pi} \frac{\, e^{z  \text{Ei}(-\omega)}}{\omega} e^{-\omega} 
	=\lim_{\omega_0\to0^+} \frac{1}{2\pi} \frac{
	1-e^{z  \text{Ei}(-\omega_0 )} 
	}{z}={1\over2\pi z}\,.
\end{equation}
Putting everything together, we arrive at 
\beq
\tilde{c}_0(z) = \frac{e^{-z \gamma_E}}{\Gamma(1+z)}\, ,
\eeq
for $z>0$. Analytically continuing $z$ to $-\lambda$ gives (\ref{c0value}).

The ordered integral when $\alpha = -1$ in \eqref{eqn-Rec-solSca01} can be converted to the same form as \eqref{eq:intToBeProved} by changing $t_j \rightarrow 1- t_{L+1-j}$. 
Its value is $\tilde{c}_0(\lambda)$, and no analytical continuation is needed.

\subsection{Estimate of Small $\te$ Expansion} \label{eqn-apd-DiffEqn-cVal0}

We would like to justify the small $\te$ expansion in \eqref{eto0}. It is equivalent to showing that difference between the full expression in (\ref{1tod}) and the first term of \eqref{eqn-apd-DiffEqn-cVal} vanishes as $\te\to0$,
\beq\la{differenceeq}
\lim_{\te\to0}\mathfrak{D}(\te)=0\,,\qquad\text{where}\qquad\mathfrak{D}(\te)\equiv
{\te}^{-z} \int\limits_0^\infty \frac{\dd \omega}{2\pi \omega} e^{-\omega} \Big(e^{z\,\text{Ei}(-\omega \te)} - (\omega \te)^z e^{\gamma_E z}\Big)\,.
\eeq
Here, we have deformed the contour and taken discontinuities as before, as $\mathfrak{D}(\te)$ is well-behaved near the origin. We have stripped out an overall factor of $\te^{-z}$, so that each of the two integrals in (\ref{differenceeq}) is finite in the $\te\to0$ limit. 

Recall that at the end, we wish to analytic continue $\tilde{c}_0(z)$ to negative $z$. One may worry that this analytic continuation would not commute with the $\te\to0$ limit. For that aim, we prove (\ref{differenceeq}) for both, $z>0$ and $z<0$. For $z<0$ the integrals in (\ref{eqn-apd-DiffEqn-cVal}) and (\ref{backtoc}) no longer converge. Instead, they should be understood as keyhole integrations around the cut. The keyhole contribution, however, cancels between the two terms in $\mathfrak{D}(\te)$, so we can keep using the definition (\ref{differenceeq}) for $z<0$ as well.

To prove (\ref{differenceeq}) we note that for $z>0$ ($z<0$), the integrand in (\ref{differenceeq}) is negative (positive) along the integration domain. 
The reason for this is simple. Let us define
\beq
\mathfrak{d}(x) = \text{Ei}(-x) - \gamma_E - \log x\,, \qquad (x \geq 0 )\,,
\eeq
and recast the difference in the bracket in (\ref{differenceeq}) as
\beq \label{integrandD}
e^{z\,\text{Ei}(-\omega\, \te)} - (\omega \te)^z e^{z \gamma_E} = (\omega \te)^z e^{z \gamma_E} \Big( e^{z\, \mathfrak{d}(\te\, \omega)} -1 \Big)\, .
\eeq
The function $\mathfrak{d}(x)$ is a monotonic decreasing function since
\beq\la{dmathfrakd}
\frac{\dd}{\dd x} \mathfrak{d}(x) = \frac{e^{-x}-1}{x} <0\, .
\eeq
It implies that $\mathfrak{d}(x)$ obtains its maximal value at $x=0$. Using the small $x$ asymptotics of $\text{Ei}(-x)$, one finds that $\mathfrak{d}(0)=0$. As a consequence, $\exp(z \mathfrak{d}(\omega \te)) -1$ is negative (positive) along the integration domain. 
The measure multiplying (\ref{integrandD}) is always positive, so $\mathfrak{D}$ is bounded from above (below) as $\mathfrak{D} \leq 0$, ($\mathfrak{D} \geq 0$).

\paragraph{Lower bound for $z>0$.} To estimate the lower bound, we can use the elementary inequality 
\beq
e^x - 1 \geq x\,, \qquad \forall x \in \mathbb{R}\,,
\eeq
with $x=z\, \mathfrak{d}(\te\, \omega)$ in (\ref{integrandD}). Hence,
\beq
\begin{aligned}
\mathfrak{D} & \geq z \int\limits_0^\infty \frac{\dd \omega}{2\pi} \frac{e^{-\omega}}{\omega^{1-z}} e^{z \gamma_E} \mathfrak{d}(\omega \te)\\
& = -\frac{e^{z \gamma_E}}{2\pi}{\te}^{-z } \Gamma (z ) \left[z  {\te}^{z } (\log ({\te})+\psi ^{(0)}(z )+\gamma_E )+\, _2F_1\left(z ,z ;z +1;-\frac{1}{{\te}}\right)\right] \\
& = -\frac{z \Gamma(1+z) e^{z \gamma_E} }{2\pi} \te+ \cO(\epsilon^2)\, ,
\end{aligned}
\eeq
where we performed the integration analytically using $\mathtt{Mathematica}$.\footnote{See also equation (6.228) in \cite{gradshteyn2014table}, where a closed form formula is given as $z \int_0^\infty \dd p\, \text{Ei}(-p \te) e^{-p} p^{z-1} =- {\te}^{-z } \Gamma (z ) \, _2F_1\left(z ,z ;z +1;-\frac{1}{{\te}}\right)$.} The leading order expansion in $\te$ then follows. 

\paragraph{Upper bound for $z<0$.} To estimate the upper bound for $z<0$, we need to bound $\( e^{z \mathfrak{d}(x)} -1 \)$ from above. We start by noting that the $x$-derivative of this expression is manifstly positive, (\ref{dmathfrakd}). As a result, $\( e^{z \mathfrak{d}(x)} -1 \)$ is convex. In some neighborhoods of $x=0$,
\begin{equation}
    \( e^{z \mathfrak{d}(x)} -1 \) = -z x + \cO(x)
\end{equation}
while in some neighborhood of $x\to\infty$,
\begin{equation}
    \( e^{z \mathfrak{d}(x)} -1 \) = e^{-\gamma_E  z } x^{-z } + \cO(1/x^0)
\end{equation}
We see that in both it is bounded by the linear function $-2 z\, x$. Combining these facts implies the inequality
\beq
\Big( e^{z \mathfrak{d}(x)} -1 \Big) \le -2 z\, x\,,\qquad \forall x \ge 0 \,. 
\eeq
It then follows that
\beq
\begin{aligned}
\mathfrak{D}(\te) & \leq -2z e^{\gamma_E}\te\int\limits_{0}^{\infty} \frac{\dd \omega}{2\pi} \omega^{z} e^{-\omega} 
= - \frac{2 e^{\gamma_E}z \Gamma(1+z)}{\pi} \te \, .
\end{aligned}
\eeq
The leading order expansion in $\te$ follows as before.  

\paragraph{Summary.} We have shown that the difference $\mathfrak{D}$ is bounded as
\beq
\mathcal{O}(\te) \leq \mathfrak{D}(\te) \leq 0 \, ,
\eeq
for $z>0$ and as,
\beq
0 \leq \mathfrak{D}(\te) \leq \cO\(\te\) \, ,
\eeq
for $z<0$. Therefore, in the small $\te$ limit, the difference goes to zero.

\section{Verification of the Condensed Fermion Boundary Equation}\la{cfbeapp}

We would like to verify boundary equation for the condensed fermion mesonic line operator (\ref{becf}),  
\begin{equation}\label{eq:beqintermsofflds}
 \widetilde\cO_R^{(2,-{1\over2})}+2\delta_-\widetilde\cO_R^{(0,{1\over2})}
 = 0\,.
\end{equation}
Using the expressions of these two boundary operators in terms of fields,
\beq
\widetilde\cO_R^{(2,-{1\over2})}=
\ii \sqrt{4\pi\over k}\[D_3 \psi_- + \psi_-  \frac{\lambda}{\epsilon}\] + \frac{\lambda^2}{\epsilon^2} \[\rm{Empty}\]_- \,,
\eeq
and
\beq
\label{eq:defCond+-}
\delta_-\widetilde\cO_R^{(0,{1\over2})}=
-{\ii\over\sqrt{2}} \sqrt{4\pi\over k}D_-\psi_+ + \frac{\lambda}{2\epsilon} \frac{\lambda}{2\epsilon} \[\rm{Empty}\]_{\alpha} \[\gamma^-\gamma^+\]^{\alpha}_{~-} \,,
\eeq
we fined that
\beq
\widetilde\cO_R^{(2,-{1\over2})}+2\delta_-\widetilde\cO_R^{(0,{1\over2})}\propto  \[D_3 \psi_- + \psi_-  \frac{\lambda}{\epsilon}\]
 -\sqrt{2}D_-\psi_+ \,.
\eeq
Here, the constant terms of order $\cO(1/\te^2)$ cancel in the sum \eqref{eq:beqintermsofflds}.

\begin{figure}
    \centering
 \includegraphics[width=0.6\textwidth]{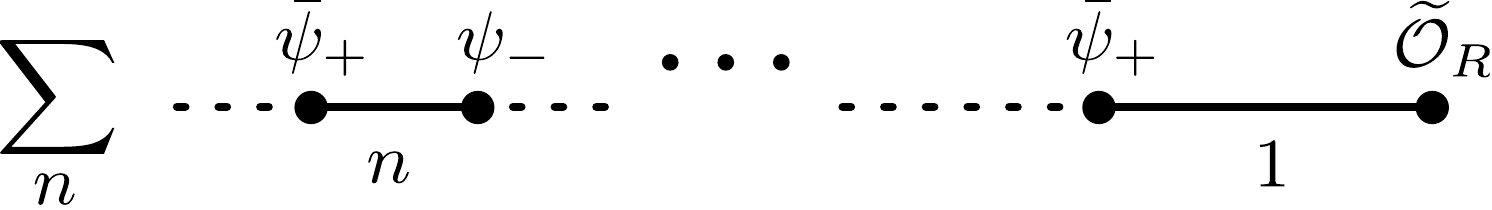}
    \caption{Inserting the boundary operator $\widetilde{\mathcal{O}}_R \in \{D_3 \psi_-, \psi_-, D_+ \psi+\}$ at the rightmost boundary $x_0$ would only affect the first ladder that is being condensed. Therefore, verifying (\ref{becf}) for the first ladder is sufficient.}
    \label{fig:condensedEOM}
\end{figure}

When inserting \eqref{eq:beqintermsofflds} at the right boundary, diagrammatically, only the first fermion antifermion pair is affected. The rest of the diagrams still end on the $\bar{\psi}_+-\psi_-$ pairs, see figure \ref{fig:condensedEOM}. It is therefore sufficient to show that \eqref{eq:beqintermsofflds} holds for the first pair.

The integrand of the first fermion anti-fermion pair satisfies the universal recursion relation. Therefore, one only needs to show that the seed, given by the $0$- and $1$- ladders, vanishes. For this aim, we apply the same logic as in the uncondensed fermion in section \ref{sec:ferevoandbdry}. Namely, we place the operator in (\ref{eq:beqintermsofflds}) and $\bar{\psi}_+$ at the right and left ends of a straight line. 
Every diagram contributing to the expectation value of this mesonic line operator is of a similar form to the diagrams in figure \ref{fig:condensed-Identity}. The only difference is that at the right end, we have to insert a superposition of the three operators $D_-\psi_+$, $D_3\psi_-$, and $\psi_-$. We consider each one of them in turn.

For $D_-\psi_+$ we have
\begin{align}
\mathtt{F}_{D_-\psi_+}(t_{1}) & = \int \frac{\dd^3 k_0}{(2\pi)^3} e^{\ii k^3_0 x t_1} [\widetilde\Delta(k_0)]_{++} (-\ii k^+_0)\,, \label{eqn-CondFer-EOMd1} \\ 
\mathtt{F}_{D_-\psi_+}(t_{1};t_2) & = \int{\dd^3 k_1\over(2\pi)^3}{\dd^3 k_0\over(2\pi)^3}\ \,\Phi^{(1)}_{-1}(t_1; t_{2}| k_0,k_1)\[{-\ii k_0^+\widetilde\Delta(k_0)}{\gamma^+ \over k_1^+}\widetilde\Delta(k_1)\]_{++}\nn \\
& = \int{\dd^3 k_1\over(2\pi)^3}{\dd^3 k_0\over(2\pi)^3}\ \,\Phi^{(1)}_{-1}(t_1; t_{2}| k_0,k_1) \frac{-\sqrt{2}\ii k_0^+}{k_1^+} [\widetilde\Delta(k_0)]_{+-}[\widetilde\Delta(k_1)]_{-+}\, , \label{eqn-CondFer-EOMd2}
\end{align}
as for the $0$- and $1$- ladder seeds.

The $\psi_-$ insertion has been considered in section \ref{sec:ferAllLoop}, and the seeds are given in \eqref{preintegrand-fer}. Applying $D_3$ derivatives to it amounts to modifying the first $k_0$ propagator in the first two recursion seeds
\begin{align}
\mathtt{F}_{D_3 \psi_-}(t_{1}) & = \int \frac{\dd^3 k_0}{(2\pi)^3} e^{\ii k^3_0 x t_1} [\mathbf{V}_{D_3\psi}(k_0) \widetilde\Delta(k_0) ]_{-+}\,, \label{eqn-CondFer-EOMd31} \\
\mathtt{F}_{D_3 \psi_-}(t_{1};t_2) & = \int{\dd^3 k_1\over(2\pi)^3}{\dd^3 k_0\over(2\pi)^3}\ \,\Phi^{(1)}_{-1}(t_1; t_{2}| k_0,k_1)\[\mathbf{V}_{D_3\psi}(k_0){\widetilde\Delta(k_0)}{\gamma^+ \over k_1^+}\widetilde\Delta(k_1)\]_{-+}\nn \\
& = \int{\dd^3 k_1\over(2\pi)^3}{\dd^3 k_0\over(2\pi)^3}\ \,\Phi^{(1)}_{-1}(t_1; t_{2}| k_0,k_1) \frac{\sqrt{2}}{k_1^+} [\mathbf{V}_{D_3\psi} \widetilde\Delta(k_0)]_{--}[\widetilde\Delta(k_1)]_{-+}\, .
\label{eqn-CondFer-EOMd32}
\end{align}
Here, the effective vertex $\mathbf{V}_{D_3\psi}(q)$ of the $D_3 \psi_-$ insertion receives contributions from two diagrams. These can be obtained by changing the endpoint covariant derivative from $D_+$ to $D_3$ in figure \ref{fig:ferderseed}. Explicitly,
\beq \label{eqn-ferdSeedVertex}
\mathbf{V}_{D_3\psi}(q) = - \ii q_3 + 2\pi \ii \lambda \int \frac{\dd^3 p}{(2\pi)^3} \Delta(p)\gamma^+  \frac{1}{(p-q)^+}\, .
\eeq
Note that the $e^{\ii p_3 \epsilon}$ in the $p$ propagator is not being stripped out.

For those seeds, the loop momentum integral over $p$ \eqref{eqn-ferdSeedVertex} can always be evaluated explicitly. Verifying \eqref{eq:beqintermsofflds} for the recursion seeds then becomes algebraic, as most of the terms, except one, cancel 
\textit{before} taking the $k_0$ and $k_1$ integrals.

Starting from the 0-ladders, the relevant matrix element in \eqref{eqn-CondFer-EOMd31} reads
\beq \label{eqn-CondFer-EOMd31-Vertex}
\begin{aligned}
[\mathbf{V}_{D_3\psi}(q) \widetilde\Delta(q) ]_{-+} & = - \ii q_3 [\widetilde\Delta(q) ]_{-+} +2\pi \ii \lambda \int \frac{\dd^3 p}{(2\pi)^3}  \frac{1}{(p-q)^+} [\Delta(p)\gamma^+ \widetilde\Delta(q)]_{-+} \\
& = - \ii q_3 [\widetilde\Delta(q) ]_{-+} +2\pi \ii \lambda \sqrt{2} \int \frac{\dd^3 p}{(2\pi)^3}  \frac{1}{(p-q)^+} [\Delta(p)]_{--} [\widetilde\Delta(q)]_{-+} \\
& = \Big(- \ii q_3  + \lambda \frac{e^{-\epsilon |q_\perp|}}{\epsilon}\Big) [\widetilde\Delta(q) ]_{-+}\, . \\
\end{aligned}
\eeq
We can perform the loop integration over $p$ because in our scheme the $--$ element of the propagator $\Delta(p)$, apart from an exponential factor, does not receive loop corrections. Importantly, we notice that apart from a counterterm, the term in the bracket is proportional to the $-+$ component of the exact propagator, as we have shown in red below,
\beq
- \ii q_3  + \lambda \frac{e^{-\epsilon |q_\perp|}}{\epsilon} = - \ii q_3  + \lambda \frac{e^{-\epsilon |q_\perp|} \textcolor{red}{-1}}{\epsilon} + \textcolor{red}{\frac{\lambda}{\epsilon}} = q^2 [\widetilde\Delta(q)]_{+-}+ \frac{\lambda}{\epsilon}\, .
\eeq
Moreover, using \eqref{eq:propExactProp}, the product of two components of the exact propagator can be further converted to an equivalent product that is proportional to \eqref{eqn-CondFer-EOMd1}, 
\beq
\begin{aligned}
[\mathbf{V}_{D_3\psi}(q) \widetilde\Delta(q) ]_{-+} & = \frac{\lambda}{\epsilon} [\widetilde\Delta(q) ]_{-+} +  q^2 [\widetilde\Delta(q)]_{+-}[\widetilde\Delta(q) ]_{-+} \\
& = \frac{\lambda}{\epsilon} [\widetilde\Delta(q) ]_{-+} +  q^2 [\widetilde\Delta(q) ]_{--} [\widetilde\Delta(q)]_{++} +1 \\
& = \frac{\lambda}{\epsilon} [\widetilde\Delta(q) ]_{-+} + \sqrt{2} (-\ii q^+) [\Delta(q)]_{++} +1 \, .
\end{aligned}
\eeq
The first term cancels against the $0$-ladder contribution of the boundary operator $\frac{\lambda}{\epsilon}\psi_-$. The second term cancels ($\sqrt{2}$ times) \eqref{eqn-CondFer-EOMd1}. Therefore, the last term is the only contribution to \eqref{eq:beqintermsofflds} at the level of 0-ladders. This term, when integrated against $e^{\ii q_3 t_1 x}$, yields a contact term $\delta(x \, t_1)$, which has no support for strictly positive $t_1$.

The computation for the one-ladder follows the same logic and is even simpler. The momentum loop integral of the vertex in \eqref{eqn-CondFer-EOMd32} can also be done and results in 
\beq
\begin{aligned}
[\mathbf{V}_{D_3\psi}(q) \widetilde\Delta(q) ]_{--} & = - \ii q_3 [\widetilde\Delta(q) ]_{--} +2\pi \ii \lambda \int \frac{\dd^3 p}{(2\pi)^3}  \frac{1}{(p-q)^+} [\Delta(p)\gamma^+ \widetilde\Delta(q)]_{--} \\
& =  \Big[q^2 [\widetilde\Delta(q)]_{+-}+ \frac{\lambda}{\epsilon}\Big] [\widetilde\Delta(q) ]_{--} \\
& = -\sqrt{2}\ii q^+ [\widetilde\Delta(q)]_{+-} + \frac{\lambda}{\epsilon} [\widetilde\Delta(q)]_{+-} [\widetilde\Delta(q)]_{--}\, ,
\end{aligned}
\eeq
where we substitute the expression of $[\widetilde\Delta(q)]_{--}$ for the first term.

The first term cancels ($\sqrt{2}$ times)  \eqref{eqn-CondFer-EOMd2}. The second term cancels against ($\sqrt{2}$ times) the $1$-ladder contribution to the boundary operator  $\frac{\lambda}{\epsilon}\psi_-$ insertion. There is no additional piece left, so \eqref{eq:beqintermsofflds} holds.

The full cancellation of the 0- and 1-ladder seeds shows that the quantum boundary equation of motion \eqref{eq:beqintermsofflds} holds as an operator equation. Here, each of the boundary operators acquires additive counterterms, however when we taking the sum those divergences cancel out.

\section{Supersymmetry Conventions of the ${\cal N}=2$ theory} \label{susyappendix}

\subsection{SUSY Conventions}

We shall follow the convention of \cite{Hama:2010av}. See also the lecture notes \cite{Marino:2011nm} and the references therein.

\paragraph{Spinor.}

A fundamental $SU(2)$ spinor is denoted by $\psi^a$. Its conjugate is denoted by $\bar{\psi}^a$. The spinor indices take value in $1,2$. In three dimensions, there is no chirality, so we do not have dotted indices.

\paragraph{$\gamma$ Matrices.}

The Euclidean $\gamma$ matrices are taken to be the Pauli matrices. The index structures are as follows:
\beq
\gamma_\mu = (\gamma_\mu)^a_{\ b}\,.
\eeq

For spinor bilinears, the spinor indices are contracted with the help of the antisymmetric $\varepsilon$-symbols. For instance,
\beq
\bar{\psi} \psi =\bar{\psi}^a \varepsilon_{ab} \psi^b\,, \qquad \bar{\psi} \gamma^\mu \psi = \bar{\psi}^a \varepsilon_{ab} (\gamma^\mu)^b_{\ c} \psi^c \, ,
\eeq
where $\varepsilon_{21} = - \varepsilon_{12} = 1$.

In practise, there is no need to lower the spinor indices. For definiteness, we choose the convention that all spinor indices are raised/lowered by the anti-symmetric $\varepsilon$-tensor \textit{from the left},
\beq
\psi^a = \varepsilon^{ab}\psi_b\,, \qquad \psi_a = \varepsilon_{ab} \psi^b\, .
\eeq
Consistency requires $\varepsilon^{12} = 1$.

\paragraph{Spinor Identities.}

Using the spinor conventions above, one can verify that
\beq
\bar{\epsilon} \lambda = \lambda \bar{\epsilon}\,, \qquad \bar{\epsilon}\gamma^\mu \lambda = - \lambda \gamma^\mu \bar{\epsilon}\,,
\eeq
and in particular, 
\beq \label{eqn-SUSY-spinorId2}
(\gamma^\mu \bar{\epsilon}) \lambda = - \bar{\epsilon} \gamma^\mu \lambda\, .
\eeq

\paragraph{Other $\gamma$-Matrices.}

$\gamma$-matrices with all upper and lower indices can be constructed using $\varepsilon$ as
\beq
\gamma_{ab} = \varepsilon_{ac} \gamma^c_{\ d}\,, \qquad \gamma^{ab} = \varepsilon^{bd} \gamma^c_{\ d}\,.
\eeq
Both of them are \textit{symmetric} matrices. The Hermitian conjugate of $(\gamma)^{ab}$ equals to $-(\gamma)_{ab}$.

\paragraph{R-charge.}

The $\mathcal{N} = 2$ theory has a $U(1)$ R-symmetry, so we have to choose a convention for the $R$-charges.

For matter fields, the convention we use is\footnote{This convention is the standard one where the $R$-charge of the SUSY generator $Q$ is $-1/2$. The matter $R$ charges can be then deduced from the superpotential, $W(\Phi) = \tr \Phi^4$, where $\Phi$ is the chiral superfield that encodes the boson, fermion and the auxiliary boson $F$.}
\beq
[\phi]_R = +\frac{1}{2}\,, \qquad [\psi]_R = - \frac{1}{2}\,, \qquad [F]_R = - \frac{3}{2}\,,
\eeq
while the conjugates take the negatives of these values. Here $F$ is an auxiliary scalar field in the chiral multiplet, which vanishes on-shell.

\subsection{SUSY Transformation of Fields}

The $\mathcal{N}=2$ SUSY transformations can be parameterized by two complex fundamental $SU(2)$ spinors, $\epsilon$ and $\bar{\epsilon}$. In Euclidean signature, they are independent.

In order to maintain superconformal symmetry, those spinors have to satisfy the \textit{conformal} Killing equation\footnote{Here $D_\mu$ stands for \textit{spacetime} covariant derivative.}
\beq
D_\mu \epsilon = \frac{1}{3} \gamma_\mu \epsilon_c\,, \qquad D_\mu \bar{\epsilon} = \frac{1}{3} \gamma_\mu \bar{\epsilon}_c\,,
\eeq
where $\epsilon_c$, $\bar{\epsilon}_c$ are two constant spinors. In flat spacetime, the solution is
\beq \label{eqn-SUSY-KillingVecSol}
\epsilon = \epsilon_s + x^\mu \gamma_\mu \epsilon_c, \quad \bar{\epsilon} = \bar{\epsilon}_s + x^\mu \gamma_\mu \bar{\epsilon}_c \, ,
\eeq
where $\epsilon_s$, $\bar{\epsilon}_s$ are another set of constant spinors.

\paragraph{Vector Multiplet.}

The 3d $\mathcal{N}=2$ vector superfield $V$ consists of the following component fields,
\beq
V:\quad A_\mu\,,\ \sigma\,,\ \lambda\,,\ \bar{\lambda}\,,\ D\,.
\eeq
Here, $A_\mu$ is the gauge field, $\sigma$ is a scalar field, $\lambda$ and $\bar{\lambda}$ are two complex fermions, and $D$ is an auxiliary scalar field. They satisfy the following transformation rules (for simplicity, we only list the relevant ones)
\beq \label{eqn-SUSY-vec}
\delta A_\mu = -\frac{\ii}{2} \big(\bar{\epsilon}\gamma_\mu \lambda - \bar{\lambda} \gamma_\mu \epsilon \big)\,, \quad 
\delta \sigma = +\frac{1}{2} \big(\bar{\epsilon}\lambda - \bar{\lambda} \epsilon \big)\, ,
\eeq
and
\begin{equation} \label{eqn-SUSY-vec2}
\begin{aligned}
&\delta \lambda=-\frac{1}{2} \gamma^{\mu \nu} \epsilon F_{\mu \nu}-D \epsilon+\mathrm{i} \gamma^\mu \epsilon D_\mu \sigma+\frac{2 \mathrm{i}}{3} \sigma \gamma^\mu D_\mu \epsilon\, , \\
&\delta \bar{\lambda}=-\frac{1}{2} \gamma^{\mu \nu} \bar{\epsilon} F_{\mu \nu}+D \bar{\epsilon}-\mathrm{i} \gamma^\mu \bar{\epsilon} D_\mu \sigma-\frac{2 \mathrm{i}}{3} \sigma \gamma^\mu D_\mu \bar{\epsilon}\, .
\end{aligned}
\end{equation}
Here, $\gamma^{\mu \nu} \equiv 1/2 [\gamma^\mu, \gamma^\nu]$.

\paragraph{Matter Multiplet.}

The 3d $\mathcal{N}=2$ chiral multiplet consists of a complex scalar $\phi$, a complex fermion $\psi$, and a complex auxiliary scalar field $F$.

The fields in a chiral multiplet transform as
\beq \label{eqn-SUSY-matter}
\begin{aligned}
\delta \phi & = \bar{\epsilon} \psi\,, \\
\delta \phi^\dagger & = \epsilon \bar{\psi}\,, \\
\delta \psi & = \ii \gamma^\mu \epsilon D_\mu \phi + \ii \epsilon \sigma \phi + \frac{\ii}{3}\gamma^\mu D_\mu \epsilon \phi + \bar{\epsilon} F\,, \\
\delta \bar{\psi} & = \ii \gamma^\mu \bar{\epsilon} D_\mu \phi^\dagger + \ii \phi^\dagger \sigma \bar{\epsilon} + \frac{\ii}{3} \phi^\dagger \gamma^\mu D_\mu \bar{\epsilon} + \bar{F} \epsilon\, . 
\end{aligned}
\eeq

\paragraph{Lagrangian.}

The supersymmetric Lagrangian $\mathcal{L} = \frac{\ii k}{4\pi} \mathcal{L}_{CS} + \mathcal{L}_m $ consists of the supersymmetric Chern-Simons term,
\beq
\mathcal{L}_{CS} = \tr \epsilon^{\mu\nu\lambda}\Big(A_\mu \partial_\nu A_\lambda -\frac{2\ii}{3} A_\mu A_\nu A_\lambda \Big) -\bar{\lambda}\lambda+2D\sigma\,,  
\eeq
and the matter fields Lagrangian,
\beq
\mathcal{L}_m = D_\mu \phi^\dagger D_\mu \phi + \phi^\dagger \sigma^2 \phi +\ii \phi^\dagger D \phi +\bar{F}F -\ii \bar{\psi} \gamma^\mu D_\mu \psi + \ii \bar{\psi} \sigma \psi + \ii \bar{\psi} \lambda \phi -\ii \phi^\dagger \bar{\lambda} \psi\,.
\eeq 

The auxiliary scalar fields $\sigma$ and $D$ in the vector multiplet do not have kinetic terms, so they serve as constraints.  The equation of motion of the $D$ field relates $\sigma$ to the adjoint bi-scalar pair
\beq
\frac{\delta \mathcal{L}}{\delta D} = 2\delta D \sigma (\frac{\ii k}{4\pi}) + \ii \phi^\dagger \delta D \phi = 0\,, \qquad \Rightarrow \qquad\sigma =  - \frac{2\pi}{k} \phi \phi^\dagger \, .
\eeq
Similarly, the gauginos are related to the scalar-fermion adjoint,
\beq \label{eqn-SUSY-ValGaugino}
\lambda = -\frac{4\pi}{k} \psi \phi^\dagger\,, \qquad \bar{\lambda} = \frac{4\pi}{k} \phi \bar{\psi}\, .
\eeq

\paragraph{Line Operators.}

The bi-scalar adjoint in the line operator \eqref{Wbos} can be replaced by $\sigma$ as
\beq \label{Wbos2}
\mathcal{W}_\alpha[\cC] = \mathcal{P} \exp{\ \ii \int\limits_\cC \!\! \(A\cdot \dd x-\ii \alpha \sigma |\dd x|\)}\, .
\eeq

\subsection{SUSY Algebra}

In this section, we present the SUSY algebra that is compatible with the field transformation rules. 
For definiteness, we first fix the bosonic part of the SUSY algebra, namely conformal group commutation relations.

\paragraph{Bosonic Part: Conformal Algebra.}

The $SO(4,1)$ conformal algebra reads
\beq \label{eqn-SUSY-confAlg}
\begin{aligned}
[M_{\mu \nu}, M_{\rho \sigma}] & = - \ii \left( \delta_{\mu \sigma} M_{\nu \rho} +\delta_{\nu \rho} M_{\mu \sigma} - \delta_{\mu \rho} M_{\nu \sigma} -\delta_{\nu \sigma} M_{\mu \rho }\right)\,, \\
[M_{\mu \nu}, P_\rho] & = -\ii (\delta_{\nu \rho} P_\mu - \delta_{\mu \rho} P_\nu)\,, \\
[D,P_\mu]  & = - \ii P_\mu\,, \\
[D,K_{\mu}] & = + \ii K_\mu\,, \\
[P_\mu, K_{\nu}] & = -2\ii (\delta_{\mu \nu} D + M_{\mu \nu})\,.
\end{aligned}
\eeq
The action of the dilatation operator on operators reads $[D,\mathcal{O}(0)] = -\ii \Delta \mathcal{O}(0)$.

\paragraph{Differential Operator Realization.}

All conformal generators pick a minus sign when acting on operators as differential operators. Explicitly,\footnote{This is an artifact of the coset space. See \cite{Minwalla:1997ka} for details.}
\beq \label{eqn-SUSY-confAlg-Diff}
\begin{aligned}
M_{\mu \nu} \mathcal{O}(x) & = \ii (x_\mu \partial_\nu - x_\nu \partial_\mu) \mathcal{O}(x)\,, \\
K_\mu \mathcal{O}(x) & = \ii (x^2 \partial_\mu - 2 x_\mu x\cdot \partial - 2 x_\mu \Delta) \mathcal{O}(x)\,, \\
P_\mu \mathcal{O}(x) & = \ii \partial_\mu \mathcal{O}(x)\,, \\
D \mathcal{O}(x) & = (-\ii x \cdot \partial -\ii \Delta) \mathcal{O}(x)\, .
\end{aligned}
\eeq

\paragraph{General Transformation Rules.}

The SUSY algebra can be inferred from the field transformation rules. In order to find the (anti-)commutators between supercharges, it suffices to study the commutators of two SUSY variations, parameterized by $\epsilon$ and $\bar{\epsilon}$,\footnote{ We have turned off the vector multiplet in the field transformations for notational simplicity. Resuming the gauge multiplet will modify all derivatives to the corresponding covariant ones.}
\beq \label{eqn-SUSY-doubleComm}
\begin{aligned}
[\delta_\epsilon, \delta_{\bar{\epsilon}}] \phi & = \ii \xi^\mu \partial_\mu \phi +\frac{\rho}{2} \phi - \frac{\alpha}{2} \phi\,, \\
[\delta_\epsilon, \delta_{\bar{\epsilon}}] \phi^\dagger & = \ii \xi^\mu \partial_\mu \phi^\dagger +\frac{\rho}{2} \phi^\dagger + \frac{\alpha}{2} \phi^\dagger, \\
[\delta_\epsilon, \delta_{\bar{\epsilon}}] \psi & = \ii \xi^\mu \partial_\mu \psi + \frac{\ii}{4} \Theta_{\mu \nu} \gamma^{\mu \nu} \psi +\rho \psi + \frac{\alpha}{2} \psi\,,\\
[\delta_\epsilon, \delta_{\bar{\epsilon}}] \bar{\psi} & = \ii \xi^\mu \partial_\mu \psi + \frac{\ii}{4} \Theta_{\mu \nu} \gamma^{\mu \nu} \bar{\psi} +\rho \bar{\psi} - \frac{\alpha}{2} \bar{\psi}\,,
\end{aligned}
\eeq
where
\beq\label{eqn-SUSY-doubleComm-Parameter}
\xi^\mu= \bar{\epsilon} \gamma^\mu \epsilon\,,\quad 
\Theta^{\mu\nu}= \partial^{[\mu} \xi^{\nu ]}\,,\quad
\rho= \frac{\ii}{3} \big(\bar{\epsilon} \gamma^\mu D_\mu \epsilon + D_\mu \bar{\epsilon} \gamma^\mu \epsilon \big)\,,\quad
\alpha= \frac{\ii}{3} \big(D_\mu \bar{\epsilon} \gamma^\mu \epsilon -  \bar{\epsilon} \gamma^\mu D_\mu \epsilon \big)\,.
\eeq
Other commutators vanish, since
\beq
[\delta_\epsilon,\delta_\epsilon] = [\delta_{\bar{\epsilon}},\delta_{\bar{\epsilon}}]= 0\, .
\eeq

On the RHS of \eqref{eqn-SUSY-doubleComm},  the quantum numbers of the respective fields are represented by the coefficients of the terms linear in $\rho$ and $\alpha$. For instance, the coefficient $1/2$ of the $\rho \phi$ term is the tree-level dimension of $\phi$, and the coefficient of $\alpha \phi$ term is minus the $R$-charge of $\phi$. For spinning particles, the terms linear in $\Theta_{\mu \nu}$ gives the spin. When the interaction is turned on, the dimension and spins are replaced by the quantum corrected values.

\paragraph{SUSY Operator: Definitions.}

The action of the supercharges can be defined as the coefficient of the SUSY variations. Explicitly, we \textit{define}
\beq
\delta_\epsilon \mathcal{O} = \epsilon_s [Q,\mathcal{O}] + \ii \epsilon_c [S,\mathcal{O}]\,, \qquad \delta_{\bar{\epsilon}} = \bar{\epsilon}_s [\bar{Q},\mathcal{O}] + \ii \bar{\epsilon}_c [\bar{S},\mathcal{O}]\,.
\eeq
The commutator of SUSY variations can then be expressed in terms of the supercharge commutators,
\beq \label{eqn-SUSY-doubleComm-QS}
[\delta_\epsilon,\delta_{\bar{\epsilon}}] = \bar{\epsilon}^b_s \{Q_a,\bar{Q}_b\} \epsilon_s^a - \bar{\epsilon}^b_c \{S_a,\bar{S}_b\} \epsilon_c^a + \ii \bar{\epsilon}^b_c \{Q_a,\bar{S}_b\} \epsilon_s^a + \ii \bar{\epsilon}^b_s \{S_a,\bar{Q}_b\} \epsilon_c^a\, .
\eeq

\paragraph{Commutation Relation Between Supercharges.}

The SUSY commutation relation directly follows from the double commutator above by comparing \eqref{eqn-SUSY-doubleComm-QS} with the RHS of \eqref{eqn-SUSY-doubleComm}. Plugging in \eqref{eqn-SUSY-KillingVecSol}, the SUSY variation parameters in \eqref{eqn-SUSY-doubleComm-Parameter} take the values
\beq
\begin{aligned}
\xi^\mu 
& = \bar{\epsilon}_s \gamma^\mu \epsilon_s + x^\mu (\bar{\epsilon}_s \epsilon_c - \bar{\epsilon}_c \epsilon_s ) + x^\sigma \big(\bar{\epsilon}_s \gamma_{\mu\sigma} \epsilon_c + \bar{\epsilon}_c \gamma_{\mu\sigma} \epsilon_s\big) -\bar{\epsilon}_c \big( 2 x^\mu x\cdot{\gamma} - x^2 \gamma^\mu \big)\epsilon_c\,, \\
\rho & = \ii (\bar{\epsilon}_s \epsilon_c - \bar{\epsilon}_c \epsilon_s ) -2\ii \bar{\epsilon}_c \gamma_\mu \epsilon_c x^\mu\,, \\
\alpha & = -\ii (\bar{\epsilon}_s \epsilon_c + \bar{\epsilon}_c \epsilon_s )\, .
\end{aligned}
\eeq
For instance, equating the $\bar{\epsilon}_s\epsilon_s$ term on both sides, we find
\beq
\bar{\epsilon}_s^b \{Q_a,\bar{Q}_b\} \epsilon^a_s = \ii \bar{\epsilon}_s \gamma^\mu \epsilon_s \partial_\mu = \bar{\epsilon}_s^b \gamma^\mu_{ba} \epsilon_s^a P_\mu\,, 
\eeq
where we identify the derivative operator with the translation generator $P_\mu$ by \eqref{eqn-SUSY-confAlg-Diff}. Since $\epsilon,\bar{\epsilon}$ are generic, we must have
\beq
\{Q_a,\bar{Q}_b\} = \gamma^\mu_{ab} P_\mu \, ,
\eeq
where we use the fact that $\gamma_{ab}$ is symmetric, $\gamma_{ab}^\mu = \gamma_{ba}^\mu$. 

Similarly, comparing the coefficients of $\bar{\epsilon}_c \epsilon_c$ and using $\Delta_\phi = 1/2$, we find\footnote{This minus sign comes from the definition of $\gamma^{ab}$: $\{S^a,\bar{S}^b\} = (\gamma^\mu)^b_{\ c}\varepsilon^{ca} K_\mu$. Sometimes people define the matrix structure above as $\gamma^{ba}$. In our definition, the Hermitian conjugate of $\gamma_{ab}$ equals to $-\gamma^{ab}$.}
\beq
\{S_a,\bar{S}_b\} = -\gamma^\mu_{ab} K_\mu \qquad \leftrightarrow \qquad \{S^a,\bar{S}^b\} = - (\gamma^\mu)^{ab} K_\mu\,.
\eeq

The most complicated anticommutations relation come form the mixing terms. Identifying\footnote{This replacement means that we choose the $\mathcal{R}$-charge generator to satisfy $[\mathcal{R},\mathcal{O}] = r_\mathcal{O} \mathcal{O}$.}
\beq
\rho \phi/2 \rightarrow \rho \Delta\,, \qquad \alpha/2 \phi \rightarrow r \alpha \simeq \mathcal{R}\,,
\eeq
we find,
\beq \label{eqn-SUSY-QScomm}
\begin{aligned}
\{Q_a,\bar{S}_b\} & = \varepsilon_{ab} \big(\ii D -\mathcal{R} \big) - \frac{1}{2} \varepsilon_{\mu \nu \rho} M^{\mu \nu} \gamma^\rho_{ab}\,, \\
\{S_a,\bar{Q}_b\} & = \varepsilon_{ba} \big(\ii D + \mathcal{R} \big) - \frac{1}{2} \varepsilon_{\mu \nu \rho} M^{\mu \nu} \gamma^\rho_{ab}\,.
\end{aligned}
\eeq
All the other anticommutators among $Q$, $\bar{Q}$, $S$, and $\bar{S}$ unlisted above vanish.

\paragraph{Commutation Relation: the Rest.}

What remains are commutation relations between supercharges and conformal generators. Those can be inferred directly from the space-time transformation properties of the supercharges, so we shall omit the details.

\paragraph{Reality Conditions.}

Recall that the Hermitian conjugate ${}^\dagger$ is defined in \textit{Lorentzian} signature.
As usual, the bosonic part satisfies
\beq
D^\dagger = - D\,, \qquad M^\dagger = M\,, \qquad P^\dagger = K\,, \qquad \mathcal{R}^\dagger = \mathcal{R}\, .
\eeq
Consider the $QQ$ anti-commutator, taking conjugates, we find
\beq
\{(Q_a)^\dagger, (\bar{Q}_b)^\dagger \} = (\gamma^\mu_{ab})^\dagger (P_\mu)^\dagger = - \gamma_\mu^{ab} K^\mu\,. 
\eeq
Hence, we should choose
\beq \label{eqn-SUSY-RC-QS}
(Q_a)^\dagger = \bar{S}^a\,, \qquad (\bar{Q}_a)^\dagger = S^a \, . 
\eeq
This choice is unique since $Q^\dagger$ has an opposite $R$ charge as compared to $Q$. The non-vanishing anticommutators \eqref{eqn-SUSY-QScomm} that involve the uncharged bosonic generators on the RHS must be identified with $\{Q,Q^\dagger\}$.

\section{Boundary Operators of the $1/2$-BPS Condensed Fermion Line} \la{cfboapp}

In this appendix we analyze the superconformal properties of the boundary operators of the $1/2$-BPS condensed fermion line in (\ref{bocondenced}). 
The action of the line superconformal symmetries on these boundary operators is composed of two contributions. One is the direct action on the boundary fields. The other is a boundary term that comes about due to the fact that the symmetries action on the line connection is a total derivative. It is convenient to split the boundary operators in (\ref{bocondenced}) into two groups. 

Consider first the operators $\{\widetilde\cO_R^{(0,-{1\over2})},\widetilde\cO_L^{(0,{1\over2})}\}$. These operators do not have fields in them, so the charges act on them only through line boundary terms. 
The action of superconformal charges $\mathcal{S}^\pm$ on the line leads to a boundary term that vanishes at the origin (\ref{eqn-SUSY-condensedF-actionS}). From (\ref{eqn-SUSY-condensedF-BdrTerm}) we see that they are also annihilated by either $Q_2$ or $\bar Q_1$. For example, 
the action of $\bar{Q}_1=(\mathcal{Q}_+ - \mathcal{Q}_-)/2$ on $\widetilde{\cO}_R^{(0,-\frac{1}{2})}$ leads to a boundary term
\beq
\begin{pmatrix}
0 & 0 \\
u \phi^\dagger & 0 \\ 
\end{pmatrix} \widetilde{\cO}_R^{(0,-\frac{1}{2})} = 
\begin{pmatrix}
0 & 0 \\
u \phi^\dagger & 0 \\ 
\end{pmatrix}
\begin{pmatrix}
0 \\
1 \\
\end{pmatrix} = 0\, .
\eeq
Hence, these boundary operators are superconformal primaries. It follows that the dimension of these BPS boundary operators is related to their transverse spin as in \eqref{eqn-SUSY-BPSalphappP1}, which leads to \eqref{eqn-SUSY-CondFer-Half}. 
The factors $\pm 1/2$ in \eqref{eqn-SUSY-CondFer-Half} correspond to the $R$-charges of the empty endpoints, which must be also transported along the empty segments with the spins. Since $R$-charge is a global symmetry that does not depend on the shape of the path, transporting it is trivial.\footnote{Compared with the normal $\alpha = 1$ line, the $1/2$ BPS condition for the condensed line changes. For instance, the uncondensed $\alpha =1$ right operator ${\cO}_R^{(0,0)}$ is annihilated by $Q_2$. On the contrary, for condensed line the right operator $\widetilde{\cO}_R^{(0,-\frac{1}{2})}$ is annihilated by the other supercharge $\bar{Q}_1$. From this perspective, the right operator in the condensed theory behaves like the left operator in the uncondensed theory. }

The analysis of the opertors in the second group, $\{\widetilde\cO_R^{(0,-{3\over2})},\widetilde\cO_L^{(0,{3\over2})}\}$, is 
more involved because these operators are \textit{not} superconformal primaries.
\footnote{One way to motivate this is to use the duality dictionary, see table (\ref{tab_comp_cond}). For instance, $\widetilde{\cO}_R^{(0,-\frac{3}{2})}$ is dual to $D_-\phi$, which is a decedent field of $\psi^2$ for the $\alpha = -1$ case \eqref{eqn-SUSY-a2}. Consequently $\widetilde{\cO}_R^{(0,-\frac{3}{2})}$ is expected to be a decedent field. }
For instance, since $\psi^2$ is the descendent of $\phi$, \eqref{eqn-SUSY-a1}, the action of $S^1 = 1/2(\mathcal{S}^+ -\mathcal{S}^-) $ on $\widetilde{\cO}_R^{(0,-\frac{3}{2})}$ is non-trivial. It is given by
\beq\la{S1ong2}
S^1 \widetilde{\cO}_R^{(0,-\frac{3}{2})} \propto 
\begin{pmatrix}
D_ - \phi \\
0 \\
\end{pmatrix}\, ,
\eeq
where we use the fact that the superconformal variation of $A_-$ vanishes at the origin. 

The resulting scalar derivative operator on the right hand side of (\ref{S1ong2}) 
is ignorant about the fermion condensation. Hence, the SUSY algebra of the line attaching to it effectively reduces to the normal $\alpha =1$ uncondenced SUSY algebra \eqref{eqn-SUSY-a1}. 
From this algebra, we see that $D_-\phi$ is not a superconformal descendent of the primary fermions, so it must be a superconformal primary on its own. Indeed, $\phi$ is invariant under $S^1$ at any point and the origin in particular. Moreover, the action of $\bar{S}^2$ on $D_-\phi$ reads
\beq
\bar{S}^2D_- \phi \Big|_{x=0}=\bar{S}^2 \partial_- \phi \Big|_{x=0}\propto (\partial_- x^-) (\gamma_- \bar{\epsilon}_c) \psi \Big|_{x=0} = (\gamma_- \bar{\epsilon}_c) \psi = 0\,,
\eeq
where the non-trivial contribution originated from the action of the $\d_-$ derivative on the position dependent part of the SUSY parameter. In the last step we used the fact that the only non-vanishing $\bar{\epsilon}_c$ component is $\bar{\epsilon}^1_c$ but $(\gamma_-)^{\bullet}_{\ 1} = 0$.

We can also directly check the BPS condition for $D_-\phi$. Since the scalar field, $\phi$ is invariant under the action of $Q_2$, we would like to verify the action of this supercharge on the covariant derivative and on the line. The SUSY variation of the line is still orthogonal to this boundary primary,
\beq
Q_2 \Big[e^{\int \mathcal{L}_u} \Big] \cdot
\begin{pmatrix}
D_- \phi \\
0 \\
\end{pmatrix}
\propto 
\begin{pmatrix}
0 & u \phi \\
0 & 0 \\
\end{pmatrix}
\begin{pmatrix}
D_- \phi \\
0 \\
\end{pmatrix}
= 0 \, .
\eeq
The non-trivial SUSY variation of the covariant derivative comes from the gauge field $A_-$. Using \eqref{eqn-SUSY-vec}, we find that the action of $Q_2$ on $A_-$ is non-zero. It is given by
\beq \label{eqn-SUSY-variationAMCondensed}
\delta A_- = \frac{\ii}{2} \Big[ \bar{\lambda}^2 (\gamma_-)^1_{\ 2} \epsilon^2 \Big] = \frac{\ii}{\sqrt{2}}  \bar{\lambda}^2 \epsilon^2 \quad \Rightarrow \quad Q_2 \circ A_- \propto \bar{\lambda}^2. 
\eeq
On the other hand, $\bar{\lambda}^2$ is invariant under $\bar{S}^2$ 
\beq
\delta \bar{\lambda}^2 \propto \sigma (\gamma^\mu D_\mu \bar{\epsilon})^2 \propto \sigma \bar{\epsilon}_c^2 = 0\,,
\eeq
where the non-trivial contribution at the origin comes from the last term of \eqref{eqn-SUSY-vec2}. 

We conclude that this operator is still being annihilated by the commutator $\{Q_2,\bar{S}^2\}$. This leads to BPS-type condition 
\beq
\Delta(D_- \phi) = r(D_-\phi) - \mathfrak{s}(D_-\phi) = \frac{1}{2} - \mathfrak{s}(D_-\phi)\, .
\eeq
The descendent field therefore satisfies the relation
\beq
\Delta(D_- \psi^2) = \Delta(D_- \phi) + \frac{1}{2}\,, \qquad \mathfrak{s}(D_-\psi^2) = \mathfrak{s}(D_-\phi) - \frac{1}{2}\, .
\eeq
By combining these relations, we arrive at the first equation of \eqref{eqn-SUSY-CondFer-3Half}. 

In order to be consistent with supersymmetric duality, where the dual operator $\psi^2$ is BPS, the operator $(D_-\phi \ 0)$ must also be lifted to a BPS operator. This lift can be achieved by modifying the lower component of the scalar operator with a color singlet term of tree-level dimension $3/2$,
\beq
\begin{pmatrix}
D_-\phi \\
0 \\
\end{pmatrix}\quad \rightarrow \quad
\begin{pmatrix}
D_-\phi \\
\frac{\ii u}{\sqrt{2}} \bar{\psi}^2 \phi \\
\end{pmatrix}\,.
\eeq
The lower component singlet operator is invariant under $Q_2$ since its constituents are invariant. However, it modifies the first component when combined with the line boundary term,
\beq
Q_2 \Big[e^{\int \mathcal{L}_u} \Big] \cdot
\begin{pmatrix}
D_- \phi \\
\frac{\ii u}{\sqrt{2}} \bar{\psi}^2 \phi \\
\end{pmatrix}
=
\begin{pmatrix}
0 & u \phi \\
0 & 0 \\
\end{pmatrix}
\begin{pmatrix}
D_- \phi \\
\frac{\ii u}{\sqrt{2}} \bar{\psi}^2 \phi \\
\end{pmatrix}
= \begin{pmatrix}
\frac{\ii u^2}{\sqrt{2}} \phi (\bar{\psi^2} \phi) \\
0 \\
\end{pmatrix}
= \begin{pmatrix}
- \frac{1}{\sqrt{2}} \bar{\lambda}^2 \phi \\
0 \\
\end{pmatrix} \, ,
\eeq
where we used \eqref{eqn-SUSY-ValGaugino} in the very last step. It cancels the variation of the gauge field \eqref{eqn-SUSY-variationAMCondensed}. From this operator, we can then find the corrected $\widetilde{O}_R^{(0,-\frac{3}{2})}$ as its $\bar{Q}_1$ descendent,
\beq
\widetilde{O}_R^{(0,-\frac{3}{2})} = - \delta_{\bar{Q}_1} \begin{pmatrix}
D_- \phi \\
\frac{\ii u}{\sqrt{2}} \bar{\psi}^2 \phi \\
\end{pmatrix} =
\begin{pmatrix}
D_- \psi^2 \\
u D_-(\phi^\dagger \cdot \phi) 
+\frac{\ii u}{\sqrt{2}} \bar{\psi}^2 \cdot \psi^2 \\
\end{pmatrix}\, .
\eeq
The lower component is a color singlet, which decouples from the upper component in the planar limit.

Similarly, we find that the boundary operator $\widetilde\cO_L^{(0,{3\over2})}$ is a superconformal descendent of $(D_+ \phi^\dagger\ 0)$. The latter is a superconformal primary but not a BPS operator. Nevertheless, it is being annihilated by the anticommutator $\{\bar{Q}_1,\bar{S}^1\}$, so it also satisfies the BPS-type condition. This condition leads to the second equation of \eqref{eqn-SUSY-CondFer-3Half}. 

Modifying the second component of $(D_+\phi^\dagger\,,\ 0)$ will make it BPS. Explicitly, the supersymmetric BPS lift of $(D_+\phi^\dagger\,,\ 0)$ is given by
\beq
(D_+\phi^\dagger\,,\  0)\quad\rightarrow\quad(D_+\phi^\dagger\,,\  \frac{\ii u}{\sqrt{2}} \phi^\dagger \cdot \psi^1)\, .
\eeq
Correspondingly, its $Q_2$ descendent boundary operator $\widetilde\cO_L^{(0,{3\over2})}$ is lifted to
\beq
\big(D_+ \bar{\psi}^1\,,\ 
    0\big)\quad\rightarrow\quad\big(D_+ \bar{\psi}^1\,,\ 
    u\, D_+(\phi^\dagger \cdot \phi) + \frac{\ii u}{\sqrt{2}} \bar{\psi}^1 \cdot \psi^1\big)\,.
\eeq

\section{Fermion Self-energy: the Differential Equation Approach}
\label{sec:ferselfEnergyDiff}

In this appendix we present a differential equation 
method of solving the Schwinger-Dyson equation of the fermionic self-energy \eqref{SDeq}. Recall that in matrix components, this equation takes the form
\beq
\Sigma_- = \Sigma_3 = 0\,,
\eeq
and 
\beq \label{eqn-apd-SDfer}
\begin{aligned}
\Sigma_+(p) & = - 4\pi \lambda \int \frac{\dd^3 q}{(2\pi)^3} \frac{\Sigma_0}{(q_\mu + \Sigma_\mu)^2 + \Sigma_0^2} \frac{e^{\ii \epsilon q_3}}{(p-q)^+}\,,  \\
\Sigma_0(p) - M_{c.t.} & =  4\pi \lambda \int \frac{\dd^3 q}{(2\pi)^3} \frac{1}{(q_\mu + \Sigma_\mu)^2 + \Sigma_0^2} \frac{q^+}{(p-q)^+} e^{\ii \epsilon q_3}\, .
\end{aligned}
\eeq

Our regulator preserves the rotational symmetry in the transverse plane. Therefore, we can assume the non-vanishing components of the self-energy take the following form,
\beq
\Sigma_0(p) = f(\lambda,|p_\perp|)\, |p_\perp| e^{-b \epsilon |p_\perp|}\,, \qquad \Sigma_+(p)= g(\lambda,|p_\perp|)\, p^- e^{-a \epsilon |p_\perp|}\,,
\eeq
where $f$, $g$ are two unknown functions and $a$, $b$ are two unknown constants. For simplicity, we will suppress the $\lambda$ dependence of $f$ and $g$.

\paragraph{Differential Equation.}

Using the identity
\beq
\frac{\partial}{\partial p^-} \frac{1}{p^+} = 2\pi \delta^{(2)}(p_\perp),
\eeq
we can convert \eqref{eqn-apd-SDfer} to a system of differential equations,
\beq \label{eqn-FerSelfEnergy-DiffEqnFG}
\begin{aligned}
\frac{\partial}{\partial p^-} \Big( p^- g e^{-a\epsilon |p_\perp|} \Big) & = -2\lambda f |p_\perp| e^{-(1+b)\epsilon |p_\perp|} \int \frac{\dd q_3}{2\pi} \frac{1}{q_3^2 + p_\perp^2 (1+f^2 e^{-2b\epsilon |p_\perp|} +g e^{-a\epsilon |p_\perp|} )}\,, \\
& = -\lambda f e^{-(1+b)\epsilon |p_\perp|} \frac{1}{\sqrt{1+f^2 e^{-2b\epsilon |p_\perp|} +g e^{-a\epsilon |p_\perp|} }}, \\
\frac{\partial}{\partial p^-} \Big( f |p_\perp| e^{-b\epsilon |p_\perp|} \Big) & = \lambda \frac{p^+}{|p_\perp|} e^{-\epsilon |p_\perp|}\frac{1}{\sqrt{1+f^2 e^{-2b\epsilon |p_\perp|} +g e^{-a\epsilon |p_\perp|} }}\,.
\end{aligned}
\eeq

Comparing the exponential term on both sides, we find 
\beq
a = 1+b\,, \quad b =1  \qquad \Rightarrow \qquad a = 2\,, \quad b=1\,.
\eeq

\paragraph{Differential Equations for $f^2 +g$.}

In order for the exact propagator to be consistent with conformal symmetry, it must have a pole at $p=0$. Consequently, 
\beq
(p_\mu + \Sigma_\mu(p))^2 + \Sigma_0^2(p)-p^2\ \propto\ f^2 e^{-2b\epsilon |p_\perp|} + g e^{-a\epsilon |p_\perp|}=0\,.
\eeq
As $a =2 = 2b$, the exponential term factors out, and $f^2+g =0$ suffices. We would like to show first that this choice is indeed compatible with the structures of the equations in (\ref{eqn-FerSelfEnergy-DiffEqnFG}).

By plugging $a=2, b=1$ into \eqref{eqn-FerSelfEnergy-DiffEqnFG}, we find,
\beq
\begin{aligned}
\frac{\partial}{\partial p^-} \Big( p^- g e^{-2\epsilon |p_\perp|} \Big) & = -\frac{\lambda f e^{-2\epsilon |p_\perp|}}{\sqrt{1+(f^2+g) e^{-2b\epsilon |p_\perp|}}}\,,\\
\frac{\partial}{\partial p^-} \Big( f |p_\perp|e^{-\epsilon |p_\perp|} \Big) & = \frac{\lambda p^+}{|p_\perp|} e^{-\epsilon |p_\perp|}\frac{1}{\sqrt{1+(f^2+g) e^{-2b\epsilon |p_\perp|}}}\, .
\end{aligned}
\eeq
Taking linear combinations of these two equations to cancel the RHS, we end with
\beq
e^{\epsilon |p_\perp|} p^+ \frac{\partial}{\partial p^-} \Big( p^- g e^{-2\epsilon |p_\perp|} \Big) + f |p_\perp| \frac{\partial}{\partial p^-}\Big( f |p_\perp|e^{-\epsilon |p_\perp|} \Big) = 0\,.
\eeq
After some straightforward algebra, this equation becomes a first order differential equation for the combination $f^2+g$,
\beq
\Big(1-\epsilon |p_\perp|+ \frac{1}{2} |p_\perp| \frac{\partial}{\partial |p_\perp|}\Big)(g+f^2)=0\qquad \Rightarrow\qquad g+f^2 = \mathtt{d} \frac{e^{2\epsilon |p_\perp|}}{p_\perp^2}\, ,
\eeq
where $\mathtt{d}$ is a constant of integration. Choosing this constant to be zero gives us the solution needed.\footnote{The constant $\mathtt{d}$ determines the pole of the exact propagator, whose denominator takes the form of $p^2 +\mathtt{d}^2$.}

\paragraph{Differential Equations for $f,g$.}

After setting $f^2+g=0$, \eqref{eqn-FerSelfEnergy-DiffEqnFG} simplifies to the following set of first order differential equations, 
\beq
g + |p_\perp| \Big(\frac{1}{2}g'-\epsilon g\Big) = -\lambda f\,,\qquad |p_\perp| \Big(f' |p_\perp| + f-\epsilon f |p_\perp|\Big) = \lambda |p_\perp|\,.
\eeq
The second equation only involves $f$. It an be solved as
\beq
f(|q_\perp|) = -\frac{\lambda }{\epsilon |p_\perp|}+\mathtt{c}_f\frac{ e^{\epsilon |p_\perp|}}{|p_\perp|},
\eeq
where $\mathtt{c}_f$ is the constant of integration. The function $g$ is then determined by $g = -f^2$. Notably, the first differential equation above is obeyed for any value of 
$\mathtt{c}_f$.

\paragraph{Final Solution.}

The original integral equation \eqref{eqn-apd-SDfer} can be used to calculate the value of the constant $\mathtt{c}_f$. Upon reinserting the solution for $f$ and $g$, we find that this integral equation fixes both $\mathtt{c}_f$ and the mass counterterm $M_{c.t.}$ to
\beq
\mathtt{c}_f = \frac{\lambda}{\epsilon}\,, \qquad \quad M_{c.t.} = \mathtt{c}_f\,.
\eeq
The final non-vanishing components of the self energy read
\beq
\Sigma_0 = \frac{\lambda}{\epsilon}\Big(1-e^{-\epsilon |p_\perp|} \Big)\,, \qquad \Sigma_+ = -\lambda^2 \frac{(1-e^{-\epsilon |p_\perp|})^2}{\epsilon^2 |p_\perp|^2} p^- \, ,
\eeq
in agreement with the solution that we have found in the main text, \eqref{eq:exactprop}.

\section{Perturbative Analysis of the Framing Factor} \label{apd:FramingFactor}

The dependence of the mesonic line operators on the framing of the line is given by the framing factor in (\ref{ff}). In section \ref{sec:susy} we have presented an indirect derivation of this factor. In this appendix we perform further perturbative checks of this result in Lorentz and lightcone gauges.




\subsection{Lorentz Gauge}

In Appendix \ref{apd:oneLoopGen}, we computed the one-loop expectation value of the mesonic line operator along an arbitrary smooth path. At this order, we have found that the framing factor is given by \eqref{eqn-apd-relTorsionLorentz}, in agreement with the expansion of \eqref{ff}.
As for closed loops, this framing dependence originated solely from the gluon-gluon integration on the line.\footnote{In particular, the one-loop gluon-matter interaction in the Lorentz gauge does not lead to a framing-dependent contribution.} It quantifies the total angle in which the framing vector rotates along the path. 


\subsection{Lightcone Gauge}

In the lightcone gauge, the planar gluon-gluon self-interaction 
factors out of the expectation values of the mesonic lines, allowing us to analyze it separately.

For lines in which the endpoints point in the third direction, i.e., $e_1 = e_0 \propto (0,0,1)$, these are the only framing-dependent contributions, as shown below. They exponentiate into the framing factor \eqref{ff}. On the other hand, for lines in which at least one of the endpoints is not pointing in the third direction, we find that both of these properties no longer hold. At one-loop order, the matter-gluon interactions depend on the framing. It combines with the gluon-gluon ones into a non-Lorentz invariant result that differs from the correct one we obtained in the Lorentz gauge. Similar issues with lightcone gauge has arisen before, see for example \cite{Labastida:1997uw}. We will not attempt to resolve this puzzle at present and leave it for future study.

\paragraph{Gluon-Gluon Interaction.}

Consider any diagram with a gluon exchange between two points on the Wilson line, (see figure \ref{fig:Aspin} for an example)
\beq \la{gluonexchange}
\int \dd x^\mu\int \dd y^\nu\<\ldots \wick{\c1 A_\mu(x)\c1 A_\nu(y)}\ldots\>\,.
\eeq
The gluon propagator in light-cone gauge is localized in the third direction, with the only non-zero components being
\eqref{eqn-LC-gluonProp},
\beq\la{gluonprop}
\wick{\c1 A_3(x)\,\c1 A_+(0)}=-\wick{\c1 A_+(x)\,\c1 A_3(0)}={2\over k}{\delta(x^3)\over x^+}\, .
\eeq
Hence, for lines that can be uniquely parameterized by their projection on the third direction, the gluon exchange (\ref{gluonexchange}) is local. Moreover, the gluon rainbow diagrams with more than one propagator do not contribute. The reason is that after dressing out the center of mass integration, the total number of delta functions is less than the remaining number of integrations. For instance, consider the two-rainbow diagram given by
\beq
\int\limits_{s_1\leq t_1 \leq t_2 \leq s_2}\!\!\!\!\!\!\!\!\!\! \dd s_1\,\dd s_2\,\dd t_1\,\dd t_2\,G(t_1,t_2) G(s_1,s_2)\,,
\eeq
where $G(t_1,t_2)\propto\delta(t_2-t_1)$ represents the gluon propagator contracted between two points on the line. 
Because of the path ordering, this integral is non-vanishing only for $s_1 = t_1=t_2 = s_2$. However, one is still left with an integral over $t_1$ between $s_1$ and $s_2$, so the final result vanishes.

We conclude that the gluon exchange diagrams factor out of any planar diagram. They give an overall factor that can be dressed out of the mesonic line expectation value. To compute it, we follow the framing regularization prescription. It amounts to plugging $y=x+\epsilon\, n$ and taking the $\epsilon\to0$ limit. We find  
\begin{align} \label{eqn-Bos-ASpinFactor}
\int \dd x^\mu\int \dd y^\nu\<\ldots \wick{\c1 A_\mu(x)\c1 A_\nu(y)}\ldots\>=&\<\ldots\>\times{\lambda\over2}\iint \(\dd x^+\dd y^3-\dd x^3\dd y^+\){\delta(x^3-y^3)\over x^+-y^+}\nn\\
=&\<\ldots\>\times{\lambda\over2}\int \dd \log (y^+_s-x_s^+)\,.
\end{align}
where the parametrization of the two lines is chosen based on the value of the third component, $s = x_s^3$. In this parametrization, $\dd x_3 = \dd y_3$, which simplifies the double integral.

Let $y(x)=x+\epsilon\,n$ be a framed point along the line, where $n$ is a unit normal vector at $x$. This parametrization of the point $y$ is different from the one in (\ref{eqn-Bos-ASpinFactor}), which is defined by its projection on the third direction as $y_s^3=x_s^3$. Consequently, the difference inside the logarithm in (\ref{eqn-Bos-ASpinFactor}) is different from the plus component of the framing vector. It is given by 
\beq
y_s^+ - x_s^+ =\epsilon ({\bf n}_s^+ -{\bf n}_s^3 \dot{x}_s^+) + \cO(\epsilon^2)\,.
\eeq
Here, ${\bf n}^+$ is in the $x^1\!\!-\!x^2$ plane, while $n^+$ in (\ref{eqn-ff-int}) is in the normal plane to the path at $x_s$. These are two different coordinate systems, (unless the path is pointing in the third direction). 

Consider first a path in which the two endpoints are pointing in the third direction, $e_0=e_1\propto(0,0,1)$. In this case, at $L$-loops we have $L$ such ordered contractions 
\beq
{\lambda^L\over2^L}\int\!\!\prod_{s_1<\ldots<s_L}\!\!\dd \log ({\bf n}_{s_i}^+ e_{s_i}^3 -{\bf n}_{s_i}^3 e_{s_i}^+)=
{1\over L!}\({\lambda\over2} \log {n_1^+\over n_0^+}\)^L\,,
\eeq
where we have used that $e_1^+=e_0^+=0$ and $e_1^3=e_0^3$. Hence, these contributions therefore simply exponentiate to the overall framing factor (\ref{ff}). 

On the other hand, if one of the line endpoints is not pointing in the third direction then already at one loop-order we get that the gluon exchange diagram in figure \ref{fig:1loop}.a yields
\beq
W_a^\text{1-loop}/W^\text{tree}={\lambda\over2} \log {{\bf n}_1^+ e_1^3 -{\bf n}_1^3 e_1^+\over{\bf n}_0^+ e_0^3 -{\bf n}_0^3 e_0^+}\,.
\eeq
In this generic case also the gluon-scalar diagram in figure \ref{fig:1loop}.c gives a framing dependence contribution. The two however do not combine into a Lorentz invariant one-loop result.

\bibliography{bib}

\providecommand{\href}[2]{#2}\begingroup\raggedright\begin{thebibliography}{100}

\bibitem{Sezgin:2002rt}
E.~Sezgin and P.~Sundell, {\it {Massless higher spins and holography}},  {\em
  Nucl. Phys. B} {\bf 644} (2002) 303--370,
  [\href{http://arxiv.org/abs/hep-th/0205131}{{\tt hep-th/0205131}}]. [Erratum:
  Nucl.Phys.B 660, 403--403 (2003)].

\bibitem{Klebanov:2002ja}
I.~R. Klebanov and A.~M. Polyakov, {\it {AdS dual of the critical O(N) vector
  model}},  {\em Phys. Lett. B} {\bf 550} (2002) 213--219,
  [\href{http://arxiv.org/abs/hep-th/0210114}{{\tt hep-th/0210114}}].

\bibitem{Giombi:2009wh}
S.~Giombi and X.~Yin, {\it {Higher Spin Gauge Theory and Holography: The
  Three-Point Functions}},  {\em JHEP} {\bf 09} (2010) 115,
  [\href{http://arxiv.org/abs/0912.3462}{{\tt arXiv:0912.3462}}].

\bibitem{Benini:2011mf}
F.~Benini, C.~Closset, and S.~Cremonesi, {\it {Comments on 3d Seiberg-like
  dualities}},  {\em JHEP} {\bf 10} (2011) 075,
  [\href{http://arxiv.org/abs/1108.5373}{{\tt arXiv:1108.5373}}].

\bibitem{Giombi:2011kc}
S.~Giombi, S.~Minwalla, S.~Prakash, S.~P. Trivedi, S.~R. Wadia, and X.~Yin,
  {\it {Chern-Simons Theory with Vector Fermion Matter}},  {\em Eur. Phys. J.
  C} {\bf 72} (2012) 2112, [\href{http://arxiv.org/abs/1110.4386}{{\tt
  arXiv:1110.4386}}].

\bibitem{Aharony:2011jz}
O.~Aharony, G.~Gur-Ari, and R.~Yacoby, {\it {d=3 Bosonic Vector Models Coupled
  to Chern-Simons Gauge Theories}},  {\em JHEP} {\bf 03} (2012) 037,
  [\href{http://arxiv.org/abs/1110.4382}{{\tt arXiv:1110.4382}}].

\bibitem{Maldacena:2011jn}
J.~Maldacena and A.~Zhiboedov, {\it {Constraining Conformal Field Theories with
  A Higher Spin Symmetry}},  {\em J. Phys. A} {\bf 46} (2013) 214011,
  [\href{http://arxiv.org/abs/1112.1016}{{\tt arXiv:1112.1016}}].

\bibitem{Maldacena:2012sf}
J.~Maldacena and A.~Zhiboedov, {\it {Constraining conformal field theories with
  a slightly broken higher spin symmetry}},  {\em Class. Quant. Grav.} {\bf 30}
  (2013) 104003, [\href{http://arxiv.org/abs/1204.3882}{{\tt
  arXiv:1204.3882}}].

\bibitem{Chang:2012kt}
C.-M. Chang, S.~Minwalla, T.~Sharma, and X.~Yin, {\it {ABJ Triality: from
  Higher Spin Fields to Strings}},  {\em J. Phys. A} {\bf 46} (2013) 214009,
  [\href{http://arxiv.org/abs/1207.4485}{{\tt arXiv:1207.4485}}].

\bibitem{Jain:2012qi}
S.~Jain, S.~P. Trivedi, S.~R. Wadia, and S.~Yokoyama, {\it {Supersymmetric
  Chern-Simons Theories with Vector Matter}},  {\em JHEP} {\bf 10} (2012) 194,
  [\href{http://arxiv.org/abs/1207.4750}{{\tt arXiv:1207.4750}}].

\bibitem{Aharony:2012nh}
O.~Aharony, G.~Gur-Ari, and R.~Yacoby, {\it {Correlation Functions of Large N
  Chern-Simons-Matter Theories and Bosonization in Three Dimensions}},  {\em
  JHEP} {\bf 12} (2012) 028, [\href{http://arxiv.org/abs/1207.4593}{{\tt
  arXiv:1207.4593}}].

\bibitem{Yokoyama:2012fa}
S.~Yokoyama, {\it {Chern-Simons-Fermion Vector Model with Chemical Potential}},
   {\em JHEP} {\bf 01} (2013) 052, [\href{http://arxiv.org/abs/1210.4109}{{\tt
  arXiv:1210.4109}}].

\bibitem{Gur-Ari:2012lgt}
G.~Gur-Ari and R.~Yacoby, {\it {Correlators of Large N Fermionic Chern-Simons
  Vector Models}},  {\em JHEP} {\bf 02} (2013) 150,
  [\href{http://arxiv.org/abs/1211.1866}{{\tt arXiv:1211.1866}}].

\bibitem{Aharony:2012ns}
O.~Aharony, S.~Giombi, G.~Gur-Ari, J.~Maldacena, and R.~Yacoby, {\it {The
  Thermal Free Energy in Large N Chern-Simons-Matter Theories}},  {\em JHEP}
  {\bf 03} (2013) 121, [\href{http://arxiv.org/abs/1211.4843}{{\tt
  arXiv:1211.4843}}].

\bibitem{Jain:2013py}
S.~Jain, S.~Minwalla, T.~Sharma, T.~Takimi, S.~R. Wadia, and S.~Yokoyama, {\it
  {Phases of large $N$ vector Chern-Simons theories on $S^2 \times S^1$}},
  {\em JHEP} {\bf 09} (2013) 009, [\href{http://arxiv.org/abs/1301.6169}{{\tt
  arXiv:1301.6169}}].

\bibitem{Takimi:2013zca}
T.~Takimi, {\it {Duality and higher temperature phases of large N Chern-Simons
  matter theories on $S^2$ x $S^1$}},  {\em JHEP} {\bf 07} (2013) 177,
  [\href{http://arxiv.org/abs/1304.3725}{{\tt arXiv:1304.3725}}].

\bibitem{Jain:2013gza}
S.~Jain, S.~Minwalla, and S.~Yokoyama, {\it {Chern Simons duality with a
  fundamental boson and fermion}},  {\em JHEP} {\bf 11} (2013) 037,
  [\href{http://arxiv.org/abs/1305.7235}{{\tt arXiv:1305.7235}}].

\bibitem{Yokoyama:2013pxa}
S.~Yokoyama, {\it {A Note on Large N Thermal Free Energy in Supersymmetric
  Chern-Simons Vector Models}},  {\em JHEP} {\bf 01} (2014) 148,
  [\href{http://arxiv.org/abs/1310.0902}{{\tt arXiv:1310.0902}}].

\bibitem{Bardeen:2014paa}
W.~A. Bardeen and M.~Moshe, {\it {Spontaneous Breaking of Scale Invariance in a
  D=3 U(N ) Model with Chern-Simons Gauge Fields}},  {\em JHEP} {\bf 06} (2014)
  113, [\href{http://arxiv.org/abs/1402.4196}{{\tt arXiv:1402.4196}}].

\bibitem{Jain:2014nza}
S.~Jain, M.~Mandlik, S.~Minwalla, T.~Takimi, S.~R. Wadia, and S.~Yokoyama, {\it
  {Unitarity, Crossing Symmetry and Duality of the S-matrix in large N
  Chern-Simons theories with fundamental matter}},  {\em JHEP} {\bf 04} (2015)
  129, [\href{http://arxiv.org/abs/1404.6373}{{\tt arXiv:1404.6373}}].

\bibitem{Bardeen:2014qua}
W.~A. Bardeen, {\it {The Massive Fermion Phase for the U(N) Chern-Simons Gauge
  Theory in D=3 at Large N}},  {\em JHEP} {\bf 10} (2014) 039,
  [\href{http://arxiv.org/abs/1404.7477}{{\tt arXiv:1404.7477}}].

\bibitem{Gurucharan:2014cva}
V.~Gurucharan and S.~Prakash, {\it {Anomalous dimensions in non-supersymmetric
  bifundamental Chern-Simons theories}},  {\em JHEP} {\bf 09} (2014) 009,
  [\href{http://arxiv.org/abs/1404.7849}{{\tt arXiv:1404.7849}}]. [Erratum:
  JHEP 11, 045 (2017)].

\bibitem{Dandekar:2014era}
Y.~Dandekar, M.~Mandlik, and S.~Minwalla, {\it {Poles in the $S$-Matrix of
  Relativistic Chern-Simons Matter theories from Quantum Mechanics}},  {\em
  JHEP} {\bf 04} (2015) 102, [\href{http://arxiv.org/abs/1407.1322}{{\tt
  arXiv:1407.1322}}].

\bibitem{Frishman:2014cma}
Y.~Frishman and J.~Sonnenschein, {\it {Large N Chern-Simons with massive
  fundamental fermions - A model with no bound states}},  {\em JHEP} {\bf 12}
  (2014) 165, [\href{http://arxiv.org/abs/1409.6083}{{\tt arXiv:1409.6083}}].

\bibitem{Moshe:2014bja}
M.~Moshe and J.~Zinn-Justin, {\it {3D Field Theories with Chern--Simons Term
  for Large $N$ in the Weyl Gauge}},  {\em JHEP} {\bf 01} (2015) 054,
  [\href{http://arxiv.org/abs/1410.0558}{{\tt arXiv:1410.0558}}].

\bibitem{Aharony:2015pla}
O.~Aharony, P.~Narayan, and T.~Sharma, {\it {On monopole operators in
  supersymmetric Chern-Simons-matter theories}},  {\em JHEP} {\bf 05} (2015)
  117, [\href{http://arxiv.org/abs/1502.00945}{{\tt arXiv:1502.00945}}].

\bibitem{Inbasekar:2015tsa}
K.~Inbasekar, S.~Jain, S.~Mazumdar, S.~Minwalla, V.~Umesh, and S.~Yokoyama,
  {\it {Unitarity, crossing symmetry and duality in the scattering of $
  \mathcal{N}=1 $ susy matter Chern-Simons theories}},  {\em JHEP} {\bf 10}
  (2015) 176, [\href{http://arxiv.org/abs/1505.06571}{{\tt arXiv:1505.06571}}].

\bibitem{Bedhotiya:2015uga}
A.~Bedhotiya and S.~Prakash, {\it {A test of bosonization at the level of
  four-point functions in Chern-Simons vector models}},  {\em JHEP} {\bf 12}
  (2015) 032, [\href{http://arxiv.org/abs/1506.05412}{{\tt arXiv:1506.05412}}].

\bibitem{Gur-Ari:2015pca}
G.~Gur-Ari and R.~Yacoby, {\it {Three Dimensional Bosonization From
  Supersymmetry}},  {\em JHEP} {\bf 11} (2015) 013,
  [\href{http://arxiv.org/abs/1507.04378}{{\tt arXiv:1507.04378}}].

\bibitem{Minwalla:2015sca}
S.~Minwalla and S.~Yokoyama, {\it {Chern Simons Bosonization along RG Flows}},
  {\em JHEP} {\bf 02} (2016) 103, [\href{http://arxiv.org/abs/1507.04546}{{\tt
  arXiv:1507.04546}}].

\bibitem{Radicevic:2015yla}
D.~Radi\v{c}evi\'c, {\it {Disorder Operators in Chern-Simons-Fermion
  Theories}},  {\em JHEP} {\bf 03} (2016) 131,
  [\href{http://arxiv.org/abs/1511.01902}{{\tt arXiv:1511.01902}}].

\bibitem{Geracie:2015drf}
M.~Geracie, M.~Goykhman, and D.~T. Son, {\it {Dense Chern-Simons Matter with
  Fermions at Large N}},  {\em JHEP} {\bf 04} (2016) 103,
  [\href{http://arxiv.org/abs/1511.04772}{{\tt arXiv:1511.04772}}].

\bibitem{Aharony:2015mjs}
O.~Aharony, {\it {Baryons, monopoles and dualities in Chern-Simons-matter
  theories}},  {\em JHEP} {\bf 02} (2016) 093,
  [\href{http://arxiv.org/abs/1512.00161}{{\tt arXiv:1512.00161}}].

\bibitem{Yokoyama:2016sbx}
S.~Yokoyama, {\it {Scattering Amplitude and Bosonization Duality in General
  Chern-Simons Vector Models}},  {\em JHEP} {\bf 09} (2016) 105,
  [\href{http://arxiv.org/abs/1604.01897}{{\tt arXiv:1604.01897}}].

\bibitem{Gur-Ari:2016xff}
G.~Gur-Ari, S.~A. Hartnoll, and R.~Mahajan, {\it {Transport in
  Chern-Simons-Matter Theories}},  {\em JHEP} {\bf 07} (2016) 090,
  [\href{http://arxiv.org/abs/1605.01122}{{\tt arXiv:1605.01122}}].

\bibitem{Karch:2016sxi}
A.~Karch and D.~Tong, {\it {Particle-Vortex Duality from 3d Bosonization}},
  {\em Phys. Rev. X} {\bf 6} (2016), no.~3 031043,
  [\href{http://arxiv.org/abs/1606.01893}{{\tt arXiv:1606.01893}}].

\bibitem{Murugan:2016zal}
J.~Murugan and H.~Nastase, {\it {Particle-vortex duality in topological
  insulators and superconductors}},  {\em JHEP} {\bf 05} (2017) 159,
  [\href{http://arxiv.org/abs/1606.01912}{{\tt arXiv:1606.01912}}].

\bibitem{Seiberg:2016gmd}
N.~Seiberg, T.~Senthil, C.~Wang, and E.~Witten, {\it {A Duality Web in 2+1
  Dimensions and Condensed Matter Physics}},  {\em Annals Phys.} {\bf 374}
  (2016) 395--433, [\href{http://arxiv.org/abs/1606.01989}{{\tt
  arXiv:1606.01989}}].

\bibitem{Giombi:2016ejx}
S.~Giombi, {\it {Higher Spin \textemdash{} CFT Duality}},  in {\em {Theoretical
  Advanced Study Institute in Elementary Particle Physics}: {New Frontiers in
  Fields and Strings}}, pp.~137--214, 2017.
\newblock \href{http://arxiv.org/abs/1607.02967}{{\tt arXiv:1607.02967}}.

\bibitem{Hsin:2016blu}
P.-S. Hsin and N.~Seiberg, {\it {Level/rank Duality and Chern-Simons-Matter
  Theories}},  {\em JHEP} {\bf 09} (2016) 095,
  [\href{http://arxiv.org/abs/1607.07457}{{\tt arXiv:1607.07457}}].

\bibitem{Radicevic:2016wqn}
D.~Radi\v{c}evi\'c, D.~Tong, and C.~Turner, {\it {Non-Abelian 3d Bosonization
  and Quantum Hall States}},  {\em JHEP} {\bf 12} (2016) 067,
  [\href{http://arxiv.org/abs/1608.04732}{{\tt arXiv:1608.04732}}].

\bibitem{Karch:2016aux}
A.~Karch, B.~Robinson, and D.~Tong, {\it {More Abelian Dualities in 2+1
  Dimensions}},  {\em JHEP} {\bf 01} (2017) 017,
  [\href{http://arxiv.org/abs/1609.04012}{{\tt arXiv:1609.04012}}].

\bibitem{Giombi:2016zwa}
S.~Giombi, V.~Gurucharan, V.~Kirilin, S.~Prakash, and E.~Skvortsov, {\it {On
  the Higher-Spin Spectrum in Large N Chern-Simons Vector Models}},  {\em JHEP}
  {\bf 01} (2017) 058, [\href{http://arxiv.org/abs/1610.08472}{{\tt
  arXiv:1610.08472}}].

\bibitem{Wadia:2016zpd}
S.~R. Wadia, {\it {Chern\textendash{}Simons theories with fundamental matter: A
  brief review of large $N$ results including Fermi\textendash{}Bose duality
  and the S-matrix}},  {\em Int. J. Mod. Phys. A} {\bf 31} (2016), no.~32
  1630052.

\bibitem{Aharony:2016jvv}
O.~Aharony, F.~Benini, P.-S. Hsin, and N.~Seiberg, {\it {Chern-Simons-matter
  dualities with $SO$ and $USp$ gauge groups}},  {\em JHEP} {\bf 02} (2017)
  072, [\href{http://arxiv.org/abs/1611.07874}{{\tt arXiv:1611.07874}}].

\bibitem{Giombi:2017rhm}
S.~Giombi, V.~Kirilin, and E.~Skvortsov, {\it {Notes on Spinning Operators in
  Fermionic CFT}},  {\em JHEP} {\bf 05} (2017) 041,
  [\href{http://arxiv.org/abs/1701.06997}{{\tt arXiv:1701.06997}}].

\bibitem{Benini:2017dus}
F.~Benini, P.-S. Hsin, and N.~Seiberg, {\it {Comments on global symmetries,
  anomalies, and duality in (2 + 1)d}},  {\em JHEP} {\bf 04} (2017) 135,
  [\href{http://arxiv.org/abs/1702.07035}{{\tt arXiv:1702.07035}}].

\bibitem{Sezgin:2017jgm}
E.~Sezgin, E.~D. Skvortsov, and Y.~Zhu, {\it {Chern-Simons Matter Theories and
  Higher Spin Gravity}},  {\em JHEP} {\bf 07} (2017) 133,
  [\href{http://arxiv.org/abs/1705.03197}{{\tt arXiv:1705.03197}}].

\bibitem{Nosaka:2017ohr}
T.~Nosaka and S.~Yokoyama, {\it {Complete factorization in minimal $
  \mathcal{N}=4 $ Chern-Simons-matter theory}},  {\em JHEP} {\bf 01} (2018)
  001, [\href{http://arxiv.org/abs/1706.07234}{{\tt arXiv:1706.07234}}].

\bibitem{Komargodski:2017keh}
Z.~Komargodski and N.~Seiberg, {\it {A symmetry breaking scenario for
  QCD$_{3}$}},  {\em JHEP} {\bf 01} (2018) 109,
  [\href{http://arxiv.org/abs/1706.08755}{{\tt arXiv:1706.08755}}].

\bibitem{Giombi:2017txg}
S.~Giombi, {\it {Testing the Boson/Fermion Duality on the Three-Sphere}},
  \href{http://arxiv.org/abs/1707.06604}{{\tt arXiv:1707.06604}}.

\bibitem{Gaiotto:2017tne}
D.~Gaiotto, Z.~Komargodski, and N.~Seiberg, {\it {Time-reversal breaking in
  QCD$_{4}$, walls, and dualities in 2 + 1 dimensions}},  {\em JHEP} {\bf 01}
  (2018) 110, [\href{http://arxiv.org/abs/1708.06806}{{\tt arXiv:1708.06806}}].

\bibitem{Jensen:2017dso}
K.~Jensen and A.~Karch, {\it {Bosonizing three-dimensional quiver gauge
  theories}},  {\em JHEP} {\bf 11} (2017) 018,
  [\href{http://arxiv.org/abs/1709.01083}{{\tt arXiv:1709.01083}}].

\bibitem{Jensen:2017xbs}
K.~Jensen and A.~Karch, {\it {Embedding three-dimensional bosonization
  dualities into string theory}},  {\em JHEP} {\bf 12} (2017) 031,
  [\href{http://arxiv.org/abs/1709.07872}{{\tt arXiv:1709.07872}}].

\bibitem{Gomis:2017ixy}
J.~Gomis, Z.~Komargodski, and N.~Seiberg, {\it {Phases Of Adjoint QCD$_3$ And
  Dualities}},  {\em SciPost Phys.} {\bf 5} (2018), no.~1 007,
  [\href{http://arxiv.org/abs/1710.03258}{{\tt arXiv:1710.03258}}].

\bibitem{Inbasekar:2017ieo}
K.~Inbasekar, S.~Jain, P.~Nayak, and V.~Umesh, {\it {All tree level scattering
  amplitudes in Chern-Simons theories with fundamental matter}},  {\em Phys.
  Rev. Lett.} {\bf 121} (2018), no.~16 161601,
  [\href{http://arxiv.org/abs/1710.04227}{{\tt arXiv:1710.04227}}].

\bibitem{Inbasekar:2017sqp}
K.~Inbasekar, S.~Jain, S.~Majumdar, P.~Nayak, T.~Neogi, T.~Sharma, R.~Sinha,
  and V.~Umesh, {\it {Dual superconformal symmetry of $ \mathcal{N} $ = 2
  Chern-Simons theory with fundamental matter at large N}},  {\em JHEP} {\bf
  06} (2019) 016, [\href{http://arxiv.org/abs/1711.02672}{{\tt
  arXiv:1711.02672}}].

\bibitem{Cordova:2017vab}
C.~Cordova, P.-S. Hsin, and N.~Seiberg, {\it {Global Symmetries, Counterterms,
  and Duality in Chern-Simons Matter Theories with Orthogonal Gauge Groups}},
  {\em SciPost Phys.} {\bf 4} (2018), no.~4 021,
  [\href{http://arxiv.org/abs/1711.10008}{{\tt arXiv:1711.10008}}].

\bibitem{GuruCharan:2017ftx}
V.~Guru~Charan and S.~Prakash, {\it {On the Higher Spin Spectrum of
  Chern-Simons Theory coupled to Fermions in the Large Flavour Limit}},  {\em
  JHEP} {\bf 02} (2018) 094, [\href{http://arxiv.org/abs/1711.11300}{{\tt
  arXiv:1711.11300}}].

\bibitem{Benini:2017aed}
F.~Benini, {\it {Three-dimensional dualities with bosons and fermions}},  {\em
  JHEP} {\bf 02} (2018) 068, [\href{http://arxiv.org/abs/1712.00020}{{\tt
  arXiv:1712.00020}}].

\bibitem{Aitken:2017nfd}
K.~Aitken, A.~Baumgartner, A.~Karch, and B.~Robinson, {\it {3d Abelian
  Dualities with Boundaries}},  {\em JHEP} {\bf 03} (2018) 053,
  [\href{http://arxiv.org/abs/1712.02801}{{\tt arXiv:1712.02801}}].

\bibitem{Argurio:2018uup}
R.~Argurio, M.~Bertolini, F.~Bigazzi, A.~L. Cotrone, and P.~Niro, {\it {QCD
  domain walls, Chern-Simons theories and holography}},  {\em JHEP} {\bf 09}
  (2018) 090, [\href{http://arxiv.org/abs/1806.08292}{{\tt arXiv:1806.08292}}].

\bibitem{Jensen:2017bjo}
K.~Jensen, {\it {A master bosonization duality}},  {\em JHEP} {\bf 01} (2018)
  031, [\href{http://arxiv.org/abs/1712.04933}{{\tt arXiv:1712.04933}}].

\bibitem{Chattopadhyay:2018wkp}
A.~Chattopadhyay, P.~Dutta, and S.~Dutta, {\it {From Phase Space to Integrable
  Representations and Level-Rank Duality}},  {\em JHEP} {\bf 05} (2018) 117,
  [\href{http://arxiv.org/abs/1801.07901}{{\tt arXiv:1801.07901}}].

\bibitem{Turiaci:2018nua}
G.~J. Turiaci and A.~Zhiboedov, {\it {Veneziano Amplitude of Vasiliev Theory}},
   {\em JHEP} {\bf 10} (2018) 034, [\href{http://arxiv.org/abs/1802.04390}{{\tt
  arXiv:1802.04390}}].

\bibitem{Choudhury:2018iwf}
S.~Choudhury, A.~Dey, I.~Halder, S.~Jain, L.~Janagal, S.~Minwalla, and
  N.~Prabhakar, {\it {Bose-Fermi Chern-Simons Dualities in the Higgsed Phase}},
   {\em JHEP} {\bf 11} (2018) 177, [\href{http://arxiv.org/abs/1804.08635}{{\tt
  arXiv:1804.08635}}].

\bibitem{Karch:2018mer}
A.~Karch, D.~Tong, and C.~Turner, {\it {Mirror Symmetry and Bosonization in 2d
  and 3d}},  {\em JHEP} {\bf 07} (2018) 059,
  [\href{http://arxiv.org/abs/1805.00941}{{\tt arXiv:1805.00941}}].

\bibitem{Aharony:2018npf}
O.~Aharony, L.~F. Alday, A.~Bissi, and R.~Yacoby, {\it {The Analytic Bootstrap
  for Large $N$ Chern-Simons Vector Models}},  {\em JHEP} {\bf 08} (2018) 166,
  [\href{http://arxiv.org/abs/1805.04377}{{\tt arXiv:1805.04377}}].

\bibitem{Yacoby:2018yvy}
R.~Yacoby, {\it {Scalar Correlators in Bosonic Chern-Simons Vector Models}},
  \href{http://arxiv.org/abs/1805.11627}{{\tt arXiv:1805.11627}}.

\bibitem{Aitken:2018cvh}
K.~Aitken, A.~Baumgartner, and A.~Karch, {\it {Novel 3d bosonic dualities from
  bosonization and holography}},  {\em JHEP} {\bf 09} (2018) 003,
  [\href{http://arxiv.org/abs/1807.01321}{{\tt arXiv:1807.01321}}].

\bibitem{Aharony:2018pjn}
O.~Aharony, S.~Jain, and S.~Minwalla, {\it {Flows, Fixed Points and Duality in
  Chern-Simons-matter theories}},  {\em JHEP} {\bf 12} (2018) 058,
  [\href{http://arxiv.org/abs/1808.03317}{{\tt arXiv:1808.03317}}].

\bibitem{Dey:2018ykx}
A.~Dey, I.~Halder, S.~Jain, L.~Janagal, S.~Minwalla, and N.~Prabhakar, {\it
  {Duality and an exact Landau-Ginzburg potential for quasi-bosonic
  Chern-Simons-Matter theories}},  {\em JHEP} {\bf 11} (2018) 020,
  [\href{http://arxiv.org/abs/1808.04415}{{\tt arXiv:1808.04415}}].

\bibitem{Skvortsov:2018uru}
E.~Skvortsov, {\it {Light-Front Bootstrap for Chern-Simons Matter Theories}},
  {\em JHEP} {\bf 06} (2019) 058, [\href{http://arxiv.org/abs/1811.12333}{{\tt
  arXiv:1811.12333}}].

\bibitem{Argurio:2019tvw}
R.~Argurio, M.~Bertolini, F.~Mignosa, and P.~Niro, {\it {Charting the phase
  diagram of QCD$_{3}$}},  {\em JHEP} {\bf 08} (2019) 153,
  [\href{http://arxiv.org/abs/1905.01460}{{\tt arXiv:1905.01460}}].

\bibitem{Armoni:2019lgb}
A.~Armoni, T.~T. Dumitrescu, G.~Festuccia, and Z.~Komargodski, {\it {Metastable
  vacua in large-N QCD$_{3}$}},  {\em JHEP} {\bf 01} (2020) 004,
  [\href{http://arxiv.org/abs/1905.01797}{{\tt arXiv:1905.01797}}].

\bibitem{Chattopadhyay:2019lpr}
A.~Chattopadhyay, D.~Suvankar, and Neetu, {\it {Chern-Simons Theory on Seifert
  Manifold and Matrix Model}},  {\em Phys. Rev. D} {\bf 100} (2019), no.~12
  126009, [\href{http://arxiv.org/abs/1902.07538}{{\tt arXiv:1902.07538}}].

\bibitem{Dey:2019ihe}
A.~Dey, I.~Halder, S.~Jain, S.~Minwalla, and N.~Prabhakar, {\it {The large N
  phase diagram of $ \mathcal{N} $ = 2 SU(N) Chern-Simons theory with one
  fundamental chiral multiplet}},  {\em JHEP} {\bf 11} (2019) 113,
  [\href{http://arxiv.org/abs/1904.07286}{{\tt arXiv:1904.07286}}].

\bibitem{Halder:2019foo}
I.~Halder and S.~Minwalla, {\it {Matter Chern Simons Theories in a Background
  Magnetic Field}},  {\em JHEP} {\bf 11} (2019) 089,
  [\href{http://arxiv.org/abs/1904.07885}{{\tt arXiv:1904.07885}}].

\bibitem{Aharony:2019mbc}
O.~Aharony and A.~Sharon, {\it {Large N renormalization group flows in 3d $
  \mathcal{N} $ = 1 Chern-Simons-Matter theories}},  {\em JHEP} {\bf 07} (2019)
  160, [\href{http://arxiv.org/abs/1905.07146}{{\tt arXiv:1905.07146}}].

\bibitem{Li:2019twz}
Z.~Li, {\it {Bootstrapping conformal four-point correlators with slightly
  broken higher spin symmetry and $3D$ bosonization}},  {\em JHEP} {\bf 10}
  (2020) 007, [\href{http://arxiv.org/abs/1906.05834}{{\tt arXiv:1906.05834}}].

\bibitem{Jain:2019fja}
S.~Jain, V.~Malvimat, A.~Mehta, S.~Prakash, and N.~Sudhir, {\it {All order
  exact result for the anomalous dimension of the scalar primary in
  Chern-Simons vector models}},  {\em Phys. Rev. D} {\bf 101} (2020), no.~12
  126017, [\href{http://arxiv.org/abs/1906.06342}{{\tt arXiv:1906.06342}}].

\bibitem{Inbasekar:2019wdw}
K.~Inbasekar, S.~Jain, V.~Malvimat, A.~Mehta, P.~Nayak, and T.~Sharma, {\it
  {Correlation functions in ${\cal N}=2$ Supersymmetric vector matter
  Chern-Simons theory}},  {\em JHEP} {\bf 04} (2020) 207,
  [\href{http://arxiv.org/abs/1907.11722}{{\tt arXiv:1907.11722}}].

\bibitem{Inbasekar:2019azv}
K.~Inbasekar, L.~Janagal, and A.~Shukla, {\it {Mass-deformed $N=3$
  supersymmetric Chern-Simons-matter theory}},  {\em Phys. Rev. D} {\bf 100}
  (2019), no.~8 085008, [\href{http://arxiv.org/abs/1908.08119}{{\tt
  arXiv:1908.08119}}].

\bibitem{Jensen:2019mga}
K.~Jensen and P.~Patil, {\it {Chern-Simons dualities with multiple flavors at
  large $N$}},  {\em JHEP} {\bf 12} (2019) 043,
  [\href{http://arxiv.org/abs/1910.07484}{{\tt arXiv:1910.07484}}].

\bibitem{Kalloor:2019xjb}
R.~R. Kalloor, {\it {Four-point functions in large $N$ Chern-Simons fermionic
  theories}},  {\em JHEP} {\bf 10} (2020) 028,
  [\href{http://arxiv.org/abs/1910.14617}{{\tt arXiv:1910.14617}}].

\bibitem{Ghosh:2019sqf}
S.~Ghosh and S.~Mazumdar, {\it {Thermal Correlators and Bosonization Dualities
  in Large $N$ Chern Simons Matter Theories}},
  \href{http://arxiv.org/abs/1912.06589}{{\tt arXiv:1912.06589}}.

\bibitem{Argurio:2020her}
R.~Argurio, A.~Armoni, M.~Bertolini, F.~Mignosa, and P.~Niro, {\it {Vacuum
  structure of large $N$ $QCD_{3}$ from holography}},  {\em JHEP} {\bf 07}
  (2020) 134, [\href{http://arxiv.org/abs/2006.01755}{{\tt arXiv:2006.01755}}].

\bibitem{Inbasekar:2020hla}
K.~Inbasekar, L.~Janagal, and A.~Shukla, {\it {Scattering Amplitudes in
  $\mathcal{N} = 3$ Supersymmetric $SU(N)$ Chern-Simons-Matter Theory at Large
  $N$}},  {\em JHEP} {\bf 04} (2020) 101,
  [\href{http://arxiv.org/abs/2001.02363}{{\tt arXiv:2001.02363}}].

\bibitem{Jain:2020rmw}
S.~Jain, R.~R. John, and V.~Malvimat, {\it {Momentum space spinning correlators
  and higher spin equations in three dimensions}},  {\em JHEP} {\bf 11} (2020)
  049, [\href{http://arxiv.org/abs/2005.07212}{{\tt arXiv:2005.07212}}].

\bibitem{Minwalla:2020ysu}
S.~Minwalla, A.~Mishra, and N.~Prabhakar, {\it {Fermi seas from Bose
  condensates in Chern-Simons matter theories and a bosonic exclusion
  principle}},  {\em JHEP} {\bf 11} (2020) 171,
  [\href{http://arxiv.org/abs/2008.00024}{{\tt arXiv:2008.00024}}].

\bibitem{Jain:2020puw}
S.~Jain, R.~R. John, and V.~Malvimat, {\it {Constraining momentum space
  correlators using slightly broken higher spin symmetry}},  {\em JHEP} {\bf
  04} (2021) 231, [\href{http://arxiv.org/abs/2008.08610}{{\tt
  arXiv:2008.08610}}].

\bibitem{Mishra:2020wos}
A.~Mishra, {\it {On thermal correlators and bosonization duality in
  Chern-Simons theories with massive fundamental matter}},  {\em JHEP} {\bf 01}
  (2021) 109, [\href{http://arxiv.org/abs/2010.03699}{{\tt arXiv:2010.03699}}].

\bibitem{Jain:2021wyn}
S.~Jain, R.~R. John, A.~Mehta, A.~A. Nizami, and A.~Suresh, {\it {Momentum
  space parity-odd CFT 3-point functions}},  {\em JHEP} {\bf 08} (2021) 089,
  [\href{http://arxiv.org/abs/2101.11635}{{\tt arXiv:2101.11635}}].

\bibitem{Jain:2021vrv}
S.~Jain, R.~R. John, A.~Mehta, A.~A. Nizami, and A.~Suresh, {\it {Higher spin
  3-point functions in 3d CFT using spinor-helicity variables}},  {\em JHEP}
  {\bf 09} (2021) 041, [\href{http://arxiv.org/abs/2106.00016}{{\tt
  arXiv:2106.00016}}].

\bibitem{Gandhi:2021gwn}
Y.~Gandhi, S.~Jain, and R.~R. John, {\it {Anyonic correlation functions in
  Chern-Simons matter theories}},  {\em Phys. Rev. D} {\bf 106} (2022), no.~4
  046014, [\href{http://arxiv.org/abs/2106.09043}{{\tt arXiv:2106.09043}}].

\bibitem{Gabai:2022snc}
B.~Gabai, J.~Sandor, and X.~Yin, {\it {Anyon scattering from lightcone
  Hamiltonian: the singlet channel}},  {\em JHEP} {\bf 09} (2022) 145,
  [\href{http://arxiv.org/abs/2205.09144}{{\tt arXiv:2205.09144}}].

\bibitem{Mehta:2022lgq}
U.~Mehta, S.~Minwalla, C.~Patel, S.~Prakash, and K.~Sharma, {\it {Crossing
  Symmetry in Matter Chern-Simons Theories at finite $N$ and $k$}},
  \href{http://arxiv.org/abs/2210.07272}{{\tt arXiv:2210.07272}}.

\bibitem{Jain:2022ajd}
P.~Jain, S.~Jain, B.~Sahoo, K.~S. Dhruva, and A.~Zade, {\it {Mapping Slightly
  Broken Higher Spin (SBHS) theory correlators to Free theory correlators: A
  momentum space bootstrap using SBHS symmetry}},
  \href{http://arxiv.org/abs/2207.05101}{{\tt arXiv:2207.05101}}.

\bibitem{Vasiliev:1992av}
M.~A. Vasiliev, {\it {More on equations of motion for interacting massless
  fields of all spins in (3+1)-dimensions}},  {\em Phys. Lett. B} {\bf 285}
  (1992) 225--234.

\bibitem{Sezgin:2003pt}
E.~Sezgin and P.~Sundell, {\it {Holography in 4D (super) higher spin theories
  and a test via cubic scalar couplings}},  {\em JHEP} {\bf 07} (2005) 044,
  [\href{http://arxiv.org/abs/hep-th/0305040}{{\tt hep-th/0305040}}].

\bibitem{Leigh:2003gk}
R.~G. Leigh and A.~C. Petkou, {\it {Holography of the N=1 higher spin theory on
  AdS(4)}},  {\em JHEP} {\bf 06} (2003) 011,
  [\href{http://arxiv.org/abs/hep-th/0304217}{{\tt hep-th/0304217}}].

\bibitem{short}
B.~Gabai, A.~Sever, and D.-l. Zhong, {\it {Line Operators in
  Chern-Simons\textendash{}Matter Theories and Bosonization in Three
  Dimensions}},  {\em Phys. Rev. Lett.} {\bf 129} (2022), no.~12 121604,
  [\href{http://arxiv.org/abs/2204.05262}{{\tt arXiv:2204.05262}}].

\bibitem{Ivri}
I.~Nagar, {\it {To appear}}, .

\bibitem{Cuomo:2021rkm}
G.~Cuomo, Z.~Komargodski, and A.~Raviv-Moshe, {\it {Renormalization Group Flows
  on Line Defects}},  {\em Phys. Rev. Lett.} {\bf 128} (2022), no.~2 021603,
  [\href{http://arxiv.org/abs/2108.01117}{{\tt arXiv:2108.01117}}].

\bibitem{bootstrap}
B.~Gabai, A.~Sever, and D.-l. Zhong, ``{Line operators in Chern-Simons-Matter
  theories and Bosonization in Three Dimensions III - The Line Bootstrap, To
  appear}.'' 2022.

\bibitem{Migdal:1983qrz}
A.~A. Migdal, {\it {Loop Equations and 1/N Expansion}},  {\em Phys. Rept.} {\bf
  102} (1983) 199--290.

\bibitem{Witten:1988hf}
E.~Witten, {\it {Quantum Field Theory and the Jones Polynomial}},  {\em Commun.
  Math. Phys.} {\bf 121} (1989) 351--399.

\bibitem{Bray:1977tk}
A.~J. Bray and M.~A. Moore, {\it {Critical Behavior of a Semiinfinite System: n
  Vector Model in the Large n Limit}},  {\em Phys. Rev. Lett.} {\bf 38} (1977)
  735--738.

\bibitem{Cuomo:2021cnb}
G.~Cuomo, M.~Mezei, and A.~Raviv-Moshe, {\it {Boundary conformal field theory
  at large charge}},  {\em JHEP} {\bf 10} (2021) 143,
  [\href{http://arxiv.org/abs/2108.06579}{{\tt arXiv:2108.06579}}].

\bibitem{Padayasi:2021sik}
J.~Padayasi, A.~Krishnan, M.~A. Metlitski, I.~A. Gruzberg, and M.~Meineri, {\it
  {The extraordinary boundary transition in the 3d O(N) model via conformal
  bootstrap}},  {\em SciPost Phys.} {\bf 12} (2022), no.~6 190,
  [\href{http://arxiv.org/abs/2111.03071}{{\tt arXiv:2111.03071}}].

\bibitem{Herzog:2017xha}
C.~P. Herzog and K.-W. Huang, {\it {Boundary Conformal Field Theory and a
  Boundary Central Charge}},  {\em JHEP} {\bf 10} (2017) 189,
  [\href{http://arxiv.org/abs/1707.06224}{{\tt arXiv:1707.06224}}].

\bibitem{Cuomo:2022xgw}
G.~Cuomo, Z.~Komargodski, M.~Mezei, and A.~Raviv-Moshe, {\it {Spin Impurities,
  Wilson Lines and Semiclassics}},  \href{http://arxiv.org/abs/2202.00040}{{\tt
  arXiv:2202.00040}}.

\bibitem{Aharony:2022ntz}
O.~Aharony, G.~Cuomo, Z.~Komargodski, M.~Mezei, and A.~Raviv-Moshe, {\it
  {Phases of Wilson Lines in Conformal Field Theories}},
  \href{http://arxiv.org/abs/2211.11775}{{\tt arXiv:2211.11775}}.

\bibitem{Makeenko:1979pb}
Y.~M. Makeenko and A.~A. Migdal, {\it {Exact Equation for the Loop Average in
  Multicolor QCD}},  {\em Phys. Lett. B} {\bf 88} (1979) 135. [Erratum:
  Phys.Lett.B 89, 437 (1980)].

\bibitem{Bianchi:2016vvm}
M.~S. Bianchi, L.~Griguolo, M.~Leoni, A.~Mauri, S.~Penati, and D.~Seminara,
  {\it {The quantum 1/2 BPS Wilson loop in ${\cal N}=4$ Chern-Simons-matter
  theories}},  {\em JHEP} {\bf 09} (2016) 009,
  [\href{http://arxiv.org/abs/1606.07058}{{\tt arXiv:1606.07058}}].

\bibitem{Agmon:2020pde}
N.~B. Agmon and Y.~Wang, {\it {Classifying Superconformal Defects in Diverse
  Dimensions Part I: Superconformal Lines}},
  \href{http://arxiv.org/abs/2009.06650}{{\tt arXiv:2009.06650}}.

\bibitem{Kapustin:2009kz}
A.~Kapustin, B.~Willett, and I.~Yaakov, {\it {Exact Results for Wilson Loops in
  Superconformal Chern-Simons Theories with Matter}},  {\em JHEP} {\bf 03}
  (2010) 089, [\href{http://arxiv.org/abs/0909.4559}{{\tt arXiv:0909.4559}}].

\bibitem{Hama:2010av}
N.~Hama, K.~Hosomichi, and S.~Lee, {\it {Notes on SUSY Gauge Theories on
  Three-Sphere}},  {\em JHEP} {\bf 03} (2011) 127,
  [\href{http://arxiv.org/abs/1012.3512}{{\tt arXiv:1012.3512}}].

\bibitem{Giveon:2008zn}
A.~Giveon and D.~Kutasov, {\it {Seiberg Duality in Chern-Simons Theory}},  {\em
  Nucl. Phys. B} {\bf 812} (2009) 1--11,
  [\href{http://arxiv.org/abs/0808.0360}{{\tt arXiv:0808.0360}}].

\bibitem{Kapustin:2011gh}
A.~Kapustin, {\it {Seiberg-like duality in three dimensions for orthogonal
  gauge groups}},  \href{http://arxiv.org/abs/1104.0466}{{\tt
  arXiv:1104.0466}}.

\bibitem{Willett:2011gp}
B.~Willett and I.~Yaakov, {\it {$\mathcal N$ = 2 dualities and
  $Z$-extremization in three dimensions}},  {\em JHEP} {\bf 10} (2020) 136,
  [\href{http://arxiv.org/abs/1104.0487}{{\tt arXiv:1104.0487}}].

\bibitem{Minwalla:1997ka}
S.~Minwalla, {\it {Restrictions imposed by superconformal invariance on quantum
  field theories}},  {\em Adv. Theor. Math. Phys.} {\bf 2} (1998) 783--851,
  [\href{http://arxiv.org/abs/hep-th/9712074}{{\tt hep-th/9712074}}].

\bibitem{Drukker:2009hy}
N.~Drukker and D.~Trancanelli, {\it {A Supermatrix model for N=6 super
  Chern-Simons-matter theory}},  {\em JHEP} {\bf 02} (2010) 058,
  [\href{http://arxiv.org/abs/0912.3006}{{\tt arXiv:0912.3006}}].

\bibitem{Ouyang:2015iza}
H.~Ouyang, J.-B. Wu, and J.-j. Zhang, {\it {Novel BPS Wilson loops in
  three-dimensional quiver Chern\textendash{}Simons-matter theories}},  {\em
  Phys. Lett. B} {\bf 753} (2016) 215--220,
  [\href{http://arxiv.org/abs/1510.05475}{{\tt arXiv:1510.05475}}].

\bibitem{Lee:2010hk}
K.-M. Lee and S.~Lee, {\it {1/2-BPS Wilson Loops and Vortices in ABJM Model}},
  {\em JHEP} {\bf 09} (2010) 004, [\href{http://arxiv.org/abs/1006.5589}{{\tt
  arXiv:1006.5589}}].

\bibitem{Mauri:2018fsf}
A.~Mauri, H.~Ouyang, S.~Penati, J.-B. Wu, and J.~Zhang, {\it {BPS Wilson loops
  in $ \mathcal{N} $ \ensuremath{\geq} 2 superconformal Chern-Simons-matter
  theories}},  {\em JHEP} {\bf 11} (2018) 145,
  [\href{http://arxiv.org/abs/1808.01397}{{\tt arXiv:1808.01397}}].

\bibitem{Drukker:2019bev}
N.~Drukker et~al., {\it {Roadmap on Wilson loops in 3d
  Chern\textendash{}Simons-matter theories}},  {\em J. Phys. A} {\bf 53}
  (2020), no.~17 173001, [\href{http://arxiv.org/abs/1910.00588}{{\tt
  arXiv:1910.00588}}].

\bibitem{Drukker:2020opf}
N.~Drukker, {\it {BPS Wilson loops and quiver varieties}},  {\em J. Phys. A}
  {\bf 53} (2020), no.~38 385402, [\href{http://arxiv.org/abs/2004.11393}{{\tt
  arXiv:2004.11393}}].

\bibitem{Guadagnini:1989kr}
E.~Guadagnini, M.~Martellini, and M.~Mintchev, {\it {Perturbative Aspects of
  the Chern-Simons Field Theory}},  {\em Phys. Lett. B} {\bf 227} (1989)
  111--117.

\bibitem{Chen:1992ee}
W.~Chen, G.~W. Semenoff, and Y.-S. Wu, {\it {Two loop analysis of nonAbelian
  Chern-Simons theory}},  {\em Phys. Rev. D} {\bf 46} (1992) 5521--5539,
  [\href{http://arxiv.org/abs/hep-th/9209005}{{\tt hep-th/9209005}}].

\bibitem{Guadagnini:1989am}
E.~Guadagnini, M.~Martellini, and M.~Mintchev, {\it {Wilson Lines in
  Chern-Simons Theory and Link Invariants}},  {\em Nucl. Phys. B} {\bf 330}
  (1990) 575--607.

\bibitem{Polyakov:1988md}
A.~M. Polyakov, {\it {Fermi-Bose Transmutations Induced by Gauge Fields}},
  {\em Mod. Phys. Lett. A} {\bf 3} (1988) 325.

\bibitem{gradshteyn2014table}
I.~S. Gradshteyn and I.~M. Ryzhik, {\em Table of integrals, series, and
  products}.
\newblock Academic press, 2014.

\bibitem{Marino:2011nm}
M.~Marino, {\it {Lectures on localization and matrix models in supersymmetric
  Chern-Simons-matter theories}},  {\em J. Phys. A} {\bf 44} (2011) 463001,
  [\href{http://arxiv.org/abs/1104.0783}{{\tt arXiv:1104.0783}}].

\bibitem{Labastida:1997uw}
J.~M.~F. Labastida and E.~Perez, {\it {Kontsevich integral for Vassiliev
  invariants from Chern-Simons perturbation theory in the light cone gauge}},
  {\em J. Math. Phys.} {\bf 39} (1998) 5183--5198,
  [\href{http://arxiv.org/abs/hep-th/9710176}{{\tt hep-th/9710176}}].

\end{thebibliography}\endgroup


\providecommand{\href}[2]{#2}\begingroup\raggedright\endgroup
\bibliographystyle{JHEP}

\end{document}